\documentclass[twocolumn]{aastex631}

\usepackage{gensymb}

\newcommand{\fdeg}{\mbox{$.\!\!{\degree}$}}
\newcommand{\oiii}{[O\,{\sc iii}]}
\newcommand{\oii}{[O\,{\sc ii}]}
\newcommand{\cii}{[C\,{\sc ii}]}
\newcommand{\lya}{Ly$\alpha$}

\received{2025 April 4}
\revised{2025 October 23}
\accepted{2025 November 5}
\shorttitle{Himiko and CR7 with JWST and ALMA}
\shortauthors{Kiyota et al.}
\graphicspath{{./}{figures/}}
\usepackage{amsmath}

\begin{document}

\title{
Comprehensive JWST+ALMA Study on the Extended Ly$\alpha$ Emitters, Himiko and CR7 at $z\sim 7$:\\
Blue Major Merger Systems in Stark Contrast to Submillimeter Galaxies
}
\correspondingauthor{Tomokazu Kiyota}
\email{tomokazu.kiyota@grad.nao.ac.jp}

\author[0009-0004-4332-9225]{Tomokazu Kiyota}
\affiliation{Department of Astronomical Science, The Graduate University for Advanced Studies, SOKENDAI, 2-21-1 Osawa, Mitaka, Tokyo, 181-8588, Japan}
\affiliation{National Astronomical Observatory of Japan, 2-21-1 Osawa, Mitaka, Tokyo, 181-8588, Japan}

\author[0000-0002-1049-6658]{Masami Ouchi}
\affiliation{National Astronomical Observatory of Japan, 2-21-1 Osawa, Mitaka, Tokyo, 181-8588, Japan}
\affiliation{Institute for Cosmic Ray Research, The University of Tokyo, 5-1-5 Kashiwanoha, Kashiwa, Chiba 277-8582, Japan}
\affiliation{Department of Astronomical Science, The Graduate University for Advanced Studies, SOKENDAI, 2-21-1 Osawa, Mitaka, Tokyo, 181-8588, Japan}
\affiliation{Kavli Institute for the Physics and Mathematics of the Universe (WPI), University of Tokyo, Kashiwa, Chiba 277-8583, Japan}

\author[0000-0002-5768-8235]{Yi Xu}
\affiliation{Institute for Cosmic Ray Research, The University of Tokyo, 5-1-5 Kashiwanoha, Kashiwa, Chiba 277-8582, Japan}
\affiliation{Department of Astronomy, Graduate School of Science, the University of Tokyo, 7-3-1 Hongo, Bunkyo, Tokyo 113-0033, Japan} 

\author[0000-0002-0984-7713]{Yurina Nakazato}
\affiliation{Department of Physics, The University of Tokyo, 7-3-1 Hongo, Bunkyo, Tokyo 113-0033, Japan}
\affiliation{Center for Computational Astrophysics, Flatiron Institute, 162 5th Avenue, New York, NY 10010}

\author{Kenta Soga} 
\affiliation{Center for Computational Sciences, University of Tsukuba, Ten-nodai, 1-1-1 Tsukuba, Ibaraki 305-8577, Japan} 

\author[0000-0002-1319-3433]{Hidenobu Yajima}
\affiliation{Center for Computational Sciences, University of Tsukuba, Ten-nodai, 1-1-1 Tsukuba, Ibaraki 305-8577, Japan} 

\author[0000-0001-7201-5066]{Seiji Fujimoto}
\affiliation{David A. Dunlap Department of Astronomy and Astrophysics, University of Toronto, 50 St. George Street, Toronto, Ontario, M5S 3H4, Canada}
\affiliation{Dunlap Institute for Astronomy and Astrophysics, 50 St. George Street, Toronto, Ontario, M5S 3H4, Canada}

\author[0000-0002-6047-430X]{Yuichi Harikane}
\affiliation{Institute for Cosmic Ray Research, The University of Tokyo, 5-1-5 Kashiwanoha, Kashiwa, Chiba 277-8582, Japan} 

\author[0000-0003-2965-5070]{Kimihiko Nakajima}
\affiliation{National Astronomical Observatory of Japan, 2-21-1 Osawa, Mitaka, Tokyo, 181-8588, Japan}
\affiliation{Institute of Liberal Arts and Science, Kanazawa University, Kakuma-machi, Kanazawa, Ishikawa, 920-1192, Japan} 

\author[0000-0001-9011-7605]{Yoshiaki Ono}
\affiliation{Institute for Cosmic Ray Research, The University of Tokyo, 5-1-5 Kashiwanoha, Kashiwa, Chiba 277-8582, Japan}

\author[0000-0002-1199-6523]{Dongsheng Sun}
\affiliation{Institute for Cosmic Ray Research, The University of Tokyo, 5-1-5 Kashiwanoha, Kashiwa, Chiba 277-8582, Japan}

\author[0000-0002-3801-434X]{Haruka Kusakabe}
\affiliation{National Astronomical Observatory of Japan, 2-21-1 Osawa, Mitaka, Tokyo, 181-8588, Japan}
\affiliation{Department of General Systems Studies, Graduate School of Arts and Sciences,  The University of Tokyo, 3-8-1 Komaba, Meguro-ku, Tokyo 153-8902, Japan
} 

\author[0000-0002-8680-248X]{Daniel Ceverino}
\affiliation{Departamento de Fisica Teorica, Modulo 8, Facultad de Ciencias, Universidad Autonoma de Madrid, 28049 Madrid, Spain}
\affiliation{CIAFF, Facultad de Ciencias, Universidad Autonoma de Madrid, E-28049 Madrid, Spain}

\author[0000-0001-6469-8725]{Bunyo Hatsukade}
\affiliation{National Astronomical Observatory of Japan, 2-21-1 Osawa, Mitaka, Tokyo, 181-8588, Japan}
\affiliation{Department of Astronomical Science, The Graduate University for Advanced Studies, SOKENDAI, 2-21-1 Osawa, Mitaka, Tokyo, 181-8588, Japan}
\affiliation{Department of Astronomy, Graduate School of Science, the University of Tokyo, 7-3-1 Hongo, Bunkyo, Tokyo 113-0033, Japan} 

\author[0000-0002-2364-0823]{Daisuke Iono}
\affiliation{National Astronomical Observatory of Japan, 2-21-1 Osawa, Mitaka, Tokyo, 181-8588, Japan}
\affiliation{Department of Astronomical Science, The Graduate University for Advanced Studies, SOKENDAI, 2-21-1 Osawa, Mitaka, Tokyo, 181-8588, Japan}

\author[0000-0002-4052-2394]{Kotaro Kohno}
\affiliation{Institute of Astronomy, Graduate School of Science, The University of Tokyo, 2-21-1 Osawa, Mitaka, Tokyo 181-0015, Japan}
\affiliation{Research Center for the Early Universe, Graduate School of Science, The University of Tokyo, 7-3-1 Hongo, Bunkyo-ku, Tokyo 113-0033, Japan}

\author[0000-0002-6939-0372]{Koichiro Nakanishi}
\affiliation{National Astronomical Observatory of Japan, 2-21-1 Osawa, Mitaka, Tokyo, 181-8588, Japan}
\affiliation{Department of Astronomical Science, The Graduate University for Advanced Studies, SOKENDAI, 2-21-1 Osawa, Mitaka, Tokyo, 181-8588, Japan}

\begin{abstract}
We present various properties of two bright extended Ly$\alpha$ objects, Himiko and CR7, at $z=6.6$ thoroughly investigated with JWST/NIRCam photometry, NIRSpec-IFU spectroscopy, and ALMA data, uncovering their physical origins. Himiko (CR7) shows at least five (four) clumps with small separations of 2.4--7.3 kpc and velocity offsets of $\Delta v<220~\mathrm{km~s^{-1}}$ in the [O\,{\sc iii}]$\lambda\lambda4959,5007$ line maps, three of which exhibit stellar components with comparable stellar masses ranging in $\log{(M_*/M_\odot)}=8.4$--$9.0$ ($8.3$--$8.8$), indicative of major merger systems that are consistent with our numerical simulations. The [C\,{\sc ii}]158$\mu$m and Ly$\alpha$ lines are found in the middle of two clumps (the brightest clump) in Himiko (CR7), suggesting that the distribution of neutral gas does not always coincide with that of ionized gas or stars in merging processes. We find that some of the clumps have broad [O\,{\sc iii}] components (250--400$~\mathrm{km~s^{-1}}$) in Himiko and CR7, likely tracing outflow and tidal features, while the central clump in Himiko presents a broad H$\alpha$ ($\sim1000~\mathrm{km~s^{-1}}$) line explained by an AGN with a low mass black hole of $M_\mathrm{BH}=10^{6.6}~M_\odot$, which contribute to the extended and bright nature of Himiko and CR7. We find low metallicities of $12+\log(\mathrm{O/H})=$7.9--8.1 in Himiko and CR7 based on auroral [O\,{\sc iii}]$\lambda4363$ and strong lines that are consistent with no 1-mm continuum detection corresponding to the dust mass limits of $M_\mathrm{dust}\lesssim 9\times 10^6 M_\odot$. Himiko and CR7 are metal- and dust-poor blue merger systems with stellar and dust masses $\gtrsim2$ orders of magnitude smaller than the massive dust-rich merger systems represented by submillimeter galaxies.

\end{abstract}


\section{Introduction} \label{sec:intro}

Recent advancements in observing the high-redshift Universe have enabled us to explore the detailed physical properties of early galaxies. 
Over the past several decades, ground-based telescopes, such as Subaru and Keck, and space telescopes, like the Hubble Space Telescope (HST), have observed thousands of galaxies in the epoch of reionization ($z>6$). 
These observations have provided deep photometric data spanning the rest-frame ultraviolet (UV) to near-infrared (NIR) wavelengths and UV spectroscopic data, which are crucial for understanding the physical conditions of these early systems (e.g., \citealt{bouwens15, finkelstein15}). 

Wide-field ground-based surveys conducted with Subaru have successfully identified bright targets, resulting in large galaxy samples, including \lya\ emitters (LAEs), even at $z>6$. 
Subaru narrow-band (NB) surveys effectively select LAEs at $z=2$--$7$ by detecting NB excess in combination with broadband imaging (e.g., \citealt{ouchi10, matthee15, ouchi18, kikuta23}). 
These surveys have been critical in identifying large populations of high-redshift galaxies and have enabled the discovery of unique and rare objects. 
Notably, such surveys have led to the identification of high-redshift \lya\ blobs (e.g., \citealt{shibuya18}; see also e.g., \citealt{steidel00, matsuda04, prescott12, cantalupo14} for \lya\ blobs or \lya\ nebulas at $z=2$--$3$ studies) and protoclusters (e.g., \citealt{toshikawa18, harikane19a}). 
The discovery of these sources provides insight into the processes that govern galaxy formation and evolution. 

The remarkable findings of these surveys are large \lya\ blobs at $z=6.6$, known as ‘Himiko’ (e.g., \citealt{ouchi09, walter12, ouchi13, hirashita14, ota14, zabl15, carniani18}) and ‘CR7’ (COSMOS Redshift 7; \citealt{matthee15, sobral15, bowler17b, matthee17, sobral19, matthee20, marconcini24b}). 
Himiko and CR7 were identified as the most luminous \lya\ emitter candidates at $z\sim6.6$ 
in the UKIDSS/UDS and COSMOS/UltraVISTA field,
respectively, using the Subaru/Suprime-Cam NB921 filter \citep{ouchi09, matthee15}. 
They are confirmed spectroscopically with Keck and the Very Large Telescope (VLT) aimed at the \lya\ emission. 
Both are significantly bright in the rest-frame UV continuum ($m_\mathrm{AB}\sim25$ or $M_{\mathrm{UV}}\sim-22$) and \lya\ emission with their \lya\ luminosity of $3.9\times10^{43}~\mathrm{erg~s^{-1}}$ and $8.9\times10^{43}~\mathrm{erg~s^{-1}}$, and their \lya\ emissions are spatially extended over 17~kpc and 16~kpc for Himiko and CR7, respectively \citep{ouchi09, sobral15}. 
The HST/Wide Field Camera 3 (WFC3) imaging revealed their clumpy UV counterparts, which are composed of at least three components within 10~kpc$\times$10~kpc, indicating potentially merging systems. 
The follow-up observations with the Atacama Large Millimeter/submillimeter Array (ALMA) Band~6 detected \cii$158\micron$ emissions but no dust continuum \citep{matthee17, carniani18}. 

In addition, NIR photometry with Spitzer/IRAC revealed a $3.6\micron$ excess, suggesting the presence of strong H$\beta$+\oiii\ lines, although the spatial resolution is lower than that of UV photometry \citep{ouchi13, matthee15, bowler17a}. 
However, these datasets alone suffer from dust-age-metallicity degeneracies and potential misinterpretation of the stellar populations due to contributions from nebular emissions.
The limited data make it difficult to clarify the physical origins of their extended \lya\ emission and their bright nature in the rest-frame UV, leaving the topic still under debate (e.g., the discussion on LAEs at $z=2$--$3$; \citealt{sobral18b}). 

The advent of the James Webb Space Telescope (JWST; \citealt{gardner23}) has changed our ability to study early galaxies with remarkable sensitivity and resolution. 
JWST/Near Infrared Spectrograph (NIRSpec; \citealt{jakobsen22})  is the first instrument to observe the rest-frame optical spectrum of Himiko and CR7, providing their detailed physical properties, such as metallicity, ionization conditions, and dynamics. 
Furthermore, JWST has an Integral Field Unit (IFU; \citealt{boker22}), allowing for high-resolution, three-dimensional spectroscopic observations. 
IFU observations enable us to investigate the morphology and kinematics of clumpy or merging galaxies like Himiko and CR7 in detail, offering a comprehensive view of their internal structure, dynamics, and interactions 
(e.g., \citealt{perna23, arribas24, ishikawa24, jones24a, jones24b, lamperti24, marconcini24b, marconcini24a, marshall24, scholtz24, morishita25, parlanti25})
. 

This paper comprehensively analyzes Himiko and CR7 using publicly available data from the JWST Near Infrared Camera (NIRCam; \citealt{rieke23}) and the NIRSpec IFU. We also include archival ALMA Band~6 data to examine their cold gas and dust characteristics. We integrate these data sets to explore their dynamics, stellar populations, ionization conditions, and dust content. 

The structure of this paper is as follows.
Section~\ref{sec:data} outlines the data processing of the JWST, ALMA, and Subaru. Section~\ref{sec:analysis} details the analysis procedures, including flux measurements and spectral energy distribution (SED) fitting. Section~\ref{sec:results} presents the results and the derived physical properties. In Section~\ref{sec:discussion}, we interpret the results and discuss their implications for the extended Ly$\alpha$ emission, UV bright nature, and galaxy mergers. 
Finally, Section~\ref{sec:conclusions} summarizes our study. 
Throughout the paper, we use the Chabrier initial mass function (\citealt{chabrier03}), and a flat $\Lambda$CDM cosmology with $H_0 = 67.7~\mathrm{km~s^{-1}~Mpc^{-1}}$, $\Omega_m = 0.3111$ and $\Omega_\Lambda=0.6889$ \citep{planck20}. 
All magnitudes are in the AB system \citep{oke83}.

\section{Data and reduction} \label{sec:data}
\subsection{JWST/NIRSpec} \label{subsec:jwst-nirspec}

We use the JWST/NIRSpec IFU data of Himiko and CR7 that were taken by the Guaranteed Time Observations (GTO) programs \#GTO-1215 and \#GTO-1217 (PI: Nora Luetzgendorf) as part of the Galaxy Assembly with NIRSpec Integral Field Spectroscopy (GA-NIFS; \citealt{perna23IAU})
in 2022 December and 2023 May, respectively.
Both programs were carried out with two sets of dispersers/filters: PRISM/CLEAR and G395H/F290LP. The spectral resolutions were $R\sim100$ and $R\sim2700$, covering 0.60--5.30\,$\micron$ and 2.87--5.27\,$\micron$, respectively. 
The observations were conducted with a medium cycling pattern and eight dithers. 
The exposure times were 3968 and 18207 seconds, respectively. 

We reduce the NIRSpec IFU data obtained from the Mikulski Archive for Space Telescopes (MAST) portal\footnote{\url{https://mast.stsci.edu/portal/Mashup/Clients/Mast/Portal.html}}. 
We use the JWST Science Calibration pipeline (version 1.14.0; \citealt{bushouse_2024_10870758}), and the calibration reference data system (CRDS) contexts are \texttt{jwst\_1227.pmap} and \texttt{jwst\_1263.pmap} for Himiko and CR7, respectively.
We mainly follow the reduction procedure presented by the team of the Targeting Extremely Magnified Panchromatic Lensed Arcs and Their Extended Star formation (TEMPLATES) that is the JWST Early Release Science (ERS) program (\#ERS-1355, PI: Jane R. Rigby; \citealt{rigby23}). The reduction procedure is coded in the script that is now publicly available (version 1.0.2; \citealt{jane_rigby_2024_10933642})\footnote{\url{https://github.com/JWST-Templates}}. 
Here, we briefly summarize the reduction procedures detailed in \cite{rigby23}. 
First, we conduct \texttt{calwebb\_detector1} as stage 1, including detector-level corrections for each exposure. We then correct $1/f$ noise of the data using \texttt{NSClean} (\citealt{rauscher24}). 
We perform \texttt{calwebb\_spec2} to the stage 1 individual exposure as a stage 2, including WCS/wavelength corrections, flat-fielding, path-loss correction, and flux calibration. 
We perform drizzle weighting that accomplishes the spaxel scale of $0\farcs06$. 
Finally, instead of stage 3, we use \texttt{reproject} (version 0.13.1; \citealt{robitaille_2024_10931886}) and \texttt{reproject\_interp} packages for the stage 2 data to obtain the median stack final 3D data cube, reducing outliers. 
We estimate a median value for each wavelength channel with masks around the object to subtract the background. 
We approximate the background as a function of wavelength 
and subtract them from each wavelength channel to obtain the final background-subtracted 3D data cube. 

Previous studies have reported astrometric offsets ($\sim0\farcs2$) between IFU data and NIRCam or HST images (e.g., \citealt{fujimoto24, jones24a, jones24b, jones24c, zamora24}). Similarly, we identify astrometric offsets between NIRCam and NIRSpec IFU data. These offsets are corrected by aligning the \oiii$\lambda$5007 (H$\alpha$) emission line peak observed with the IFU to the NIRCam F356W (F444W) image of Himiko (CR7), where the \oiii$\lambda$5007 (H$\alpha$) falls in the filter response.

\subsection{JWST/NIRCam} \label{subsec:jwst-nircam}

We utilize JWST/NIRCam images to extract the photometric properties of Himiko and CR7. 
The images of Himiko were obtained through Public Release IMaging for Extragalactic Research (PRIMER) as part of the General Observers (GO) program (\#GO-1837, PI: James S. Dunlop). 
The observations were conducted using eight filters: F090W, F115W, F150W, F200W, F277W, F356W, F410M, and F444W.
We utilize the images reduced with \texttt{grizli} (\citealt{brammer23}), available on the Dawn JWST Archive website\footnote{\url{https://dawn-cph.github.io/dja/index.html}} (version 7.0 of the UDS field; see \citealt{valentino23} for details of the reduction). 

The images of CR7 were obtained as part of the COSMOS-Web survey (\#GO-1727, PI: Jeyhan Kartaltepe and Caitlin M. Casey; for an overview, see \citealt{casey23}) using four filters: F115W, F150W, F277W, and F444W.
We employ the COSMOS-Web data presented in the COSMOS-Web Data Release (DR) 0.5\footnote{\url{https://cosmos.astro.caltech.edu/page/cosmosweb-dr}}, 
which include the NIRCam data from epochs 1 and 2 taken in 2023 January and April, respectively.
We choose images with a pixel size of $0\farcs03$/pixel provided by the COSMOS-Web DR0.5.

\subsection{ALMA} \label{subsec:alma}

We observed Himiko with ALMA Band~6 to characterize \cii158$\micron$ emission line (250.24~GHz at $z_\mathrm{Ly\alpha}=6.595$) and dust continuum. 
The observation was carried out on 2015 July 22, 23, and 24 as a Cycle 1 program (\#2012.1.00033.S, PI: M. Ouchi). 
The extended array configuration with 44 12-m antennas was used, and the baseline ranges from 15 to 1600~m. The observation was conducted in frequency division mode with four spectral windows centered at 250.262~GHz, 252.138~GHz, 265.638~GHz, and 267.513~GHz. 
A total bandwidth was 7.5~GHz and a spectral resolution would be 0.976~kHz, corresponding to $\sim1.2~\mathrm{km~s^{-1}}$. 
The on-source integration time was 3.2~hours. 
Ceres and J0238+166 were observed as flux calibrators. Bandpass calibration was performed with J2357-5311, J0334-4008, and J0423-0120. J0208-0047 was used for gain calibration or complex gain calibration. 

Himiko was also observed with the same band in 2012 July as a Cycle 0 program (\#2011.0.00115.S, PI: M. Ouchi; \citealt{ouchi13}). 
These Cycles 0 and 1 data have previously been used in \cite{carniani18}, which reported the detection of \cii\ emission from Himiko. 
CR7 was observed with the Band~6 in 2016 May and November (Cycle 3, ID: \#2015.1.00122.S, PI: D. Sobral; \citealt{matthee17}). 

We retrieve these data from the ALMA Science Archive\footnote{\url{https://almascience.org/}}. 
We calibrate the Cycles 0, 1, and 3 data using the Common Astronomy Software Applications package (CASA; \citealt{casa22}) version 3.4, 4.3.1, and 4.7.0, respectively, using the calibration script or pipeline script in the archive. 
We image from the calibrated visibility using CASA version 6.6.4. 
For Himiko data, we combine the Cycles 0 and 1 data using the CASA task \texttt{CONCAT} to obtain high-sensitivity data with a total on-source time of 6.4~hours. 
Using the CASA task \texttt{TCLEAN}, we generate both continuum images and \cii$158\micron$ line cubes with a natural weighting, a pixel scale of $0\farcs1$, and a spectral channel bin of $50~\mathrm{km~s^{-1}}$. 
Continuum maps are created by excluding spectral windows containing \cii$158\micron$ emission. 
The resulting continuum map sensitivities for Himiko and CR7 are $\sigma_\mathrm{cont} = 10.5~\mu\mathrm{Jy~beam^{-1}}$ and $6.4~\mu\mathrm{Jy~beam^{-1}}$, respectively, which is consistent with previous studies \citep{carniani18, matthee17}. 
The beam sizes of the continuum maps for Himiko and CR7 are $0\farcs38 \times 0\farcs30$ with a position angle of 66\mbox{$.\!\!{ \degree }$}9 and $0\farcs46 \times 0\farcs41$ with a position angle of $38\fdeg4$, respectively.
We also produce a uv-tapered \cii$158\micron$ line cube with a Gaussian width of $0\farcs7$ to optimize the signal-to-noise ratio (S/N). 
The uv-tapered \cii$158\micron$ line cube achieves a synthesized beam full width at half maximum (FWHM) of $0\farcs94 \times 0\farcs81$ with a position angle of 81\mbox{$.\!\!{ \degree }$}7 ($0\farcs90 \times 0\farcs87$ with a position angle of $-52\fdeg0$) for Himiko (CR7). 
We use the uv-tapered line cube to measure the spatial distributions and fluxes of \cii$158\micron$ of Himiko and CR7. 
Throughout this paper, we only show statistical errors. The absolute flux uncertainty of Band~6 is $\sim10\%$ (ALMA Cycle 1 Properser's Guide\footnote{\url{https://almascience.nao.ac.jp/documents-and-tools/cycle-1/alma-proposers-guide}}).

\subsection{Subaru}

We obtain ultra-deep images from the Hyper Suprime-Cam (HSC) Subaru Strategic Program (SSP; \citealt{aihara18a, aihara18b, aihara19, aihara22}) public data release 3 (PDR3; \citealt{aihara22}). Especially, we utilize the narrowband (NB) images taken by the NB921 filter ($\lambda_{\rm c}=9215$\AA) that covers the redshifted Ly$\alpha$ emission at $z=6.6$. We study the extended Ly$\alpha$ emission of Himiko and CR7 with these images.
The point spread function (PSF) FWHM is typically $\sim0\farcs75$ for the NB921 filter.

\section{Analysis} \label{sec:analysis} 
\subsection{NIRCam Photometry}

To extract the NIRCam photometry of these objects, we conduct PSF matching. 
The reference images are F444W, with the largest PSF ($\mathrm{FWHM}\sim0\farcs16$) among the NIRCam filters. 
We measure the size of the PSF for each filter and match the PSF using a Gaussian kernel convolution. 
When we extract the photometry, we use SEP (version 1.2.1, \citealt{barbary16}), the Python version of the SOURCE EXTRACTOR (SExtractor; \citealt{bertin96}). 
We extract the photometry of each clump in a circular aperture with a $0\farcs4$-diameter, except for CR7-A (Figure~\ref{fig:NIRCam-cutout}, see Section~\ref{sec:results}), for which a $0\farcs5$-diameter is used. 
This is because CR7-A is more extended than other clumps. 
We do not apply aperture correction because these objects have complex morphology. 
We estimate errors with 1000 circular apertures of the same size on a blank field using the Photutils package (version 2.0.2, \citealt{larry_bradley_2024_13989456}). 
No uncertainty floor (e.g., 5\%--10\%) is applyed to the photometric errors.

\subsection{SED Modeling} \label{subsec:sed_modeling}

We perform a spectral energy distribution (SED) fitting with Prospector (version 1.2.0, \citealt{leja17, johnson21}), utilizing NIRCam photometry to characterize the stellar population of these objects. 
We do not include the ALMA photometry because the ALMA data have a lower resolution ($\sim0\farcs90$) than the NIRCam photometry ($\sim0\farcs16$), and we only have the upper limit of the dust continuum. 
We note that F090W photometry is not used for the SED fitting because \lya\ emission and intergalactic medium (IGM) absorption can contaminate the photometry. 
We employ a stellar population synthesis package called Flexible Stellar Population Synthesis (FSPS; \citealt{conroy09, conroy10}). 
Nebula emission is modeled using the Cloudy photo-ionization framework (\citealt{byler17}) and is implemented in the FSPS. 
We apply the \citet{chabrier03} initial mass function, the Calzetti dust attenuation law (\citealt{calzetti00}), and the \citet{madau95} model of intergalactic medium attenuation. 
We assume a nonparametric star formation history (SFH) as described by \citet{harikane25}. 
We do not include AGN models in the SED fitting because our photometry is limited to the rest-frame UV through the optical wavelength. 
We allow varying total stellar mass ($M_*$) and optical depth in the $V$ band ($\tau_V$) to be in the range of $6<\log{(M_*/M_\odot)}<11$ and $0<\tau_V<2$ with top-hat priors and log-uniform priors, respectively. 
We set the metallicity to $Z=0.2Z_\odot$, which is indicated from the NIRSpec IFU data (see Section~\ref{subsec:emission-line}). 
We use the redshifts measured from the NIRSpec IFU data. 
We apply the Markov Chain Monte Carlo method using the \texttt{emcee} (\citealt{Foreman-Mackey13}) to estimate the best-fit parameters and their uncertainties.

\subsection{Emission-line fitting and flux measurements} 
\subsubsection{JWST NIRSpec}

We perform least-square Gaussian fitting of emission lines to measure spatially integrated emission line fluxes using the \texttt{scipy.optimize} package \citep{virtanen20}. 
We conduct the fitting with up to three Gaussian components for the high-resolution G395H data ($R\sim2700$) that show complex emission line profiles. 
We constrain the Gaussian width to be larger than the instrumental dispersion, adopting $R=2700$ for G395H and $R=100$ for PRISM.
The optimal number of components is determined using the Akaike Information Criterion (AIC; \citealt{akaike74}), with a threshold of $\Delta\mathrm{AIC}\equiv\mathrm{AIC}_{i \mathrm{\,components}} - \mathrm{AIC}_{(i+1)\mathrm{\,components}}>20$, where $\mathrm{AIC}_{i \mathrm{~components}}$ is the AIC value in the case of $i$ Gaussian components. 
A single Gaussian component is used to fit the PRISM spectra with lower spectral resolution. 
We fix the flux ratio of $f_\mathrm{[O\,III]\lambda5007}/f_\mathrm{[O\,III]\lambda4959}=2.98$ \citep{storey00} to reduce the fitting parameters. 
When calculating emission line ratios, we use flux values obtained from the G395H spectra except for metallicity measurements and some line ratios involving [O\,{\sc ii}]$\lambda$3727. 
The error spectrum is scaled to the background noise level. 
Flux errors are calculated by summing the errors of the spectral bins in quadrature over a wavelength range of $2\times\mathrm{FWHM}$, centered on the Gaussian peak. 
All rest-frame wavelengths are adopted on the vacuum wavelength scale (e.g., 5008.240\AA\ for \oiii$\lambda5007$), while we follow the conventional emission line names based on their wavelengths in the air. 

To evaluate the morphology and kinematics of the objects, we create moment maps using G395H data cubes. 
The emission lines in each spatial pixel (spaxel) have distorted shapes, which implies complex structures and kinematics. 
Therefore, we calculate the moment 0, 1, and 2 maps (flux, velocity, and velocity dispersion) using a flux-weighted average (e.g., \citealt{walter08}) instead of the Gaussian profile fitting. 
When we consider the flux ratio maps, we conduct a single Gaussian fitting for each spaxel individually. 
The velocity dispersion is corrected for instrumental broadening assuming $\sigma_\mathrm{int} = \sqrt{\sigma_\mathrm{obs}^2 - \sigma_\mathrm{inst}^2}$, where $\sigma_\mathrm{int}$, $\sigma_\mathrm{obs}$, and $\sigma_\mathrm{inst}$ are intrinsic, observed, and instrumental dispersions, respectively. 
We assume $R=2700$ to obtain $\sigma_\mathrm{inst}$ for the G395H data.

\subsubsection{ALMA}

We create moment 0 maps around the \cii158$\micron$ emission line and sum up data from regions of $>3\sigma$ to obtain the total spectrum of Himiko and CR7. 
\cii\ flux measurements are conducted by summing up the spectral bins because the line shapes deviate from a Gaussian profile. However, these flux values are largely consistent with the results obtained by least-squares fitting of a Gaussian.
We measure the noise by placing 1000 random circular apertures around the objects, the spatial sizes of which are the same as those used for flux measurements of Himiko and CR7. 
The flux errors are calculated using the same method as in the NIRSpec IFU analysis. 
We convert the \cii\ fluxes to \cii\ luminosities following the relation: 
\begin{equation}
    L_\mathrm{[CII]} = 1.04 \times 10^{-3} \times \left( \frac{S_\mathrm{[CII]} \Delta v}{\mathrm{Jy~km~s^{-1}}} \right) \left(\frac{D_\mathrm{L}}{\mathrm{Mpc}} \right)^2 \left(\frac{\nu_\mathrm{obs}}{\mathrm{GHz}} \right) L_\odot, 
\end{equation}
where $S_\mathrm{[CII]} \Delta v$ is the line flux in Jy~km~s$^{-1}$, $D_\mathrm{L}$ is the luminosity distance in Mpc ($D_\mathrm{L}=65862.5~\mathrm{Mpc}$ at $z=6.60$), and $\nu_\mathrm{obs}$ is the observed frequency in GHz (e.g., \citealt{solomon05, carilli13}). 

The dust continuum at a rest-frame wavelength of $\lambda=158~\micron$ is undetected for Himiko and CR7. We only put the $3\sigma$ upper limit for the dust continuum estimated from the noise of the continuum maps.

\section{Results} \label{sec:results} 
\subsection{Morphology and Dynamics} \label{subsec:morphology} 

\begin{figure*}
\plotone{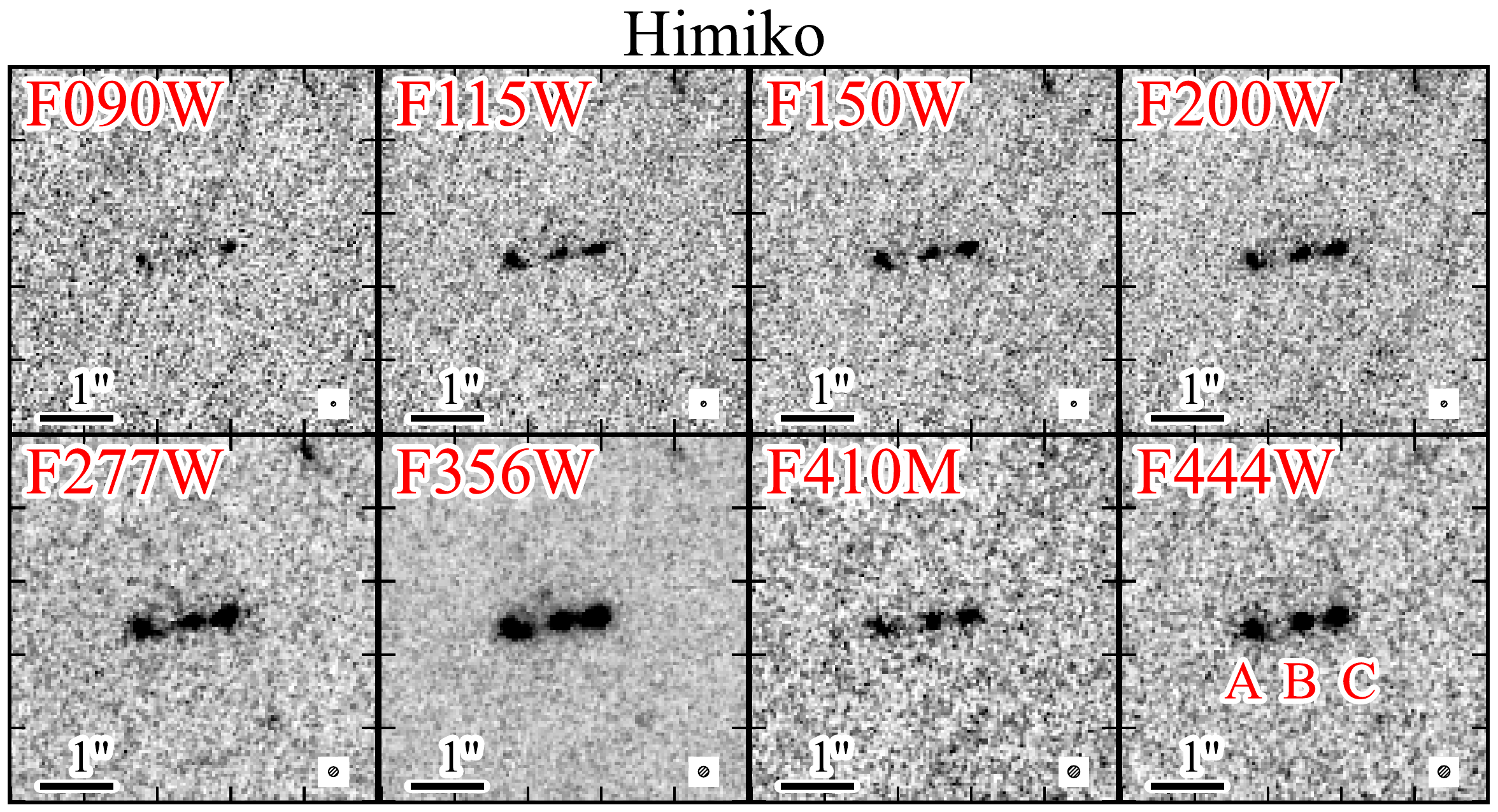}
\plotone{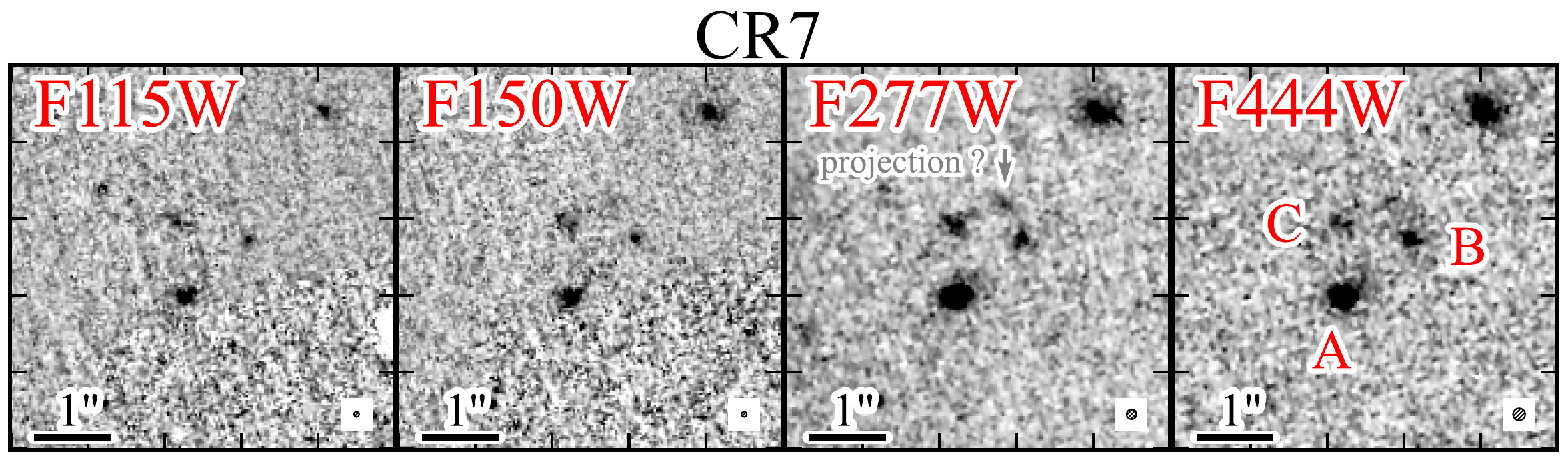}
\caption{NIRCam images of Himiko and CR7 ($5\arcsec \times 5\arcsec$). The F090W, F115W, F150W, F200W, F277W, F356W, F410M, and F444W (F115W, F150W, F277W, and F444W) images are shown for Himiko (CR7). The scale bar is $1\arcsec$ (5.5 kpc at $z=6.6$). We can see the multiple components in both Himiko and CR7. 
The circle in the bottom-right corner represents the size of the PSF FWHM. 
In the F277W image of CR7, we show a possible chance-projected object, indicated by the gray arrow (see Section~\ref{subsec:morphology}).
\label{fig:NIRCam-cutout}}
\end{figure*}

Figure~\ref{fig:NIRCam-cutout} shows $5\arcsec \times 5\arcsec$ NIRCam cutout images of Himiko (top) and CR7 (bottom). 
We identify at least three clumps within $\sim2\arcsec$ diameter ($\sim11.1$ kpc at $z=6.6$) aperture for each object in the rest-frame UV to optical images. 
Throughout the paper, we label these clumps as Himiko-A, Himiko-B, Himiko-C, and CR7-A, CR7-B, CR7-C as indicated in Figure~\ref{fig:NIRCam-cutout}, which are the same terminologies as in previous studies (e.g., \citealt{ouchi13, sobral15}). 
We note that the F356W image is significantly affected by strong emission lines, including H$\beta$ and \oiii$\lambda\lambda4959,5007$, which can enhance brightness.
The F090W image of Himiko includes the Ly$\alpha$ emission but does not reveal a clear signature of the extended Ly$\alpha$. 
This absence likely attributes to the shallower surface brightness sensitivity of JWST (6.5~m mirror) compared to that of Subaru (8~m mirror) or the difference in the filter response. 

\begin{figure*}
\gridline{\fig{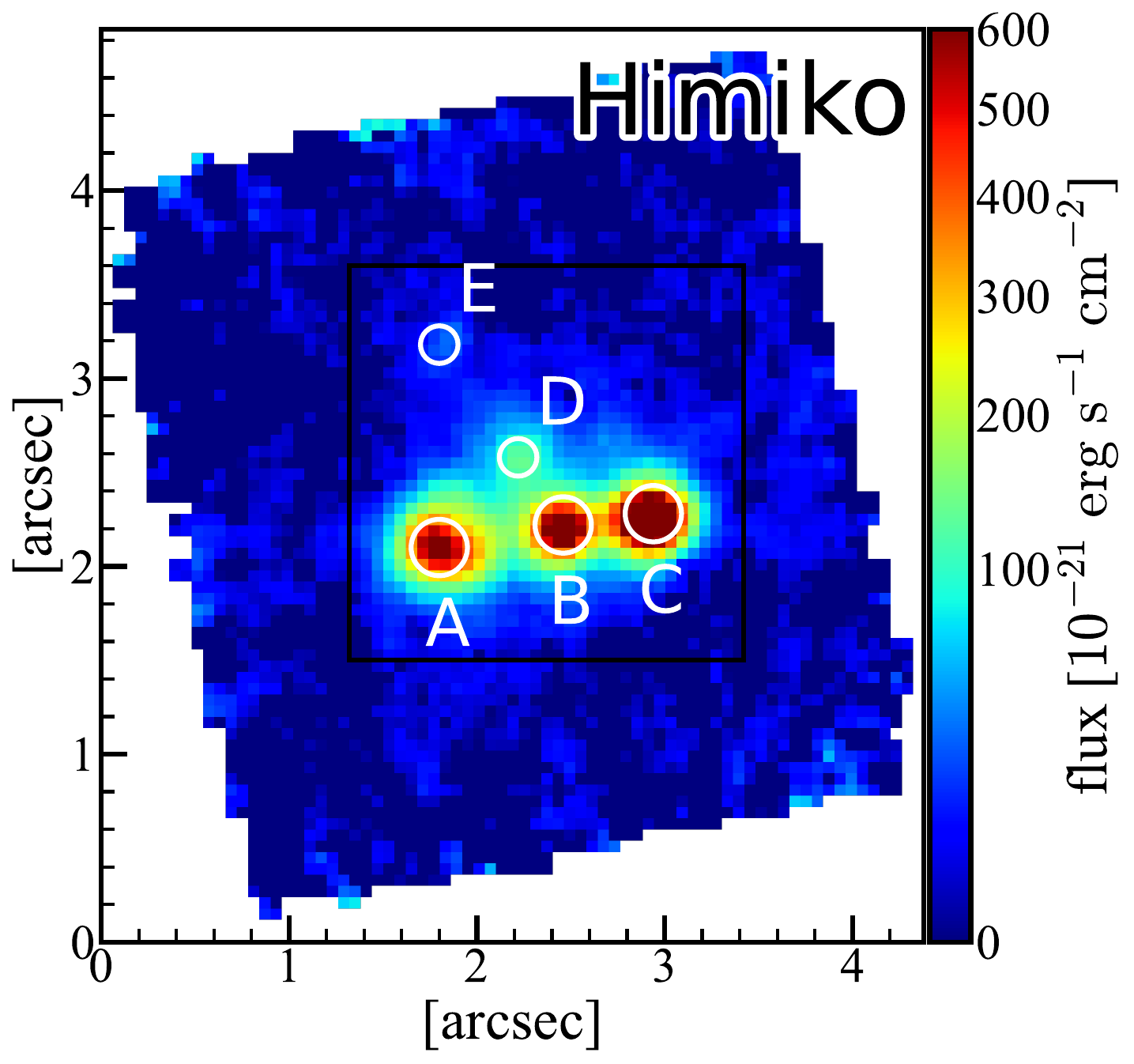}{0.42\textwidth}{}
          \fig{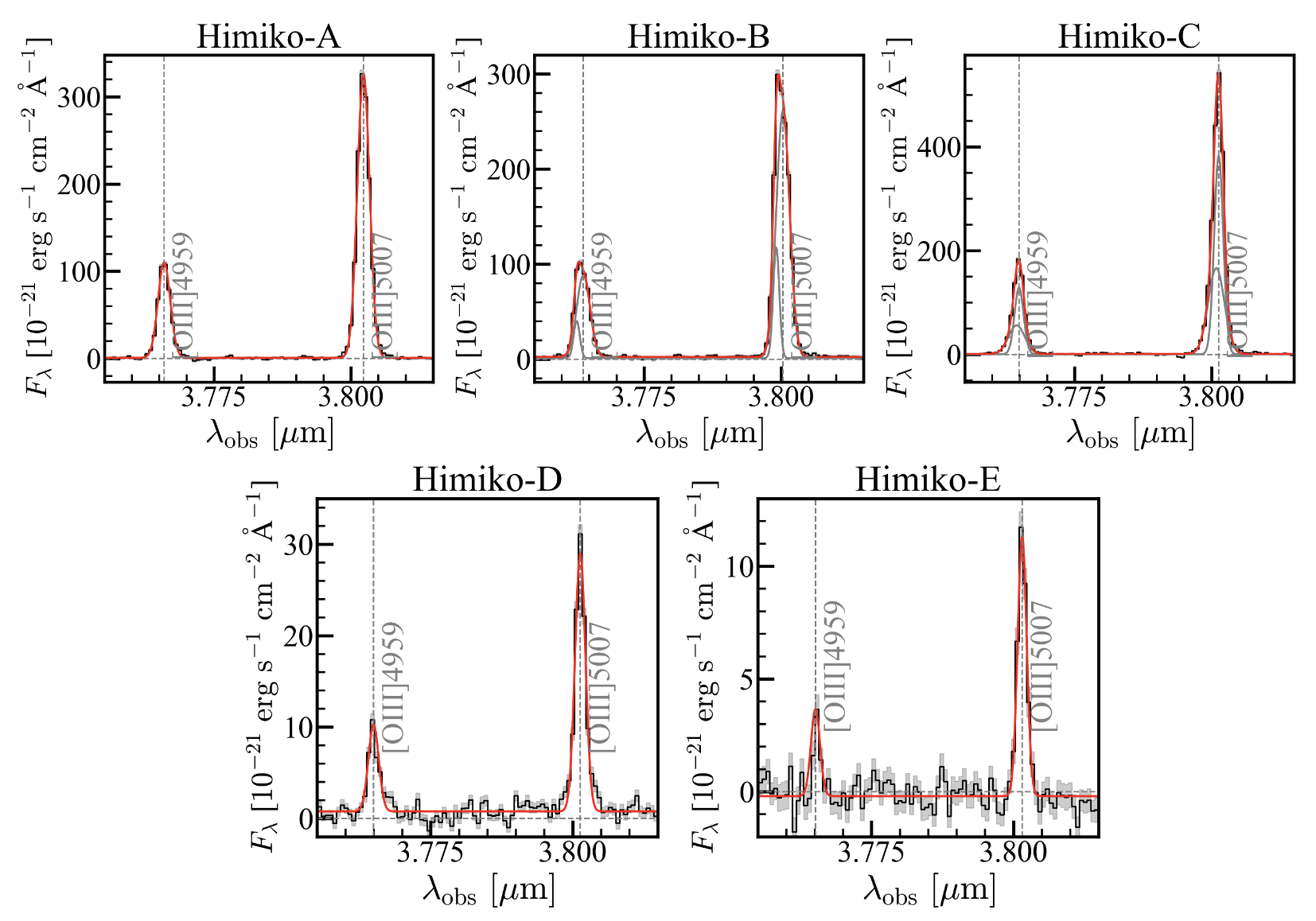}{0.60\textwidth}{}
          }
\gridline{\fig{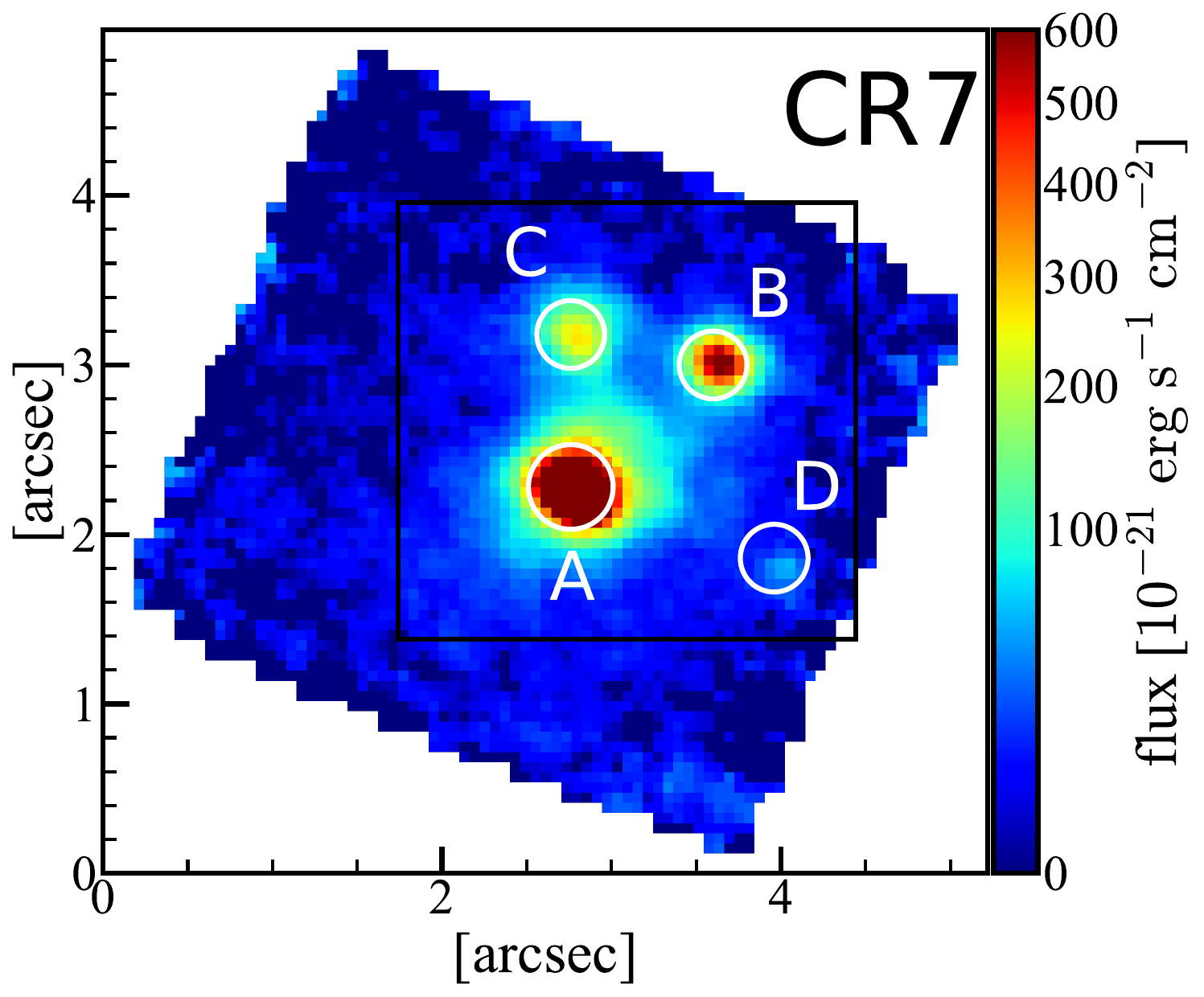}{0.42\textwidth}{}
          \fig{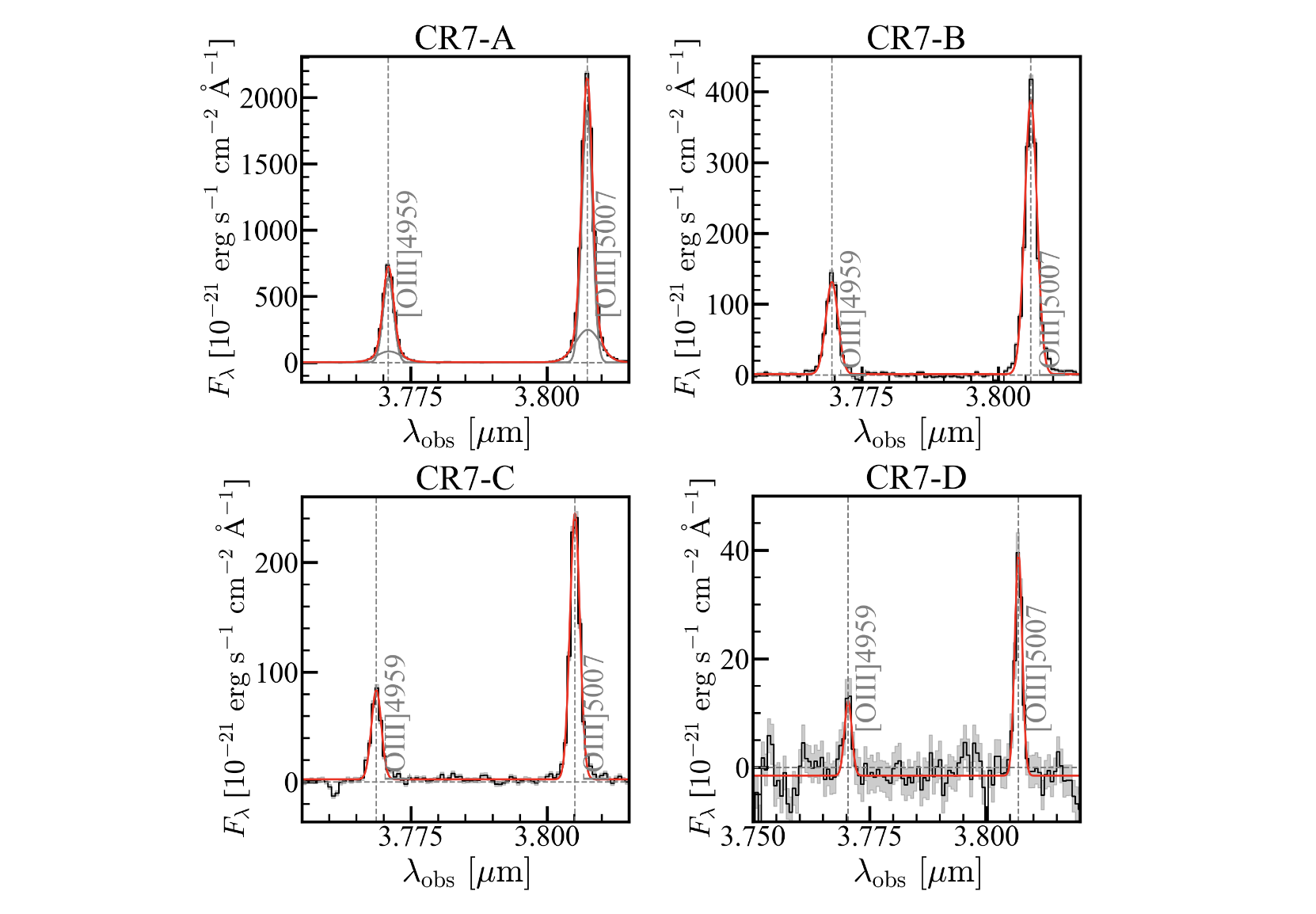}{0.60\textwidth}{}
          }
\caption{Left: JWST NIRSpec IFU G395H images collapsed around \oiii$\lambda5007$ emission line with a range of 750~km~s$^{-1}$ for Himiko (top) and CR7 (bottom). 
Black squares show the selected region for the following moment map analysis (Figure~\ref{fig:moment-maps}). 
White circles show the apertures for the extracted spectrum. 
North is up, and east is to the left side. 
We can find the five and four clumps for Himiko and CR7, respectively. 
Right: \oiii$\lambda4959, 5007$ emission lines from the individual apertures taken with NIRSpec IFU G395H, for Himiko (top) and CR7 (bottom). 
The black histograms show the spectrum, and the gray shadings show 1$\sigma$ uncertainties. 
The red and gray lines show the total Gaussian fitting results and individual Gaussian components, respectively. 
\label{fig:ifu_oiii}}
\end{figure*}

In Figure~\ref{fig:ifu_oiii}, we show the NIRSpec IFU G395H data cube collapsing around the \oiii$\lambda5007$ line, with the highest S/N among the detected emission lines. 
The spatially integrated spectrum of \oiii$\lambda\lambda4959, 5007$ for each region is also presented in Figure~\ref{fig:ifu_oiii}. 
The apertures are shown as white circles ($0\farcs3$-diameter for Himiko-A, -B, -C; $0\farcs2$-diameter for Himiko-D, -E; $0\farcs5$-diameter for CR7-A; $0\farcs4$-diameter for CR7-B, -C, -D). 
We extract the spectrum using the Photutils package, and we do not perform aperture corrections because Himiko and CR7 have complex morphology. 
Using the \oiii\ doublet, for Himiko, we find additional clumps named Himiko-D and Himiko-E with separations of 0$\farcs$43 and 1$\farcs$2 ($=2.4~\mathrm{kpc}$ and $6.6~\mathrm{kpc}$) from Himiko-B, respectively. 
We measure the redshift by performing a single Gaussian fitting for the \oiii$\lambda5007$ G395H spectrum. 
The Redshifts of Himiko-D, -E are $z=6.5902$ and $z=6.5905$, respectively, which are almost the same as those of other clumps ($z=6.5920, 6.5875, 6.5898$ for Himiko-A, -B, and -C). 
For CR7, the redshifts of CR7-A, -B, and -C are $z=6.6022$, $z=6.5994$, and $z=6.5977$. 
There is an additional clump named CR7-D (1$\farcs$3 separation from CR7-A), and the redshift is $z=6.6011$. 
These results show that Himiko and CR7 are complex systems with at least five and four clumps, respectively. 
We further discuss the origin of the multiple clumps in Section \ref{subsec:multiple-clumps}. 

We also note that in the NIRCam images of CR7 (Figure~\ref{fig:NIRCam-cutout}), there is an object in the north with a separation of $1\farcs5$ ($0\farcs5$) from CR7-A (CR7-B). 
In the F277W image of Figure~\ref{fig:NIRCam-cutout}, this object is indicated by the gray arrow. 
The object is also observed in the HST/WFC3 $YJ$ band image in Figure 6 of \citet{sobral15}. 
To check whether the object is associated with CR7, we investigate the NIRSpec IFU PRISM and G395H data at the same location as the NIRCam images. 
The object is detected with a continuum in the PRISM spectrum and an emission line at $\sim2.94~\micron$ in the G395H spectrum, which corresponds to the [Ne\,{\sc iii}]$\lambda$3869 at $z=6.6$. 
However, other emission lines are not detected in the spectrum, which might indicate that the object is merely a chance projection. 
In addition, COSMOS-Web DR1 \citep{casey23, shuntov25} conducts SED fitting for the object and reports that it has a photometric redshift of $z_\mathrm{phot}=1.8\pm 0.7$.
Further spectroscopic observations are necessary to determine its redshift. 
We assume that the object is a chance projection and do not discuss it in the latter sections of this paper. 

\begin{figure*}
\gridline{\fig{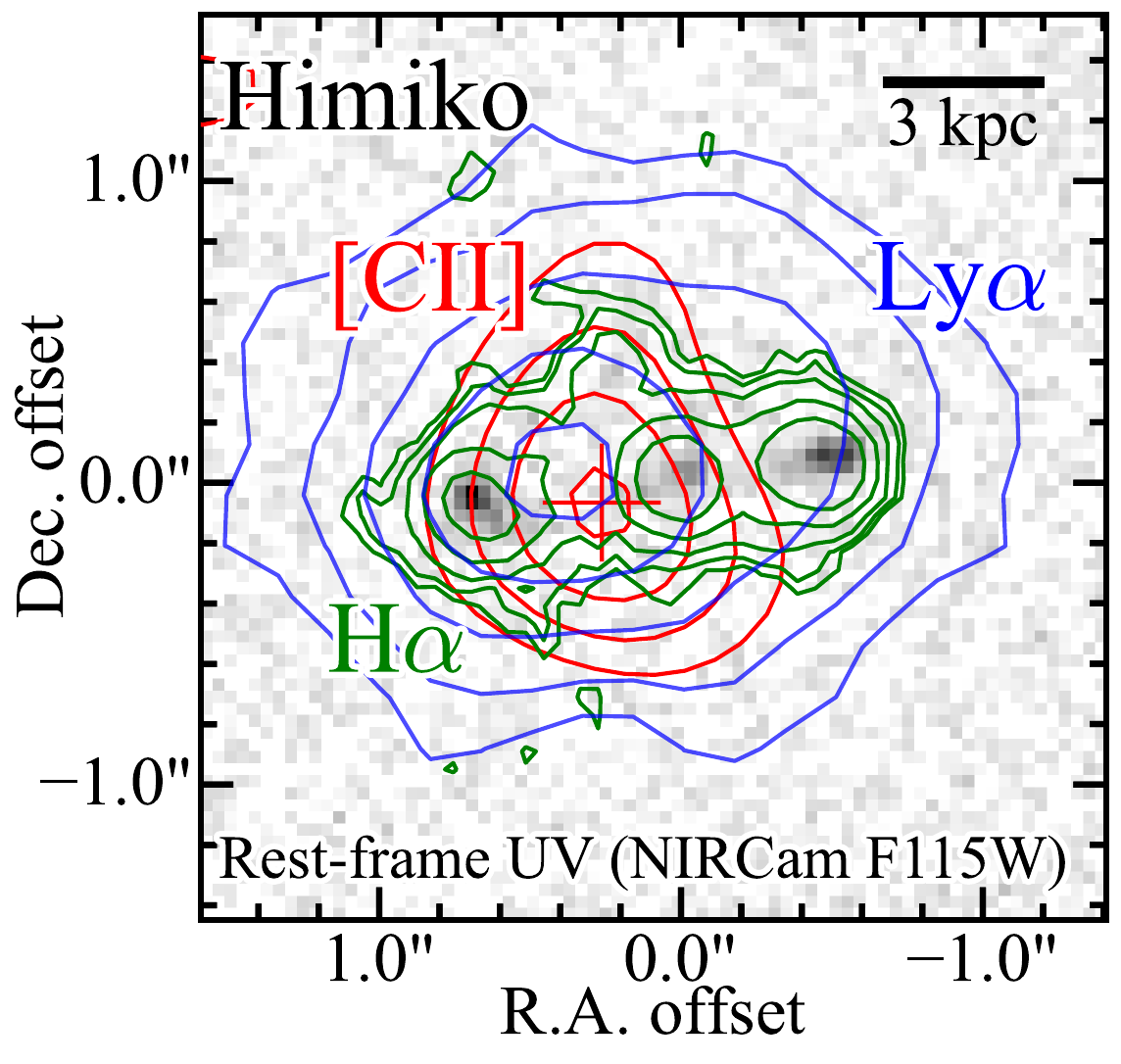}{0.45\textwidth}{}
          \fig{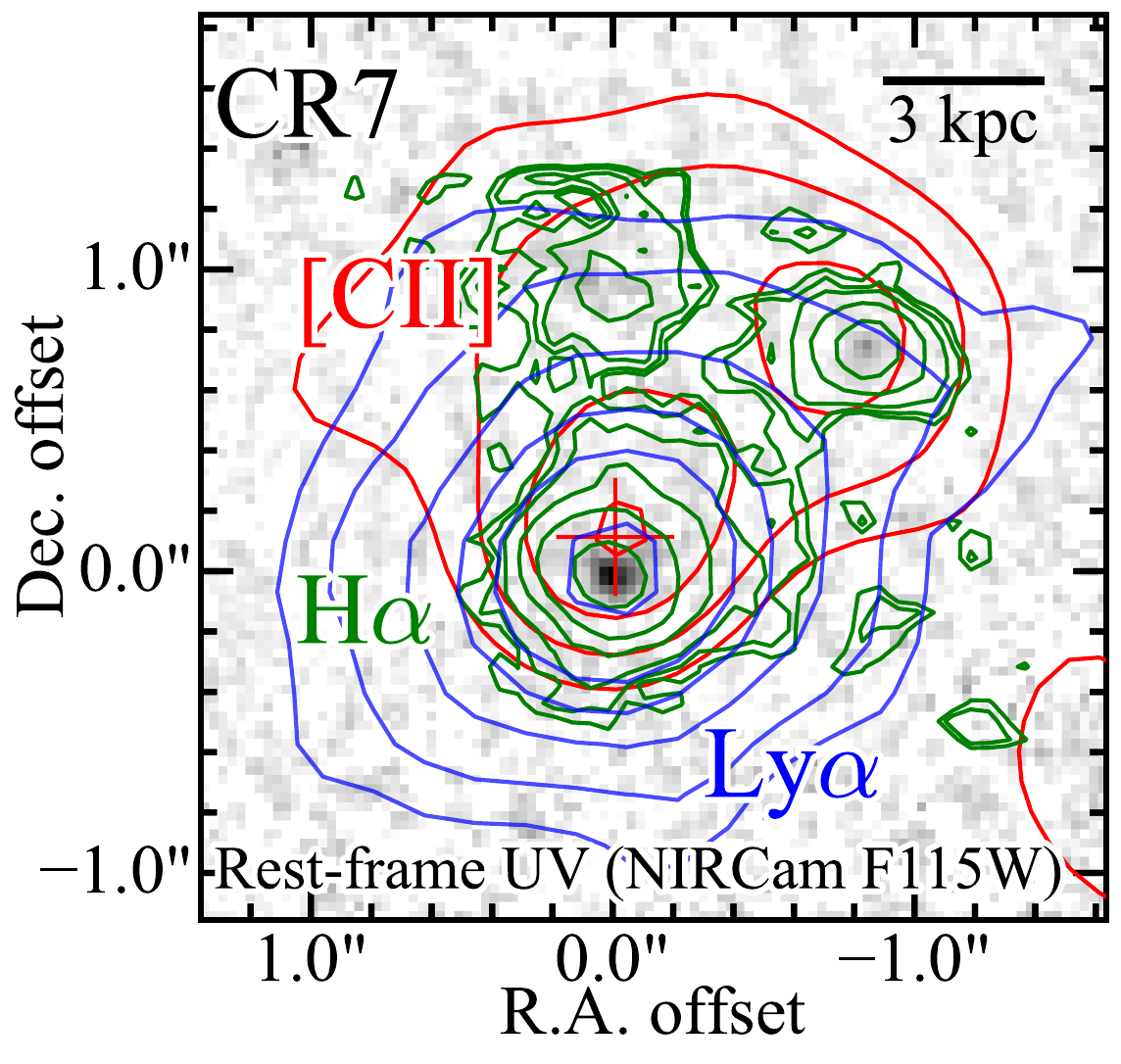}{0.45\textwidth}{}
          }
\caption{Multiwavelength images of Himiko (left) and CR7 (right). The origins of the relative coordinates of Himiko and CR7 correspond to (R.A., Decl.) = (2$^\mathrm{h}$17$^\mathrm{m}$57.571$^\mathrm{s}$, $-05$\textdegree $08\arcmin44\arcsec.816$) and (10$^\mathrm{h}$00$^\mathrm{m}$58.007$^\mathrm{s}$, $01$\textdegree $48\arcmin14\arcsec.930$), respectively (J2000.0). The gray images show the NIRCam F115W, corresponding to rest-frame UV (beam size $\sim0\farcs07$). The green contours show the H$\alpha$ emission (2$\sigma$, 3$\sigma$, 5$\sigma$, 10$\sigma$, 20$\sigma$, 60$\sigma$) taken with the NIRSpec IFU PRISM (beam size $\sim0\farcs2$). The red contours show the \cii 158$\micron$ emission (2$\sigma$, 3$\sigma$, 4$\sigma$, 5$\sigma$) obtained by the ALMA observations (beam size $\sim0\farcs9$). The red crosses indicate the \cii 158$\micron$ peak position, where the bar scales correspond to their positional uncertainty. The blue contours show the \lya\ emission (2$\sigma$, 3$\sigma$, 5$\sigma$, 7.5$\sigma$, 10$\sigma$, 15$\sigma$) taken with NB921 (beam size $\sim0\farcs8$). The scale bars show 3~kpc at $z=6.6$. In the figures, we do not conduct a PSF matching. 
\label{fig:jwst-alma-subaru-image}}
\end{figure*}

Figure~\ref{fig:jwst-alma-subaru-image} compares the spatial distribution of each emission from Himiko (left) and CR7 (right). 
The background gray images are the rest-frame UV emission (NIRCam F115W). 
The green contours show the H$\alpha$ emission taken with the NIRSpec IFU, obtained by collapsing around the H$\alpha$ wavelength. 
The red contours show the \cii\ emission obtained with ALMA Band~6, and the red cross shows the uncertainty of the ALMA positional accuracy\footnote{
The positional accuracy of ALMA imaging is roughly estimated by the relationship: $\theta_\mathrm{beam}/(\mathrm{S/N})/0.9$, where $\theta_\mathrm{beam}$ is FWHM synthesized beam size in arcsecond (see equation 10.7 of ALMA Technical Handbook: \url{https://almascience.nao.ac.jp/documents-and-tools/cycle11/alma-technical-handbook}).}. 
The blue contours show the Ly$\alpha$ distributions estimated by subtracting the F115W images (UV continuum) from the NB921 images (\lya\ and UV continuum). 
The PSFs of F115W images are matched to those of NB921 images by conducting Gaussian kernel convolutions. 
The spatial distribution of the UV continuum and H$\alpha$ emission in Himiko and CR7 aligns well. 
This indicates that H$\alpha$ emission mainly comes from the star-forming regions. 
In addition, the spatial position of the \cii\ and \lya\ peaks agree within the positional uncertainty, consistent with \cite{carniani18} and \cite{matthee17}. 
\lya\ emission of CR7 mostly aligns well with the CR7-A indicated by UV continuum and H$\alpha$ emission. 
\cite{matthee20} have observed CR7 with VLT/MUSE and reported the $\sim1~\mathrm{kpc}$ spatial offset between \lya\ and UV peak of CR7-A. 
We cannot detect such an offset of \lya, which might be due to the low angular resolution of Subaru NB921 data ($0\farcs75$). 
On the other hand, we find that the peaks of \cii\ and \lya\ from Himiko have an offset ($\sim0\farcs3$) from those of the UV continuum and H$\alpha$ peaks of Himiko-A or Himiko-B. 
This offset is beyond the positional uncertainty, and we discuss the physical origin of the offset in Section~\ref{sec:discussion}. 
\cii\ emission of CR7 is spatially resolved, and we find two flux peaks near the position of CR7-A and -B, indicating that the \cii\ is mainly associated with the stellar component of CR7 and consistent with \citet{matthee17}. 

\begin{figure*}
\gridline{\fig{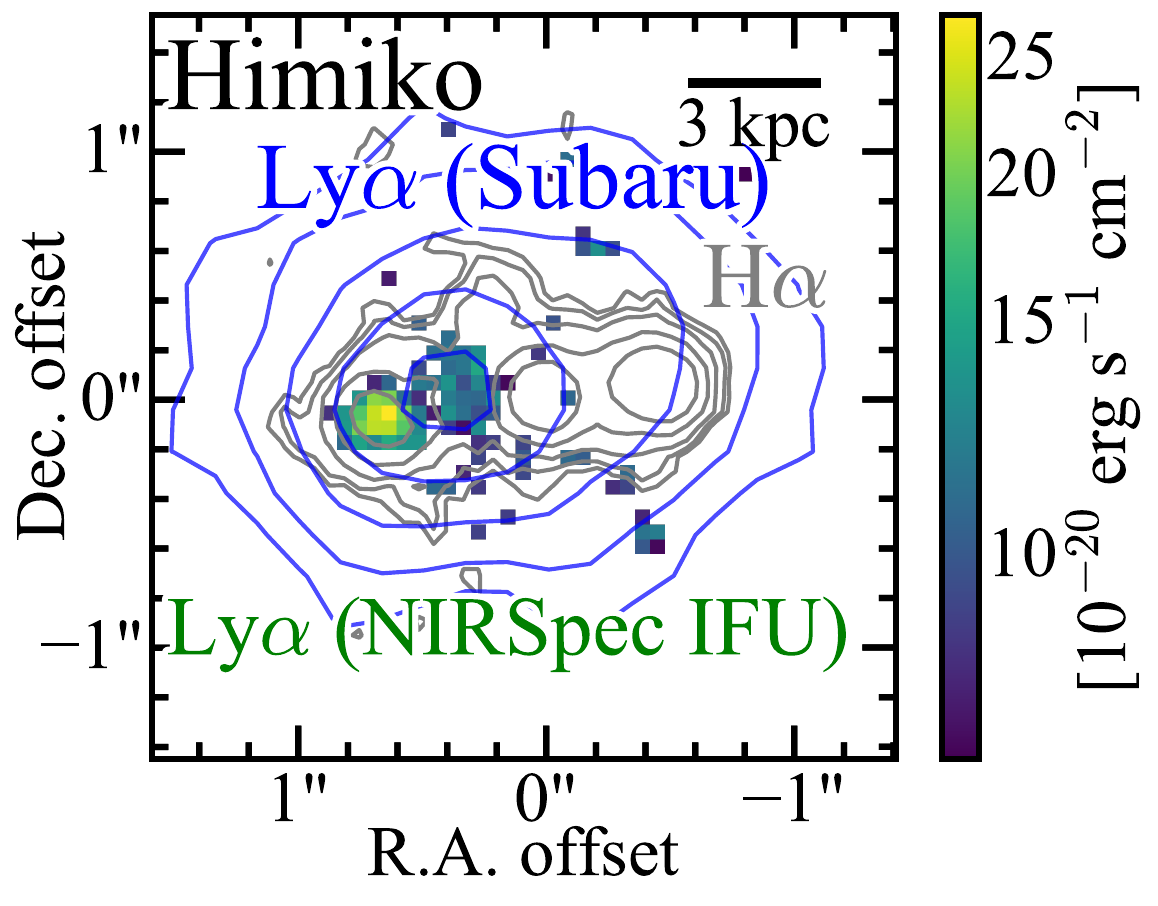}{0.50\textwidth}{}
          \fig{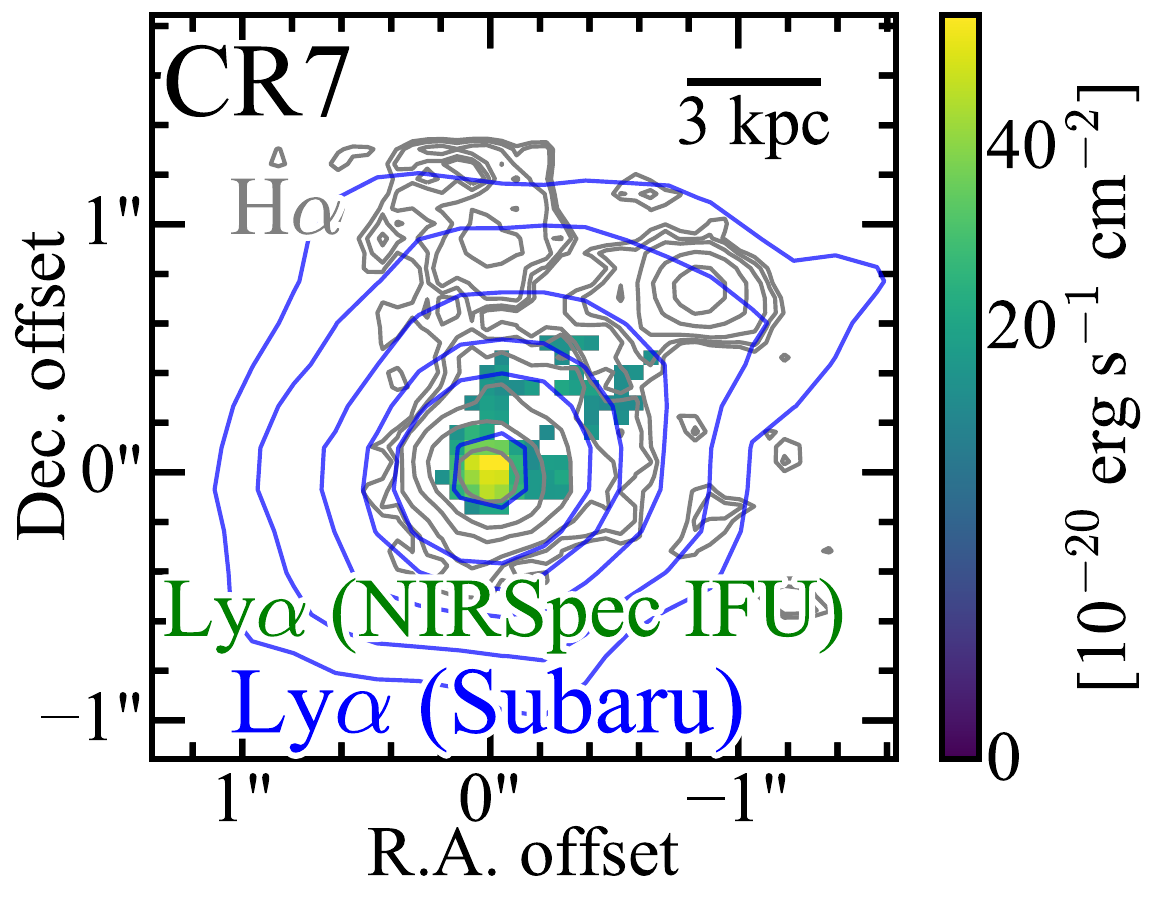}{0.50\textwidth}{}
          }
\caption{Distributions of Ly$\alpha$ and H$\alpha$ emission from Himiko (left) and CR7 (right). 
The background image represents the Ly$\alpha$ emission observed with NIRSpec IFU PRISM. 
Only spaxels with S/N $> 3$ are shown.
The blue and gray contours indicate the Ly$\alpha$ emission from NB921 and the H$\alpha$ emission from NIRSpec IFU PRISM, respectively. 
The contour levels and the relative coordinates are the same for Figure~\ref{fig:jwst-alma-subaru-image}. 
The scale bars correspond to 3~kpc at $z=6.6$. 
A PSF matching is not applied in the figures. 
\label{fig:lya-map}}
\end{figure*}

Figure~\ref{fig:lya-map} shows the spatial distribution of the Ly$\alpha$ emission from Himiko (left) and CR7 (right), observed with NIRSpec IFU (background map) and Subaru NB921 (blue contours). 
For comparison, the H$\alpha$ distribution is also shown with gray contours.
In the NIRSpec IFU images, the Ly$\alpha$ emission peaks for Himiko and CR7 are located around Himiko-A and CR7-A, respectively. 
Furthermore, Ly$\alpha$ emission can be observed between Himiko-A and Himiko-B, coinciding with the peak in the Subaru NB921 image. 
Similarly, there is Ly$\alpha$ emission between CR7-A and CR7-B. 
\cite{matthee20} have reported that the spatial distribution of \lya\ elongated from CR7-A to CR7-B, which is consistent with the JWST NIRSpec PRISM data. 

\begin{figure*}
\gridline{\fig{himiko_bkg_sbt_all_1_v6.pdf}{0.48\textwidth}{}
          \fig{himiko_bkg_sbt_all_0_v4.pdf}{0.48\textwidth}{}
          }
\caption{Integrated spectrum of Himiko taken with NIRSpec IFU PRISM ($R\sim100$; left) and G395H ($R\sim2700$; right). From top to bottom, the spectra of Himiko-A to Himiko-E are shown. The black lines show the spectra, and the blue shaded areas show their $1\sigma$ uncertainties. In the PRISM spectra of Himiko-A, there is a clear \lya\ emission line.
\label{fig:integrated-spectrum-himiko}}
\end{figure*}

\begin{figure*}
\gridline{\fig{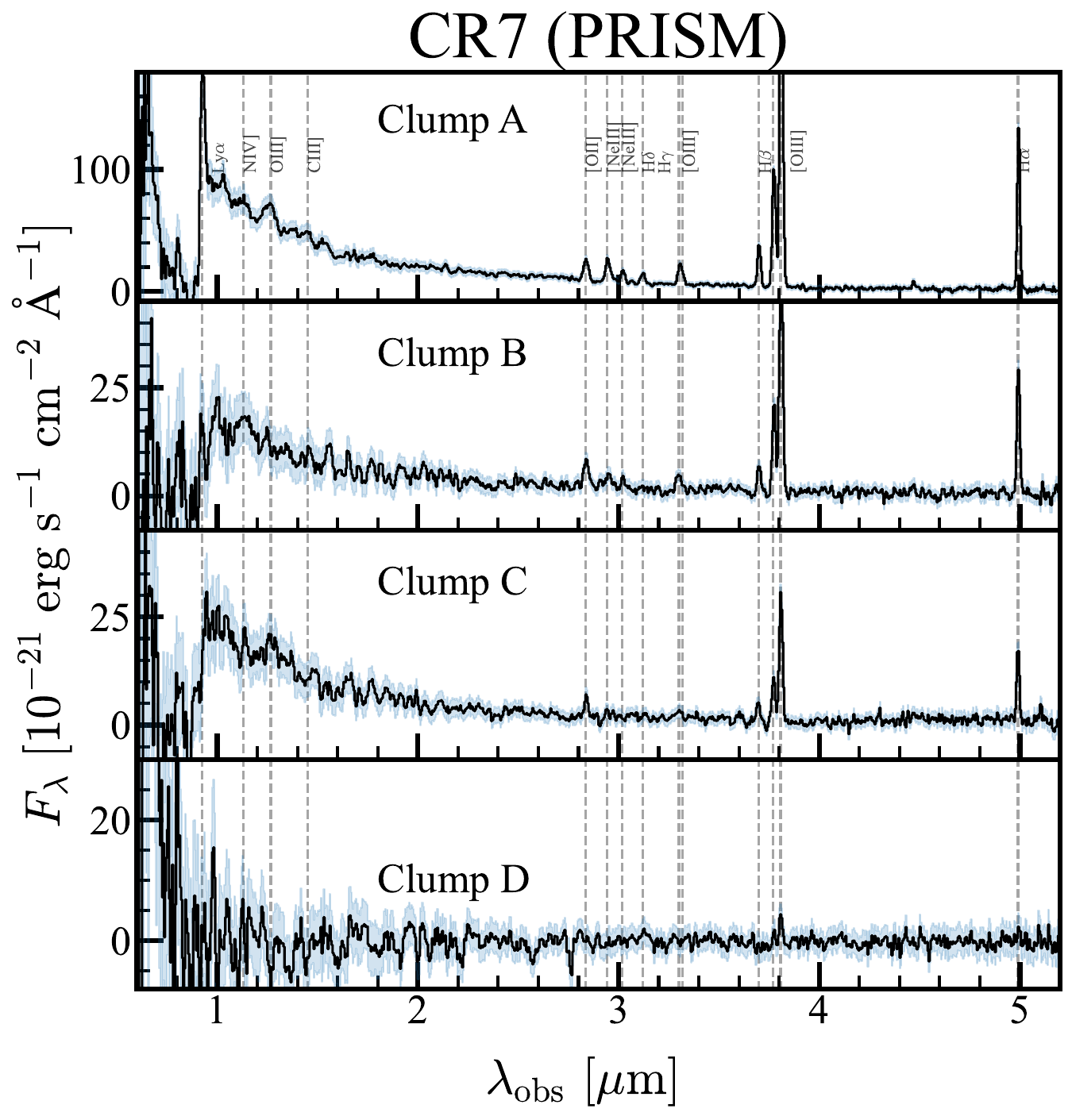}{0.48\textwidth}{}
          \fig{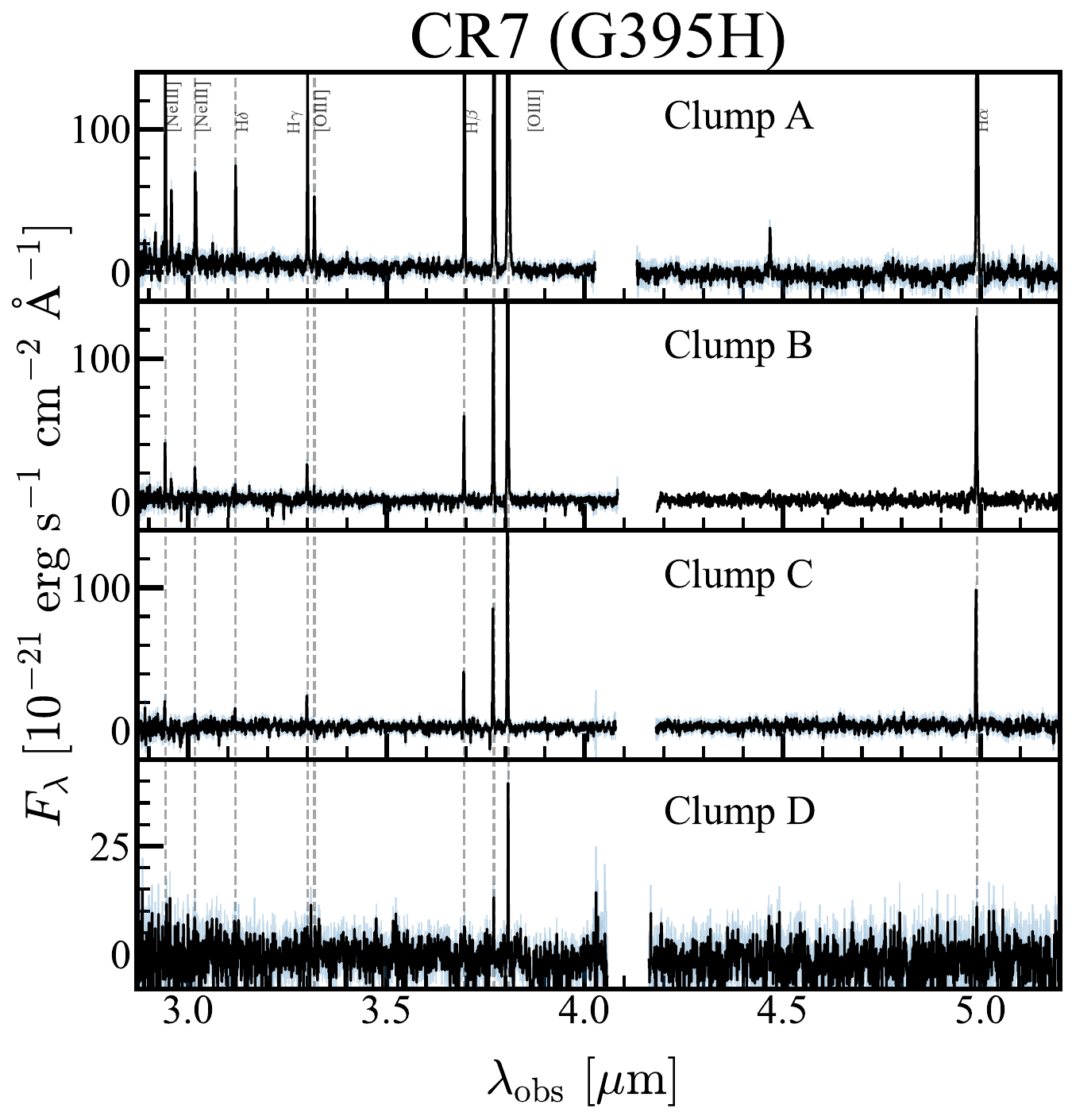}{0.48\textwidth}{}
          }
\caption{Same as in Figure~\ref{fig:integrated-spectrum-himiko} but for CR7. In the PRISM spectra of CR7-A, a clear \lya\ emission line is detected. 
\label{fig:integrated-spectrum-CR7}}
\end{figure*}

Figure~\ref{fig:integrated-spectrum-himiko} and Figure~\ref{fig:integrated-spectrum-CR7} show the spatially integrated spectrum of Himiko and CR7, respectively. 
The apertures to extract the spectrum are shown in Figure~\ref{fig:ifu_oiii}, and we conduct no aperture corrections. 
We find clear \lya\ emission lines in Himiko-A and CR7-A, although other clumps do not. We utilize these spatially integrated spectra to measure the fluxes of each clump. 
Table~\ref{tab:flux-measurement} summarizes the flux measurements for each clump. 
We also list the values of CR7 from the literature \citep{marconcini24b}. 
The flux measurements of CR7 are almost in agreement with those of \citet{marconcini24b} within 2$\sigma$, despite the slight differences in the apertures (see Figure 1 of \citealt{marconcini24b}). 

\begin{deluxetable*}{lcccccccccc}
    \tablecaption{Flux measurements of Himiko and CR7. \label{tab:flux-measurement}}
    \tablewidth{0pt}
    \tablehead{
    \colhead{Object} & \colhead{\oii} & \colhead{[Ne\, {\sc iii}]} & \colhead{H$\delta$} & \colhead{H$\gamma$} & \colhead{\oiii}& \colhead{H$\beta$} & \colhead{\oiii} & \colhead{H$\alpha$} & \colhead{reference} \\
    \colhead{} & \colhead{$\lambda$3727} & \colhead{$\lambda$3869} & \colhead{} & \colhead{} & \colhead{$\lambda$4363} & \colhead{} & \colhead{$\lambda$5007} & \colhead{} & \colhead{}
    } 
    \startdata
    Himiko-A & 0.6$\pm$0.2 & 0.6$\pm$0.1 & 0.3$\pm$0.1 & 0.6$\pm$0.1 & 0.2$\pm$0.1 & 1.2$\pm$0.1 & 9.2$\pm$0.1 & 3.2$\pm$0.1 & This work \\
    Himiko-B & 1.0$\pm$0.2 & 0.7$\pm$0.1 & 0.4$\pm$0.1 & 0.7$\pm$0.1 & 0.2$\pm$0.1 & 1.2$\pm$0.1 & 9.8$\pm$0.1 & 4.4$\pm$0.1 & This work \\
    Himiko-C & 1.5$\pm$0.2 & 1.1$\pm$0.1 & 0.5$\pm$0.1 & 0.9$\pm$0.1 & 0.2$\pm$0.1 & 1.8$\pm$0.1 & 13.6$\pm$0.1 & 4.9$\pm$0.1 & This work \\
    Himiko-D & \nodata & \nodata & \nodata & \nodata & \nodata & \nodata & 0.63$\pm$0.04 & 0.3$\pm$0.1 & This work \\
    Himiko-E & \nodata & \nodata & \nodata & \nodata & \nodata & \nodata & 0.24$\pm$0.04 & \nodata & This work \\
    \hline
    CR7-A & 3.8$\pm$0.5 & 4.0$\pm$0.2 & 1.7$\pm$0.1 & 3.3$\pm$0.1 & 1.3$\pm$0.1 & 6.8$\pm$0.2 & 56.5$\pm$0.4 & 19.1$\pm$0.3 & This work \\
          & 5.5$\pm$1.6 & 4.6$\pm$0.7 & 1.8$\pm$0.5 & 3.2$\pm$0.4 & 1.15$\pm$0.4 & 7.3$\pm$0.4 & 62.6$\pm$0.4 & 21.5$\pm$0.7 & \citet{marconcini24b} \\
    CR7-B & 1.8$\pm$0.4 & 0.8$\pm$0.1 & 0.2$\pm$0.1 & 0.6$\pm$0.1 & 0.2$\pm$0.1 & 1.5$\pm$0.1 & 10.7$\pm$0.1 & 4.1$\pm$0.2 & This work \\
          & 1.8$\pm$0.4 & 0.8$\pm$0.3 & 0.3$\pm$0.2 & 0.3$\pm$0.2 & 0.1$\pm$0.1 & 1.5$\pm$0.3 & 12.5$\pm$0.3 & 4.3$\pm$0.5 & \citet{marconcini24b} \\
    CR7-C & 1.1$\pm$0.4 & 0.5$\pm$0.1 & 0.2$\pm$0.1 & 0.5$\pm$0.1 & \nodata & 0.9$\pm$0.1 & 5.4$\pm$0.1 & 2.5$\pm$0.1 & This work \\
          & 1.4$\pm$0.4 & 0.4$\pm$0.3 & 0.2$\pm$0.2 & \nodata & \nodata & 0.9$\pm$0.3 & 5.8$\pm$0.3 & 2.3$\pm$0.3 & \citet{marconcini24b} \\
    CR7-D & \nodata & \nodata & \nodata & \nodata & \nodata & \nodata &  0.8$\pm$0.1 & \nodata & This work \\
          & \nodata & \nodata & \nodata & \nodata & \nodata & \nodata & \nodata & \nodata & \citet{marconcini24b} \\
    \enddata
    \tablecomments{
    Integrated emission line fluxes ($>$3$\sigma$) that are estimated by a single Gaussian fitting. Each aperture is shown in Figure~\ref{fig:ifu_oiii}. 
    The fluxes are not corrected for dust extinction and units of $10^{-18}$~erg~s$^{-1}$~cm$^{-2}$. 
    All flux values are measured from the G395H spectra except for \oii$\lambda3727$, which is derived from the PRISM spectra. 
    \citet{marconcini24b} have used slightly different apertures compared to this work (see Figure 1 of \citealt{marconcini24b}), but the flux measurements of CR7 align well. 
    }
\end{deluxetable*}

\begin{figure*}
\gridline{\fig{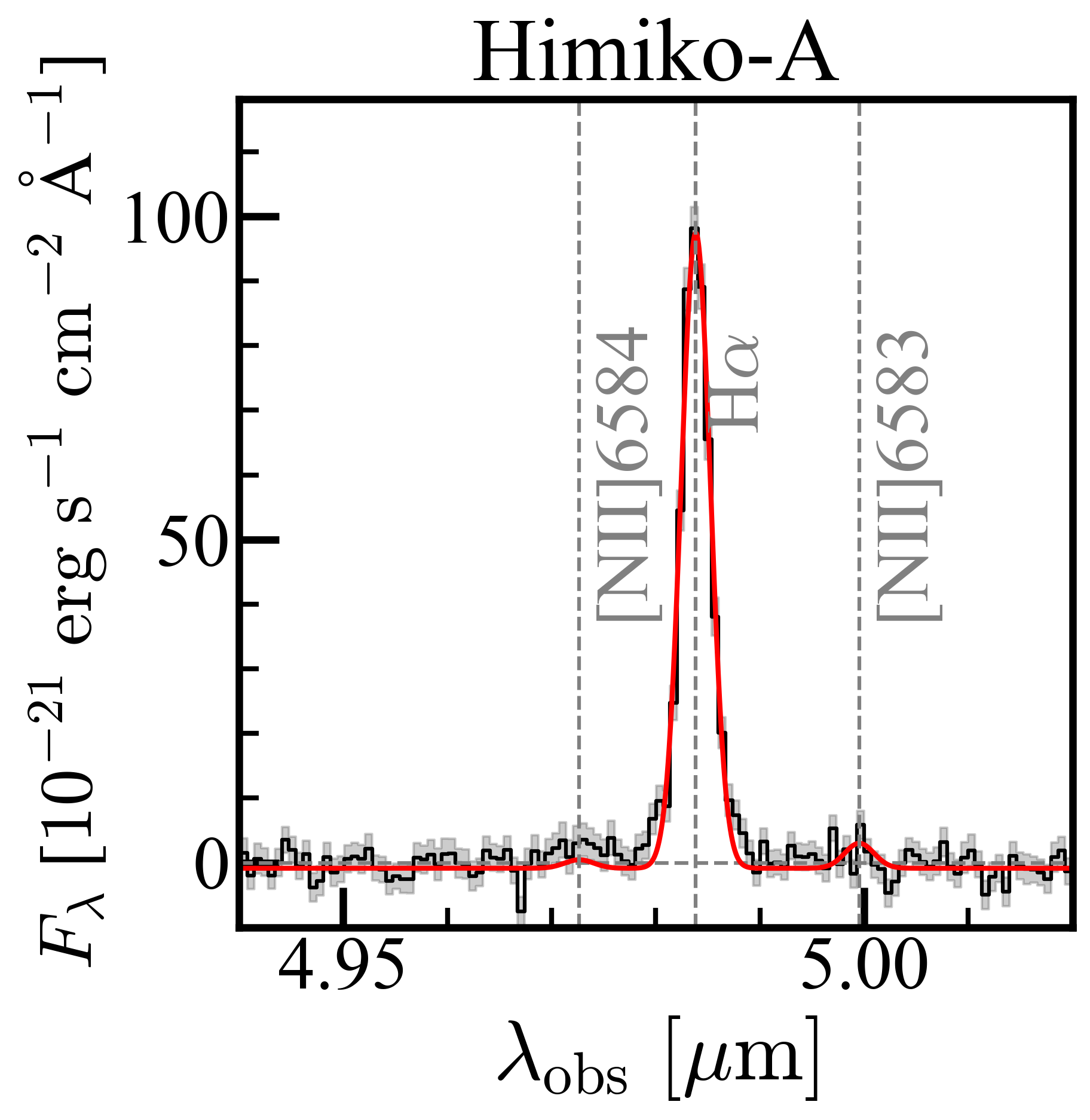}{0.28\textwidth}{}
          \fig{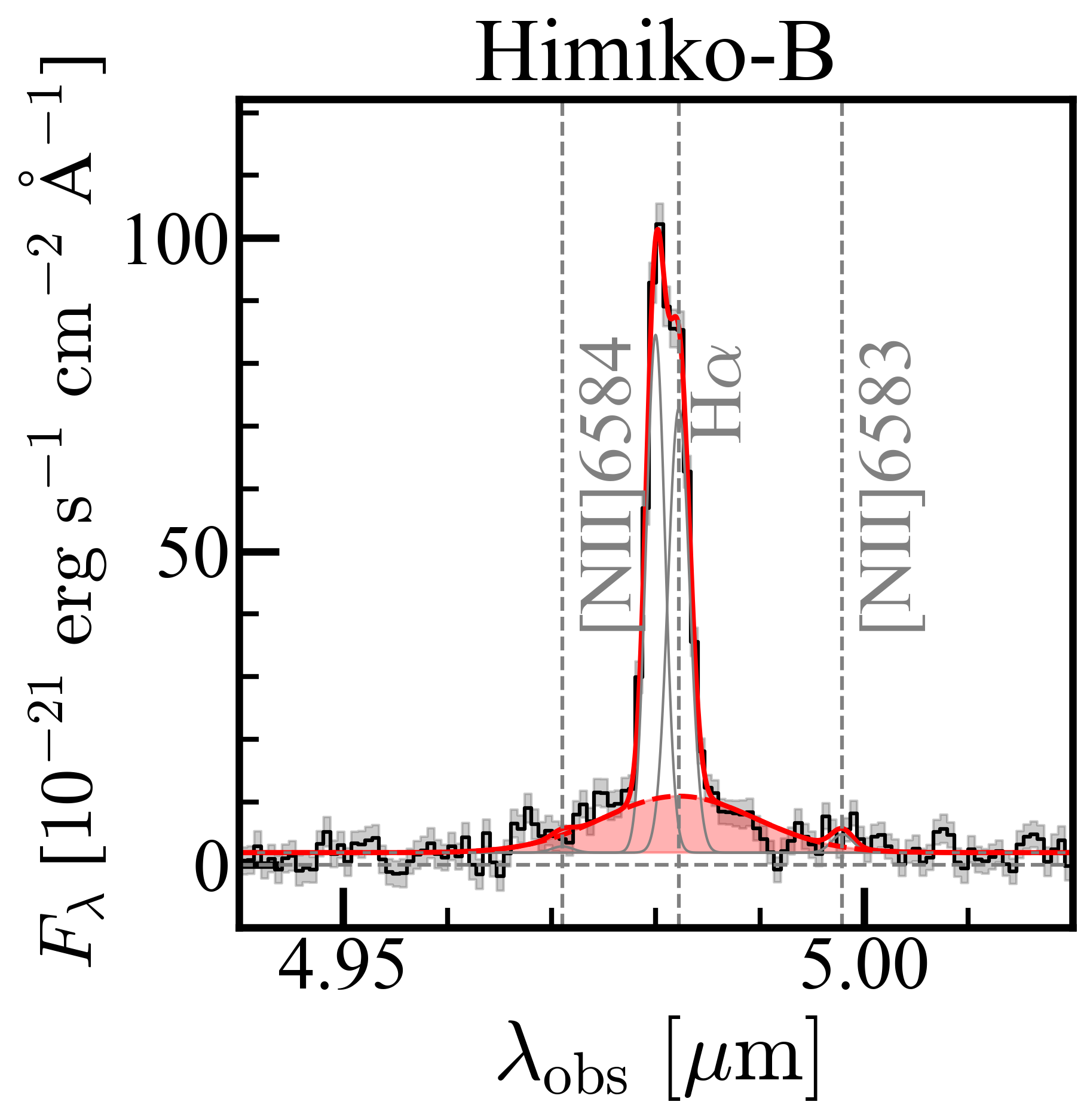}{0.28\textwidth}{}
          \fig{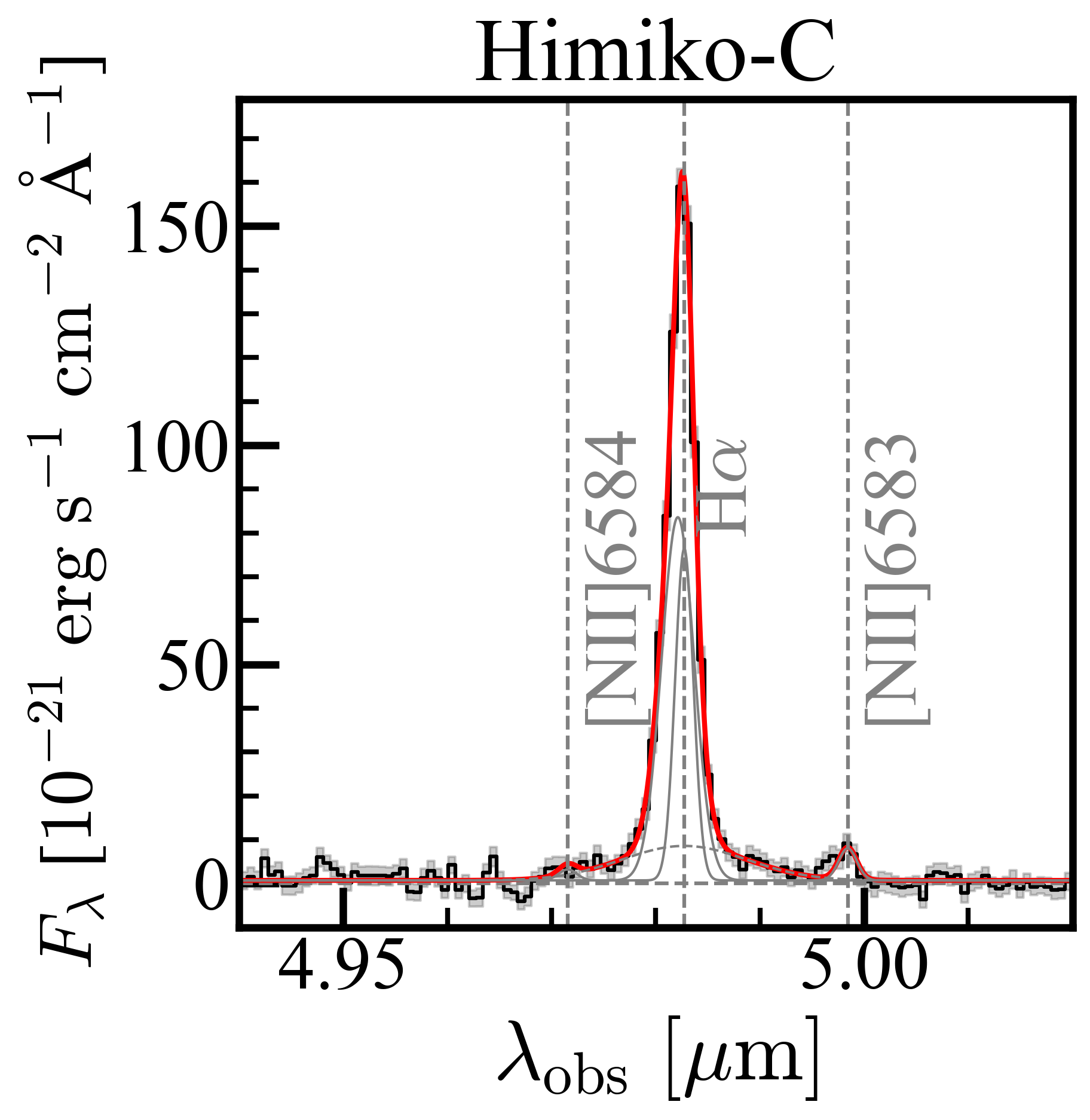}{0.28\textwidth}{}
          }
\gridline{\fig{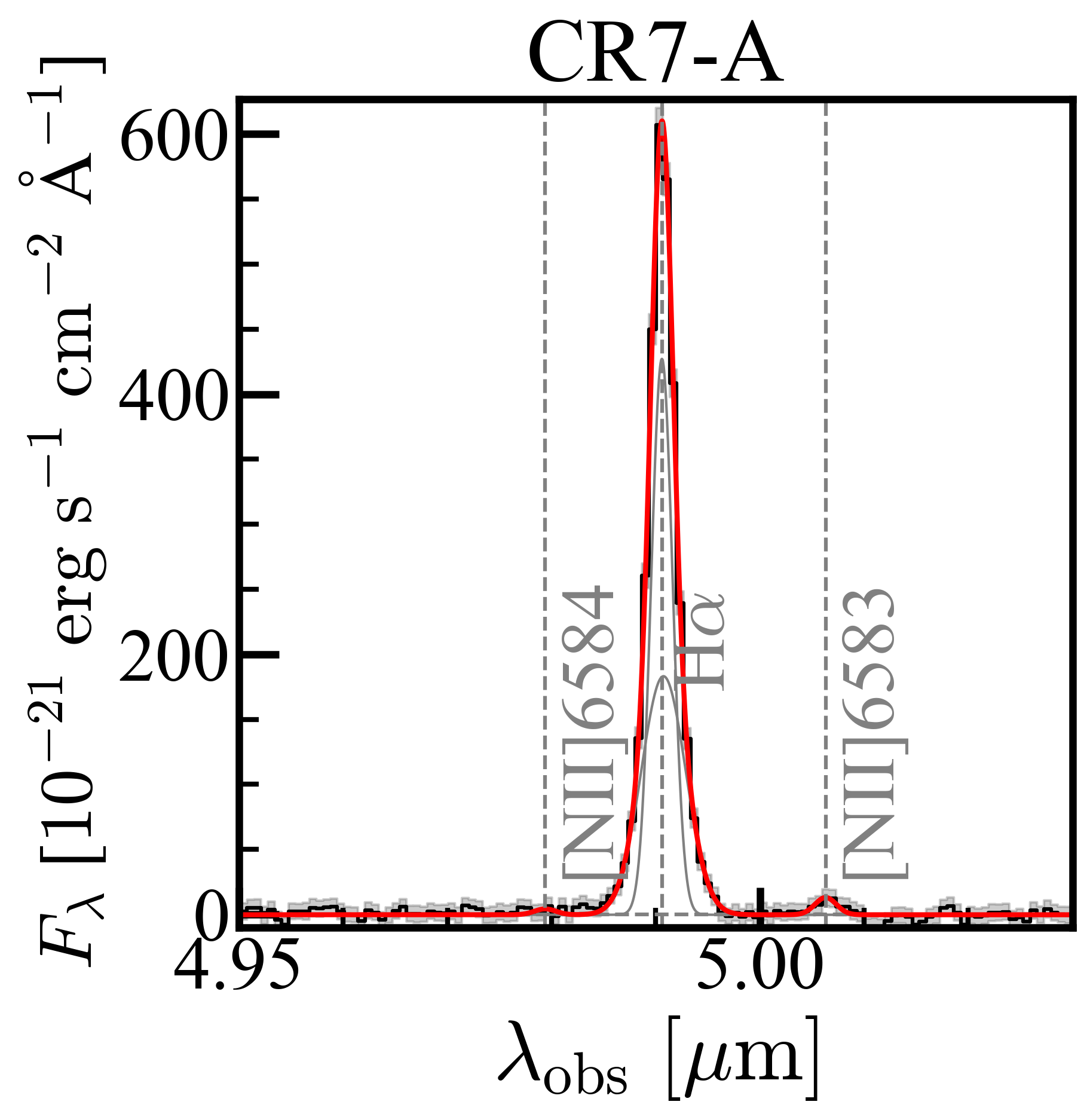}{0.28\textwidth}{}
          \fig{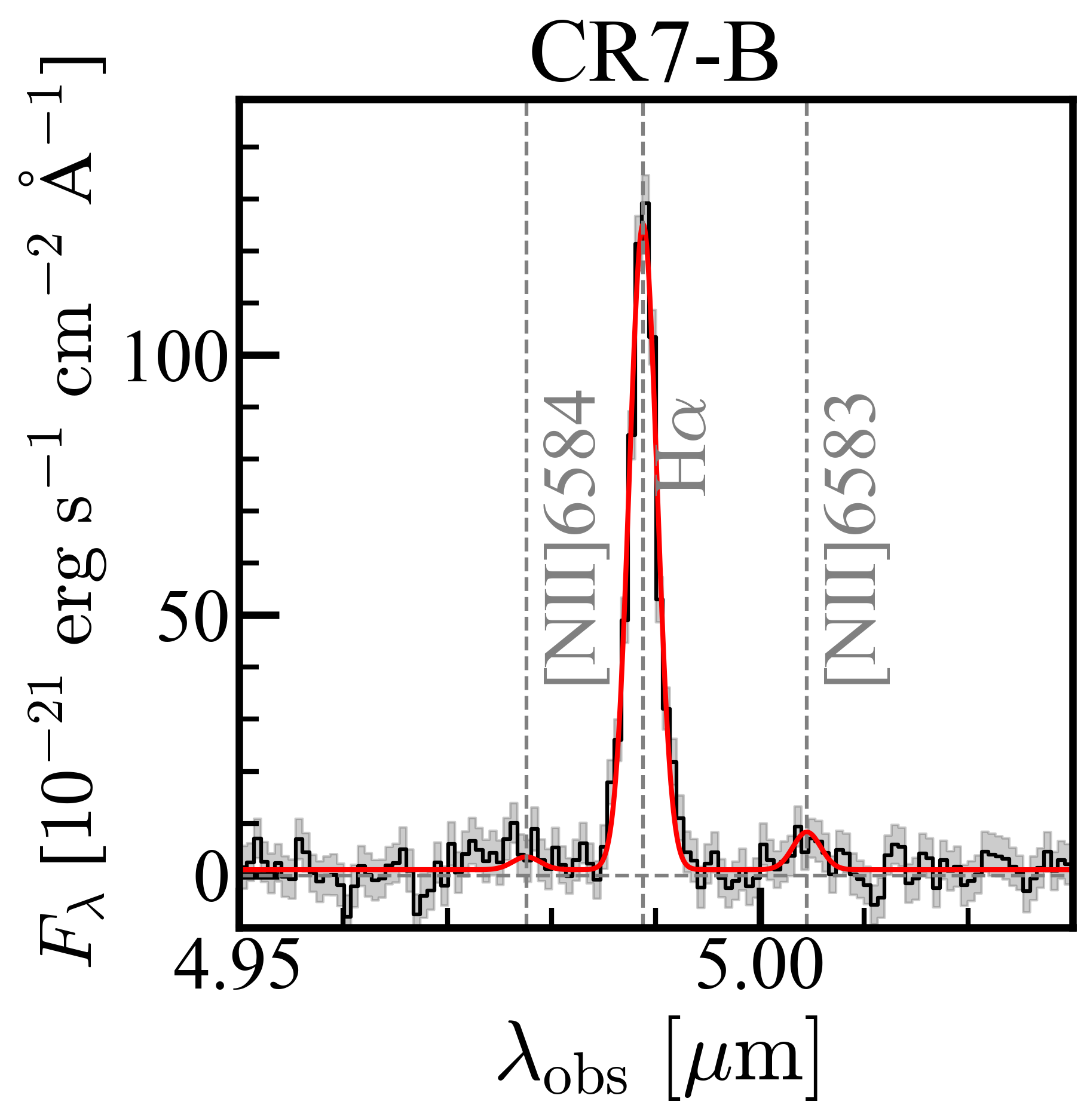}{0.28\textwidth}{}
          \fig{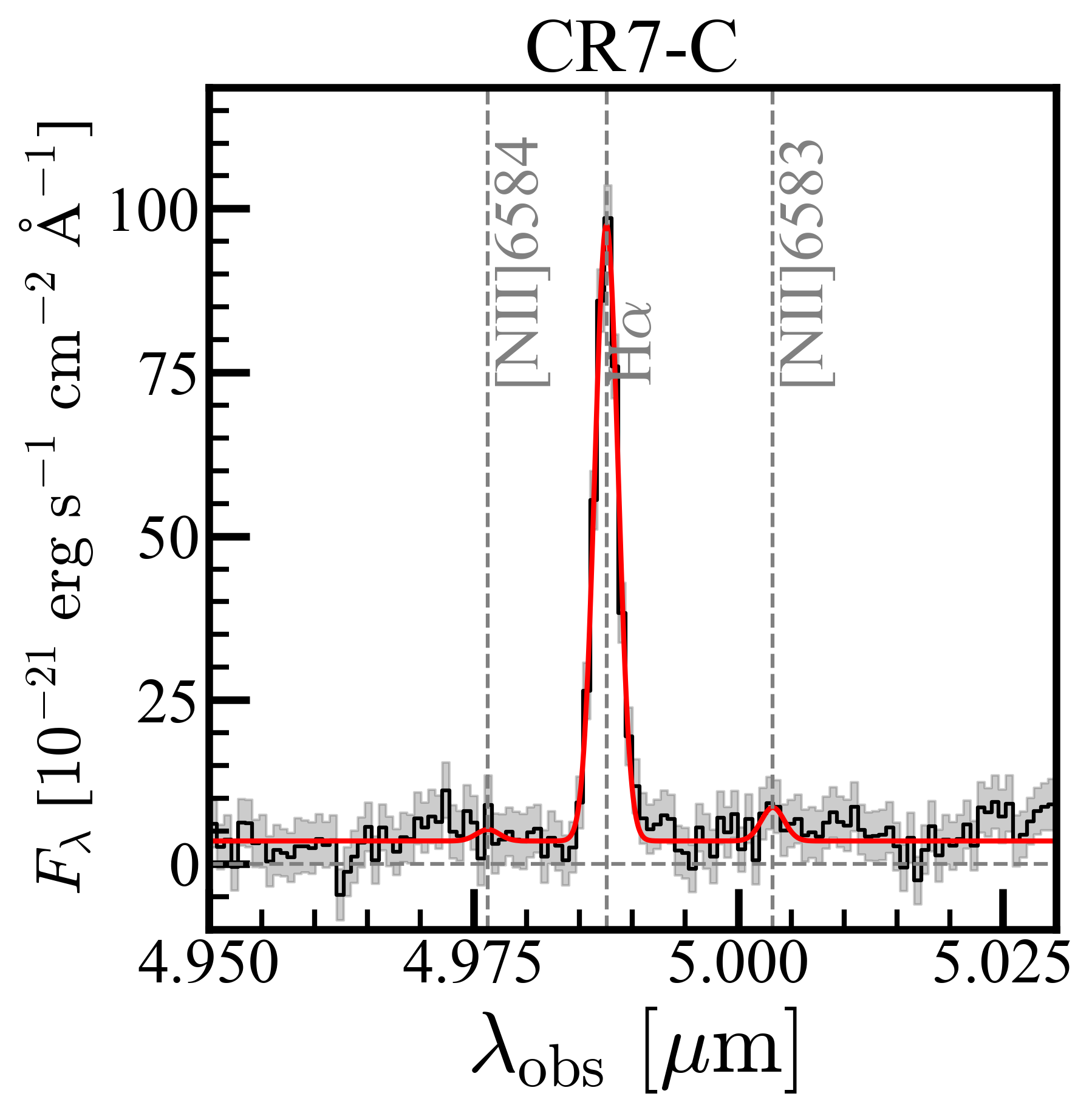}{0.28\textwidth}{}
          }
\caption{
H$\alpha$ emission lines of Himiko (top) and CR7 (bottom), observed with NIRSpec IFU G395H. 
Clumps A to C are displayed from left to right. 
The black histograms represent the spectra, with gray shading indicating 1$\sigma$ uncertainties. 
The red lines show the total Gaussian fit, while the gray lines represent individual Gaussian components. 
For Himiko-B, the broad (FWHM$\sim1000~\mathrm{km~s^{-1}}$) H$\alpha$ component is highlighted with the red shading (see Section~\ref{subsec:AGN}). 
\label{fig:Ha}}
\end{figure*}

Figure~\ref{fig:Ha} shows the spectrum of H$\alpha$ emission from Himiko-A, -B, -C, and CR7-A, -B, -C. 
Thanks to the high-resolution data from NIRSpec IFU, we can investigate the multiple components of the H$\alpha$ emission lines (see Section~\ref{sec:analysis}). 
The right panel of Figure~\ref{fig:ifu_oiii} also shows the \oiii\ emission lines from the individual clumps. 
The \oiii$\lambda$5007 emission lines have the highest S/N, allowing us to detect both narrow and broad components. 
Especially, Himiko-C and CR7-A have broad components of $\mathrm{FWHM=250}$, and $400~\mathrm{km~s^{-1}}$ with $\Delta\mathrm{AIC}>20$, respectively. 
The broad components in the \oiii\ emission lines may originate from outflows or tidal features. 
We discuss the connection between this outflow or tidal features with the origin of the extended \lya\ emission in Section~\ref{subsec:extended-Lya}. 
Additionally, we conduct the Gaussian fitting to H$\alpha$ emission lines independently of \oiii\ emission lines. 
For H$\alpha$ emission lines, there are also several components in one emission line. 
Himiko-B shows a distorted H$\alpha$ emission line profile, which is best characterized by two narrow components ($\mathrm{FWHM}\sim100~\mathrm{km~s^{-1}}$) and one broad component. 
The two narrow components have similar redshifts compared to their \oiii\ emission line components. 
The two narrow components remain even when using smaller apertures, suggesting that they may originate from gas clouds in front or behind Himiko-B. 
H$\alpha$ emission line in Himiko-B also has a broad component with FWHM$=1030\pm110~\mathrm{km~s^{-1}}$, which cannot be observed in the \oiii\ emission line. 
This broad permitted line (H$\alpha$) and the narrow forbidden line (\oiii) can indicate the possibility of active galactic nuclei (AGN). 
We further characterize the AGN signature in Section~\ref{subsec:AGN}. 
We note that no significant broad H$\beta$ component is detected from Himiko-B, possibly due to the lower S/N than H$\alpha$. 
Similarly, Himiko-C has a broad component in the H$\alpha$ emission line with $\mathrm{FWHM}=840\pm140\mathrm{~km~s^{-1}}$, which could be attributed to either stellar outflows or AGN activity. 
Since a threshold value of $\mathrm{FWHM=1000~km~s^{-1}}$ is often adopted to identify AGNs (e.g., \citealt{harikane23b, matthee24}), the nature of this broad component remains uncertain. 
We do not further investigate the broad H$\alpha$ component of Himiko-C in this study. 

\begin{figure*}
\gridline{\fig{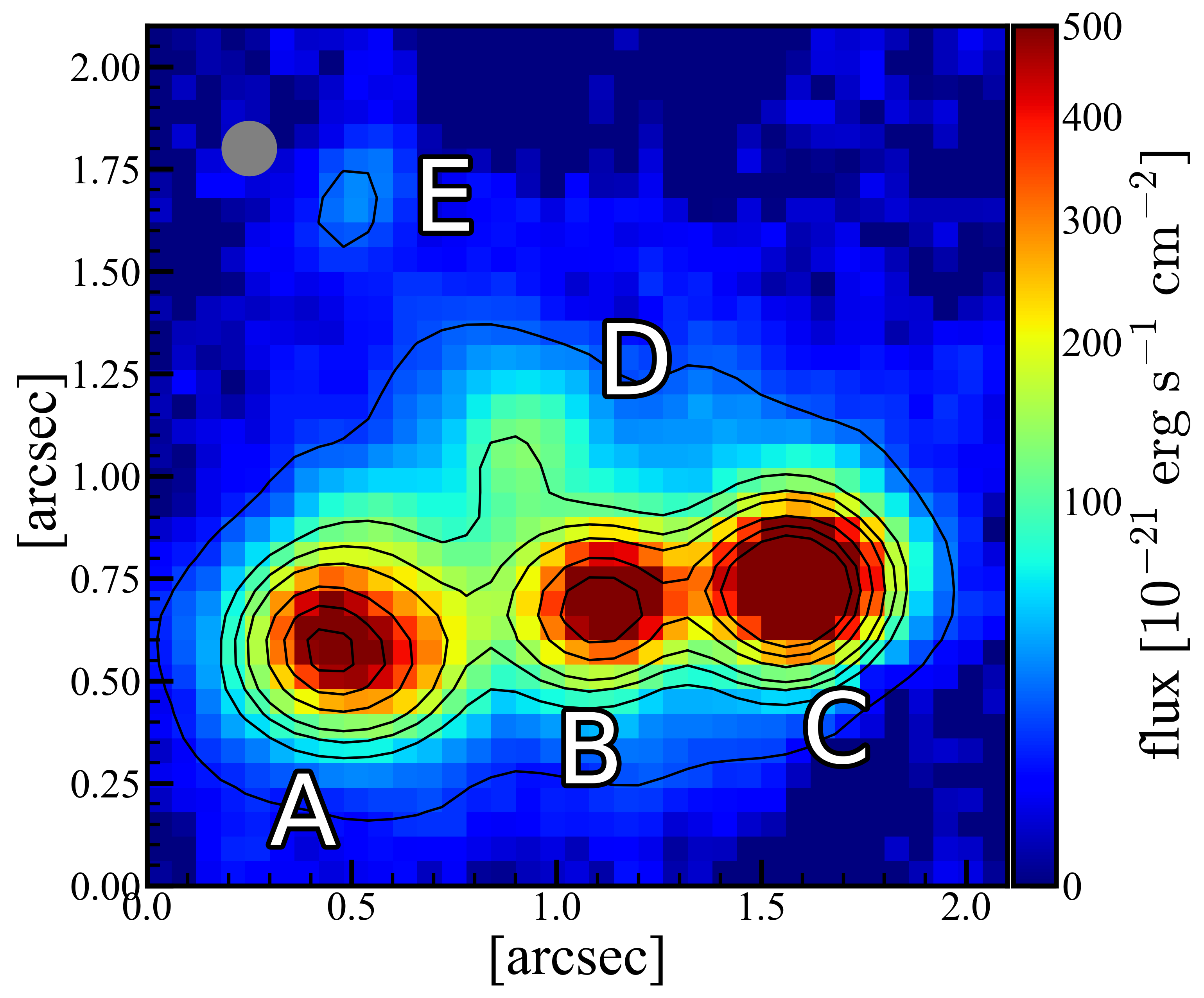}{0.30\textwidth}{}
          \fig{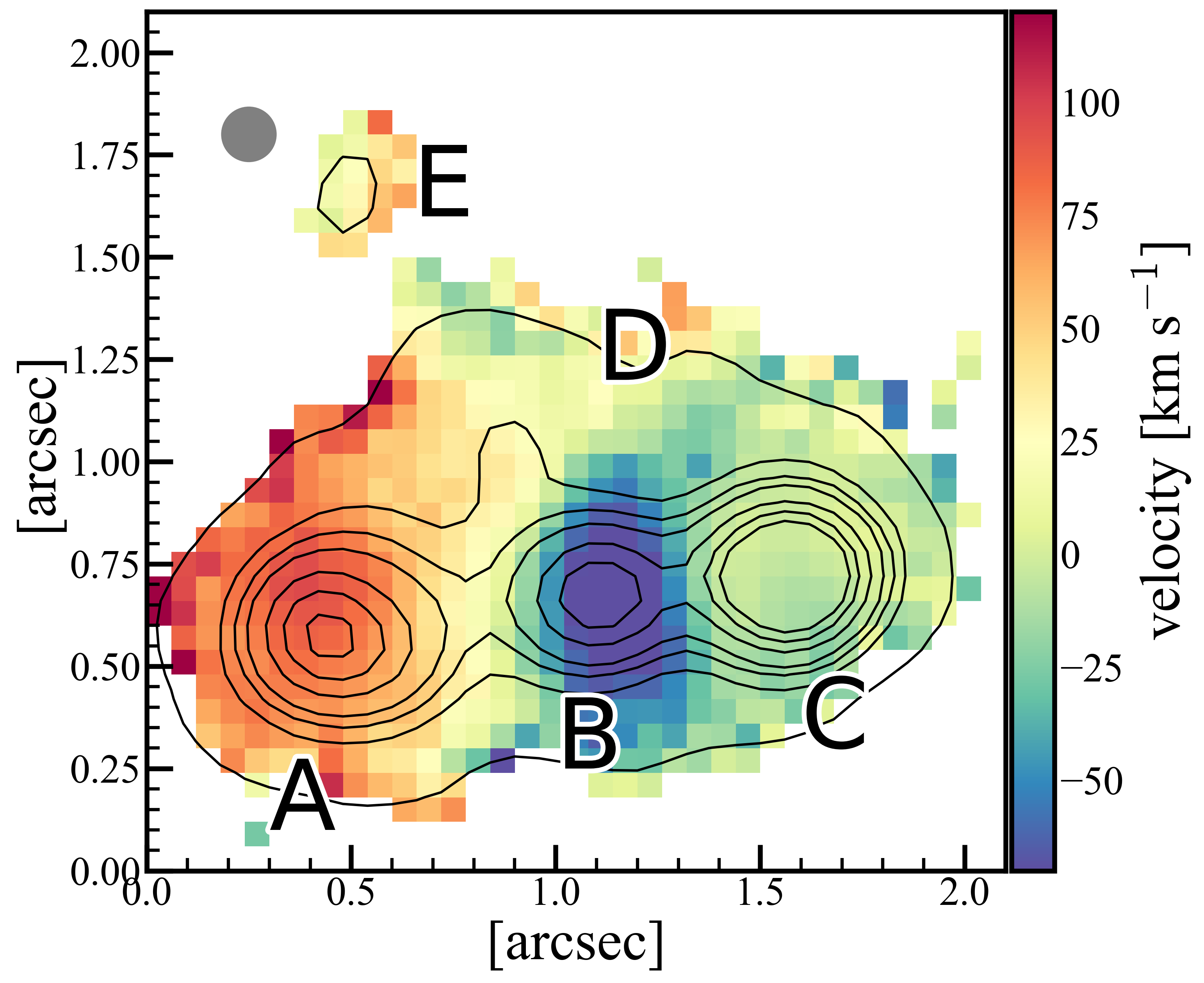}{0.30\textwidth}{}
          \fig{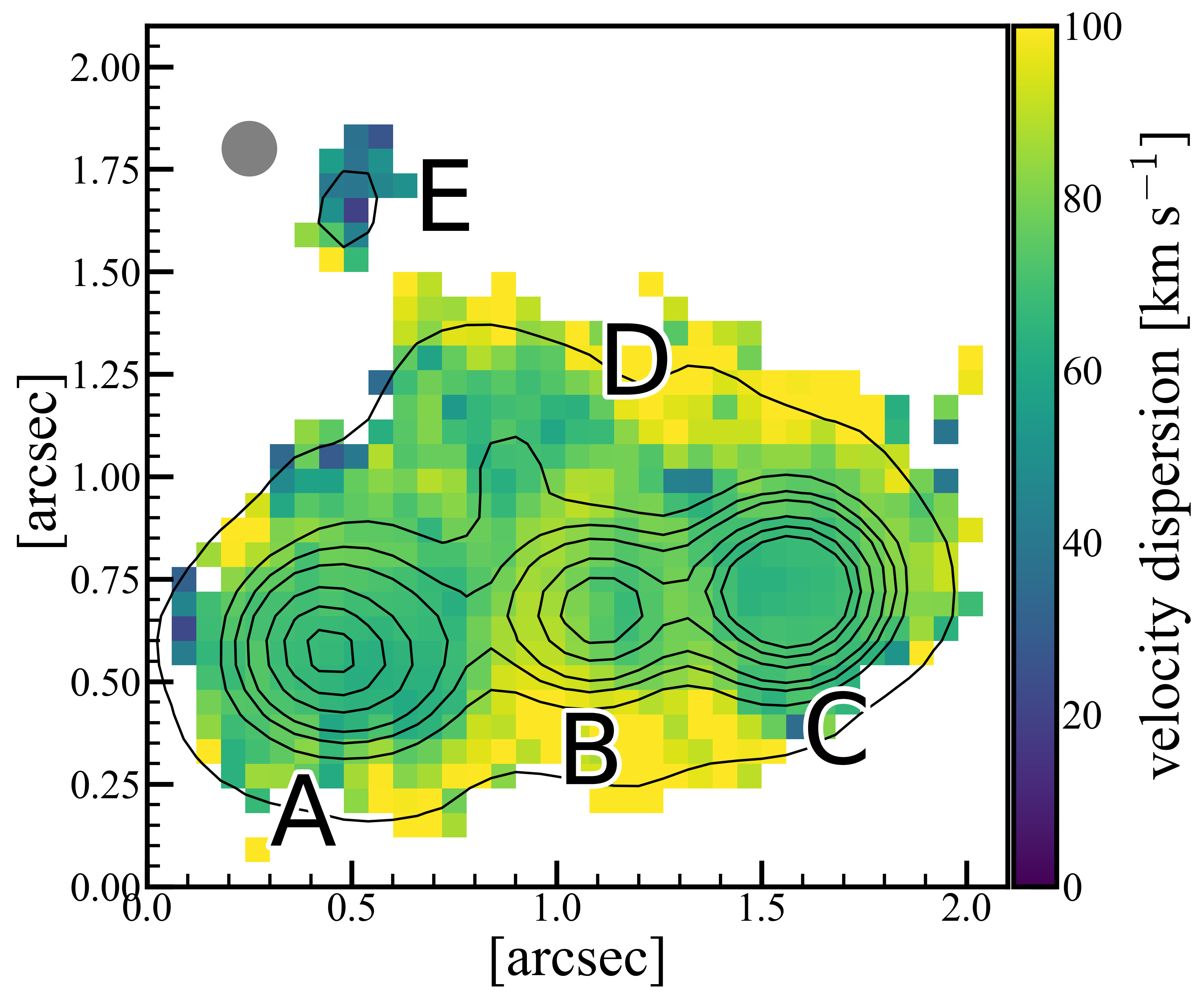}{0.30\textwidth}{}
          }
\gridline{\fig{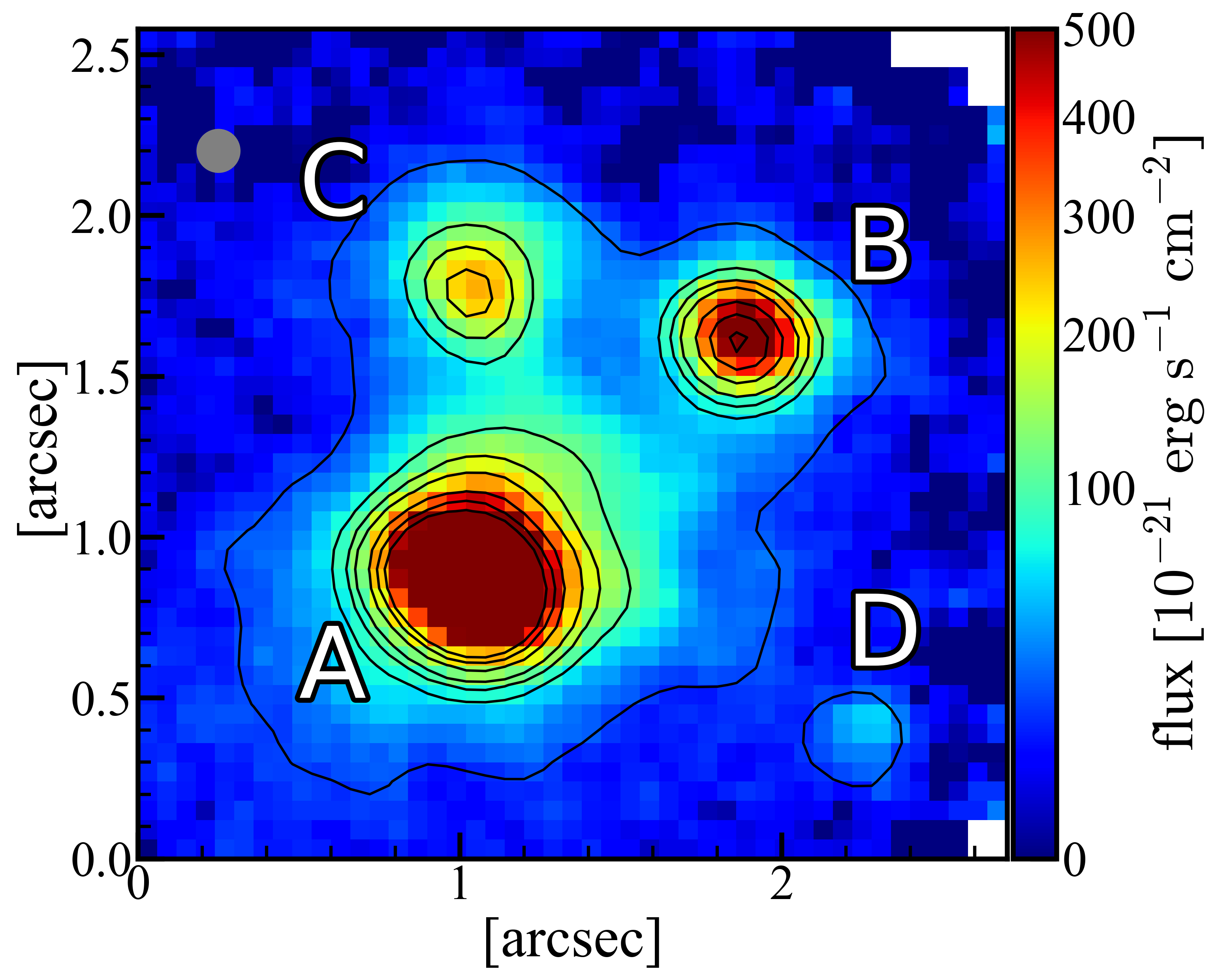}{0.30\textwidth}{}
          \fig{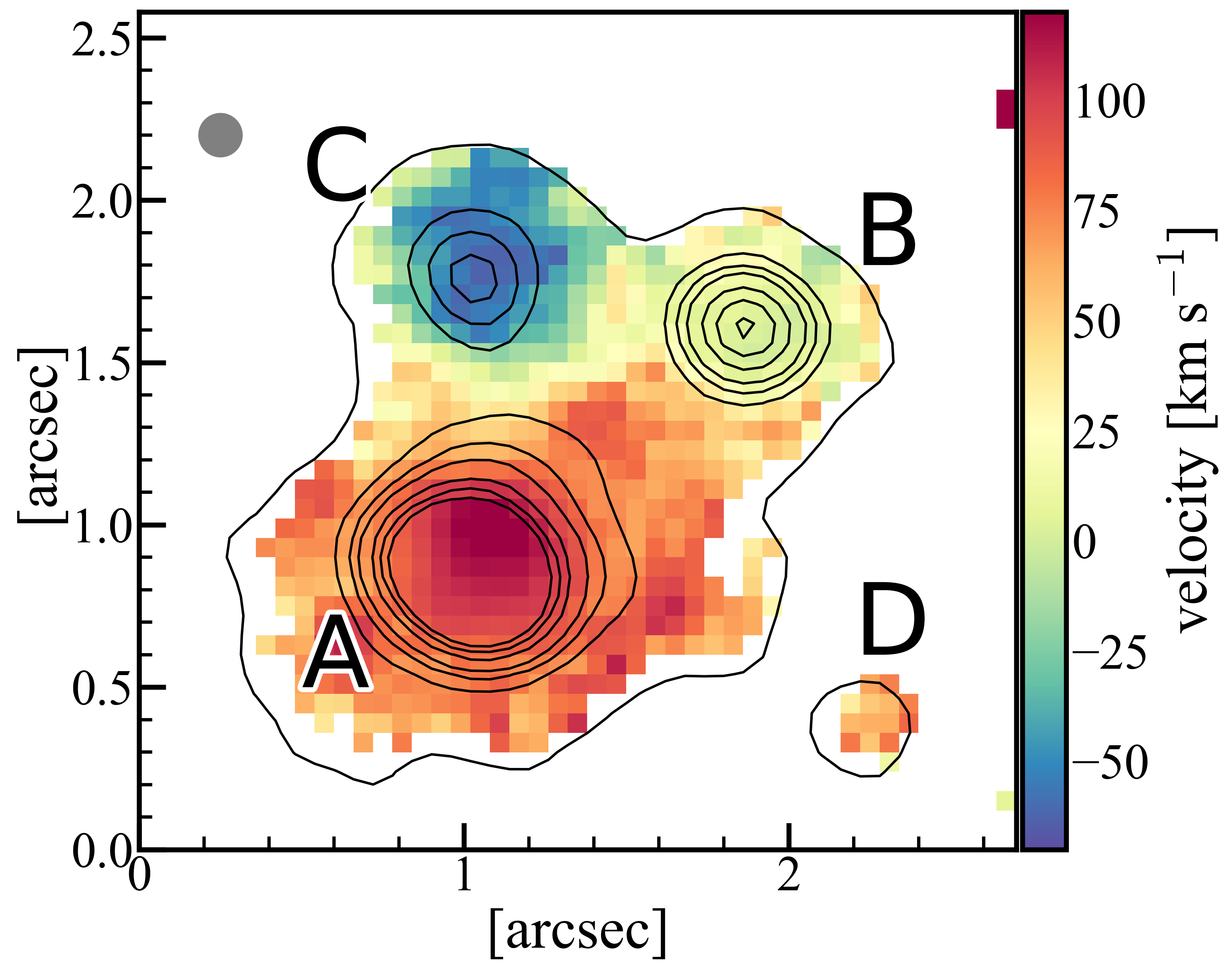}{0.30\textwidth}{}
          \fig{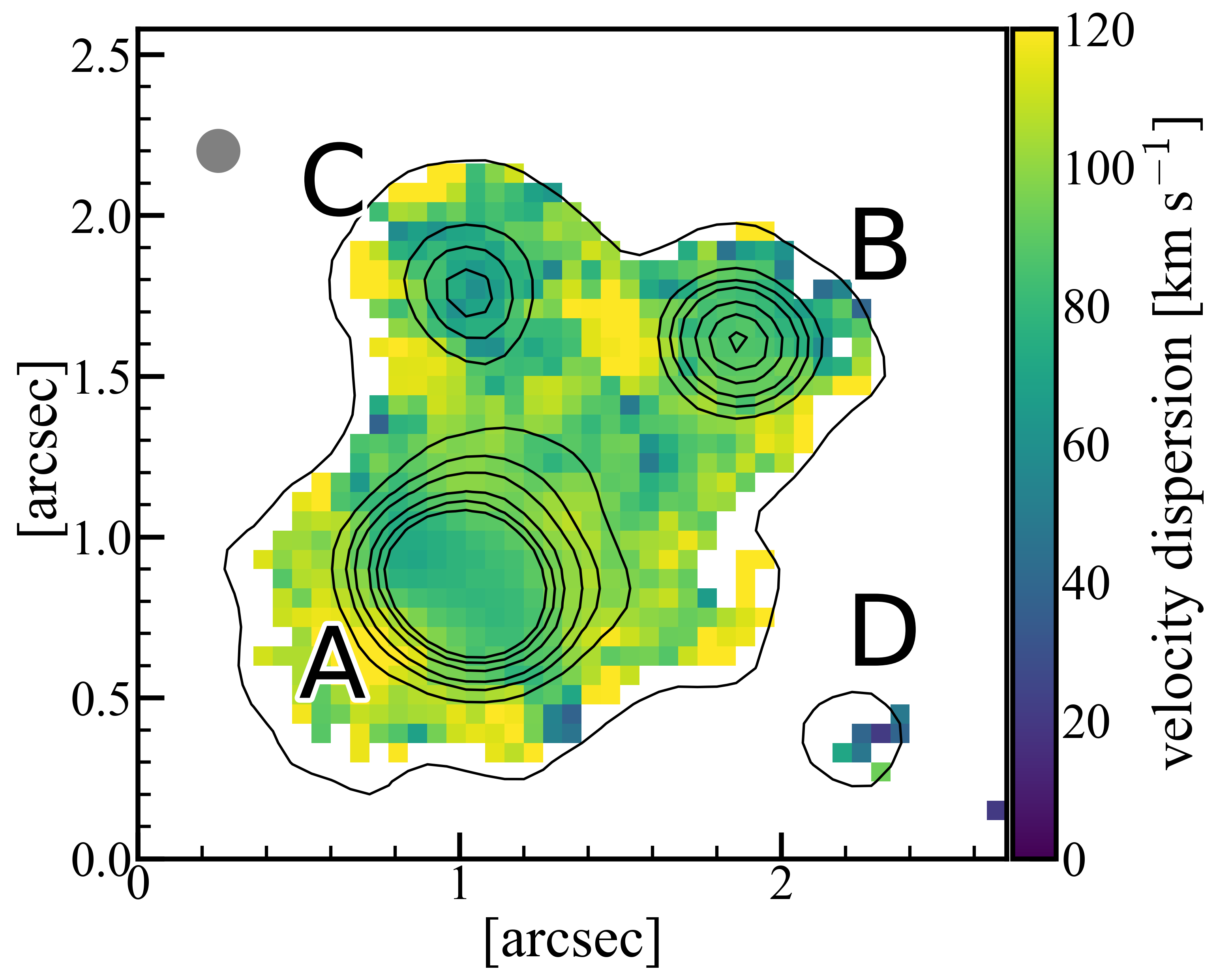}{0.30\textwidth}{}
          }
\caption{
\oiii$\lambda5007$ moment maps of Himiko (top) and CR7 (bottom) obtained by NIRSpec IFU G395H. 
Moment 0, 1, and 2 maps are shown from left to right. 
A fiducial velocity corresponds to the redshift of Himiko-C or CR7-B. 
We only use the spaxels of $\mathrm{S/N>5}$. 
Velocity dispersions are corrected for instrumental broadening. 
No dust extinction correction is applied. 
Black contours show arbitrary \oiii$\lambda5007$ flux levels. 
The upper left gray circle shows the PSF FWHM around the \oiii$\lambda5007$ emission line wavelength. 
\label{fig:moment-maps}}
\end{figure*}

We also investigate moment maps to characterize the dynamics of these objects, which cannot be achieved from the spatially integrated spectra. 
In Figure~\ref{fig:moment-maps}, we show the [O{\sc iii}]$\lambda$5007 moment 0 (left), 1 (middle), 2 (right) maps of Himiko (top), and CR7 (bottom). 
In Figure~\ref{fig:moment-maps}, we only use the spaxels with $\mathrm{S/N}>5$ to calculate the moment maps. 
We find that the velocity (moment 1) maps of both Himiko and CR7 show complex kinematics with a velocity offset of $\Delta v \lesssim 220~\mathrm{km~s^{-1}}$, indicating that each clump has close separations between other clumps. 
In the velocity maps, neither Himiko nor CR7 shows an entire galaxy-scale, disk-like velocity gradient. This disfavors a single rotating disk and motivates a merger scenario for the origin of the clumps discussed in Sections~\ref{subsec:multiple-clumps} and \ref{subsec:jwst-clump-comparison}.
CR7-A shows a velocity gradient from north to south, which might be a signature of a rotating disk (see also \citealt{marconcini24b}). In this study, a detailed investigation of this potential rotation signature is beyond the scope, and we do not further explore it. 
\citet{matthee17} have reported the velocity map of CR7 utilizing ALMA \cii$158\micron$ emission line. 
The \cii\ velocity map has a similar velocity kinematics as that of \oiii$\lambda5007$, indicating that the \cii\ emission aligns well with the ionized gas distribution represented by \oiii\ emission. 

\begin{figure*}
\gridline{\fig{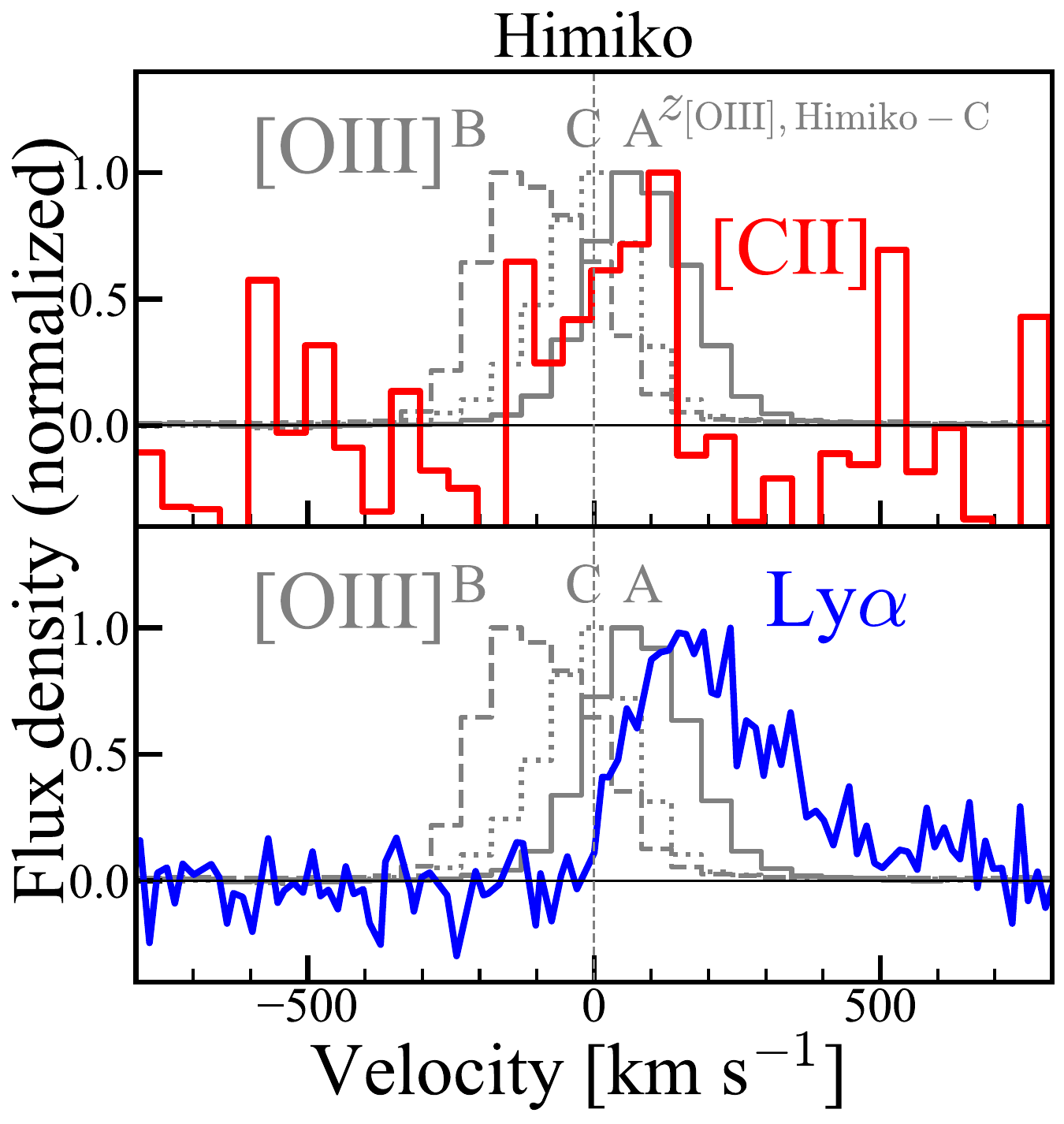}{0.4\textwidth}{}
          \fig{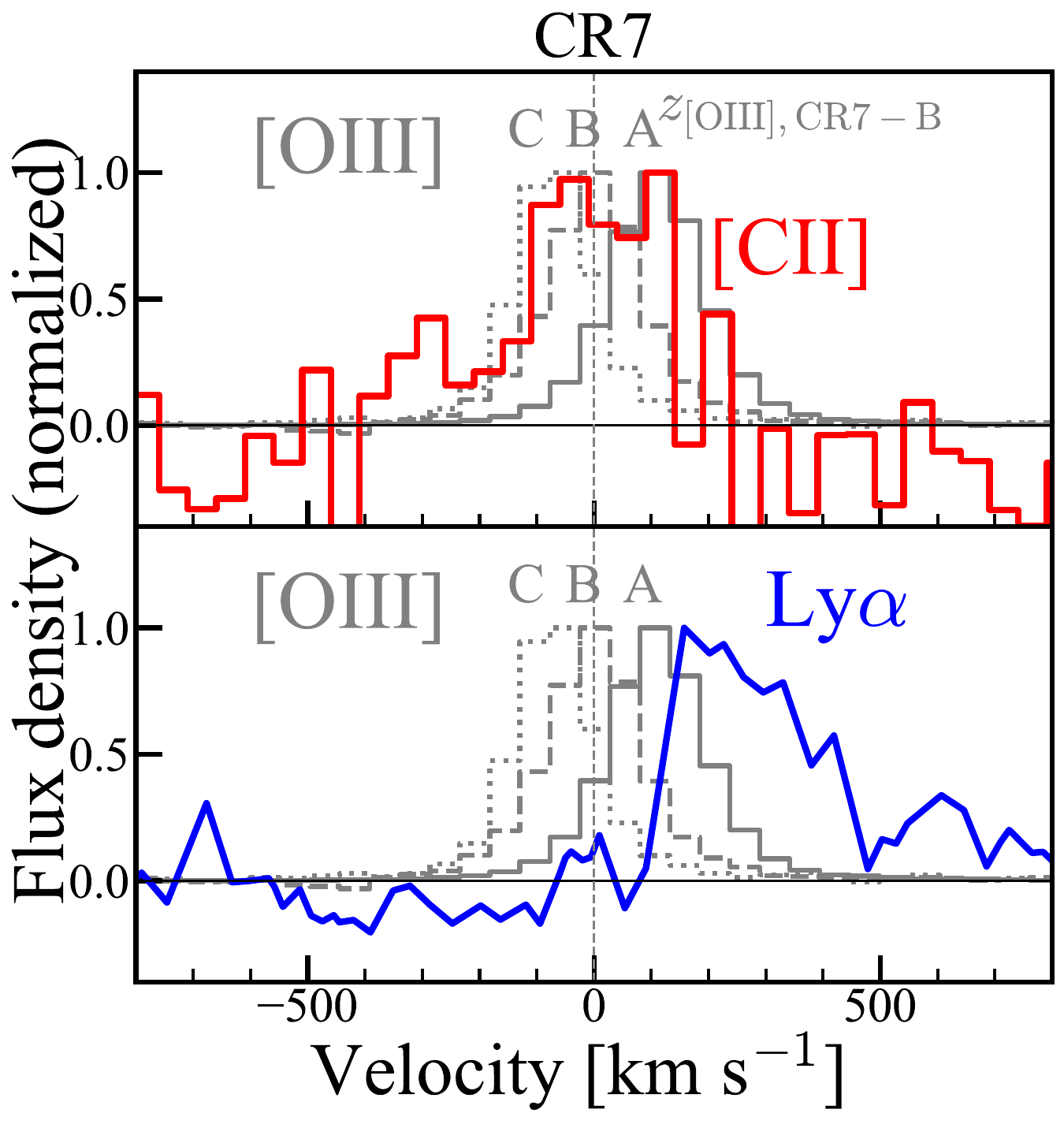}{0.4\textwidth}{}
          }
\caption{
\oiii$\lambda5007$, \cii 158$\micron$, and \lya\ emission lines of Himiko (left) and CR7 (right). 
\oiii$\lambda5007$ lines from G395H spectra are shown in gray, and clump IDs are added in the figure. 
\cii 158$\micron$ and \lya\ lines are presented in red and blue, respectively. 
\cii 158$\micron$ lines in this figure are integrated over the regions where $\mathrm{S/N}>3$. 
\lya\ spectra of Himiko and CR7 are taken from Keck/DEIMOS \citep{ouchi09} and VLT/X-SHOOTER \citep{sobral15}, respectively. 
\label{fig:spectrum-jwst-alma}}
\end{figure*}

Figure~\ref{fig:spectrum-jwst-alma} shows the comparison between the \oiii$\lambda5007$ (gray), \cii$158\micron$ (red), and Ly$\alpha$ (blue; \citealt{ouchi09, sobral15}) spectra of Himiko (left) and CR7 (right).
The \cii\ emission line of Himiko is mainly aligned with the \oiii$\lambda5007$ emission line of Himiko-A.
For CR7, the \cii\ emission is located within the \oiii$\lambda5007$ emission from CR7-A to CR7-C, as indicated by the \oiii\ and \cii\ velocity maps (see also \citealt{matthee17}). 
The peaks of the Ly$\alpha$ emission from both Himiko and CR7 are redshifted relative to the \oiii\ emission lines.
However, measuring the velocity offset of the \lya\ emission is challenging because the \lya\ emission from Himiko and CR7 is extended across multiple clumps.

\subsection{Spectral energy distribution} \label{subsec:sed} 

\begin{deluxetable*}{lcccccccc}
    \tablecaption{NIRCam photometry measurements of Himiko and CR7. \label{tab:NIRCam-phot}}
    \tablewidth{0pt}
    \tablehead{
    \colhead{Object} & \colhead{F090W} & \colhead{F115W} & \colhead{F150W} & \colhead{F200W} & \colhead{F277W} & \colhead{F356W} & \colhead{F410M} & \colhead{F444W} \\
    \colhead{} & \colhead{($\mu\mathrm{Jy}$)} & \colhead{($\mu\mathrm{Jy}$)} & \colhead{($\mu\mathrm{Jy}$)} & \colhead{($\mu\mathrm{Jy}$)} & \colhead{($\mu\mathrm{Jy}$)} & \colhead{($\mu\mathrm{Jy}$)} & \colhead{($\mu\mathrm{Jy}$)} & \colhead{($\mu\mathrm{Jy}$)}
    }
    \startdata
    Himiko-A & $0.048\pm0.007$ & $0.127\pm0.007$ & $0.101\pm0.006$ & $0.092\pm0.005$ & $0.096\pm0.003$ & $0.183\pm0.003$ & $0.066\pm0.006$ & $0.098\pm0.004$ \\
    Himiko-B & $0.025\pm0.007$ & $0.065\pm0.007$ & $0.064\pm0.006$ & $0.075\pm0.005$ & $0.070\pm0.003$ & $0.178\pm0.003$ & $0.071\pm0.006$ & $0.109\pm0.004$ \\
    Himiko-C & $0.041\pm0.007$ & $0.117\pm0.007$ & $0.117\pm0.006$ & $0.112\pm0.005$ & $0.122\pm0.003$ & $0.262\pm0.003$ & $0.078\pm0.006$ & $0.120\pm0.004$ \\
    \hline
    CR7-A & \nodata   & $0.242\pm0.013$ & $0.246\pm0.010$ & \nodata   & $0.282\pm0.006$ & \nodata   & \nodata   & $0.238\pm0.006$ \\
    CR7-B & \nodata   & $0.046\pm0.010$ & $0.048\pm0.008$ & \nodata   & $0.048\pm0.005$ & \nodata   & \nodata   & $0.056\pm0.005$ \\
    CR7-C & \nodata   & $0.051\pm0.010$ & $0.043\pm0.008$ & \nodata   & $0.040\pm0.005$ & \nodata   & \nodata   & $0.029\pm0.005$ \\
    \enddata
    \tablecomments{All Himiko clumps, CR7-B, and CR7-C are extracted with $0\farcs4$-diameter apertures, while $0\farcs5$-diameter aperture is used for CR7-A.}
\end{deluxetable*}

\begin{figure*}
\gridline{\fig{SED_plot_summary_Himiko_v2.pdf}{0.4\textwidth}{}
          \fig{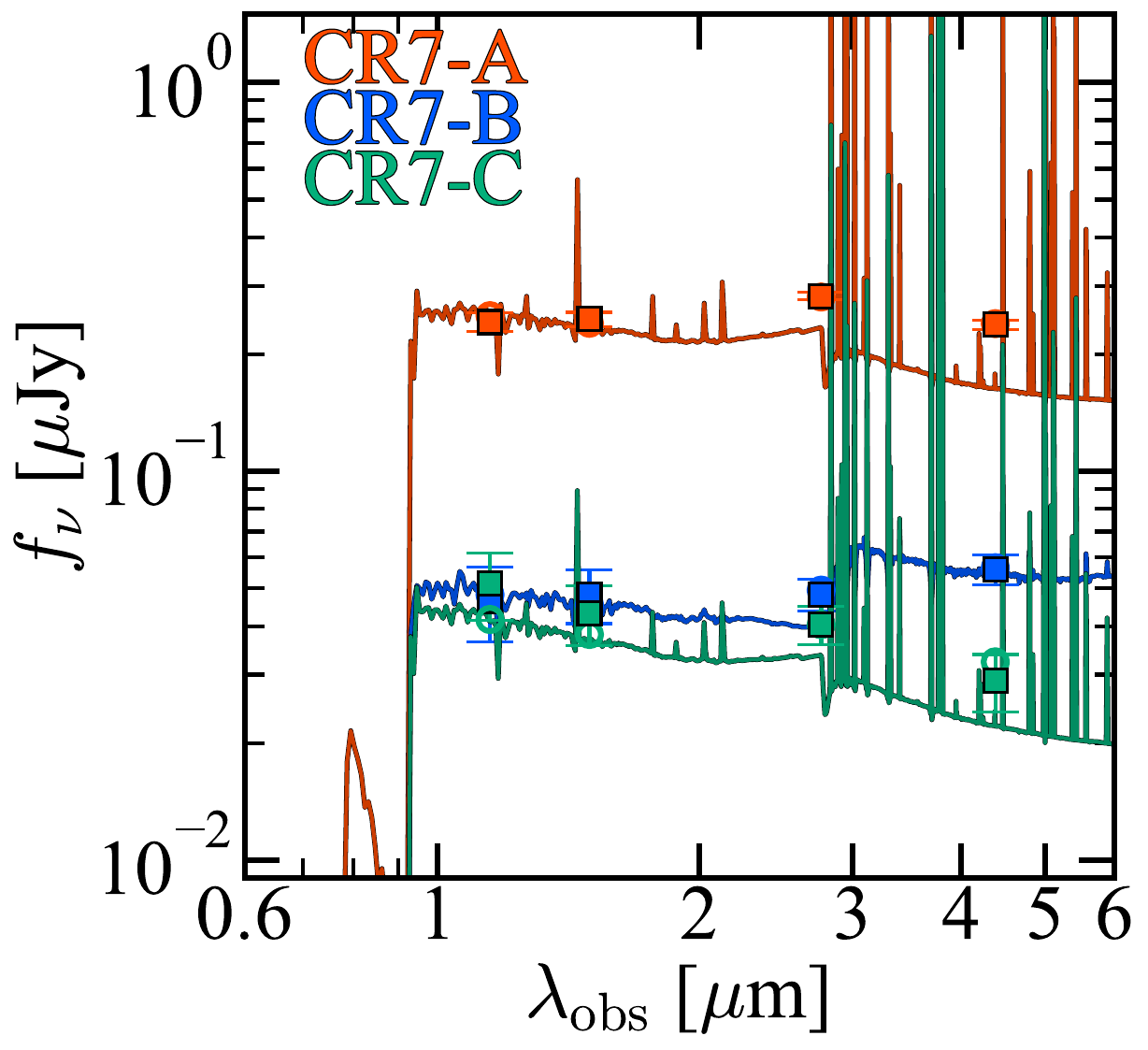}{0.4\textwidth}{}
          }
\caption{SED of Himiko (left) and CR7 (right). 
Orange, blue, and green colors represent the clumps A, B, and C, respectively. 
The solid lines show the best-fit SEDs. 
The squares and error bars show the observed fluxes and their 1$\sigma$ uncertainties, respectively. 
The circles show the best-fit photometry. 
We do not use the F090W fluxes in the SED fitting due to the contamination of Ly$\alpha$ emission and IGM absorption. 
\label{fig:sed-fitting}}
\end{figure*}

\begin{figure}
\plotone{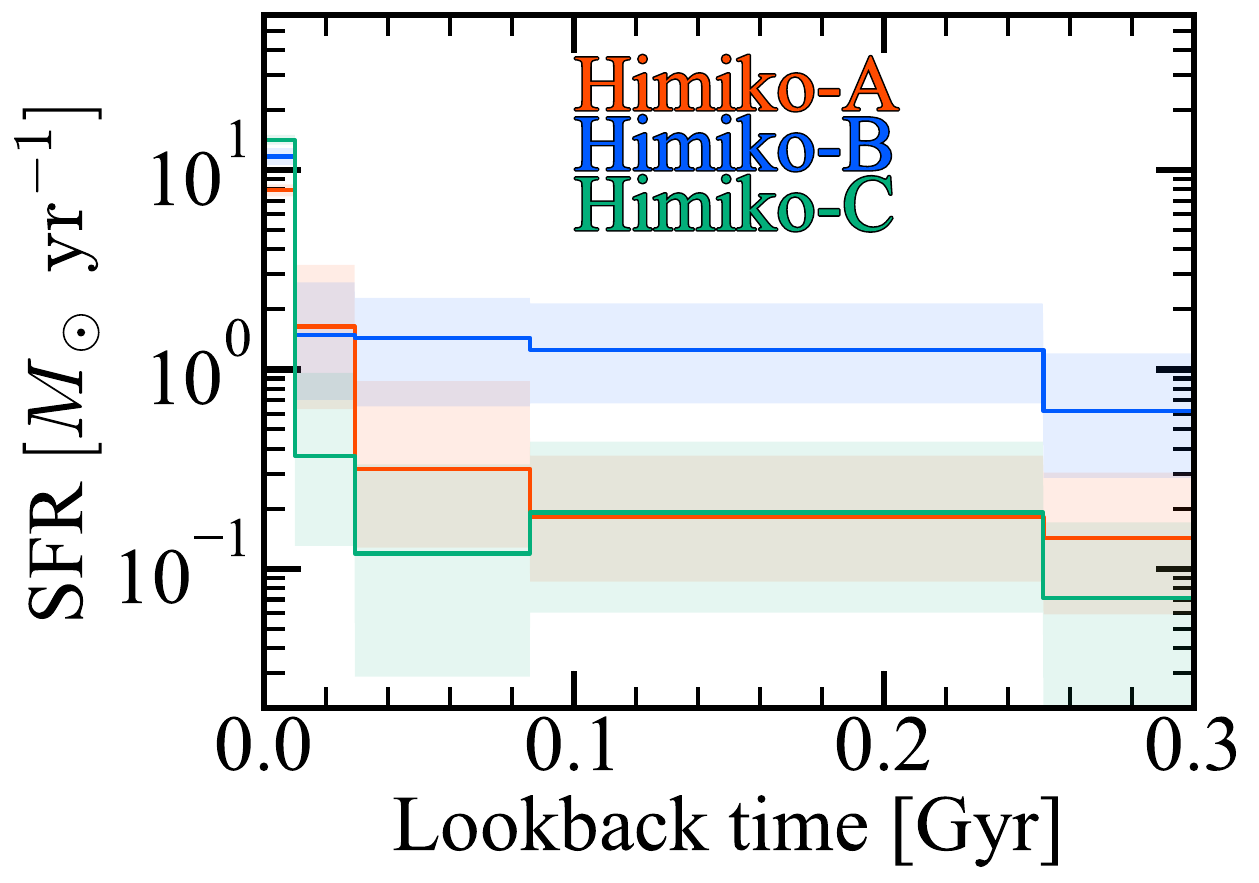}
\plotone{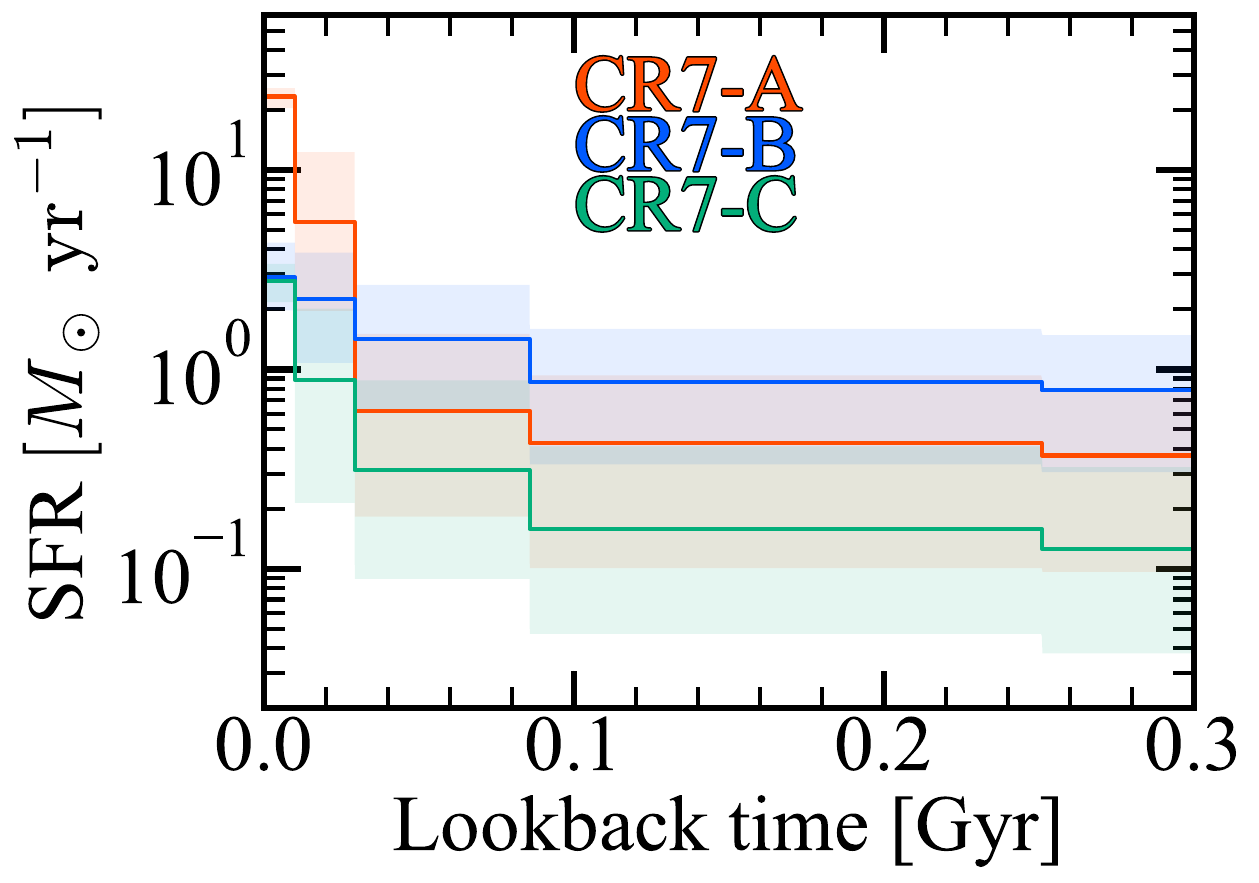}
\caption{Best-fit SFHs of Himiko (top) and CR7 (bottom) inferred from the SED fitting with NIRCam photometry. The orange, blue, and green lines show the best-fit SFHs of Clumps A, B, and C, respectively. The shaded regions show the corresponding 1$\sigma$ uncertainties.
\label{fig:sfh}}
\end{figure}

Table~\ref{tab:NIRCam-phot} summarizes the NIRCam photometry with $1\sigma$ uncertainties for Himiko and CR7, respectively. 
SED fittings are performed only for clumps detected in multiple filters with a significance greater than $5\sigma$ (specifically, Himiko-A, -B, and -C; CR7-A, -B, and -C). 
The parameter settings are described in Section~\ref{subsec:sed_modeling}. 
Table~\ref{tab:SED-fitting} shows the physical properties of Himiko and CR7 as inferred from the SED fitting. 
Figure~\ref{fig:sed-fitting} displays the best-fit SEDs for these objects. 
For Himiko, Himiko-A and -C have a similar spectral shape, while Himiko-B has a redder spectrum with $A_V=0.48^{+0.06}_{-0.06}~\mathrm{mag}$ compared to the other two clumps ($A_V\sim0.03$--$0.21$ mag). 
The stellar masses of individual Himiko clumps range $\log{(M_*/M_\odot)}=8.4$--$8.9$, with a mean stellar mass of $\log{(M_*/M_\odot)}=8.6$. 
This result is broadly consistent with that of \citet{harikane25}. 
In the case of CR7, CR7-A is the brightest clump and has the highest star formation rate (SFR) measured at $23.4^{+2.4}_{-3.0}~M_\odot~\mathrm{yr^{-1}}$, which is averaged over the past 10~Myr. 
This rate is approximately ten times higher than that of the other two clumps, indicating that CR7-A mainly dominates star formation in CR7. 
CR7-A and CR7-B have comparable stellar masses, each with $\log{(M_*/M_\odot)}=8.8$--$8.9$, while CR7-C has a lower stellar mass of $\log{(M_*/M_\odot)}=8.3^{+0.3}_{-0.3}$. 
These results align well with the stellar mass estimates of CR7 derived from NIRSpec IFU spectra by \citet{marconcini24b} within 2$\sigma$. 
Individual Himiko and CR7 clumps primarily lie on the SFR--$M_*$ main sequence at $z\sim6$ (e.g., \citealt{santini17}). 
Figure~\ref{fig:sfh} shows the inferred SFHs from the SED fitting. 
The SFHs of all the clumps show rising trends (starbursts) toward the 0--10 Myr age bins, although the uncertainties in the SFHs are large at $>10$ Myr (see also \citealt{harikane25, marconcini24b}). 
These increasing trends in SFHs could be associated with galaxy mergers (e.g., \citealt{hopkins08, asada24}). 
We further discuss the implications of galaxy mergers and their scenarios in Section \ref{sec:discussion}. 

\begin{deluxetable*}{lcccccc}
    \tablecaption{Physical properties of Himiko and CR7 (SED fitting and direct method). \label{tab:SED-fitting}}
    \tablewidth{0pt}
    \tablehead{
    \colhead{} & \multicolumn{3}{c}{SED fitting} & \multicolumn{2}{c}{direct method} \\
    \cline{2-4}
    \cline{5-6}
    \colhead{Object} & \colhead{$\log{(M_*)}$} & \colhead{SFR} & \colhead{$A_V$} & \colhead{$T_e$(\oiii)} & \colhead{$12+\log{(\mathrm{O/H})}$} & \colhead{reference}\\
    \colhead{} & \colhead{($M_\odot$)} & \colhead{($M_\odot~\mathrm{yr}^{-1}$)} & \colhead{(mag)} & \colhead{($10^4\,\mathrm{K}$)}
    }
    \decimalcolnumbers
    \startdata
    Himiko-A & $8.4^{+0.2}_{-0.1}$ & $7.9^{+0.6}_{-0.6}$ & $0.03^{+0.03}_{-0.02}$ & $1.6\pm0.3$ & $7.92\pm0.19$ & This work \\
    Himiko-B & $8.9^{+0.1}_{-0.2}$ & $11.7^{+1.2}_{-1.0}$ & $0.48^{+0.06}_{-0.06}$ & $1.6\pm0.3$ & $7.90\pm0.20$ & This work \\
    Himiko-C & $8.4^{+0.1}_{-0.1}$ & $14.1^{+0.9}_{-0.7}$ & $0.21^{+0.04}_{-0.04}$ & $1.4\pm0.2$ & $8.02\pm0.14$ & This work \\
    \hline
    CR7-A & $8.8^{+0.1}_{-0.2}$ & $23.4^{+2.4}_{-3.0}$ & $0.13^{+0.04}_{-0.05}$ & $1.6\pm0.1$ & $7.90\pm0.06$ & This work \\
          & $9.3^{+0.1}_{-0.1}$ & $25.7^{+1.2}_{-1.1}$ & $0.20^{+0.01}_{-0.01}$ & $1.6$ & $8.0$ & \citet{marconcini24b} \\
    CR7-B & $8.9^{+0.2}_{-0.2}$ & $2.9^{+1.4}_{-0.9}$ & $0.13^{+0.15}_{-0.09}$ & $1.3\pm0.2$ & $8.11\pm0.20$ & This work \\
          & $8.5^{+0.2}_{-0.2}$ & $5.9^{+0.5}_{-0.4}$ & $0.14^{+0.05}_{-0.03}$ & $0.9$ & $8.08$ & \citet{marconcini24b} \\
    CR7-C & $8.3^{+0.3}_{-0.3}$ & $2.8^{+0.6}_{-0.6}$ & $0.05^{+0.07}_{-0.04}$ & \nodata & \nodata & This work \\
          & $8.3^{+0.3}_{-0.1}$ & $3.8^{+0.7}_{-0.7}$ & $0.08^{+0.04}_{-0.03}$ & \nodata & \nodata & \citet{marconcini24b} \\
    \enddata
    \tablecomments{
    Columns: 
    (1) object name; 
    (2) stellar mass; 
    (3) star formation rate averaged over the past 10 Myr; 
    (4) $V$-band magnitude of dust attenuation; 
    (5) electron temperature from the direct method; 
    (6) gas phase metallicity $12+\log{(\mathrm{O/H})}$ from the direct method. Since \citet{marconcini24b} have not reported the electron temperature of CR7-C, the electron temperature and the metallicity of CR7-C from \citet{marconcini24b} are not listed.;
    (7) references.
    }
\end{deluxetable*}

\begin{figure*}
\plotone{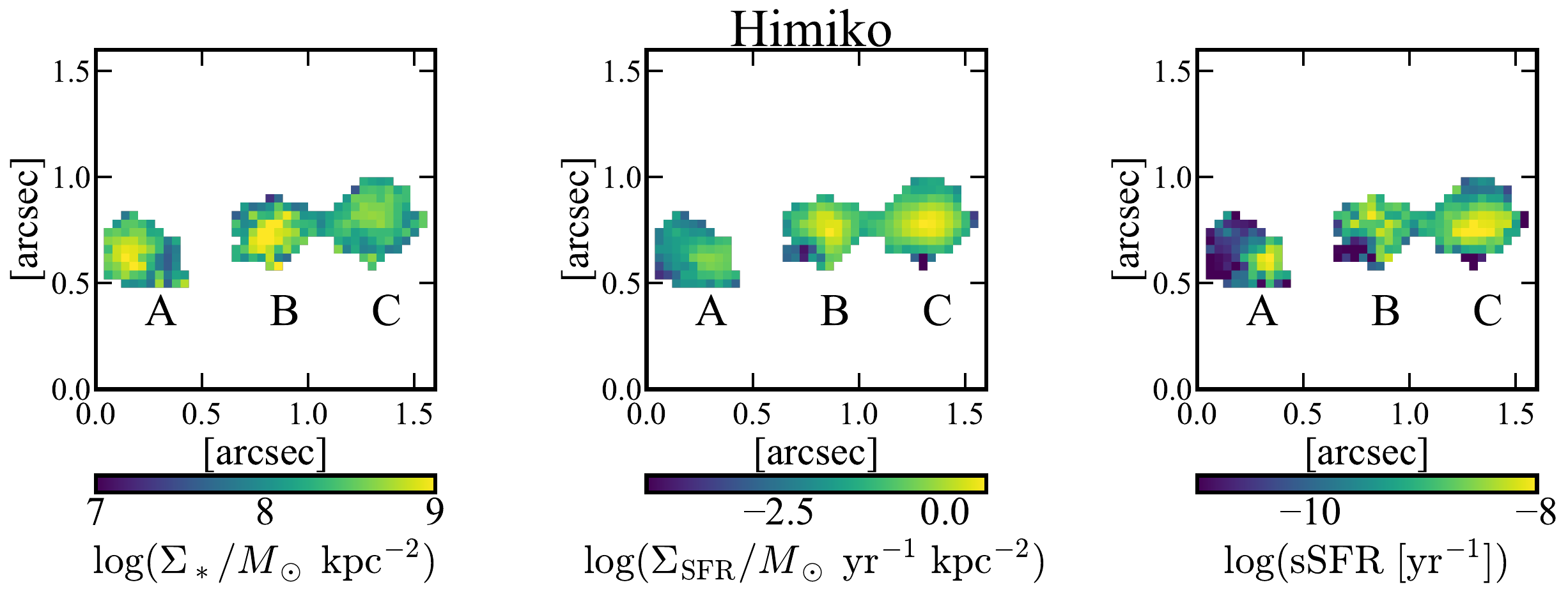}
\plotone{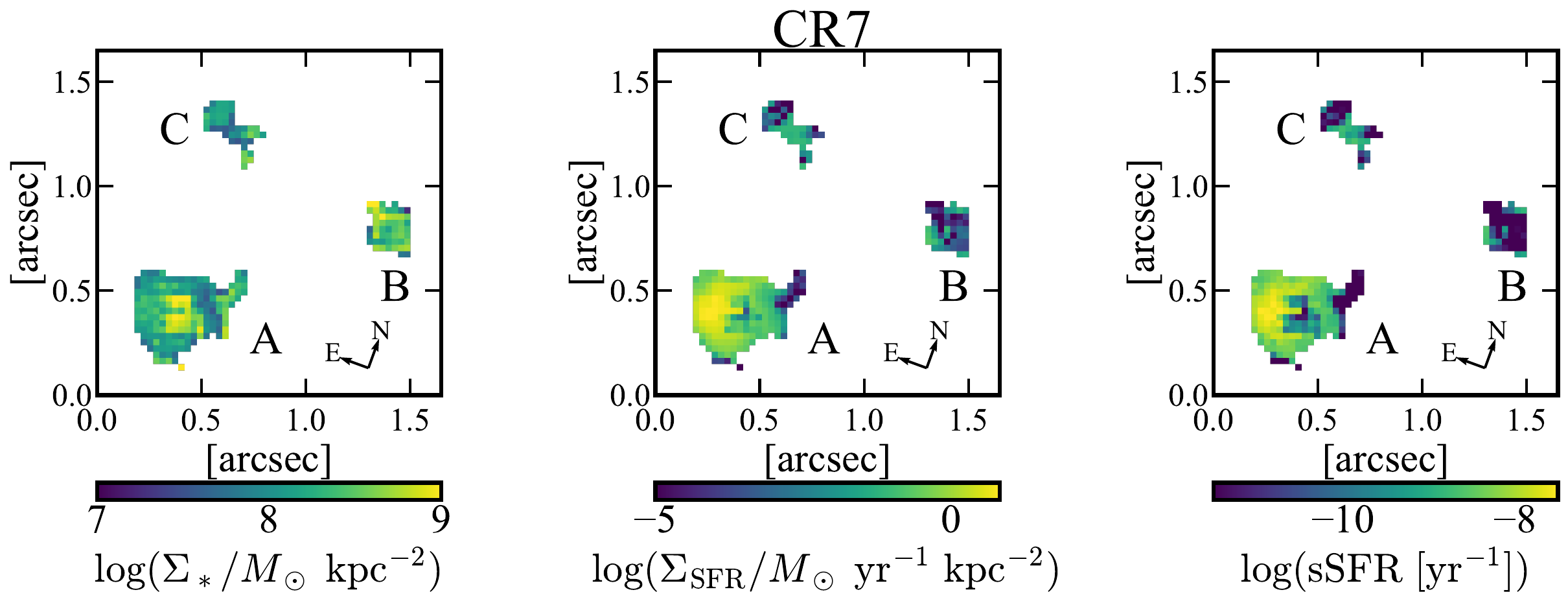}
\caption{
Results of the pixel-by-pixel SED fitting using NIRCam photometry. 
The top and bottom panels show Himiko and CR7, respectively. 
The left, middle, and right panels show the stellar mass density ($\Sigma_*$) in $M_\odot~\mathrm{kpc^{-2}}$, the star formation rate density ($\Sigma_\mathrm{SFR}$) in $M_\odot~\mathrm{yr^{-1}~kpc^{-2}}$, and the specific star formation rate (sSFR) in $\mathrm{yr^{-1}}$, respectively. 
\label{fig:pixel-by-pixel-sed}}
\end{figure*}

We also perform pixel-by-pixel SED fitting (e.g., \citealt{sorba15, sorba18, gimenez'-arteaga23}) to examine spatially resolved physical properties using the high-resolution NIRCam photometry.  
We use the NIRCam images whose PSFs are matched to F444W. 
Only pixels detected in the F444W segmentation map, constructed by SEP, are included in the fitting. 
The model settings for the pixel-by-pixel SED fitting are the same as those used for the aperture photometry, except for the stellar mass prior range, which is set to $6<\log{(M_*/M_\odot)}<10$. 
Figure~\ref{fig:pixel-by-pixel-sed} shows the stellar mass densities ($\Sigma_*$), the SFR densities ($\Sigma_\mathrm{SFR}$), and the specific star formation rate (sSFR) of Himiko (top) and CR7 (bottom) inferred from the pixel-by-pixel SED fitting. 

For Himiko, the central region of Himiko-B has high stellar mass density distributions ($\log{(\Sigma_*\,[M_\odot~\mathrm{kpc^{-2}}])}\sim 9$). 
SFR and sSFR distributions are high in the central regions of Himiko clumps ($\Sigma_{\mathrm{SFR}}\sim3~M_\odot~\mathrm{yr^{-1}~kpc^{-2}}$ at maximum). 
One interesting point is that the west side of Himiko-A has about one order of magnitude higher sSFR than the east. 
This region is near the peak position of the \lya\ emission indicated by the Subaru NB image (Figure~\ref{fig:jwst-alma-subaru-image}). 
This result could indicate that the region also contributes to the bright \lya\ emission of Himiko, although the gas distribution is also key to explaining the distribution of the \lya\ emission. 
For CR7, CR7-A has the highest stellar mass density ($\log{(\Sigma_*\,[M_\odot~\mathrm{kpc^{-2}}])}\sim 8.5$) and SFR surface density ($\Sigma_{\mathrm{SFR}}\sim5~M_\odot~\mathrm{yr^{-1}~kpc^{-2}}$), which is consistent with the SED fitting using the total photometry of CR7-A. 
\citet{marconcini24b} have conducted spatially resolved SED fittings of CR7 using NIRSpec IFU data, and their results are consistent with ours.

\subsection{Emission line diagnostics} \label{subsec:emission-line} 

We focus on the IFU data to characterize the ionizing conditions of Himiko and CR7. 
Using the Balmer lines (H$\alpha$ and H$\beta$), we assess dust extinction, assuming an electron density of $n_e = 100~\mathrm{cm^{-3}}$, an electron temperature of $T_e = 15000~\mathrm{K}$, and case B recombination \citep{osterbrock06}. We use the \citet{calzetti00} extinction law and obtain a nebular attenuation of $A_V = 0.00, 0.73, 0.02$ ($A_V = 0.01, 0.00, 0.02$) for Himiko-A, -B, and -C (CR7-A, -B, and -C), respectively. 
These values are consistent with the SED fitting results using NIRCam photometry (Section~\ref{subsec:sed}). 
These low values are also in line with the low dust content indicated by the non-detection of dust continuum through ALMA Band~6 observations (see also \citealt{matthee17, carniani18}). Further discussion of this low dust condition and its relation to merger systems are provided in Section~\ref{subsec:comparison}. 

We calculate the standard line ratios: \oiii$\lambda5007$/H$\beta$ (R3), 
(\oiii$\lambda\lambda5007, 4959$ + \oii$\lambda3727$)/H$\beta$ (R23), and \oiii$\lambda5007$/\oii$\lambda3727$ (O32). 
We also detect the \oiii$\lambda4363$ auroral line that enables us to measure an electron temperature and metallicity using the direct method. 
When calculating the line ratios and metallicity, we do not apply dust extinction corrections, as Himiko and CR7 have low dust contents, and the procedure does not change the overall results.
We estimate the electron temperature of O$^{2+}$ zone $T_e$(\oiii) using the \oiii$\lambda4363$/\oiii$\lambda5007$ ratio with the Pyneb \citep{luridiana15} package \texttt{getTemDen}. 
We assume an electron number density of $n_e=100~\mathrm{cm^{-3}}$ because we do not resolve the \oii$\lambda3726, 3729$ lines in the PRISM spectra ($R\sim100$). 
The O$^{+}$ zone temperature $T_e$(\oii) is estimated from $T_e$(\oiii) \citep{garnett92}. 
We obtain the abundance of O$^{2+}$/H$^+$ with the \oiii$\lambda4959, 5007$, H$\beta$, and $T_e$(\oiii), and O$^{+}$/H$^{+}$ with the \oii$\lambda3727$, H$\beta$ and $T_e$(\oii) using the Pyneb package \texttt{getIonAbundance}. 
We ignore a higher ionization abundance of O$^{3+}$/H$^{+}$ \citep{izotov06}. 

\begin{figure*}
    \plotone{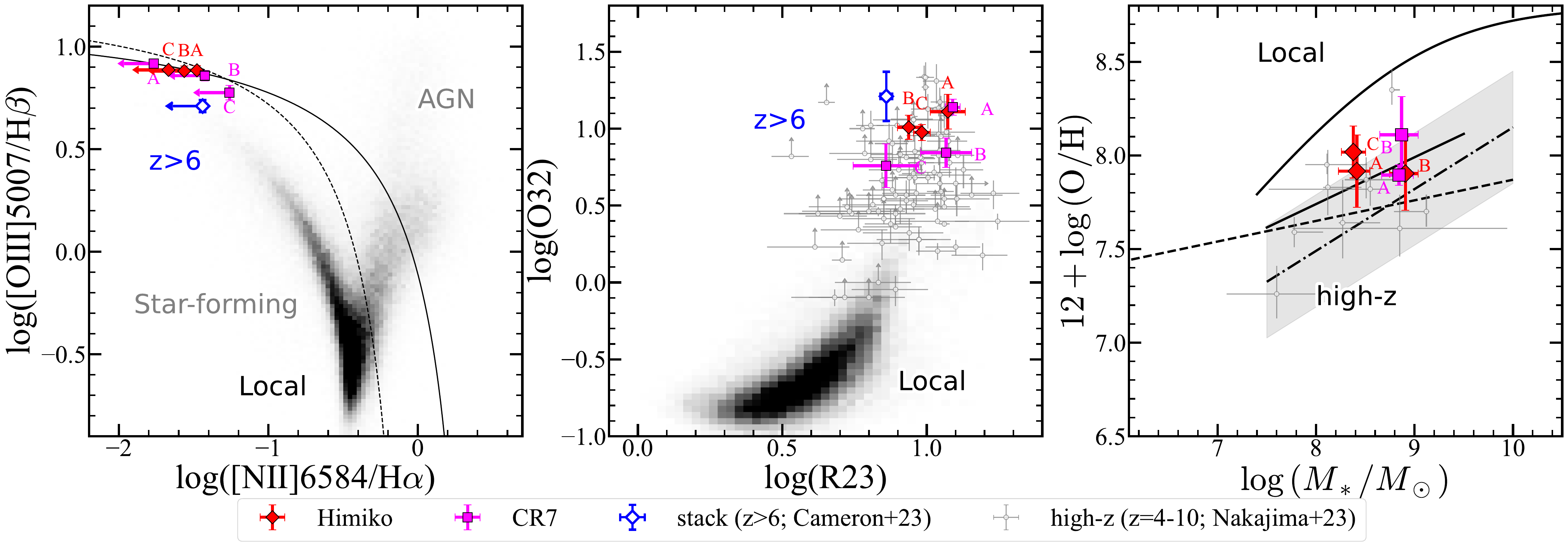}
    \caption{
    Left: [N\,{\sc ii}] BPT diagram. 
    Himiko and CR7 are plotted as red diamonds and magenta squares, respectively. Labels indicate clump names. 
    Blue plots show JADES stacked results at $z\gtrsim6$ \citep{cameron23}.
    Two demarcation curves between AGNs and star-forming galaxies are also shown as solid \citep{kewley01} and dashed \citep{kauffmann03} lines. 
    The gray regions show the SDSS galaxy distributions (SDSS DR16; \citealt{ahumada20}). 
    Middle: O32-R23 diagram. 
    The gray open circles show the JWST samples at $z=4$--$10$ \citep{nakajima23}. 
    Others are the same as the left panel. 
    Right: gas-phase metallicities as a function of stellar mass. 
    The solid, dashed, and dot-dashed lines show the mass metallicity relation based on the JWST samples at $z>4$ presented in \citet{nakajima23}, \citet{curti24a}, and \citet{heintz23}, respectively. 
    The gray shade represents the uncertainty of the \citet{heintz23} relation. 
    The solid black curve shows the relationship in the local Universe \citep{andrews13}. 
    Others are the same as the two panels. 
\label{fig:line_ratio}}
\end{figure*}

Figure~\ref{fig:line_ratio} shows the line ratio diagnostics (left: [N\,{\sc ii}] BPT diagram; \citealt{baldwin81}, middle: O32-R23) and mass-metallicity relation (right) of Himiko (red squares) and CR7 (magenta squares). 
In the [N\,{\sc ii}] BPT diagram (left panel of Figure~\ref{fig:line_ratio}), all clumps of Himiko and CR7 have high \oiii$\lambda5007$/H$\beta$ values, indicating the high ionization parameters. 
We cannot find an AGN signature only in the [N\,{\sc ii}] BPT diagram, although they are located at the edge of the demarcation curves between star-forming and AGN \citep{kewley01, kauffmann03}. 
This might be due to the low metallicity or high ionization states (e.g., \citealt{nakajima22}). 

In the O32-R23 diagram (middle panel of Figure~\ref{fig:line_ratio}), the Himiko and CR7 clumps show higher O32 values than the Sloan Digital Sky Survey (SDSS) galaxies. 
When we compare Himiko and CR7 clumps with the other high redshift galaxies observed with JWST (e.g., \citealt{nakajima23, cameron23}), they both show high O32 values, which can be explained by the hard ionizing radiation field and a low metallicity (e.g., \citealt{nakajima16}). 

From the direct method, we obtain the electron temperature $T_e$(\oiii) of $(1.6\pm0.3)\times10^4$\,K, $(1.6\pm0.3)\times10^4$\,K, $(1.4\pm0.2)\times10^4$\,K, $(1.6\pm0.1)\times10^4$\,K, and $(1.3\pm0.2)\times10^4$\,K for Himiko-A, -B, -C, CR7-A, and -B, respectively. 
The metallicities are estimated as $12+\log{(\mathrm{O/H})}=7.92\pm0.19, 7.90\pm0.20, 8.02\pm0.14, 7.90\pm0.06$, and $8.11\pm0.20$ for Himiko-A, -B, -C, CR7-A, and -B, respectively (see Table~\ref{tab:SED-fitting}). 
For CR7-C, we do not detect the \oiii$\lambda$4363 line.
In the relation between metallicity and stellar mass (right panel of Figure~\ref{fig:line_ratio}), the positions of Himiko and CR7 broadly align with the mass-metallicity relation established by JWST-observed galaxies at $z>4$ (e.g., \citealt{heintz23, curti24a, nakajima23}). 
The caveat is that Himiko-B shows the broad H$\alpha$ emission line explained by AGN activity (see Section~\ref{subsec:AGN}). The metallicity measurements might be affected by the AGN.

\begin{figure}
\plotone{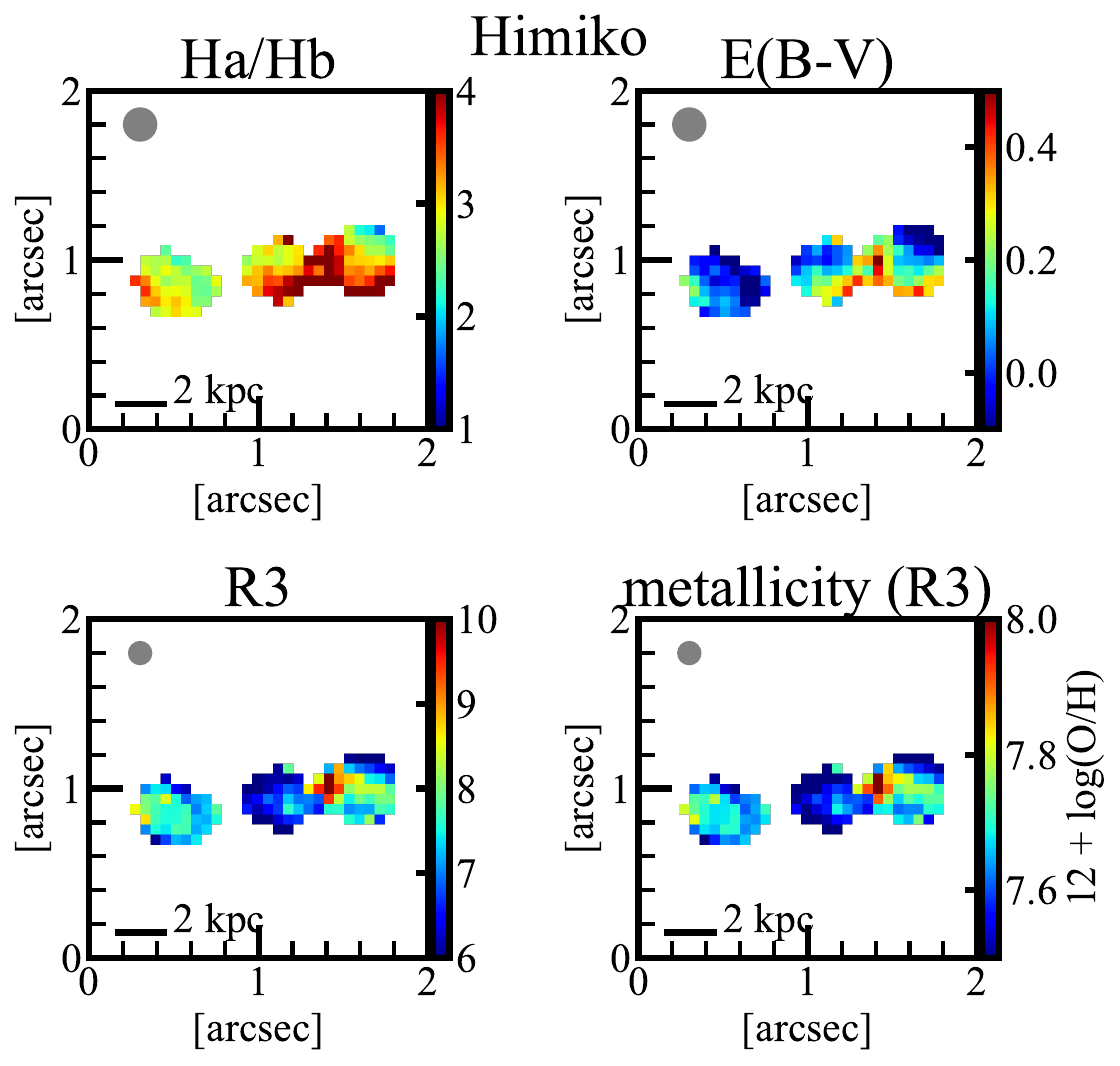}
\plotone{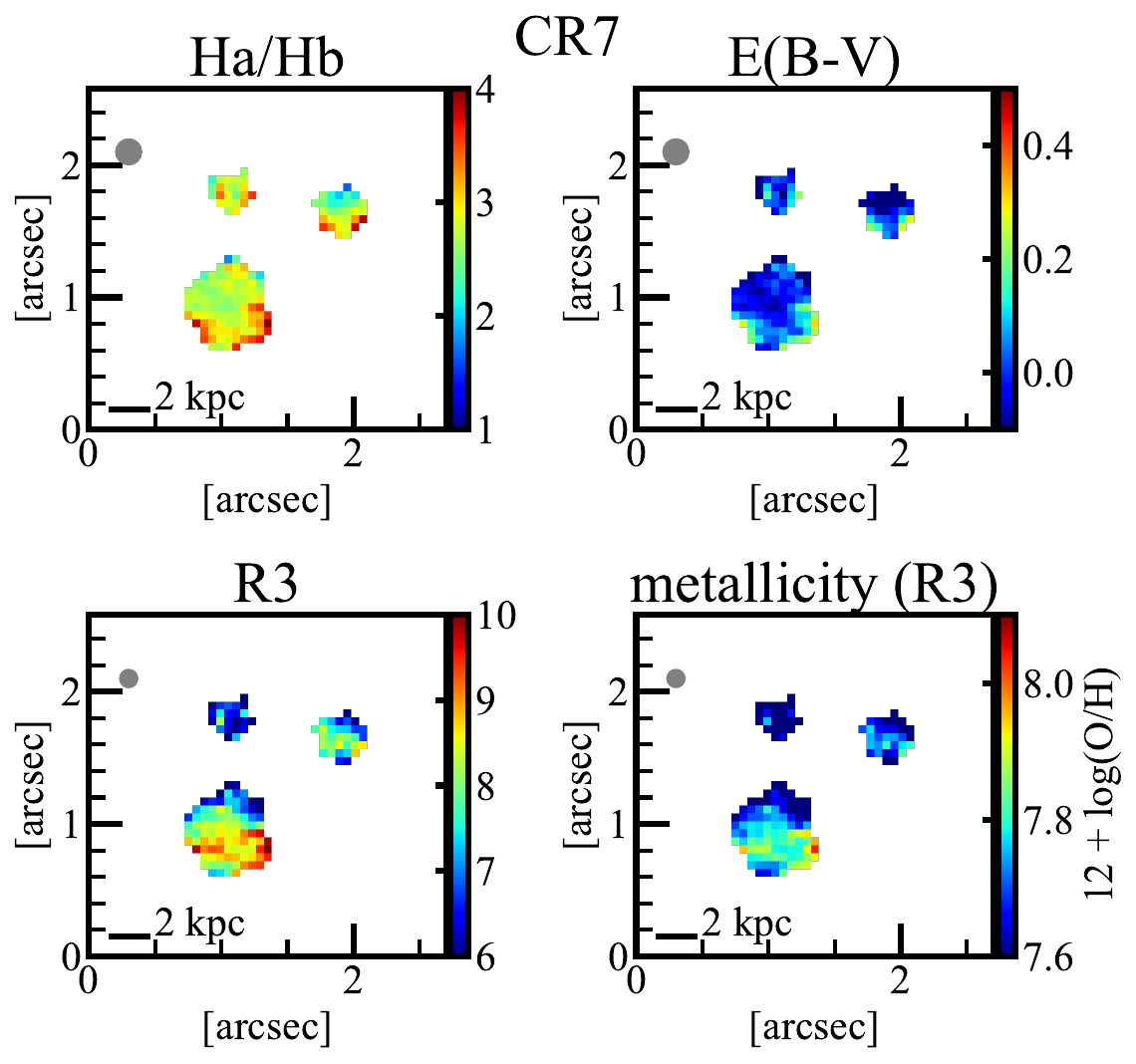}
\caption{Line ratio maps of H$\alpha$/H$\beta$, $E(B-V)$ derived from H$\alpha$/H$\beta$, R3, and metallicity derived from R3. 
The top (bottom) four panels show Himiko (CR7). 
The metallicity is estimated using the relation from \citet{hirschmann23} and is broadly consistent with the direct method applied to the spatially integrated spectra of individual clumps (see Figure~\ref{fig:line_ratio}).
The gray circle in the upper left corner indicates the PSF FWHM size, and the scale bar shows 2~kpc at $z=6.6$.
\label{fig:line_ratio_map}}
\end{figure}

Additionally, we can explore the spatially resolved line ratio properties, thanks to the high-resolution spectroscopy obtained with JWST. Figure~\ref{fig:line_ratio_map} presents the line ratio maps (H$\alpha$/H$\beta$, $E(B-V)$ derived from H$\alpha$/H$\beta$, R3, and metallicity derived from R3) for Himiko (top) and CR7 (bottom). 
We do not correct the dust extinction for the R3 maps. 
From the H$\alpha$/H$\beta$ and $E(B-V)$ maps of Himiko, we observe that Himiko-A has low dust extinctions ($E(B-V)\sim0$). The $E(B-V)$ values are slightly higher ($\gtrsim0.2$) in the region between Himiko-B and Himiko-C compared to Himiko-A or the central region of Himiko-B and -C. 
The region between Himiko-B and Himiko-C has smaller fluxes than the central areas of the clumps, and the overall dust extinction of Himiko is low ($E(B-V)\lesssim0.18$). 
In the case of CR7, all clumps show low dust extinctions ($E(B-V)\sim0$). 
When examining the R3 maps, we find high R3 values ($\mathrm{R3}\sim10$) in CR7-A. 
Notably, CR7-A has a spatial gradient of R3; the north is low ($\mathrm{R3}\sim7$) while the south is high ($\mathrm{R3}\sim9$), which aligns with \citet{marconcini24b}. 
The metallicities derived from R3 using the relation presented in \citet{hirschmann23} range from $12+\log(\mathrm{O/H})=7.5$–$8.0$, which generally agrees with those measured by the direct method (see Figure~\ref{fig:line_ratio}). 

\subsection{Signature of AGN} \label{subsec:AGN}

Himiko-B has the broad permitted H$\alpha$ emission line ($\mathrm{FWHM}=1030\pm110~\mathrm{km~s^{-1}}$) with narrow forbidden \oiii\ emission (Section~\ref{subsec:morphology}). 
The broad H$\alpha$ emission feature suggests that the broad component is powered by AGN activity, especially from its broad line region (e.g., \citealt{kocevski23, harikane23b, maiolino23,  juodvzbalis25}). 

Table~\ref{tab:BH-properties} summarizes the black hole properties of Himiko-B. 
We estimate the black hole mass ($M_\mathrm{BH}$) following the equation 5 of \citet{reines13}:
\begin{equation}
\begin{split}
& \log{\left(\frac{M_\mathrm{BH}}{M_\odot} \right)} = \log{(\epsilon)} + 6.57 \notag \\
& +0.47\log{\left( \frac{L_\mathrm{H\alpha}}{10^{42}~\mathrm{erg~s^{-1}}} \right)} + 2.06 \log{\left( \frac{\mathrm{FWHM_{H\alpha}}}{10^3~\mathrm{km~s^{-1}}} \right)}, 
\end{split}
\end{equation}
where $\epsilon$ is a scaling factor related to the geometry and kinematics of the broad line region. We adopt $\epsilon=1.075$ following \citet{reines15}. 
Here, $L_\mathrm{H\alpha}$ and $\mathrm{FWHM_{H\alpha}}$ are the broad-line H$\alpha$ luminosity and the FWHM, respectively (see also \citealt{green05}). 
We measure a black hole mass of $\log{(M_\mathrm{BH}/M_\odot)}=6.6^{+0.1}_{-0.2}$, and its uncertainties are based on the variation of $\epsilon$ ($\epsilon=0.75$--$1.4$; \citealt{onken04, reines13}). 
The corresponding Eddington luminosity:
\begin{equation}
    L_{\mathrm{Edd}}=1.3\times10^{38}\left( \frac{M_{\mathrm{BH}}}{M_\odot} \right)\,\mathrm{erg~s^{-1}}
\end{equation}
is $5.1^{+1.6}_{-1.5}\times10^{44}$~erg~s$^{-1}$. 
The black hole mass remains unchanged even when we reduce the size of the aperture to $0\farcs2$--$0\farcs3$, confirming that the broad H$\alpha$ line comes from the central region of Himiko-B. 
The black hole mass estimations have a systematic uncertainty of 0.5~dex \citep{reines15}. 

\begin{deluxetable}{lc}
    \tablecaption{AGN properties of Himiko-B. \label{tab:BH-properties}}
    \tablewidth{0pt}
    \tablehead{
    \colhead{Measurements} & \colhead{Value}
    } 
    \startdata
    $\log{(M_\mathrm{BH}/M_\odot)_\mathrm{H\alpha}}$ & $6.6^{+0.1}_{-0.2}$ \\
    $L_\mathrm{bol}$ & $5.7^{+19.0}_{-3.8}\times10^{44}$~erg~s$^{-1}$ \\
    $L_\mathrm{Edd}$ & $5.1^{+1.6}_{-1.5}\times10^{44}$~erg~s$^{-1}$ \\
    $\lambda_\mathrm{Edd}$ & $1.1^{+3.7}_{-0.8}$ \\
    \enddata
    \tablecomments{ 
    These properties are estimated using a $0\farcs3$ circular aperture.
    }
\end{deluxetable}

\begin{figure}
\plotone{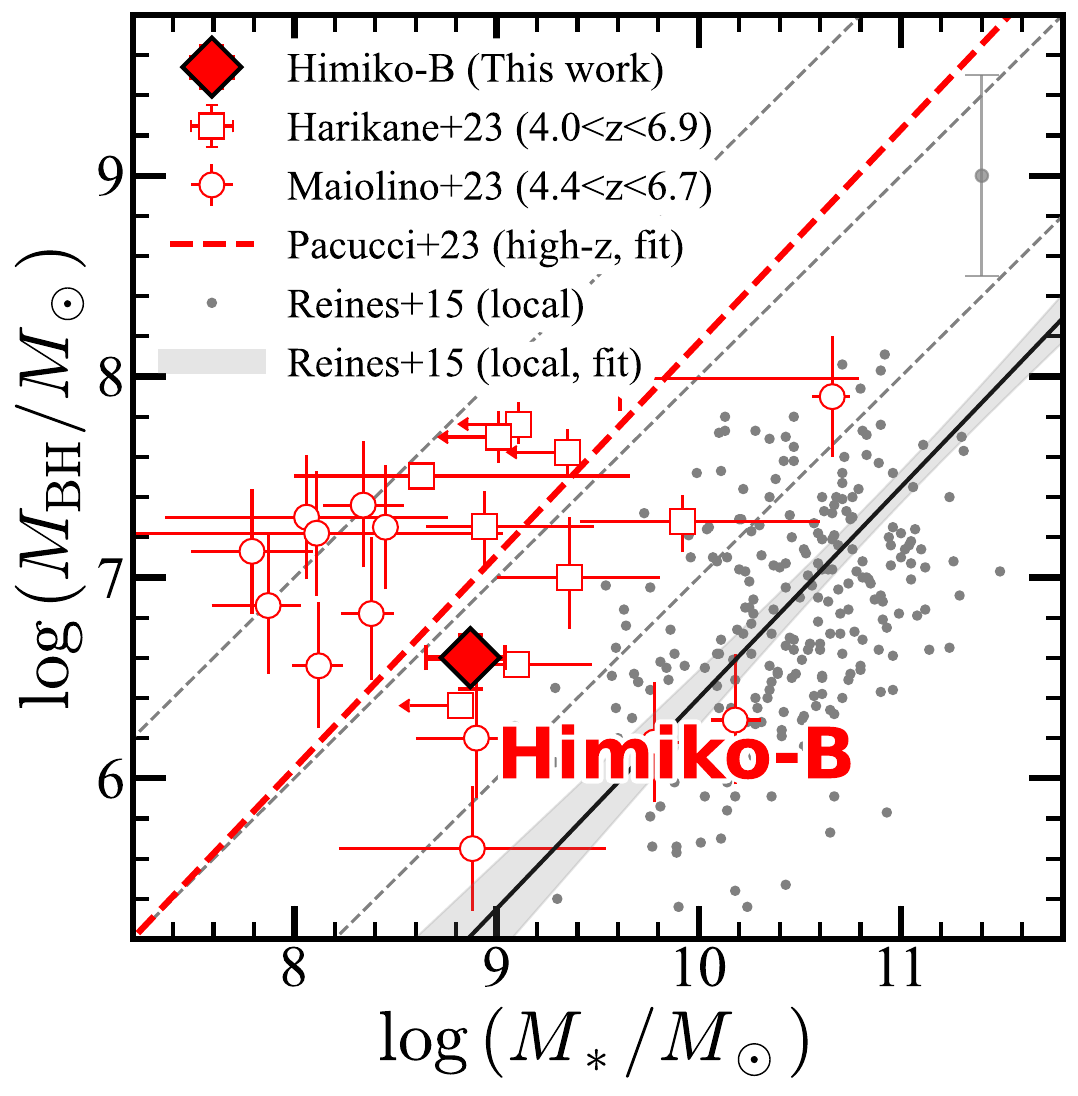}
\caption{Relation between the black hole mass and the stellar mass. The red-filled diamond shows the Himiko Clump B. The red open squares and circles show AGN samples found with JWST/NIRSpec utilizing broad emission lines at $z=4.0$--$6.9$ (\citealt{harikane23b}) and $z=4.4$--$6.7$ (\citealt{maiolino23}), respectively. The red dashed line shows the fitting result of AGN samples discovered with JWST presented in \citet{pacucci23}. 
The grey dots show the AGNs at $z\sim0$ presented in \citet{reines15}. The solid black line and grey shade show the fitting result and its error of AGNs at $z\sim0$ (\citealt{reines15}). The gray dashed lines show the black hole mass to stellar mass ratio of 0.1, 0.01, and 0.001. In the top right, the grey error bar shows the systematic uncertainty of the black hole mass measurement ($\pm$0.5~dex; \citealt{reines15}). 
\label{fig:MBH-Mstar}}
\end{figure}

Figure \ref{fig:MBH-Mstar} shows the relation between black hole mass and the stellar mass of Himiko-B.
Himiko-B is located in a region similar to the high-$z$ AGN samples identified with JWST (red plots; e.g., \citealt{harikane23b, maiolino23}). 
Compared to the local AGN samples (gray plots and line; \citealt{reines15}), Himiko-B shows a higher black hole mass (or lower stellar mass) by an order of magnitude, although the local and high-$z$ AGN samples have large scatters. 

We note that the stellar mass of Himiko-B is derived from the SED fitting using all fluxes (i.e., without decomposition into a host galaxy and a point source). 
Therefore, the actual stellar mass can be lower than our estimate due to contributions from the AGN. 
However, since we see the extended morphology of Himiko-B in the rest-frame UV to optical image (Figure \ref{fig:NIRCam-cutout}), we can consider that its stellar components mainly dominate the photometry. 

We calculate the bolometric luminosity ($L_\mathrm{bol}$) of Himiko-B following the procedure of \citet{harikane23b}. 
The best value of the bolometric luminosity is derived from the relation between H$\alpha$ luminosity and 5100\AA\ luminosity in \citet{green05}:
\begin{equation}
    L_{5100} = 10^{44} \times \left( \frac{L_\mathrm{H\alpha}}{5.25 \times10^{42}~\mathrm{erg~s^{-1}}} \right)^{1/1.157}~\mathrm{erg~s~^{-1}}, \label{eq:L5100}
\end{equation}
where $L_\mathrm{H\alpha}$ is the sum of the narrow and broad-line H$\alpha$ luminosity. 
We convert $L_{5100}$ to $L_\mathrm{bol}$ using the relation of $L_\mathrm{bol}=10.33 \times L_{5100}$ \citep{richards06}. 
For the lower limit, we use Equation (\ref{eq:L5100}) with only broad-line H$\alpha$ luminosity and calculate the bolometric luminosity using the relation of $L_\mathrm{bol}=9.8\times L_{5100}$ \citep{mclure04}.
For the upper limit, we follow the calibration in \citet{netzer09}:
\begin{equation}
\begin{split}
\log{\left ( \frac{L_\mathrm{bol}}{\mathrm{erg~s^{-1}}} \right)} = & \log{\left ( \frac{L_\mathrm{H\beta}}{\mathrm{erg~s^{-1}}} \right)} + 3.48 \notag \\ 
& + \mathrm{max[0.0, 0.31(\log{([\textsc{O\,iii}]/H\beta)} - 0.6)]}, 
\end{split}
\end{equation} 
where $L_\mathrm{H\beta}$ is the H$\beta$ luminosity. 
We obtain the bolometric luminosity of Himiko-B as $L_\mathrm{bol}=5.7^{+19.0}_{-3.8}\times10^{44}$~erg~s$^{-1}$. 
We do not apply dust-extinction corrections to the black hole mass and bolometric luminosity values because a broad H$\beta$ component is not detected and the dust extinction of the broad line is unclear.

\begin{figure}
\plotone{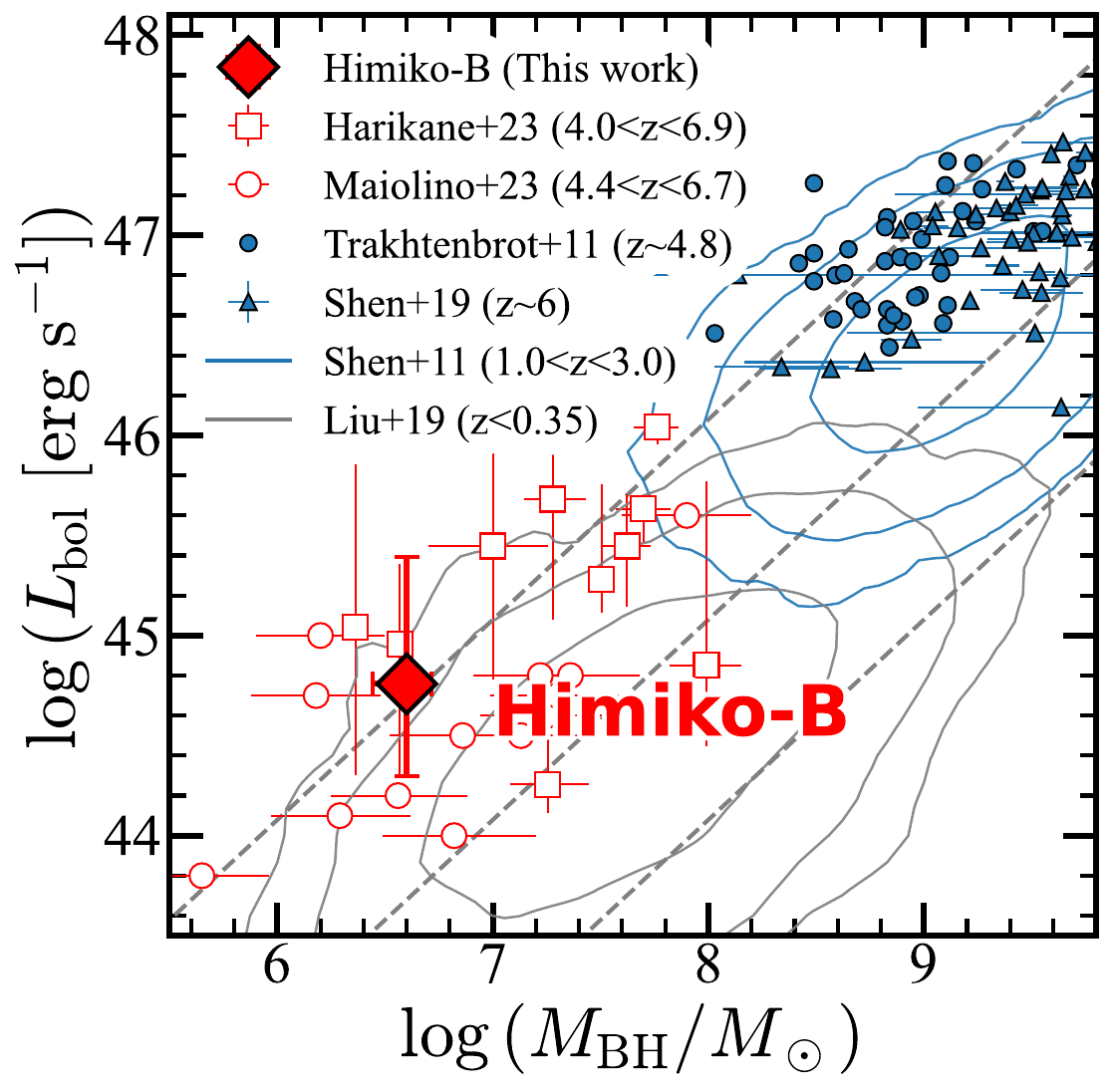}
\caption{Relation between the bolometric luminosity and the black hole mass. The red-filled diamond shows the Himiko Clump B. 
The red open squares and circles show AGNs found with JWST/NIRSpec utilizing broad emission lines at $z=4.0$--$6.9$ (\citealt{harikane23b}) and $z=4.4$--$6.7$ (\citealt{maiolino23}), respectively. 
The blue plots show the high-$z$ quasars (circle: \citealt{trakhtenbrot11}, triangle: \citealt{shen19}). 
The gray and blue contours show the distribution of broad-line AGNs at $z<0.35$ \citep{liu19} and SDSS DR7 quasars at $z=1$--$3$ \citep{shen11}, respectively. 
The gray dashed lines show 100\%, 10\%, and 1\% Eddington luminosity. 
\label{fig:Lbol-MBH}}
\end{figure}

Figure~\ref{fig:Lbol-MBH} shows the relation between the bolometric luminosity and the black hole mass. 
As a comparison, we also plot other broad-line AGNs found by JWST \citep{harikane23b, maiolino23} and high-$z$ quasars at $z\sim4$--$6$ \citep{trakhtenbrot11, shen19} in red and blue, respectively. We also show the low-$z$ AGN distributions \citep{shen11, liu19} in blue and gray contours. 
Himiko-B has a higher bolometric luminosity than the low-$z$ AGNs and a similar $M_\mathrm{BH}$--$L_\mathrm{bol}$ relation compared to the samples found by JWST, although the uncertainty of bolometric luminosity is large. 

Numerical simulations have suggested that a galaxy merger is a trigger to cause AGN activity (e.g., \citealt{hopkins08, zana22}). 
Additionally, recent JWST or ALMA observations have revealed the merging phase AGNs or quasars in the high redshift Universe (see, e.g., \citealt{perna23, ubler24, harikane23b, maiolino23, izumi24, perna25}). 
Given that Himiko is in a merging phase (Section~\ref{subsec:multiple-clumps}), we might observe the growth of a supermassive black hole induced by the merger. 

It is worth noting that there are other possibilities of producing a broad H$\alpha$ emission line, such as massive star outflow, Type IIn supernovae, and tidal disruption events (e.g., \citealt{kokubo24}). 
These mechanisms broaden both permitted H$\alpha$ line and forbidden \oiii\ line. However, Himiko has the broad ($\mathrm{FWHM>1000~\mathrm{km~s^{-1}}}$) component only in the H$\alpha$ line and remains \oiii\ line narrow ($\mathrm{FWHM\lesssim400~\mathrm{km~s^{-1}}}$), which favors emission from a high-density broad line region around a supermassive black hole (see also e.g., \citealt{maiolino23, greene24}).
We do not detect any other AGN signatures, such as high-excitation UV emission lines (e.g., C\,{\sc iv}]) in the JWST/NIRSpec data. 
Further observations are needed to obtain other AGN signs from Himiko. 

We also note that all CR7 clumps do not show any H$\alpha$ components with $\mathrm{FWHM>1000~\mathrm{km~s^{-1}}}$ that are broader than the \oiii\ components ($\mathrm{FWHM\lesssim400~\mathrm{km~s^{-1}}}$). 
\citet{marconcini24b} have argued the tentative evidence of AGN ionization for CR7-A and CR7-B using the excitation diagnostics of \oiii$\lambda$/H$\gamma$ vs \oiii$\lambda5007$/\oiii$\lambda4363$ \citep{mazzolari24} and the high \oiii$\lambda5007$/H$\beta$ values (\oiii$\lambda5007$/H$\beta\sim10$; Figure~\ref{fig:line_ratio_map}).

\subsection{[CII] and IR luminosity} \label{subsec:cii_LIR}

\begin{figure*}
\plotone{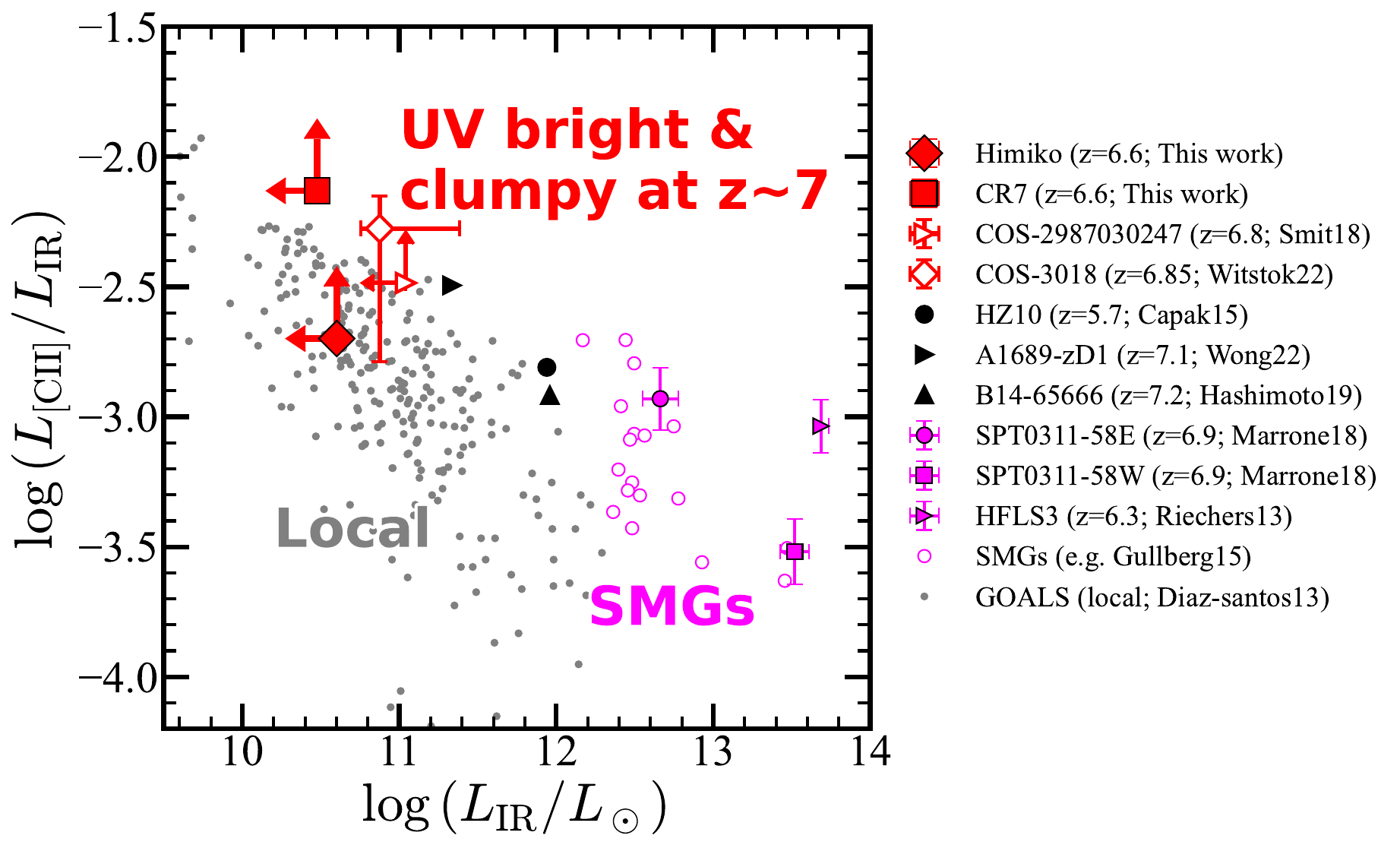}
\caption{
Relation between [C\,{\sc ii}]-IR luminosity ratio and IR luminosity. 
The red-filled diamond and square show Himiko and CR7 (total [C\,{\sc ii}] and IR luminosity), respectively. 
The red-open symbols show the UV bright ($M_\mathrm{UV} \lesssim -21.5$) and clumpy galaxies in the literature: COS-298703024 \citep{smit18} and COS-3018555981 \citep{witstok22}, as indicated in the legend. 
The black symbols show the high redshift dusty merger galaxies: HZ10 \citep{capak15}, A1689-zD1 \citep{wong22}, and B14-65666 \citep{hashimoto19}. 
The magenta-filled symbols show the clumpy submillimeter galaxies at $z=6$--$7$ (SPT0311-58E: \citealt{marrone18}, SPT0311-58W: \citealt{marrone18}, HFLS3: \citealt{riechers13}). 
The open-magenta and filled-gray plots show the submillimeter galaxies at $2<z<6$ (SPT samples: \citealt{gullberg15}) and the local galaxies from GOALS  (\citealt{diaz-santos13}), respectively. 
From the literature, we convert the IR luminosity $L_{\mathrm{IR}}$ by multiplying the factor of 1.7 for $L_{\mathrm{42.5-122.5\,\mu m}}$, and 1.3 for $L_{\mathrm{42-500\,\mu m}}$. 
\label{fig:CII_IRluminosity}
}
\end{figure*}

Figure~\ref{fig:CII_IRluminosity} shows the relation between [\textsc{C\,ii}]-IR luminosity ratio ($L_\mathrm{[\textsc{C\,ii}]}/L_\mathrm{IR}$) and IR luminosity ($L_\mathrm{IR}$; 8--1000~$\micron$). 
Their \cii\ and IR luminosities are also summarized in Table~\ref{tab:ALMA-measurement}. 
Himiko and CR7 are not detected in the dust continuum with the 3$\sigma$ upper limits of $S_\mathrm{158\,\mu m}<31~\mathrm{\mu Jy}$ and $S_\mathrm{158\,\mu m}<19~\mathrm{\mu Jy}$, respectively. 
These upper limits correspond to the IR luminosity of $<4\times10^{10}~L_\odot$ and $<3\times10^{10}~L_\odot$ for Himiko and CR7, respectively, assuming the dust temperature of $T_\mathrm{dust}=40~\mathrm{K}$ and the dust emissivity index of $\beta_\mathrm{IR}=1.5$. 
The IR luminosity depends on the $T_\mathrm{dust}$ and $\beta_\mathrm{IR}$, but the trends of $L_\mathrm{[\textsc{C\,ii}]}/L_\mathrm{IR}$ or $L_\mathrm{IR}$ for Himiko and CR7 still hold if we adopt other $T_\mathrm{dust}$ or $\beta_\mathrm{IR}$ values ($30~\mathrm{K}<T_\mathrm{dust}<50~\mathrm{K}$ or $1.5<\beta_\mathrm{IR}<2.0$). 
For comparison, we overlay some UV bright ($M_\mathrm{UV} \lesssim -21.5$) and clumpy galaxies at $z\sim7$ \citep{smit18, witstok22} in Figure~\ref{fig:CII_IRluminosity}, showing similar [C\,{\sc ii}]-IR ratio limits as those of Himiko and CR7. 
On the other hand, we can see a distinct difference compared to other dusty merger systems (\citealt{capak15, wong22, hashimoto19}) and submillimeter galaxy (SMG) samples (\citealt{marrone18, gullberg15}), indicating the low dust content of Himiko and CR7. 
It is worth noting that SPT0311 ($z=6.9$; \citealt{marrone18}) and HFLS3 ($z=6.3$; \citealt{riechers13}) are the clumpy merging galaxies at redshifts similar to those of Himiko and CR7, but they show high IR luminosity, indicating their rich dust content. 
Even at $z\sim6$--$7$, there is a variety of IR luminosity or dust content in clumpy merging galaxies. 
We discuss details about the dust content of Himiko/CR7 and the comparison with other merger galaxies in Section~\ref{subsec:comparison}. 

\begin{deluxetable*}{ccccc}
    \tablecaption{FIR properties of Himiko and CR7. \label{tab:ALMA-measurement}}
    \tablewidth{0pt}
    \tablehead{
    \colhead{} & \multicolumn{2}{c}{Himiko-Total} & \multicolumn{2}{c}{CR7-Total} \\
    \cline{2-3}
    \cline{4-5}
    \colhead{} & \colhead{This work} & \colhead{\citet{carniani18}} & \colhead{This work} & \colhead{\citet{matthee17}}
    } 
    \startdata
    $z_{[\textsc{C\,ii}]}$                                    &6.5913$\pm$0.0006 & 6.5913$\pm$0.0004 & 6.600$\pm$0.001 & 6.600$\pm$0.001 \\
    $\mathrm{FWHM_{[\textsc{C\,ii}]}}$~($\mathrm{km~s^{-1}}$)          &184$\pm$62 & 180$\pm$50 & 298$\pm$112 & 299$\pm$26 \\
    $S_\mathrm{[\textsc{C\,ii}]} \Delta\nu$~($\mathrm{mJy~km~s^{-1}}$) &70.9$\pm$15.9 & 108$\pm$13 & 196.9$\pm$32.5 & 185.2$\pm$39.0 \\
    $L_\mathrm{[\textsc{C\,ii}]}$~($10^{8}\,L_\odot$)                  & 0.80$\pm$0.17 & 1.2$\pm$0.2 & 2.22$\pm$0.37 & 2.17$\pm$0.36 \\
    $S_{158\,\mu \mathrm{m}}$~($\mu$Jy)                     & $<31$ & $<27$ & $<19$ & $<21$ \\
    $L_\mathrm{IR}$~($10^{10}\,L_\odot$) & $<4$ & $<3$ & $<3$ & $<3$ \\
    $M_\mathrm{dust}$~($10^6~M_\odot$; $T_\mathrm{d}=35\,\mathrm{K}$, $\beta_\mathrm{IR}=1.5$) & $<13$ & \nodata & $<8$ & $<8$ \\
    $M_\mathrm{dust}$~($10^6~M_\odot$; $T_\mathrm{d}=40\,\mathrm{K}$, $\beta_\mathrm{IR}=1.5$) & $<9$ & \nodata & $<5$ & \nodata \\
    $M_\mathrm{dust}$~($10^6~M_\odot$; $T_\mathrm{d}=45\,\mathrm{K}$, $\beta_\mathrm{IR}=1.5$) & $<6$ & \nodata & $<4$ & \nodata \\
    $M_\mathrm{gas}$~($10^9~M_\odot$) & 3.1 & \nodata & 8.9 & \nodata \\
    $M_\mathrm{metal}$~($10^6~M_\odot$) & $<16$ & \nodata & $<33$ & \nodata \\
    \enddata
    \tablecomments{
    We measure the \cii\ and dust properties using the spaxels above 3$\sigma$ levels. \citet{carniani18} and \citet{matthee17} have used the spaxels above 2$\sigma$ and 3$\sigma$ levels, respectively. 
    Gas mass and metal mass are estimated from the \cii158$\micron$ luminosity and metallicity using the conversion given in \citet{zanella18} and \citet{clark16} (see Section~\ref{subsec:comparison}). 
    All upper limits are given at the 3$\sigma$ levels. 
    }
\end{deluxetable*}

\section{Discussion} \label{sec:discussion}
\subsection{Origins of the multiple clumps} \label{subsec:multiple-clumps}

In Section~\ref{subsec:morphology}, we find multiple clumps of Himiko and CR7 using \oiii$\lambda\lambda4959, 5007$ emission lines. 
In this section, we discuss the origin of the multiple clumps. 

\begin{figure}
\plotone{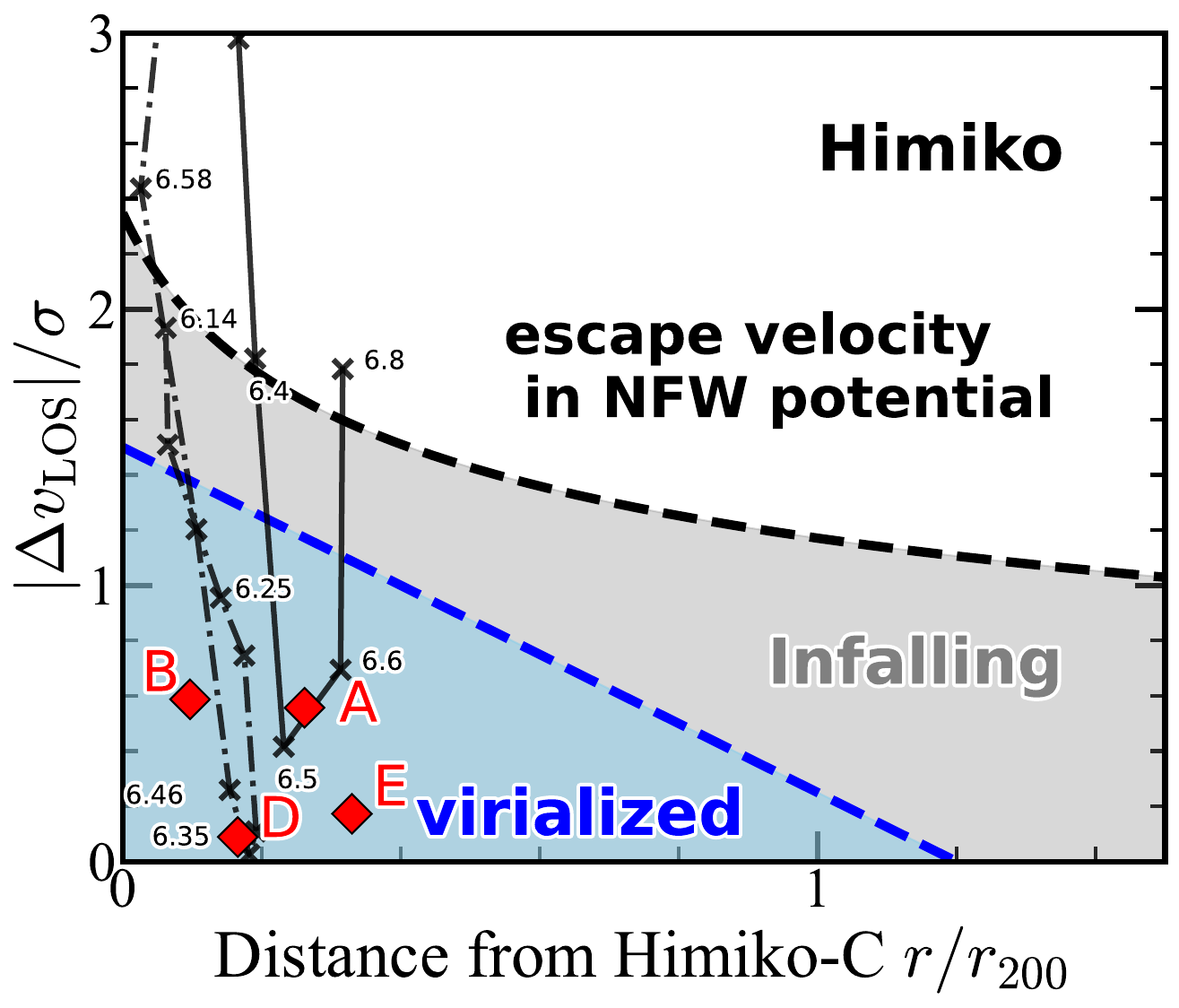}
\plotone{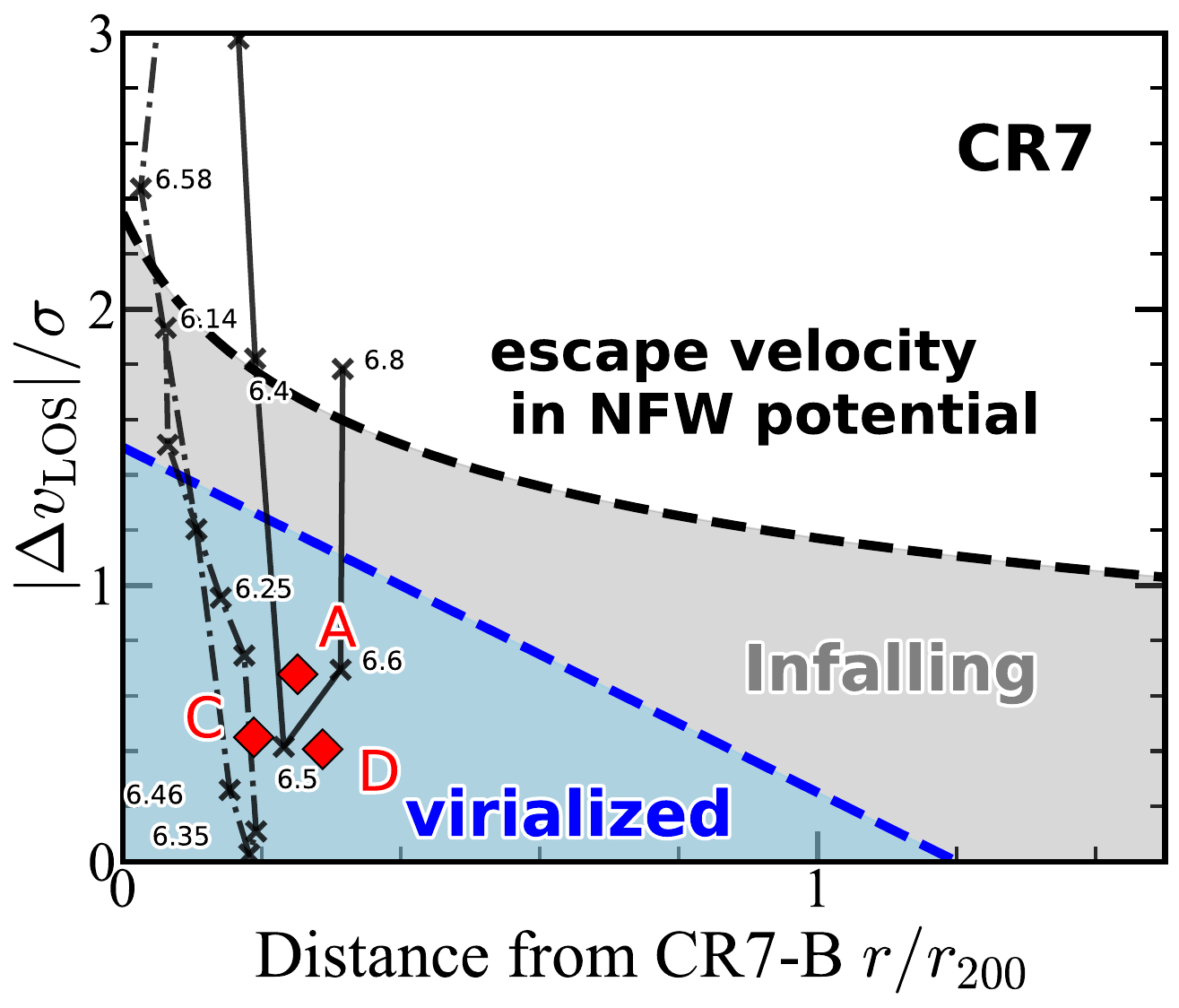}
\caption{
Projected phase space diagrams for Himiko (top) and CR7 (bottom). 
The horizontal axis shows the projected distance from Himiko-B (CR7-B) normalized to $r_{200}$.
The vertical axis is the absolute line-of-sight velocity from the fiducial clump (Himiko-B, CR7-C) normalized by the velocity dispersion of the halo. 
The red diamonds show the clumps of Himiko or CR7 with clump IDs. 
The black dashed line shows the escape velocity in the NFW potential of $\log{(M_{200}/M_\odot)}=11.3$ ($11.4$) for Himiko (CR7). 
The gray region is below the escape velocity (infalling region). 
The blue region shows the virialized region. 
The solid black line and the dash-dotted line show the simulated tracks of merging galaxies from FOREVER22 (PCR0-ID32) and FirstLight (FL956), respectively (see Sections~\ref{subsec:multiple-clumps} and \ref{subsec:extended-Lya}). The redshift of each point is indicated by the labels. 
\label{fig:phase-space}}
\end{figure}

Figure~\ref{fig:phase-space} shows the phase-space diagram (e.g., \citealt{jaffe15}) of Himiko (top) and CR7 (bottom). 
Although such phase-space diagrams are typically used at the galaxy cluster scale, we adapt and apply the same approach to the halo of a single galaxy to investigate the kinematics and membership of its clumpy substructures. 
The horizontal axis is the projected distance from Himiko-C or CR7-B normalized to a virial radius $r_{200}$. 
The vertical axis is the absolute line-of-sight velocity normalized to the line-of-sight velocity dispersion $\sigma$. 
The line-of-sight velocities of Himiko and CR7 are estimated from the \oiii$\lambda5007$ emission lines. 
The fiducial clumps are set to Himiko-C or CR7-B, located at the center along the velocity axis. 
If we set another clump as a fiducial value, the line-of-sight velocity and projected distance differ only by a factor of two, which does not change our following discussions. 
The virial mass $M_{200}$ is derived from the stellar-to-halo mass relation (SHMR) in \citet{behroozi19}. 
We assume the stellar mass as the total stellar mass of the clumps in the system, resulting in $\log{(M_{200}/M_\odot)}=11.3$ ($11.4$) for Himiko (CR7). 
We estimate the virial radius $r_{200}$ from $M_{200}$ using the relation presented in \citet{mo02}:
\begin{equation}
    r_{200} = \left( \frac{GM_{200}}{100\Omega_m H_0^2} \right)^{1/3} (1 + z)^{-1}, 
\end{equation}
where $G$ is the gravitational constant. 
The virial radius $r_{200}$ of Himiko and CR7 is $\sim24~\mathrm{kpc}$. 
The line-of-sight velocity dispersion $\sigma$ is calculated using the virial theorem: 
\begin{equation}
    M_{200} = \frac{3}{2}\frac{r_{200} \sigma^2}{G}.  
\end{equation} 
In Figure~\ref{fig:phase-space}, the black dashed line shows the escape velocity from a Navarro–Frenk–White (NFW; \citealt{navarro96}) potential of the dark matter halo (see, \citealt{jaffe15, rhee17}). 
The gray region indicates where the line-of-sight velocity is below the escape velocity, corresponding to the infalling region.
The blue region represents the virialized area as indicated by \citet{rhee17} and \citet{yoon17}.
We find that all clumps are located either in the infalling or virialized regions, indicating that they will be in equilibrium or virialized within the dark matter halo.
This result suggests that the clumps can possibly merge in the future. 

We note some caveats when estimating halo mass. We choose the SHMR of \citet{behroozi19} because the model is calibrated to a broad set of observables at $0<z<10$, such as UV luminosity function and stellar mass function, providing consistency with the averaged galaxy population. When we choose another model, for example, \citet{moster13, moster18}, the stellar to halo mass ratio is consistent with the \citet{behroozi19} relation. In addition, we need to be careful about the scatter of the SHMR. In \citet{behroozi19}, the scatter of SHMR is $\sim0.3$\,dex, but this scatter does not change our discussion. We also mention that Himiko and CR7 are merging systems, and it is difficult to estimate their exact halo profile. But still, we can roughly estimate the escape velocity (black dashed line in Figure~\ref{fig:phase-space}) through a typical assumption of a dark matter halo profile. For example, Figure 2 of \citet{dutton14} shows the differences in the circular velocity depending on dark matter halo profiles. They indicate that the circular velocity does not largely change around $r=0.2r_{200}$, where Himiko and CR7 clumps are located. The threshold of the escape velocity in Figure~\ref{fig:phase-space} (black dashed lines) remains similar, even when choosing other dark matter halo profiles. 

\begin{figure*}
\gridline{
          \fig{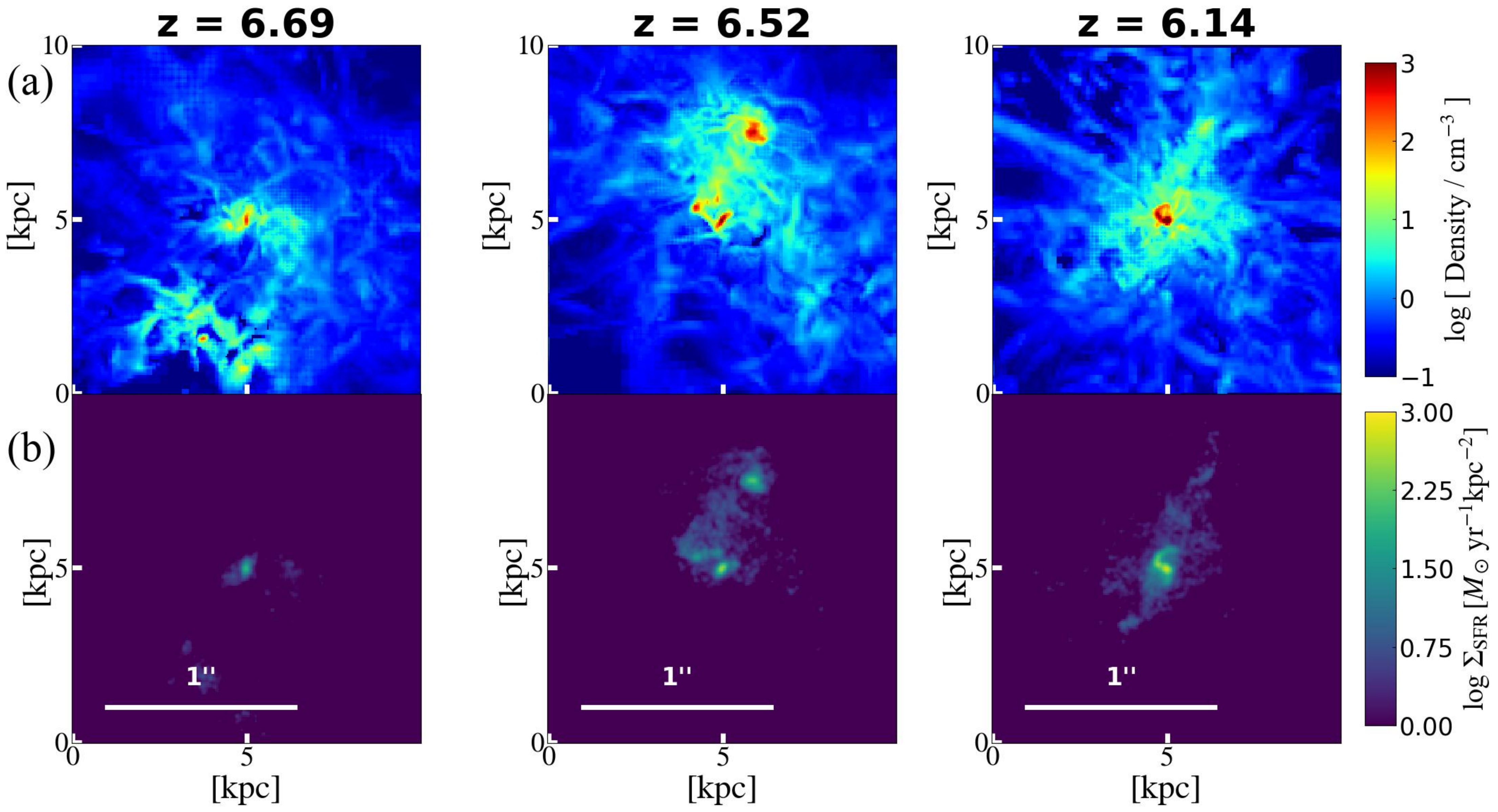}{0.65\textwidth}{}
          \fig{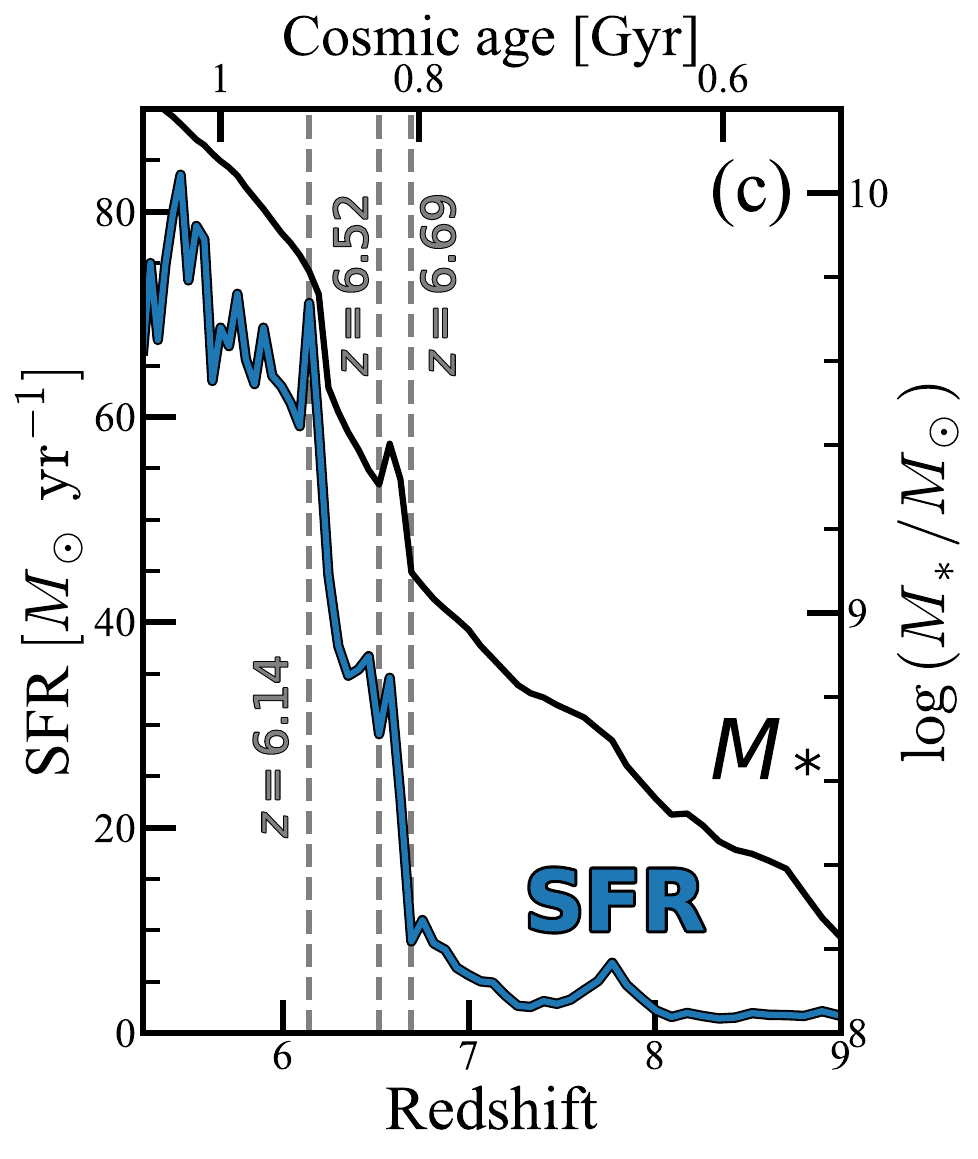}{0.33\textwidth}{}
          }
\caption{
FirstLight simulation \citep{ceverino17, nakazato24} results for a galaxy sample FL956. 
Panels (a) and (b) show the time evolution of gas density distribution and SFR surface density from $z=6.69$ to $z=6.14$, respectively. 
Each panel shows a region of 10 kpc\,$\times$\,10 kpc, corresponding to $1\farcs8 \times 1\farcs8$ at $z=6.6$. Panel (c) shows the star formation history (blue) and stellar mass evolution (black) from $z=9.0$ to $z=5.3$. 
In panel (c), $z=6.69$, $z=6.52$, and $z=6.14$ are indicated by gray dashed lines, which correspond to panels (a) and (b). 
\label{fig:firstlight}
}
\end{figure*}

It is interesting to compare these clumps with numerical simulations to check whether these systems can be observed in simulations. 
We use FirstLight simulations \citep{ceverino17}, which are zoom-in cosmological simulations with maximum resolutions of 17-32 proper pc. 
The simulation has a box size of 60 comoving Mpc (cMpc). 
Following the clump-identification methods in \citet{nakazato24}, we search for a clumpy galaxy with a total stellar mass of $\log{(M_*/M_\odot)}\sim9$ and SFR of $\mathrm{SFR}\sim30~M_\odot~\mathrm{yr^{-1}}$ at $z\sim6.5$, which are similar properties of Himiko and CR7. 
We successfully find an object similar to Himiko and CR7, named FL956. 
The FL956 track ($z=6.6$--$6.1$) in the phase space diagram is indicated in Figure~\ref{fig:phase-space} by dash-dotted lines. 
The track shows that FL956 is experiencing a similar dynamical phase at $z\sim6.5$ to those of Himiko and CR7. 
Panels (a) and (b) of Figure~\ref{fig:firstlight} show the gas density distribution and SFR surface density of FL956 at $z=6.69$ to $z=6.14$, respectively. 
At $z=7.0$, the system contains one central galaxy. 
At $z=6.69$, the counterpart galaxy enters the 10 kpc\,$\times$\,10 kpc region (left side of panel (a) in Figure~\ref{fig:firstlight}) and experiences pericenter passage at $z=6.58$. 
The surrounding gas is perturbed, leading to the formation of new clumps in a tidal tail at $z=6.52$ (middle side of panel (a) in Figure~\ref{fig:firstlight}). 
There are at least three clumps at $z=6.52$ within a $10~\mathrm{kpc}\times10~\mathrm{kpc}$ region, and the separation of each clump is a few kpc, which is also similar to CR7 and Himiko. 
The SFR surface density distribution aligns well with the gas density distribution. 

Panel (c) of Figure~\ref{fig:firstlight} displays the star formation history and stellar mass evolution of FL956 from $z=9.0$ to $z=5.3$. 
This system merges into one galaxy at $z\sim6.1$ (right side of panel (a) in Figure~\ref{fig:firstlight}), and the total stellar mass (SFR) reaches $\log{(M_*/M_\odot)=9.6}$ ($65~M_\odot~\mathrm{yr^{-1}}$). 
The stellar mass in the system increases by approximately $1~\mathrm{dex}$ during the merging phase from $z=7$ to $z=6$ within just 170~Myr. 
The star formation rate rises at $z\sim6.6$, likely driven by merger-induced star formation. 
The trend of simulated star formation history is consistent with those of Himiko and CR7 derived from the SED fittings (Figure~\ref{fig:sfh} and Section~\ref{subsec:sed}). 
Additionally, the number density of the clumpy galaxy in the FirstLight simulation is broadly consistent with the number density of Himiko and CR7 observed by Subaru ($\sim10^{-5}$\,cMpc$^{-3}$ to $10^{-6}$\,cMpc$^{-3}$; \citealt{ouchi09, sobral15}), further supporting the merger scenario (see Figures 7 and 8 of \citealt{nakazato24}). 
These results support the claim that multiple components of Himiko and CR7 are attributed to merger systems (see also \citealt{nakazato24} and \citealt{harikane25} for the formation of clumpy galaxies through mergers). 
Also, given that each clump of Himiko and CR7 has comparable stellar mass (within 1:4; see Section~\ref{subsec:sed}), Himiko and CR7 are experiencing major mergers. 

We observe the \oiii-detected clumps that are not detected in the NIRCam images (Himiko-D, Himiko-E, and CR7-D).
The origins of these clumps are an interesting topic for further investigation.
One possible explanation for these clumps is galaxy mergers.
With gravitational interactions, the gas could be spatially distorted, forming tidal features, and the ionized gas clumps could be located near the stellar clumps (Himiko-A to Himiko-C and CR7-A to CR7-C). 
In panel (a) of Figure~\ref{fig:firstlight}, we find a complex spatial gas distribution in the snapshots, which could be observed as \oiii-detected clumps like Himiko-D, Himiko-E, and CR7-D. 
Additionally, these clumps can also be explained by low-mass satellite galaxies, which fall into the same dark matter halo and will merge into one galaxy. 
In this case, small clumps also have stellar populations, but they are dark below the sensitivity limit of NIRCam. 
While galaxy mergers can explain these clumps, outflows from the stellar clumps are another plausible interpretation. Stellar- or AGN-driven winds could ionize the surrounding gas, making it visible as \oiii-detected clumps that are undetectable in the NIRCam images.
There is limited information on Himiko-D, Himiko-E, and CR7-D, as they are only detected through strong emission lines. 
Deep follow-up observations will be necessary to reveal the detailed physical conditions and origins of these clumps.

\subsection{Comparisons with JWST studies of clumpy galaxies} \label{subsec:jwst-clump-comparison}
With JWST’s high spatial resolution, numerous clumpy galaxies at high redshift have recently been identified. 
Here we briefly place Himiko and CR7 in the context of recent JWST studies of clumpy substructure.

\citet{chen23} report multiple substructures in clumpy galaxies at $z=6$--$8$, with clump stellar masses $\log(M_*/M_\odot)=7$--$9$ on scales of several kpc, comparable to those inferred for Himiko and CR7. Their analysis is based solely on NIRCam imaging, and future IFU spectroscopy will provide dynamical information. \citet{claeyssens23}, \citet{adamo24}, and \citet{mowla24} examine gravitationally lensed systems at $z=1$--$10$. Thanks to the magnification, they probe intrinsically smaller clumps with $\log(M_*/M_\odot)=6$--$8$, below the clump masses of Himiko and CR7. \citet{fujimoto24_cosmicgrape} identify $>15$ clumps in one lensed galaxy at $z=6.072$ (Cosmic Grapes), with $\log(M_*/M_\odot)=6$--$8$ and effective radii of 10--60\,pc. The moment-1 (velocity) map of the total system shows a clear gradient with Toomre $Q\sim0.2$--$0.3$ (i.e. $Q<1$), indicative of a gravitationally unstable rotating disk. From this result, they argue that these clumps form via disk instabilities rather than galaxy mergers. 
By contrast, Himiko and CR7 show no galaxy-wide, disk-like velocity gradient (Section~\ref{subsec:morphology}), favoring a merger interpretation for their clumps (see also Section~\ref{subsec:multiple-clumps}). 

These studies place Himiko and CR7 at the high-mass clumps observed at high redshift, while gravitational lensing extends to intrinsically lower masses. Spatially resolved IFU kinematics for these targets reported in these studies will be decisive for understanding their dynamical structures.

\subsection{Illuminating sources} \label{subsec:illuminating-source}
In Sections~\ref{subsec:sed} and \ref{subsec:AGN}, we discuss the recent star formation ($\sim10~\mathrm{Myr}$) in Himiko/CR7 and the AGN activity of Himiko-B. 
\cite{marconcini24b} have reported tentative evidence of AGN from the line diagnostics. 
These aspects are examined with the sources illuminating Himiko and CR7. 

Recent star formation, particularly massive O- and B-type stars, emits stronger UV radiation than older stellar populations. 
The recent increase in the star formation rate likely contributes to the strong UV emission (Figure~\ref{fig:sfh}). 
Under these conditions, these objects appear as bright UV sources.
Additionally, Himiko-B shows a broad H$\alpha$ component, indicating AGN activity. 
CR7 shows the tentative evidence of AGNs \citep{marconcini24b}. 
These AGNs also contribute to the UV-bright nature of Himiko and CR7. 
However, rest-frame UV images of Himiko and CR7 (Figure~\ref{fig:NIRCam-cutout}) reveal spatially extended stellar components instead of a point source, suggesting that the extended UV emission is primarily dominated by stellar light, although we cannot rule out the possibility that UV emission in the central region originates from the AGN.

Finally, we briefly compare LAEs at $z=2$--$3$ with Himiko and CR7. While the physical conditions might not be the same owing to the redshift difference, this comparison provides insight into the properties of Himiko and CR7.
\citet{sobral18b} have observed 21 luminous LAEs at $z=2$--$3$ with the William Herschel Telescope/ Intermediate dispersion Spectrograph and Imaging System, Keck/DEIMOS, and VLT/X-SHOOTER to study high ionization UV lines. 
They suggest that 60\% of the sources show the AGN signatures, while others lacking AGN signatures (40\%) are likely powered by low-metallicities and dust-poor unobscured starbursts. 
They also report that luminous LAEs of $M_\mathrm{UV}<-21.5$ at $z=2$--$3$ are all AGNs. 
In comparison, Himiko and CR7 ($M_\mathrm{UV}\sim-22$) appear consistent with this study, as they show unobscured starbursts, which may be triggered by galaxy mergers involving low-metallicity, dust-poor clumps.

\subsection{Origins of the extended \lya} \label{subsec:extended-Lya}

A spatially extended \lya\ emission is one of the most distinctive features of Himiko and CR7. 
In this subsection, we explore the physical origins of the extended \lya\ emission based on our analysis. 
Using NIRSpec IFU data, we identified multiple clumps, at least five for Himiko and four for CR7. 
The broad components observed in the \oiii\ emission suggest the presence of outflows or tidal features. The clumps in both Himiko and CR7 are involved in major mergers with small separations, indicating strong gravitational interactions among the clumps. 

Several physical mechanisms are proposed to explain the extended Ly$\alpha$ emission: 
1) the circumgalactic medium (CGM) scattering of Ly$\alpha$ photons produced by star-forming regions or AGNs, 
2) the cold gas inflow (cold stream, gravitational cooling), 
3) satellite galaxies, 
4) Ly$\alpha$ fluorescence (photoionization by LyC radiation escaped from the host galaxy, generated by nearby objects, or by UV background) 
5) galaxy mergers or shocks 
(e.g., \citealt{dijkstra09, ouchi09, hayes11, ouchi13, yajima13, momose14, lake15, momose16, leclercq17, kusakabe19, herenz20, kimock21}). 
The observed features of Himiko and CR7 are all related to the spatially extended gas distributions. 
Simulations also suggest that gas around clumps is spatially extended in a merger phase (Panel (a) of Figure~\ref{fig:firstlight}). 
Neutral hydrogen (H\,{\sc i}) in such extended gas around the clumps scatters Ly$\alpha$ photons produced by young stars and AGNs in the clumps. 
The scattered light of \lya\ photons by H\,{\sc i} gas can be observed as the extended morphology. 
Our analysis results (multiple clumps, major mergers, outflows) are the direct evidence for such an extended gas distribution beyond clumps. 
Scenario 1) is closely related to our analysis results, and multiple clumps (potentially Himiko-D, Himiko-E, and CR7-D) might also contribute to Scenario 3).
We cannot obtain direct evidence of scenario 2) from the data. 
However, given the fact that Himiko and CR7 have massive halos at $z\sim7$ ($\log{(M_{200}/M_\odot)}\sim11.3$) indicated by the stellar-to-halo mass relation (Section~\ref{subsec:multiple-clumps}), we cannot exclude the cold gas inflow to their dark matter halos. 
The cold gas inflow might also contribute to the extended \lya\ emission of Himiko and CR7. 
Given that Himiko-B has AGN, fluorescence (scenario 4) is also a possible mechanism (e.g., \citealt{kimock21}). 

As related to the scenario 5), some hydrodynamical simulations have suggested that a gas-rich galaxy merger is a key mechanism of the extended \lya\ emission. 
\citet{yajima13} have conducted the hydrodynamical simulations and radiative transfer calculations for the binary gas-rich major merger system at $z=3$--$7$ with a halo mass range of $3$--$7\times10^{12}~M_\odot$, which are one order of magnitude higher halo mass than Himiko or CR7. 
They have found that shocked gas regions and star formation induced by gravitational interaction produce strong \lya\ emission. 
The \lya\ emission is extended due to the gas distribution with a luminosity of $L_\mathrm{Ly\alpha}\sim10^{43}$--$10^{44}~\mathrm{erg~s^{-1}}$ and a size of $\sim10$--$20$~kpc at $z>6$. 
These results also support the physical origin of the extended \lya\ that can be explained by galaxy mergers like Himiko and CR7. 

\begin{figure*}
\plotone{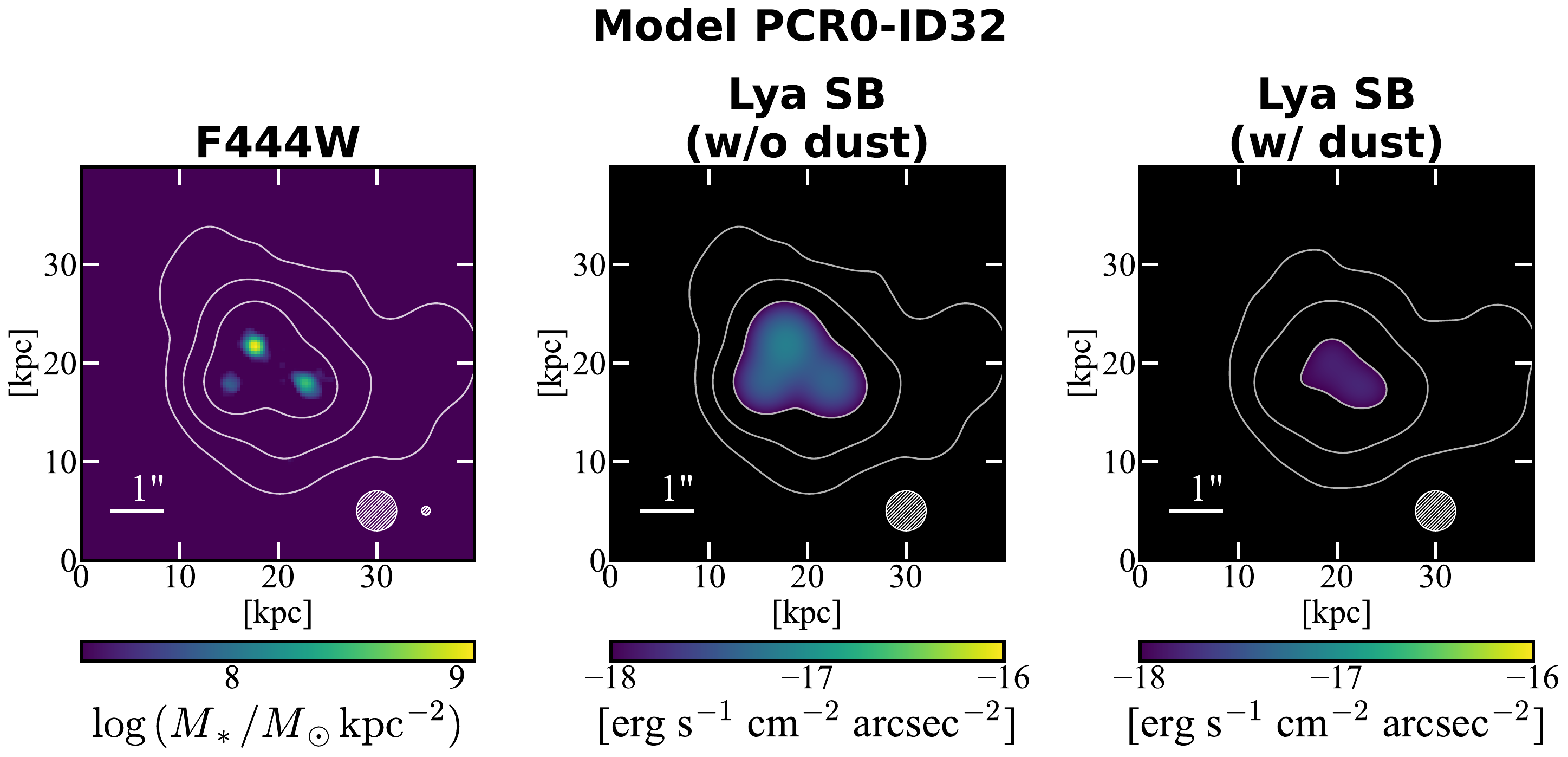}
\caption{
FOREVER22 simulation \citep{yajima22, yajima23} results for a galaxy sample, PCR0-ID32, at $z=6.8$. Each panel represents a physical scale of 40 kpc $\times$ 40 kpc. Left: stellar distribution. The simulated map is smoothed to match the spatial resolution of NIRCam F444W ($0\farcs16$). The contours show the Ly$\alpha$ surface brightness levels of $10^{-20}$, $10^{-19}$, and $10^{-18}$\,erg\,s$^{-1}$\,cm$^{-2}$\,arcsec$^{-2}$ in the dust-free model. The scale bar shows the $1\arcsec$. The circles at the bottom right show the PSF FWHM of JWST/NIRCam F444W ($\mathrm{PSF\,FWHM}\sim0\farcs16$, right) and Subaru NB921 ($0\farcs75$, left). 
Middle (Right): Ly$\alpha$ surface brightness with dust-free (dust) model. The simulated map is smoothed to match the spatial resolution of Subaru NB921 ($\mathrm{PSF\,FWHM}\sim0\farcs75$). The area brighter than the Subaru surface brightness limit ($\sim10^{-18}$\,erg\,s$^{-1}$\,cm$^{-2}$\,arcsec$^{-2}$) is indicated by the colorbar, while regions approximately fainter than the limit are shown in black. The scalebar and circles are the same as in the left panel. 
\label{fig:Lya-simulation}}
\end{figure*}

To further characterize merger galaxies with extended Ly$\alpha$ emission, such as Himiko and CR7, we investigate Himiko-type objects in the FOREVER22 simulation \citep{yajima22, yajima23}. 
FOREVER22 is a zoom-in cosmological hydrodynamics simulation of protocluster regions with a volume of (28.6 cMpc)$^3$ and an initial gas particle mass of $2.9\times10^6~M_\odot~h^{-1}$. 
\citet{yajima22} carry out post-processing radiative transfer calculations, including Ly$\alpha$ emission, using the ART$^2$ code \citep{li08, li20a, yajima12c}. 

We search for a galaxy with properties similar to those of Himiko and CR7 ($z\sim7$, $M_*\sim10^9~M_\odot$, clumpy morphology) in protocluster region 0 (PCR0) of the simulation and find a similar object named PCR0-ID32 at $z=6.8$. 
The left panel of Figure~\ref{fig:Lya-simulation} shows the stellar distribution of PCR0-ID32 at $z=6.8$. 
The simulated map is smoothed with a Gaussian kernel to match the spatial resolution of NIRCam F444W ($0\farcs16$), and the pixel scale is matched to that of Himiko's NIRCam image ($0\farcs04$; Figure~\ref{fig:NIRCam-cutout}). 
By observing down to about one third of the JWST's observational limit measured with the Himiko data ($M_*\sim10^6~M_\odot\mathrm{/pixel}$ or $M_*\sim10^{7.3}~M_\odot\mathrm{\,kpc^{-2}}$), we can detect three clumps of PCR0-ID32 within $\sim10~\mathrm{kpc}\times10~\mathrm{kpc}$. 

In Figure~\ref{fig:phase-space}, we plot the PCR0-ID32 track from $z=6.8$ to $z=6.3$. 
We find that PCR0-ID32 at $z\sim6.6$ is in a region comparable to Himiko or CR7, indicating that it shares a similar dynamical merger phase. 
The middle and right panels of Figure~\ref{fig:Lya-simulation} show Ly$\alpha$ surface brightness (SB) of PCR0-ID32 with dust-free and dust, respectively. 
The Ly$\alpha$ SB map is smoothed with a Gaussian kernel to match the spatial resolution of Subaru NB921 ($0\farcs75$). 
The surface brightness limit of Subaru NB921 is $\sim10^{-18}$\,erg\,s$^{-1}$\,cm$^{-2}$\,arcsec$^{-2}$, and only the color map above this limit is shown. 
The dust-free model shows extended Ly$\alpha$ emission ($\sim 10~\mathrm{kpc}$ scale), which coincides with the stellar distribution (middle panel of Figure~\ref{fig:Lya-simulation}). 
The Ly$\alpha$ luminosity of the object is $1.2\times10^{43}$\,erg\,s$^{-1}$, which is consistent with Himiko and CR7. 
This result shows that Himiko-type objects with extended Ly$\alpha$ emission exist in the simulation. 
In contrast, with dust extinction and obscuration (right panel of Figure~\ref{fig:Lya-simulation}), the Ly$\alpha$ emission is detected only at the PSF scale, comparable to the Subaru NB921 PSF. 
In that case, the Ly$\alpha$ luminosity is $7.4\times10^{41}$\,erg\,s$^{-1}$, which is lower than those of Himiko and CR7. 
These results suggest that stellar/gas distributions due to mergers and the presence or absence of dust play key roles in extended Ly$\alpha$ emission (see also \citealt{witten24}). 
It is possible that the dust in Himiko and CR7 has not yet grown sufficiently and/or has undergone efficient destruction processes, such as those caused by supernovae (e.g., sputtering by supernova shock waves). 
This could result in a high escape fraction, and the low dust content is consistent with the non-detection of their dust continuum in ALMA observations (see also \citealt{hirashita14}). 
Alternatively, other ionizing sources (e.g., hidden AGN) could also enhance the Ly$\alpha$ emission of Himiko and CR7, which are related to scenarios 1) and 4).

\subsection{Physical interpretations of the [CII] offset} \label{subsec:cii-offset}

In Section~\ref{subsec:morphology} and Figure~\ref{fig:jwst-alma-subaru-image}, we find that Himiko has the spatial offset between UV, H$\alpha$ (Himiko-A, -B) and [C\,{\sc ii}], Ly$\alpha$ beyond the astrometric uncertainty. 
We discuss the physical origins of the offset (see e.g., \citealt{carniani20}). 
The possible origins of the offset are 1) the photodissociation region \citep{hollenbach99} between Himiko-A and -B, and 2) shock heating by galaxy mergers. 
In scenario 1), ionizing photons caused by the star formation in Himiko-A, Himiko-B, and AGN activity of Himiko-B are the possible illuminating sources discussed in Section \ref{subsec:illuminating-source}. 
For scenario 2), gravitational interactions cause the shock heating, thus \cii\ can be emitted in a region between Himiko-A and -B. 
Some local merging galaxies tend to have a large \cii\ luminosity caused by shock heating. 
For example, \citet{appleton13} have reported the \cii\ emission from the shocked intergalactic filament in Stephan's Quintet 
(see also \citealt{umehata21} for the Ly$\alpha$ blob at $z=3.1$ case). 
Considering that Himiko is the merger system, shock heating can be a possible scenario to explain the offset. 
The presence of such gas beyond the stellar clumps can lead to resonant scattering of Ly$\alpha$ photons, thereby producing the observed extended Ly$\alpha$ emission (Section~\ref{subsec:extended-Lya}). 

It is also worth noting that the beam size of ALMA ($\sim0\farcs9$) is larger than JWST/NIRCam and NIRSpec IFU ($\sim0\farcs1$--$0\farcs2$). 
We can not exclude the possibility that the \cii\ emission from Himiko-A and Himiko-B can contribute to the total \cii\ emission, which has the offset from each clump. 
As a test, we examine the smoothed F115W image of Himiko, matched to the PSF size of Subaru or ALMA. We find that the peak flux position remains at Himiko-A, preferring that the \cii\ and Ly$\alpha$ offset in Himiko is not due to the lower angular resolution. 
Additional high-resolution ALMA data are necessary to comprehend the \cii\ distribution of Himiko, which will also enable us to explore the \cii\ kinematics.

\subsection{Comparisons with other merger systems: `blue mergers' and `red mergers'} \label{subsec:comparison}

This section compares these unique triple merger systems with other galaxy populations.
This comparison provides valuable insight into whether they share similarities with other galaxy populations. 
We focus on the dust, metal, and gas mass content of Himiko and CR7. 

We finally compare Himiko and CR7 with the dusty merger systems, including SMGs (see also Section\,\ref{subsec:cii_LIR}). 
SMGs trace heavily obscured, short-lived starburst phases with high SFRs ($10^2$--$10^3$ $M_\odot\,\mathrm{yr^{-1}}$) and large dust/gas reservoirs (e.g.,\citealt{casey14}). Their properties and number densities suggest that a subset evolves into massive quiescent/early-type galaxies at later times (e.g., \citealt{hopkins08, toft14, dudzevivciute20}). Comparisons with these distinct populations highlight the diversity in the merger phase at high redshift and provide valuable insights for a deeper understanding of both Himiko/CR7 and dusty star-forming galaxies. In addition, the comparison highlights the mechanisms of the extended Ly$\alpha$ emission.

\begin{figure*}
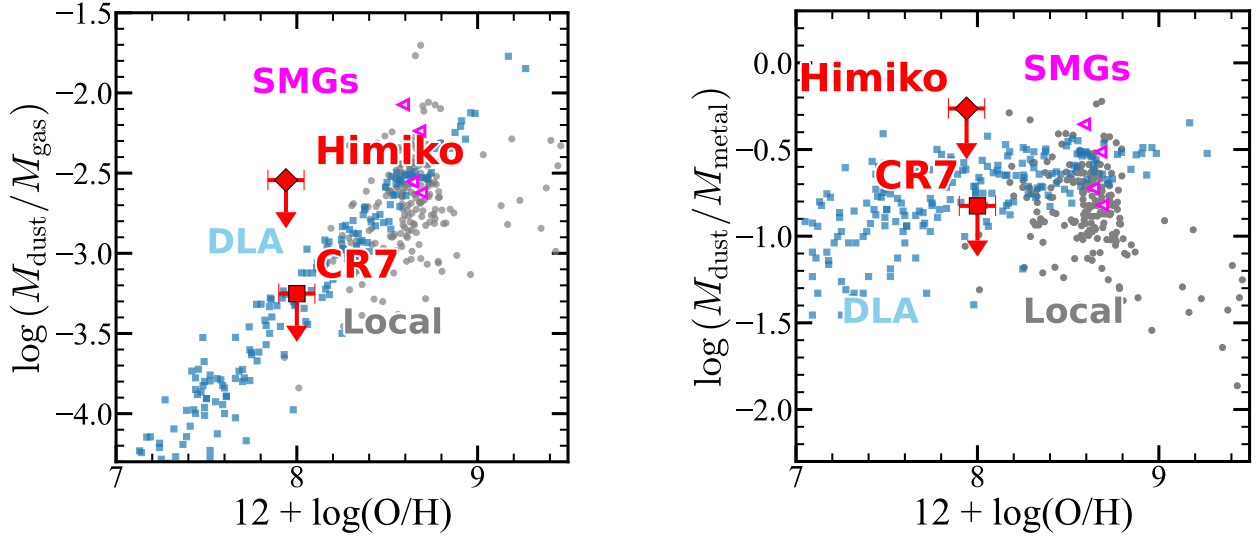

\gridline{\fig{DTG_metallicity_v3.pdf}{0.42\textwidth}{}
          \fig{DTM_metallicity_v3.pdf}{0.42\textwidth}{}
          }
\caption{Relation between DTG (DTM) and metallicity shown in left (right). 
The red diamond and square show Himiko and CR7, respectively. 
The blue, magenta, and gray plots show the DLA \citep{peroux20}, SMGs at $z\sim2$ \citep{shapley20}, and local galaxies (DustPedia; \citealt{devis19, casasola20}), respectively. 
\label{fig:DTG-DTM}}
\end{figure*}

The left (right) panel of Figure~\ref{fig:DTG-DTM} shows the relation between the dust mass-to-gas mass ratio (DTG) and metallicity (dust mass-to-metal mass ratio; DTM). 
Himiko and CR7 are colored red. 
Damped Ly$\alpha$ system (DLA; \citealt{peroux20}), local galaxies (dustpedia; \citealt{devis19, casasola20}), and submillimeter galaxies at $z\sim2$ whose metallicities are estimated by N2 ([N\,{\sc ii}]$\lambda6584$/H$\alpha$) index (\citealt{shapley20}) are plotted in blue, gray, and magenta, respectively. 

We calculate the dust masses of Himiko and CR7 using the following formula:
\begin{equation}
    M_\mathrm{dust} = \frac{D_\mathrm{L}^2 S_{\nu, \mathrm{obs}}}{(1+z) \kappa_\mathrm{d}(\nu_\mathrm{rest}) [B_\nu(T_\mathrm{dust})-B_\nu(T_\mathrm{CMB}(z))]}, 
\end{equation}
where $D_\mathrm{L}$ is the luminosity distance, $S_{\nu, \mathrm{obs}}$ is the observed flux, $\kappa_\mathrm{d}(\nu_\mathrm{rest})$ is the rest-frame dust mass absorption coefficient, and $B_\nu(T_{\mathrm{dust}})$ is the Planck function at a rest-frame frequency $\nu_\mathrm{rest}$ and dust temperature $T_\mathrm{dust}$. 
In this equation, $B_\nu(T_\mathrm{CMB}(z))$ is used to correct for the cosmic microwave background (CMB) effects (e.g., \citealt{ota14}). 
We assume $\kappa_\mathrm{d}(\nu_\mathrm{rest})=0.77(850~\mathrm{\micron}/\lambda_\mathrm{rest})^{\beta_\mathrm{IR}}~\mathrm{cm^2~g^{-1}}$ \citep{dunne00} with the rest-frame wavelength of $\lambda_\mathrm{rest}=158~\mathrm{\micron}$. 
We use the 3$\sigma$ upper limit of the dust continuum of Himiko (CR7) and obtain a dust mass upper limit of $9\times10^6~M_\odot$ ($5\times10^6~M_\odot$), assuming $\beta_\mathrm{IR}=1.5$ and $T_\mathrm{dust}=40~\mathrm{K}$ (Table~\ref{tab:ALMA-measurement}). 

The gas mass estimate has a large uncertainty, but the most promising way to obtain the gas mass at $z\gtrsim6$ is to use a cold neutral gas tracer \cii158$\micron$. 
We generally follow the estimation procedures outlined in \citet{algera25} (see also \citealt{palla24}). 
The molecular H$_2$ gas mass $M_\mathrm{H_2}$ is estimated using the following relation presented in \citet{zanella18}: 
\begin{equation}
    \log{\left( \frac{L_\mathrm{[\textsc{C\,ii}]}}{L_\odot} \right)} = -1.28 + 0.98\log{\left( \frac{M_\mathrm{H_2}}{M_\odot} \right)}, 
\end{equation}
where, $M_\mathrm{H_2}$ and $L_\mathrm{[\textsc{C\,ii}]}$ are in the units of $M_\odot$ and $L_\odot$, respectively. 
We can use the $L_\mathrm{[\textsc{C\,ii}]}/L_\odot$ taken from the ALMA observations. 
Finally, the total gas mass ($M_\mathrm{gas}$) is obtained by following the relation presented in \citet{devis19}: $M_\mathrm{gas} = \xi (M_\mathrm{HI} + M_\mathrm{H_2})$, 
where $\xi$ is a correction factor ($\sim1.3$--$1.4$) depending on the metallicity and $M_\mathrm{HI}$ is the atomic H\,{\sc i} gas mass.  
We follow \citet{clark16} and define $\xi$ as 
\begin{equation}
    \xi = \frac{1}{1-\left(f_\mathrm{He_p} + f_Z [\frac{\Delta f_\mathrm{He_p}}{\Delta Z}] \right) - f_Z},
\end{equation}
where $f_\mathrm{He_p}$ is the primordial Helium mass fraction of 0.2485 \citep{aver11}, $f_Z = Z  \times f_{Z_{\odot}}$ is the fraction by mass using $f_Z=27.36 \times 10^{(12+\log{(\mathrm{O/H)}} -12)}$. The factor of 27.36 comes from the assumption of $12+\log{(\mathrm{O/H})_\odot}=8.69$ and a solar metal mass fraction of 0.0134 \citep{asplund09}. 
Here, $\Delta f_\mathrm{He_p}/\Delta Z=1.41$ \citep{balser06} is the metallicity-dependent evolution of the helium mass fraction. 
Depending on the metallicity, $\xi$ varies from 1.33 (zero metallicity) to 1.39 (solar metallicity). 
In this paper, we ignore atmic H\,{\sc i} gas mass because the atomic gas mass estimate has large uncertainties and [C\,{\sc ii}]-to-$\mathrm{H_2}$ calibration might already trace the full gas content in galaxies at the epoch of reionization (see detailed discussions in \citealt{algera25}). 
We note that the estimate of molecular gas mass also has uncertainty ($\sim0.3$\,dex; see also \citealt{palla24, devis19}), but subsequent discussions and observed trends remain valid. 
Using the gas mass, we can also estimate the metal mass ($M_\mathrm{metal}$) of Himiko and CR7 following the relation \citep{devis19}: 
\begin{equation}
    M_\mathrm{metal}= f_Z \times M_\mathrm{gas} + M_\mathrm{dust}. 
\end{equation}
The gas and metal mass of Himiko and CR7 are summarized in Table~\ref{tab:ALMA-measurement}. 

Himiko and CR7 only have the upper limit of dust mass, so we can only show the upper limit of DTG and DTM in Figure~\ref{fig:DTG-DTM}. 
The DTG and DTM upper limits of Himiko and CR7 are consistent with the DLA samples that have similar metallicities, indicating that Himiko and CR7 are mainly in the sequence of stellar mass or dust mass-metallicity relation (see also mass-metallicity relation of Figure~\ref{fig:line_ratio}). 
However, it is important to note that Himiko and CR7 might still have lower DTG or DTM than other galaxy populations. 
We need deeper dust continuum observations aimed at Himiko and CR7 to constrain these properties. 

\begin{figure*}
\plotone{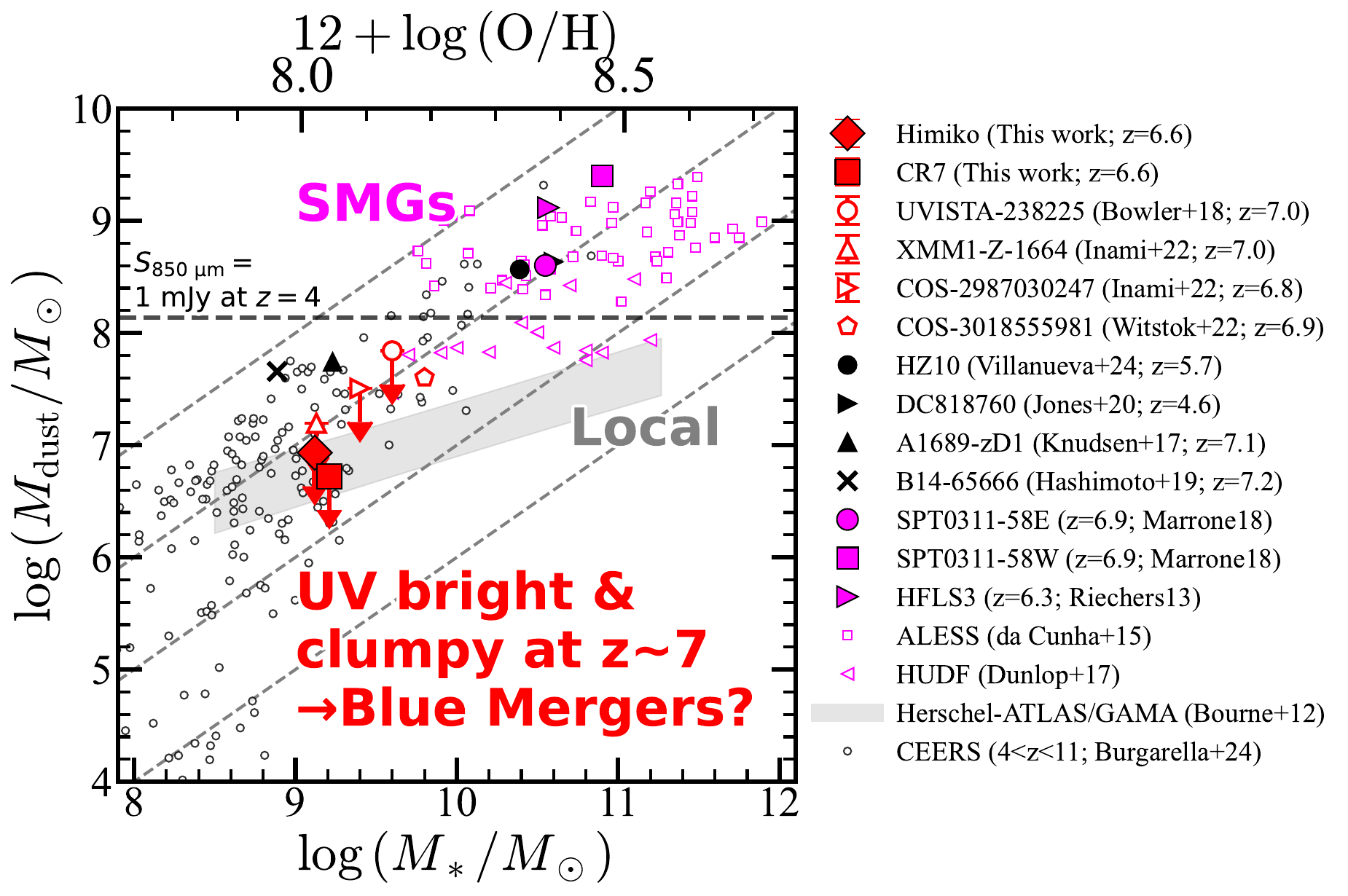}
\caption{Relation between dust mass and stellar mass. The red-filled diamonds represent Himiko and CR7, whose total stellar and dust masses are estimated in this study. 
The red-open symbols show the UV bright ($M_\mathrm{UV}\lesssim-21.5$) and clumpy galaxies in the literature: UVISTA-238225 \citep{bowler18}, XMM1-Z-1664 \citep{inami22, bouwens22}, COS-2987030247 \citep{inami22, harikane25}, COS-3018555981 \citep{inami22, witstok22, harikane25}, as indicated in the legend. 
The upper limits are $3\sigma$ levels. 
The black symbols show the other high-redshift merger galaxies (HZ10 \citep{villanueva24, capak15}, DC818760 \citep{jones20}, A1689-zD1 \citep{knudsen17, watson15}, and B14-65666 \citep{hashimoto19}. 
The magenta-filled symbols show high-redshift ($z=6$--$7$) clumpy SMGs (SPT0311-58E: \citealt{marrone18}, SPT0311-58W: \citealt{marrone18}, HFLS3: \citealt{riechers13}). 
The magenta-open symbols show large SMG samples (ALESS: \citealt{dacunha15}, HUDF: \citealt{dunlop17}). 
The gray-shaded region shows local galaxy samples from the Herschel-ATLAS/GAMA survey \citep{bourne12}. 
The black open circles show the JWST CEERS samples at $4<z<11$ \citep{burgarella24}. 
The black dashed line shows the $S_{850\micron}=1~\mathrm{mJy}$ at $z=4$, which corresponds to SCUBA's deepest detection limit \citep{holland99}. 
This threshold remains largely unchanged with redshift due to the negative-$k$ correction (e.g., \citealt{blain93, blain02}). 
The gray dashed lines correspond to the dust-to-stellar mass ratio from $10^{-4}$ to $10^{-1}$ from bottom to top. 
The ticks at the top represent metallicities, derived from stellar mass, using the mass-metallicity relation at $z=4$--$10$ as presented in \citet{nakajima23}. 
\label{fig:Mdust_Mstar}}
\end{figure*}

Figure~\ref{fig:Mdust_Mstar} shows the relation between the dust and the stellar mass. 
Himiko and CR7 are indicated in red-filled plots. 
We also present a compilation of UV bright ($M_\mathrm{UV}\lesssim-21.5$) and clumpy galaxies at $z\sim7$ with open-red symbols (e.g., \citealt{bowler18, inami22}). 
As a comparison, we display submillimeter galaxies: ALESS \citep{dacunha15}, HUDF \citep{dunlop17}, SPT0311-58 at $z=6.9$ \citep{marrone18}, HFLS3 at $z=6.3$ \citep{riechers13} with magenta plots (see also Figure~\ref{fig:CII_IRluminosity}). 
Similarly, dusty merger systems at $z\sim5$--$7$ (e.g., \citealt{villanueva24, jones20, knudsen17, hashimoto19}) are shown in black filled plots. 
The gray belt shows the local galaxy samples \citep{bourne12}. 
The JWST CEERS samples ($4<z<11$; \citealt{burgarella24}) are shown by the black open circles. 

Himiko and CR7 are consistent with other UV-bright and clumpy galaxies at $z\sim7$, suggesting that these sources might have dust properties similar to those of Himiko and CR7. 
Compared to dusty merger systems at $z\sim5$--$7$, Himiko and CR7 have lower dust mass, even in a similar stellar mass range. 
This trend can be explained by the difference in their SFR and merger phase. 
For example, B14-65666 at $z=7.15$ has $\mathrm{SFR\sim200}~M_\odot~\mathrm{yr^{-1}}$ (\citealt{hashimoto19, jones24b, sugahara25}), which is more than five times higher than that of Himiko or CR7. 
The star formation history from our simulation shows that Himiko-like objects are experiencing early-phase mergers because their SFR is still lower ($\sim30~M_\odot~\mathrm{yr^{-1}}$) than the peak SFR ($\sim80~M_\odot~\mathrm{yr^{-1}}$) at $z\sim5.4$ (Figure~\ref{fig:firstlight}). 
The dust construction of Himiko and CR7 might still be ongoing, and the dust mass remains small compared to the other dusty merger systems. 

Himiko-type objects (UV-bright, clumpy galaxies) contrast sharply with dusty mergers, including some SMGs. 
Previous observations have revealed that some SMGs are merging galaxies with distorted morphologies (e.g., \citealt{tacconi08}), while some others are experiencing disk instability (e.g., \citealt{tadaki18}). 
Himiko-type objects are metal-poor ($Z\sim0.1$--$0.2Z_\odot$), dust-poor ($M_\mathrm{dust}\lesssim9\times10^6~M_\odot$), and low-mass ($M_*\sim10^9~M_\odot$), while dusty mergers are metal-rich ($Z\sim Z_\odot$), dust-rich ($M_\mathrm{dust}\gtrsim10^8~M_\odot$), and massive ($M_*\gtrsim10^{10}~M_*$), aligning with the mass-metallicity and dust-mass-metallicity relations. 
Himiko-type objects show self-similar properties in both stellar mass and dust mass compared to SMGs. 
Himiko-type objects appear as `blue mergers', whereas dusty mergers, heavily affected by dust extinction, appear as `red mergers'. 
Interestingly, we now have two blue mergers, Himiko and CR7, showing spatially extended \lya\ emission.
Furthermore, in Figure~\ref{fig:Mdust_Mstar}, COS-2987030247 is also a clumpy merger galaxy at $z=6.8$ \citep{harikane25}. COS-2987030247 has been detected in \lya\ emission with VLT/X-SHOOTER \citep{laporte17}\footnote{\citet{laporte17} have referred to COS-2987030247 as `COSz2'.}, although with limited spatial information. 
Some other UV bright ($M_{\mathrm{UV}}\sim-21.5$) and clumpy merger galaxies reported in \citet{harikane25} are also detected in Ly$\alpha$ (e.g., COS-788571 at $z=6.9$, COS-1009842 at $z=6.8$; \citealt{endsley21}). 
These samples indicate that the `blue mergers' with extended gas distributions triggered by mergers are important physical mechanisms for emitting extended \lya\ emission, which is also suggested from the FOREVER22 simulation described in Section~\ref{subsec:extended-Lya}. Their low dust content allows \lya\ photons to escape easily. In contrast, although the `red merger' population can also exhibit extended gas distributions due to gravitational interactions, their high dust content prevents \lya\ photons from escaping, making extended \lya\ emission difficult to observe.

\section{Summary and conclusions} \label{sec:conclusions}

We investigate the two bright extended \lya\ objects, Himiko and CR7, at $z=6.6$ using publicly available JWST/NIRCam photometry, NIRSpec IFU spectroscopy, ALMA Band~6, and Subaru NB921 data. 
We also incorporate cosmological simulations, providing information on the connection between galaxy mergers and extended \lya\ emission. 
Our findings and discussions are summarized as follows: 
\begin{enumerate}
    \item 
    Himiko (CR7) is composed of at least five (four) clumps with separations of 2.4--7.3~kpc found by the \oiii$\lambda\lambda4959, 5007$ emission lines in the JWST NIRSpec IFU data. Individual clumps have small velocity offsets with $\Delta v\lesssim 220~\mathrm{km~s^{-1}}$. The stellar masses of Himiko (CR7) estimated by the SED fitting using the NIRCam photometry range in $\log{(M_*/M_\odot)}=8.4$--$9.0$ ($8.3$--$8.8$). Because of the small velocity offsets and equal stellar masses, Himiko and CR7 are major merger systems, which is also suggested by the FirstLight simulations. 
    \item 
    The star formation histories of Himiko and CR7, derived from the SED fitting using NIRCam photometry, indicate an increasing trend over the past $\sim 10~\mathrm{Myr}$. These starbursts may be triggered by mergers and are connected to the UV-bright nature of both Himiko and CR7.
    \item 
    The spatial distributions of UV and H$\alpha$ or \cii$158\mu\mathrm{m}$ and \lya\ align well in Himiko and CR7. Between Himiko-A and B, there is a spatial offset between UV, H$\alpha$, and \cii, \lya ($\sim0\farcs3$), suggesting that the distribution of neutral gas does not coincide with that of ionized gas or stellar clumps in the merging phase. However, there is no significant offset in CR7. 
    \item 
    \oiii$\lambda\lambda4959, 5007$ emission lines of Himiko-C and CR7-A show broad components ($\mathrm{FWHM}\sim250$--$400~\mathrm{km~s^{-1}}$) that can be explained by outflows or tidal features. This signature is related to the spatially extended gas distribution that can be related to the extended Ly$\alpha$ emission of Himiko and CR7. 
    \item 
    Himiko-B shows the broad H$\alpha$ emission line ($\mathrm{FWHM}\sim1000~\mathrm{km\,s^{-1}}$). This broad component is not observed in the forbidden \oiii$\lambda5007$ emission line of Himiko-B, indicating AGN activity with a low-mass black hole of $M_\mathrm{BH}\sim10^{6.6}~M_\odot$. In the $M_\mathrm{BH}$--$M_*$ plane, Himiko-B is located in the higher black hole mass region than those of local galaxies, which is similar to the recent JWST AGN samples. The AGN could also contribute to the bright UV emission from Himiko. 
    \item 
    Himiko and CR7 are not detected in 1 mm dust continuum in the ALMA Band~6 observations with $3\sigma$ upper limits of $S_{\mathrm{158~\micron}}<31~\mu\mathrm{Jy}$ and $S_{\mathrm{158~\micron}}<19~\mu\mathrm{Jy}$, respectively. From these upper limits, we can derive the dust mass upper limits of $M_{\mathrm{dust}}\lesssim9\times10^6~M_\odot$ and $M_{\mathrm{dust}}\lesssim5\times10^6~M_\odot$ for Himiko and CR7, respectively. 
    This result indicates that Himiko and CR7 are dust-poor systems. 
    \item 
    We interpret high-redshift galaxy merger systems with two classifications: `blue mergers' and `red mergers.' 
    Blue mergers, such as Himiko and CR7, are characterized by low stellar mass, low dust mass, and low metallicity. In contrast, red mergers represent the opposite population, including dusty mergers and some submillimeter galaxies.
    Blue mergers show self-similarity in stellar mass and dust mass compared to red mergers and are one of the key mechanisms for explaining the extended Ly$\alpha$ emission, like Himiko and CR7. 
\end{enumerate}

\section{acknowledgments} \label{ack}
We thank the anonymous referee for careful reading and valuable comments that improved our manuscript.
We thank Rychard Bouwens, Rebecca A. A. Bowler, Takuya Hashimoto, Jacqueline Hodge, Ryota Ikeda, Yuta Kageura, Kentaro Nagamine, Takayuki Saito, Yoichi Tamura, Daisuke Toyouchi, Takafumi Tsukui, Hideki Umehata, Joris Witstok, and Jorge A. Zavala for the valuable comments and discussions. 
We are grateful to the ALMA Help Desk for calibrating the Cycle 1 archival data and for answering our questions. 

This work is based in part on observations made with the NASA/ESA/CSA James Webb Space Telescope. The data were obtained from the Mikulski Archive for Space Telescopes at the Space Telescope Science Institute, which is operated by the Association of Universities for Research in Astronomy, Inc., under NASA contract NAS 5-03127 for JWST. These observations are associated with programs GTO-1215, GTO-1217 (GA-NIFS), GO-1727 (COSMOS-Web), and GO-1837 (PRIMER). 
The authors acknowledge the GA-NIFS, COSMOS-Web, and PRIMER teams led by Santiago Arribas \& Roberto Maiolino, Jeyhan Kartaltepe \& Caitlin M. Casey, and James S. Dunlop, respectively, for developing their observation programs. 
The authors also acknowledge the TEMPLATE team led by Jane R. Rigby for establishing the JWST/NIRSpec IFU reduction procedures. 
The JWST/NIRSpec IFU data presented in this article were obtained from the Mikulski Archive for Space Telescopes (MAST) at the Space Telescope Science Institute. The specific observations analyzed can be accessed via \dataset[doi: 10.17909/twz5-hd10]{https://doi.org/10.17909/twz5-hd10}.

Some of the data products presented herein were retrieved from the Dawn JWST Archive (DJA). DJA is an initiative of the Cosmic Dawn Center (DAWN), which is funded by the Danish National Research Foundation under grant DNRF140. 

This paper makes use of the following ALMA data: ADS/JAO.ALMA\#2011.0.00115.S, \#2012.1.00033.S, and \#2015.1.00122.S. ALMA is a partnership of ESO (representing its member states), NSF (USA) and NINS (Japan), together with NRC (Canada), NSTC and ASIAA (Taiwan), and KASI (Republic of Korea), in cooperation with the Republic of Chile. The Joint ALMA Observatory is operated by ESO, AUI/NRAO and NAOJ. 

The Hyper Suprime-Cam (HSC) collaboration includes the astronomical communities of Japan and Taiwan, and Princeton University. The HSC instrumentation and software were developed by the National Astronomical Observatory of Japan (NAOJ), the Kavli Institute for the Physics and Mathematics of the Universe (Kavli IPMU), the University of Tokyo, the High Energy Accelerator Research Organization (KEK), the Academia Sinica Institute for Astronomy and Astrophysics in Taiwan (ASIAA), and Princeton University. Funding was contributed by the FIRST program from the Japanese Cabinet Office, the Ministry of Education, Culture, Sports, Science and Technology (MEXT), the Japan Society for the Promotion of Science (JSPS), Japan Science and Technology Agency (JST), the Toray Science Foundation, NAOJ, Kavli IPMU, KEK, ASIAA, and Princeton University. 

This paper makes use of software developed for Vera C. Rubin Observatory. We thank the Rubin Observatory for making their code available as free software at \url{http://pipelines.lsst.io/}.

This paper is based on data collected at the Subaru Telescope and retrieved from the HSC data archive system, which is operated by the Subaru Telescope and Astronomy Data Center (ADC) at NAOJ. Data analysis was in part carried out with the cooperation of Center for Computational Astrophysics (CfCA), NAOJ. We are honored and grateful for the opportunity of observing the Universe from Maunakea, which has the cultural, historical and natural significance in Hawaii. 

The Pan-STARRS1 Surveys (PS1) and the PS1 public science archive have been made possible through contributions by the Institute for Astronomy, the University of Hawaii, the Pan-STARRS Project Office, the Max Planck Society and its participating institutes, the Max Planck Institute for Astronomy, Heidelberg, and the Max Planck Institute for Extraterrestrial Physics, Garching, The Johns Hopkins University, Durham University, the University of Edinburgh, the Queen’s University Belfast, the Harvard-Smithsonian Center for Astrophysics, the Las Cumbres Observatory Global Telescope Network Incorporated, the National Central University of Taiwan, the Space Telescope Science Institute, the National Aeronautics and Space Administration under grant No. NNX08AR22G issued through the Planetary Science Division of the NASA Science Mission Directorate, the National Science Foundation grant No. AST-1238877, the University of Maryland, Eotvos Lorand University (ELTE), the Los Alamos National Laboratory, and the Gordon and Betty Moore Foundation. 

This publication is based upon work supported by the World Premier International Research Center Initiative (WPI Initiative), MEXT, Japan, KAKENHI (20H00180, 23KJ2148, 21H04489, 22H04939, 23K20035, 24H00004, 23KJ0728) through the Japan Society for the Promotion of Science, and JST FOREST Program (JP-MJFR202Z). 
This work was supported by the joint research program of the Institute for Cosmic Ray Research (ICRR), University of Tokyo. 
Numerical analyses of FirstLight simulations were carried out on the analysis servers at the Center for Computational Astrophysics, National Astronomical Observatory of Japan. 
The English writing in this paper was improved with ChatGPT and Grammarly, while no sentences were generated by these softwares from scratch. 

\vspace{5mm}
\facilities{JWST (NIRSpec, NIRCam), ALMA (Band 6), Subaru (Hyper Suprime-Cam)}
\software{Astropy \citep{astropy:2013, astropy:2018, astropy:2022},  
          CASA \citep{casa22}, 
          Matplotlib \citep{hunter07},
          NumPy \citep{harris20}, 
          pandas \citep{mckinney2010data}
          Prospector \citep{leja17, johnson21}, 
          PyNeb \citep{luridiana15},
          SciPy \citep{virtanen20}, 
          SEP \citep{barbary16}, 
          }
\bibliography{reference}{}

\begin{thebibliography}{}
\expandafter\ifx\csname natexlab\endcsname\relax\def\natexlab#1{#1}\fi
\providecommand{\url}[1]{\href{#1}{#1}}
\providecommand{\dodoi}[1]{doi:~\href{http://doi.org/#1}{\nolinkurl{#1}}}
\providecommand{\doeprint}[1]{\href{http://ascl.net/#1}{\nolinkurl{http://ascl.net/#1}}}
\providecommand{\doarXiv}[1]{\href{https://arxiv.org/abs/#1}{\nolinkurl{https://arxiv.org/abs/#1}}}

\bibitem[{{Adamo} {et~al.}(2024){Adamo}, {Bradley}, {Vanzella}, {Claeyssens}, {Welch}, {Diego}, {Mahler}, {Oguri}, {Sharon}, {Abdurro'uf}, {Hsiao}, {Xu}, {Messa}, {Lassen}, {Zackrisson}, {Brammer}, {Coe}, {Kokorev}, {Ricotti}, {Zitrin}, {Fujimoto}, {Inoue}, {Resseguier}, {Rigby}, {Jim{\'e}nez-Teja}, {Windhorst}, {Hashimoto}, \& {Tamura}}]{adamo24}
{Adamo}, A., {Bradley}, L.~D., {Vanzella}, E., {et~al.} 2024, \nat, 632, 513, \dodoi{10.1038/s41586-024-07703-7}

\bibitem[{{Ahumada} {et~al.}(2020){Ahumada}, {Allende Prieto}, {Almeida}, {Anders}, {Anderson}, {Andrews}, {Anguiano}, {Arcodia}, {Armengaud}, {Aubert}, {Avila}, {Avila-Reese}, {Badenes}, {Balland}, {Barger}, {Barrera-Ballesteros}, {Basu}, {Bautista}, {Beaton}, {Beers}, {Benavides}, {Bender}, {Bernardi}, {Bershady}, {Beutler}, {Bidin}, {Bird}, {Bizyaev}, {Blanc}, {Blanton}, {Boquien}, {Borissova}, {Bovy}, {Brandt}, {Brinkmann}, {Brownstein}, {Bundy}, {Bureau}, {Burgasser}, {Burtin}, {Cano-D{\'\i}az}, {Capasso}, {Cappellari}, {Carrera}, {Chabanier}, {Chaplin}, {Chapman}, {Cherinka}, {Chiappini}, {Doohyun Choi}, {Chojnowski}, {Chung}, {Clerc}, {Coffey}, {Comerford}, {Comparat}, {da Costa}, {Cousinou}, {Covey}, {Crane}, {Cunha}, {Ilha}, {Dai}, {Damsted}, {Darling}, {Davidson}, {Davies}, {Dawson}, {De}, {de la Macorra}, {De Lee}, {Queiroz}, {Deconto Machado}, {de la Torre}, {Dell'Agli}, {du Mas des Bourboux}, {Diamond-Stanic}, {Dillon}, {Donor}, {Drory}, {Duckworth}, {Dwelly}, {Ebelke}, {Eftekharzadeh}, {Davis
  Eigenbrot}, {Elsworth}, {Eracleous}, {Erfanianfar}, {Escoffier}, {Fan}, {Farr}, {Fern{\'a}ndez-Trincado}, {Feuillet}, {Finoguenov}, {Fofie}, {Fraser-McKelvie}, {Frinchaboy}, {Fromenteau}, {Fu}, {Galbany}, {Garcia}, {Garc{\'\i}a-Hern{\'a}ndez}, {Garma Oehmichen}, {Ge}, {Geimba Maia}, {Geisler}, {Gelfand}, {Goddy}, {Gonzalez-Perez}, {Grabowski}, {Green}, {Grier}, {Guo}, {Guy}, {Harding}, {Hasselquist}, {Hawken}, {Hayes}, {Hearty}, {Hekker}, {Hogg}, {Holtzman}, {Horta}, {Hou}, {Hsieh}, {Huber}, {Hunt}, {Ider Chitham}, {Imig}, {Jaber}, {Jimenez Angel}, {Johnson}, {Jones}, {J{\"o}nsson}, {Jullo}, {Kim}, {Kinemuchi}, {Kirkpatrick}, {Kite}, {Klaene}, {Kneib}, {Kollmeier}, {Kong}, {Kounkel}, {Krishnarao}, {Lacerna}, {Lan}, {Lane}, {Law}, {Le Goff}, {Leung}, {Lewis}, {Li}, {Lian}, {Lin}, {Long}, {Longa-Pe{\~n}a}, {Lundgren}, {Lyke}, {Mackereth}, {MacLeod}, {Majewski}, {Manchado}, {Maraston}, {Martini}, {Masseron}, {Masters}, {Mathur}, {McDermid}, {Merloni}, {Merrifield}, {M{\'e}sz{\'a}ros}, {Miglio}, {Minniti},
  {Minsley}, {Miyaji}, {Mohammad}, {Mosser}, {Mueller}, {Muna}, {Mu{\~n}oz-Guti{\'e}rrez}, {Myers}, {Nadathur}, {Nair}, {Nandra}, {Correa do Nascimento}, {Nevin}, {Newman}, {Nidever}, {Nitschelm}, {Noterdaeme}, {O'Connell}, {Olmstead}, {Oravetz}, {Oravetz}, {Osorio}, {Pace}, {Padilla}, {Palanque-Delabrouille}, {Palicio}, {Pan}, {Pan}, {Parker}, {Paviot}, {Peirani}, {Ram{\'r}ez}, {Penny}, {Percival}, {Perez-Fournon}, {P{\'e}rez-R{\`a}fols}, {Petitjean}, {Pieri}, {Pinsonneault}, {Poovelil}, {Povick}, {Prakash}, {Price-Whelan}, {Raddick}, {Raichoor}, {Ray}, {Rembold}, {Rezaie}, {Riffel}, {Riffel}, {Rix}, {Robin}, {Roman-Lopes}, {Rom{\'a}n-Z{\'u}{\~n}iga}, {Rose}, {Ross}, {Rossi}, {Rowlands}, {Rubin}, {Salvato}, {S{\'a}nchez}, {S{\'a}nchez-Menguiano}, {S{\'a}nchez-Gallego}, {Sayres}, {Schaefer}, {Schiavon}, {Schimoia}, {Schlafly}, {Schlegel}, {Schneider}, {Schultheis}, {Schwope}, {Seo}, {Serenelli}, {Shafieloo}, {Shamsi}, {Shao}, {Shen}, {Shetrone}, {Shirley}, {Silva Aguirre}, {Simon}, {Skrutskie}, {Slosar},
  {Smethurst}, {Sobeck}, {Sodi}, {Souto}, {Stark}, {Stassun}, {Steinmetz}, {Stello}, {Stermer}, {Storchi-Bergmann}, {Streblyanska}, {Stringfellow}, {Stutz}, {Su{\'a}rez}, {Sun}, {Taghizadeh-Popp}, {Talbot}, {Tayar}, {Thakar}, {Theriault}, {Thomas}, {Thomas}, {Tinker}, {Tojeiro}, {Toledo}, {Tremonti}, {Troup}, {Tuttle}, {Unda-Sanzana}, {Valentini}, {Vargas-Gonz{\'a}lez}, {Vargas-Maga{\~n}a}, {V{\'a}zquez-Mata}, {Vivek}, {Wake}, {Wang}, {Weaver}, {Weijmans}, {Wild}, {Wilson}, {Wilson}, {Wolthuis}, {Wood-Vasey}, {Yan}, {Yang}, {Y{\`e}che}, {Zamora}, {Zarrouk}, {Zasowski}, {Zhang}, {Zhao}, {Zhao}, {Zheng}, {Zheng}, {Zhu}, \& {Zou}}]{ahumada20}
{Ahumada}, R., {Allende Prieto}, C., {Almeida}, A., {et~al.} 2020, \apjs, 249, 3, \dodoi{10.3847/1538-4365/ab929e}

\bibitem[{{Aihara} {et~al.}(2018{\natexlab{a}}){Aihara}, {Arimoto}, {Armstrong}, {Arnouts}, {Bahcall}, {Bickerton}, {Bosch}, {Bundy}, {Capak}, {Chan}, {Chiba}, {Coupon}, {Egami}, {Enoki}, {Finet}, {Fujimori}, {Fujimoto}, {Furusawa}, {Furusawa}, {Goto}, {Goulding}, {Greco}, {Greene}, {Gunn}, {Hamana}, {Harikane}, {Hashimoto}, {Hattori}, {Hayashi}, {Hayashi}, {He{\l}miniak}, {Higuchi}, {Hikage}, {Ho}, {Hsieh}, {Huang}, {Huang}, {Ikeda}, {Imanishi}, {Inoue}, {Iwasawa}, {Iwata}, {Jaelani}, {Jian}, {Kamata}, {Karoji}, {Kashikawa}, {Katayama}, {Kawanomoto}, {Kayo}, {Koda}, {Koike}, {Kojima}, {Komiyama}, {Konno}, {Koshida}, {Koyama}, {Kusakabe}, {Leauthaud}, {Lee}, {Lin}, {Lin}, {Lupton}, {Mandelbaum}, {Matsuoka}, {Medezinski}, {Mineo}, {Miyama}, {Miyatake}, {Miyazaki}, {Momose}, {More}, {More}, {Moritani}, {Moriya}, {Morokuma}, {Mukae}, {Murata}, {Murayama}, {Nagao}, {Nakata}, {Niida}, {Niikura}, {Nishizawa}, {Obuchi}, {Oguri}, {Oishi}, {Okabe}, {Okamoto}, {Okura}, {Ono}, {Onodera}, {Onoue}, {Osato}, {Ouchi},
  {Price}, {Pyo}, {Sako}, {Sawicki}, {Shibuya}, {Shimasaku}, {Shimono}, {Shirasaki}, {Silverman}, {Simet}, {Speagle}, {Spergel}, {Strauss}, {Sugahara}, {Sugiyama}, {Suto}, {Suyu}, {Suzuki}, {Tait}, {Takada}, {Takata}, {Tamura}, {Tanaka}, {Tanaka}, {Tanaka}, {Tanaka}, {Terai}, {Terashima}, {Toba}, {Tominaga}, {Toshikawa}, {Turner}, {Uchida}, {Uchiyama}, {Umetsu}, {Uraguchi}, {Urata}, {Usuda}, {Utsumi}, {Wang}, {Wang}, {Wong}, {Yabe}, {Yamada}, {Yamanoi}, {Yasuda}, {Yeh}, {Yonehara}, \& {Yuma}}]{aihara18a}
{Aihara}, H., {Arimoto}, N., {Armstrong}, R., {et~al.} 2018{\natexlab{a}}, \pasj, 70, S4, \dodoi{10.1093/pasj/psx066}

\bibitem[{{Aihara} {et~al.}(2018{\natexlab{b}}){Aihara}, {Armstrong}, {Bickerton}, {Bosch}, {Coupon}, {Furusawa}, {Hayashi}, {Ikeda}, {Kamata}, {Karoji}, {Kawanomoto}, {Koike}, {Komiyama}, {Lang}, {Lupton}, {Mineo}, {Miyatake}, {Miyazaki}, {Morokuma}, {Obuchi}, {Oishi}, {Okura}, {Price}, {Takata}, {Tanaka}, {Tanaka}, {Tanaka}, {Uchida}, {Uraguchi}, {Utsumi}, {Wang}, {Yamada}, {Yamanoi}, {Yasuda}, {Arimoto}, {Chiba}, {Finet}, {Fujimori}, {Fujimoto}, {Furusawa}, {Goto}, {Goulding}, {Gunn}, {Harikane}, {Hattori}, {Hayashi}, {He{\l}miniak}, {Higuchi}, {Hikage}, {Ho}, {Hsieh}, {Huang}, {Huang}, {Imanishi}, {Iwata}, {Jaelani}, {Jian}, {Kashikawa}, {Katayama}, {Kojima}, {Konno}, {Koshida}, {Kusakabe}, {Leauthaud}, {Lee}, {Lin}, {Lin}, {Mandelbaum}, {Matsuoka}, {Medezinski}, {Miyama}, {Momose}, {More}, {More}, {Mukae}, {Murata}, {Murayama}, {Nagao}, {Nakata}, {Niida}, {Niikura}, {Nishizawa}, {Oguri}, {Okabe}, {Ono}, {Onodera}, {Onoue}, {Ouchi}, {Pyo}, {Shibuya}, {Shimasaku}, {Simet}, {Speagle}, {Spergel}, {Strauss},
  {Sugahara}, {Sugiyama}, {Suto}, {Suzuki}, {Tait}, {Takada}, {Terai}, {Toba}, {Turner}, {Uchiyama}, {Umetsu}, {Urata}, {Usuda}, {Yeh}, \& {Yuma}}]{aihara18b}
{Aihara}, H., {Armstrong}, R., {Bickerton}, S., {et~al.} 2018{\natexlab{b}}, \pasj, 70, S8, \dodoi{10.1093/pasj/psx081}

\bibitem[{{Aihara} {et~al.}(2019){Aihara}, {AlSayyad}, {Ando}, {Armstrong}, {Bosch}, {Egami}, {Furusawa}, {Furusawa}, {Goulding}, {Harikane}, {Hikage}, {Ho}, {Hsieh}, {Huang}, {Ikeda}, {Imanishi}, {Ito}, {Iwata}, {Jaelani}, {Kakuma}, {Kawana}, {Kikuta}, {Kobayashi}, {Koike}, {Komiyama}, {Li}, {Liang}, {Lin}, {Luo}, {Lupton}, {Lust}, {MacArthur}, {Matsuoka}, {Mineo}, {Miyatake}, {Miyazaki}, {More}, {Murata}, {Namiki}, {Nishizawa}, {Oguri}, {Okabe}, {Okamoto}, {Okura}, {Ono}, {Onodera}, {Onoue}, {Osato}, {Ouchi}, {Shibuya}, {Strauss}, {Sugiyama}, {Suto}, {Takada}, {Takagi}, {Takata}, {Takita}, {Tanaka}, {Terai}, {Toba}, {Uchiyama}, {Utsumi}, {Wang}, {Wang}, \& {Yamada}}]{aihara19}
{Aihara}, H., {AlSayyad}, Y., {Ando}, M., {et~al.} 2019, \pasj, 71, 114, \dodoi{10.1093/pasj/psz103}

\bibitem[{{Aihara} {et~al.}(2022){Aihara}, {AlSayyad}, {Ando}, {Armstrong}, {Bosch}, {Egami}, {Furusawa}, {Furusawa}, {Harasawa}, {Harikane}, {Hsieh}, {Ikeda}, {Ito}, {Iwata}, {Kodama}, {Koike}, {Kokubo}, {Komiyama}, {Li}, {Liang}, {Lin}, {Lupton}, {Lust}, {MacArthur}, {Mawatari}, {Mineo}, {Miyatake}, {Miyazaki}, {More}, {Morishima}, {Murayama}, {Nakajima}, {Nakata}, {Nishizawa}, {Oguri}, {Okabe}, {Okura}, {Ono}, {Osato}, {Ouchi}, {Pan}, {Plazas Malag{\'o}n}, {Price}, {Reed}, {Rykoff}, {Shibuya}, {Simunovic}, {Strauss}, {Sugimori}, {Suto}, {Suzuki}, {Takada}, {Takagi}, {Takata}, {Takita}, {Tanaka}, {Tang}, {Taranu}, {Terai}, {Toba}, {Turner}, {Uchiyama}, {Vijarnwannaluk}, {Waters}, {Yamada}, {Yamamoto}, \& {Yamashita}}]{aihara22}
---. 2022, \pasj, 74, 247, \dodoi{10.1093/pasj/psab122}

\bibitem[{{Akaike}(1974)}]{akaike74}
{Akaike}, H. 1974, IEEE Transactions on Automatic Control, 19, 716

\bibitem[{{Algera} {et~al.}(2025){Algera}, {Rowland}, {Stefanon}, {Palla}, {Sommovigo}, {Inami}, {Bouwens}, {Aravena}, {Bowler}, {Dayal}, {De Looze}, {Ferrara}, {Fisher}, {Graziani}, {Gulis}, {Heintz}, {Hodge}, {van Leeuwen}, {Pallottini}, {Phillips}, {Schouws}, {Smit}, {Stark}, \& {van der Werf}}]{algera25}
{Algera}, H., {Rowland}, L., {Stefanon}, M., {et~al.} 2025, arXiv e-prints, arXiv:2501.10508, \dodoi{10.48550/arXiv.2501.10508}

\bibitem[{{Andrews} \& {Martini}(2013)}]{andrews13}
{Andrews}, B.~H., \& {Martini}, P. 2013, \apj, 765, 140, \dodoi{10.1088/0004-637X/765/2/140}

\bibitem[{{Appleton} {et~al.}(2013){Appleton}, {Guillard}, {Boulanger}, {Cluver}, {Ogle}, {Falgarone}, {Pineau des For{\^e}ts}, {O'Sullivan}, {Duc}, {Gallagher}, {Gao}, {Jarrett}, {Konstantopoulos}, {Lisenfeld}, {Lord}, {Lu}, {Peterson}, {Struck}, {Sturm}, {Tuffs}, {Valchanov}, {van der Werf}, \& {Xu}}]{appleton13}
{Appleton}, P.~N., {Guillard}, P., {Boulanger}, F., {et~al.} 2013, \apj, 777, 66, \dodoi{10.1088/0004-637X/777/1/66}

\bibitem[{{Arribas} {et~al.}(2024){Arribas}, {Perna}, {Rodr{\'\i}guez Del Pino}, {Lamperti}, {D'Eugenio}, {P{\'e}rez-Gonz{\'a}lez}, {Jones}, {Crespo G{\'o}mez}, {Curti}, {Lim}, {{\'A}lvarez-M{\'a}rquez}, {Bunker}, {Carniani}, {Charlot}, {Jakobsen}, {Maiolino}, {{\"U}bler}, {Willott}, {B{\"o}ker}, {Chevallard}, {Circosta}, {Cresci}, {Kumari}, {Parlanti}, {Scholtz}, {Venturi}, \& {Witstok}}]{arribas24}
{Arribas}, S., {Perna}, M., {Rodr{\'\i}guez Del Pino}, B., {et~al.} 2024, \aap, 688, A146, \dodoi{10.1051/0004-6361/202348824}

\bibitem[{{Asada} {et~al.}(2024){Asada}, {Sawicki}, {Abraham}, {Brada{\v{c}}}, {Brammer}, {Desprez}, {Estrada-Carpenter}, {Iyer}, {Martis}, {Matharu}, {Mowla}, {Muzzin}, {Noirot}, {Sarrouh}, {Strait}, {Willott}, \& {Harshan}}]{asada24}
{Asada}, Y., {Sawicki}, M., {Abraham}, R., {et~al.} 2024, \mnras, 527, 11372, \dodoi{10.1093/mnras/stad3902}

\bibitem[{{Asplund} {et~al.}(2009){Asplund}, {Grevesse}, {Sauval}, \& {Scott}}]{asplund09}
{Asplund}, M., {Grevesse}, N., {Sauval}, A.~J., \& {Scott}, P. 2009, \araa, 47, 481, \dodoi{10.1146/annurev.astro.46.060407.145222}

\bibitem[{{Astropy Collaboration} {et~al.}(2013){Astropy Collaboration}, {Robitaille}, {Tollerud}, {Greenfield}, {Droettboom}, {Bray}, {Aldcroft}, {Davis}, {Ginsburg}, {Price-Whelan}, {Kerzendorf}, {Conley}, {Crighton}, {Barbary}, {Muna}, {Ferguson}, {Grollier}, {Parikh}, {Nair}, {Unther}, {Deil}, {Woillez}, {Conseil}, {Kramer}, {Turner}, {Singer}, {Fox}, {Weaver}, {Zabalza}, {Edwards}, {Azalee Bostroem}, {Burke}, {Casey}, {Crawford}, {Dencheva}, {Ely}, {Jenness}, {Labrie}, {Lim}, {Pierfederici}, {Pontzen}, {Ptak}, {Refsdal}, {Servillat}, \& {Streicher}}]{astropy:2013}
{Astropy Collaboration}, {Robitaille}, T.~P., {Tollerud}, E.~J., {et~al.} 2013, \aap, 558, A33, \dodoi{10.1051/0004-6361/201322068}

\bibitem[{{Astropy Collaboration} {et~al.}(2018){Astropy Collaboration}, {Price-Whelan}, {Sip{\H{o}}cz}, {G{\"u}nther}, {Lim}, {Crawford}, {Conseil}, {Shupe}, {Craig}, {Dencheva}, {Ginsburg}, {Vand erPlas}, {Bradley}, {P{\'e}rez-Su{\'a}rez}, {de Val-Borro}, {Aldcroft}, {Cruz}, {Robitaille}, {Tollerud}, {Ardelean}, {Babej}, {Bach}, {Bachetti}, {Bakanov}, {Bamford}, {Barentsen}, {Barmby}, {Baumbach}, {Berry}, {Biscani}, {Boquien}, {Bostroem}, {Bouma}, {Brammer}, {Bray}, {Breytenbach}, {Buddelmeijer}, {Burke}, {Calderone}, {Cano Rodr{\'\i}guez}, {Cara}, {Cardoso}, {Cheedella}, {Copin}, {Corrales}, {Crichton}, {D'Avella}, {Deil}, {Depagne}, {Dietrich}, {Donath}, {Droettboom}, {Earl}, {Erben}, {Fabbro}, {Ferreira}, {Finethy}, {Fox}, {Garrison}, {Gibbons}, {Goldstein}, {Gommers}, {Greco}, {Greenfield}, {Groener}, {Grollier}, {Hagen}, {Hirst}, {Homeier}, {Horton}, {Hosseinzadeh}, {Hu}, {Hunkeler}, {Ivezi{\'c}}, {Jain}, {Jenness}, {Kanarek}, {Kendrew}, {Kern}, {Kerzendorf}, {Khvalko}, {King}, {Kirkby}, {Kulkarni},
  {Kumar}, {Lee}, {Lenz}, {Littlefair}, {Ma}, {Macleod}, {Mastropietro}, {McCully}, {Montagnac}, {Morris}, {Mueller}, {Mumford}, {Muna}, {Murphy}, {Nelson}, {Nguyen}, {Ninan}, {N{\"o}the}, {Ogaz}, {Oh}, {Parejko}, {Parley}, {Pascual}, {Patil}, {Patil}, {Plunkett}, {Prochaska}, {Rastogi}, {Reddy Janga}, {Sabater}, {Sakurikar}, {Seifert}, {Sherbert}, {Sherwood-Taylor}, {Shih}, {Sick}, {Silbiger}, {Singanamalla}, {Singer}, {Sladen}, {Sooley}, {Sornarajah}, {Streicher}, {Teuben}, {Thomas}, {Tremblay}, {Turner}, {Terr{\'o}n}, {van Kerkwijk}, {de la Vega}, {Watkins}, {Weaver}, {Whitmore}, {Woillez}, {Zabalza}, \& {Astropy Contributors}}]{astropy:2018}
{Astropy Collaboration}, {Price-Whelan}, A.~M., {Sip{\H{o}}cz}, B.~M., {et~al.} 2018, \aj, 156, 123, \dodoi{10.3847/1538-3881/aabc4f}

\bibitem[{{Astropy Collaboration} {et~al.}(2022){Astropy Collaboration}, {Price-Whelan}, {Lim}, {Earl}, {Starkman}, {Bradley}, {Shupe}, {Patil}, {Corrales}, {Brasseur}, {N{"o}the}, {Donath}, {Tollerud}, {Morris}, {Ginsburg}, {Vaher}, {Weaver}, {Tocknell}, {Jamieson}, {van Kerkwijk}, {Robitaille}, {Merry}, {Bachetti}, {G{"u}nther}, {Aldcroft}, {Alvarado-Montes}, {Archibald}, {B{'o}di}, {Bapat}, {Barentsen}, {Baz{'a}n}, {Biswas}, {Boquien}, {Burke}, {Cara}, {Cara}, {Conroy}, {Conseil}, {Craig}, {Cross}, {Cruz}, {D'Eugenio}, {Dencheva}, {Devillepoix}, {Dietrich}, {Eigenbrot}, {Erben}, {Ferreira}, {Foreman-Mackey}, {Fox}, {Freij}, {Garg}, {Geda}, {Glattly}, {Gondhalekar}, {Gordon}, {Grant}, {Greenfield}, {Groener}, {Guest}, {Gurovich}, {Handberg}, {Hart}, {Hatfield-Dodds}, {Homeier}, {Hosseinzadeh}, {Jenness}, {Jones}, {Joseph}, {Kalmbach}, {Karamehmetoglu}, {Ka{l}uszy{'n}ski}, {Kelley}, {Kern}, {Kerzendorf}, {Koch}, {Kulumani}, {Lee}, {Ly}, {Ma}, {MacBride}, {Maljaars}, {Muna}, {Murphy}, {Norman}, {O'Steen},
  {Oman}, {Pacifici}, {Pascual}, {Pascual-Granado}, {Patil}, {Perren}, {Pickering}, {Rastogi}, {Roulston}, {Ryan}, {Rykoff}, {Sabater}, {Sakurikar}, {Salgado}, {Sanghi}, {Saunders}, {Savchenko}, {Schwardt}, {Seifert-Eckert}, {Shih}, {Jain}, {Shukla}, {Sick}, {Simpson}, {Singanamalla}, {Singer}, {Singhal}, {Sinha}, {Sip{H{o}}cz}, {Spitler}, {Stansby}, {Streicher}, {{{S}}umak}, {Swinbank}, {Taranu}, {Tewary}, {Tremblay}, {Val-Borro}, {Van Kooten}, {Vasovi{'c}}, {Verma}, {de Miranda Cardoso}, {Williams}, {Wilson}, {Winkel}, {Wood-Vasey}, {Xue}, {Yoachim}, {Zhang}, {Zonca}, \& {Astropy Project Contributors}}]{astropy:2022}
{Astropy Collaboration}, {Price-Whelan}, A.~M., {Lim}, P.~L., {et~al.} 2022, \apj, 935, 167, \dodoi{10.3847/1538-4357/ac7c74}

\bibitem[{{Aver} {et~al.}(2011){Aver}, {Olive}, \& {Skillman}}]{aver11}
{Aver}, E., {Olive}, K.~A., \& {Skillman}, E.~D. 2011, \jcap, 2011, 043, \dodoi{10.1088/1475-7516/2011/03/043}

\bibitem[{{Baldwin} {et~al.}(1981){Baldwin}, {Phillips}, \& {Terlevich}}]{baldwin81}
{Baldwin}, J.~A., {Phillips}, M.~M., \& {Terlevich}, R. 1981, \pasp, 93, 5, \dodoi{10.1086/130766}

\bibitem[{{Balser}(2006)}]{balser06}
{Balser}, D.~S. 2006, \aj, 132, 2326, \dodoi{10.1086/508515}

\bibitem[{Barbary(2016)}]{barbary16}
Barbary, K. 2016, Journal of Open Source Software, 1, 58, \dodoi{10.21105/joss.00058}

\bibitem[{{Behroozi} {et~al.}(2019){Behroozi}, {Wechsler}, {Hearin}, \& {Conroy}}]{behroozi19}
{Behroozi}, P., {Wechsler}, R.~H., {Hearin}, A.~P., \& {Conroy}, C. 2019, \mnras, 488, 3143, \dodoi{10.1093/mnras/stz1182}

\bibitem[{{Bertin} \& {Arnouts}(1996)}]{bertin96}
{Bertin}, E., \& {Arnouts}, S. 1996, \aaps, 117, 393, \dodoi{10.1051/aas:1996164}

\bibitem[{{Blain} \& {Longair}(1993)}]{blain93}
{Blain}, A.~W., \& {Longair}, M.~S. 1993, \mnras, 264, 509, \dodoi{10.1093/mnras/264.2.509}

\bibitem[{{Blain} {et~al.}(2002){Blain}, {Smail}, {Ivison}, {Kneib}, \& {Frayer}}]{blain02}
{Blain}, A.~W., {Smail}, I., {Ivison}, R.~J., {Kneib}, J.~P., \& {Frayer}, D.~T. 2002, \physrep, 369, 111, \dodoi{10.1016/S0370-1573(02)00134-5}

\bibitem[{{B{\"o}ker} {et~al.}(2022){B{\"o}ker}, {Arribas}, {L{\"u}tzgendorf}, {Alves de Oliveira}, {Beck}, {Birkmann}, {Bunker}, {Charlot}, {de Marchi}, {Ferruit}, {Giardino}, {Jakobsen}, {Kumari}, {L{\'o}pez-Caniego}, {Maiolino}, {Manjavacas}, {Marston}, {Moseley}, {Muzerolle}, {Ogle}, {Pirzkal}, {Rauscher}, {Rawle}, {Rix}, {Sabbi}, {Sargent}, {Sirianni}, {te Plate}, {Valenti}, {Willott}, \& {Zeidler}}]{boker22}
{B{\"o}ker}, T., {Arribas}, S., {L{\"u}tzgendorf}, N., {et~al.} 2022, \aap, 661, A82, \dodoi{10.1051/0004-6361/202142589}

\bibitem[{{Bourne} {et~al.}(2012){Bourne}, {Maddox}, {Dunne}, {Auld}, {Baes}, {Baldry}, {Bonfield}, {Cooray}, {Croom}, {Dariush}, {de Zotti}, {Driver}, {Dye}, {Eales}, {Gomez}, {Gonz{\'a}lez-Nuevo}, {Hopkins}, {Ibar}, {Jarvis}, {Lapi}, {Madore}, {Micha{\l}owski}, {Pohlen}, {Popescu}, {Rigby}, {Seibert}, {Smith}, {Tuffs}, {van der Werf}, {Brough}, {Buttiglione}, {Cava}, {Clements}, {Conselice}, {Fritz}, {Hopwood}, {Ivison}, {Jones}, {Kelvin}, {Liske}, {Loveday}, {Norberg}, {Robotham}, {Rodighiero}, \& {Temi}}]{bourne12}
{Bourne}, N., {Maddox}, S.~J., {Dunne}, L., {et~al.} 2012, \mnras, 421, 3027, \dodoi{10.1111/j.1365-2966.2012.20528.x}

\bibitem[{{Bouwens} {et~al.}(2015){Bouwens}, {Illingworth}, {Oesch}, {Trenti}, {Labb{\'e}}, {Bradley}, {Carollo}, {van Dokkum}, {Gonzalez}, {Holwerda}, {Franx}, {Spitler}, {Smit}, \& {Magee}}]{bouwens15}
{Bouwens}, R.~J., {Illingworth}, G.~D., {Oesch}, P.~A., {et~al.} 2015, \apj, 803, 34, \dodoi{10.1088/0004-637X/803/1/34}

\bibitem[{{Bouwens} {et~al.}(2022){Bouwens}, {Smit}, {Schouws}, {Stefanon}, {Bowler}, {Endsley}, {Gonzalez}, {Inami}, {Stark}, {Oesch}, {Hodge}, {Aravena}, {da Cunha}, {Dayal}, {de Looze}, {Ferrara}, {Fudamoto}, {Graziani}, {Li}, {Nanayakkara}, {Pallottini}, {Schneider}, {Sommovigo}, {Topping}, {van der Werf}, {Algera}, {Barrufet}, {Hygate}, {Labb{\'e}}, {Riechers}, \& {Witstok}}]{bouwens22}
{Bouwens}, R.~J., {Smit}, R., {Schouws}, S., {et~al.} 2022, \apj, 931, 160, \dodoi{10.3847/1538-4357/ac5a4a}

\bibitem[{{Bowler} {et~al.}(2018){Bowler}, {Bourne}, {Dunlop}, {McLure}, \& {McLeod}}]{bowler18}
{Bowler}, R.~A.~A., {Bourne}, N., {Dunlop}, J.~S., {McLure}, R.~J., \& {McLeod}, D.~J. 2018, \mnras, 481, 1631, \dodoi{10.1093/mnras/sty2368}

\bibitem[{{Bowler} {et~al.}(2017{\natexlab{a}}){Bowler}, {Dunlop}, {McLure}, \& {McLeod}}]{bowler17a}
{Bowler}, R.~A.~A., {Dunlop}, J.~S., {McLure}, R.~J., \& {McLeod}, D.~J. 2017{\natexlab{a}}, \mnras, 466, 3612, \dodoi{10.1093/mnras/stw3296}

\bibitem[{{Bowler} {et~al.}(2017{\natexlab{b}}){Bowler}, {McLure}, {Dunlop}, {McLeod}, {Stanway}, {Eldridge}, \& {Jarvis}}]{bowler17b}
{Bowler}, R.~A.~A., {McLure}, R.~J., {Dunlop}, J.~S., {et~al.} 2017{\natexlab{b}}, \mnras, 469, 448, \dodoi{10.1093/mnras/stx839}

\bibitem[{Bradley {et~al.}(2024)Bradley, Sipőcz, Robitaille, Tollerud, Vinícius, Deil, Barbary, Wilson, Busko, Donath, Günther, Cara, Lim, Meßlinger, Conseil, Burnett, Bostroem, Droettboom, Bray, Bratholm, Ginsburg, Jamieson, Barentsen, Craig, Morris, Perrin, Rathi, Pascual, \& Georgiev}]{larry_bradley_2024_13989456}
Bradley, L., Sipőcz, B., Robitaille, T., {et~al.} 2024, astropy/photutils: 2.0.2, 2.0.2,  Zenodo, \dodoi{10.5281/zenodo.13989456}

\bibitem[{Brammer(2023)}]{brammer23}
Brammer, G. 2023, grizli, 1.9.11,  Zenodo, \dodoi{10.5281/zenodo.8370018}

\bibitem[{{Burgarella} {et~al.}(2024){Burgarella}, {Buat}, {Theul{\'e}}, {Zavala}, {Arrabal Haro}, {Bagley}, {Boquien}, {Cleri}, {Dewachter}, {Dickinson}, {Ferguson}, {Fern{\'a}ndez}, {Finkelstein}, {Fontana}, {Gawiser}, {Grazian}, {Grogin}, {Holwerda}, {Kartaltepe}, {Kewley}, {Kirkpatrick}, {Kocevski}, {Koekemoer}, {Long}, {Lotz}, {Lucas}, {Mobasher}, {Papovich}, {P{\'e}rez-Gonz{\'a}lez}, {Pirzkal}, {Ravindranath}, {Rodighiero}, {Roehlly}, {Rose}, {Seill{\'e}}, {Somerville}, {Wilkins}, {Yang}, \& {Yung}}]{burgarella24}
{Burgarella}, D., {Buat}, V., {Theul{\'e}}, P., {et~al.} 2024, arXiv e-prints, arXiv:2410.23959, \dodoi{10.48550/arXiv.2410.23959}

\bibitem[{Bushouse {et~al.}(2024)Bushouse, Eisenhamer, Dencheva, Davies, Greenfield, Morrison, Hodge, Simon, Grumm, Droettboom, Slavich, Sosey, Pauly, Miller, Jedrzejewski, Hack, Davis, Crawford, Law, Gordon, Regan, Cara, MacDonald, Bradley, Shanahan, Jamieson, Teodoro, Williams, \& Pena-Guerrero}]{bushouse_2024_10870758}
Bushouse, H., Eisenhamer, J., Dencheva, N., {et~al.} 2024, JWST Calibration Pipeline, 1.14.0,  Zenodo, \dodoi{10.5281/zenodo.10870758}

\bibitem[{{Byler} {et~al.}(2017){Byler}, {Dalcanton}, {Conroy}, \& {Johnson}}]{byler17}
{Byler}, N., {Dalcanton}, J.~J., {Conroy}, C., \& {Johnson}, B.~D. 2017, \apj, 840, 44, \dodoi{10.3847/1538-4357/aa6c66}

\bibitem[{{Calzetti} {et~al.}(2000){Calzetti}, {Armus}, {Bohlin}, {Kinney}, {Koornneef}, \& {Storchi-Bergmann}}]{calzetti00}
{Calzetti}, D., {Armus}, L., {Bohlin}, R.~C., {et~al.} 2000, \apj, 533, 682, \dodoi{10.1086/308692}

\bibitem[{{Cameron} {et~al.}(2023){Cameron}, {Saxena}, {Bunker}, {D'Eugenio}, {Carniani}, {Maiolino}, {Curtis-Lake}, {Ferruit}, {Jakobsen}, {Arribas}, {Bonaventura}, {Charlot}, {Chevallard}, {Curti}, {Looser}, {Maseda}, {Rawle}, {Rodr{\'\i}guez Del Pino}, {Smit}, {{\"U}bler}, {Willott}, {Witstok}, {Egami}, {Eisenstein}, {Johnson}, {Hainline}, {Rieke}, {Robertson}, {Stark}, {Tacchella}, {Williams}, {Willmer}, {Bhatawdekar}, {Bowler}, {Boyett}, {Circosta}, {Helton}, {Jones}, {Kumari}, {Ji}, {Nelson}, {Parlanti}, {Sandles}, {Scholtz}, \& {Sun}}]{cameron23}
{Cameron}, A.~J., {Saxena}, A., {Bunker}, A.~J., {et~al.} 2023, \aap, 677, A115, \dodoi{10.1051/0004-6361/202346107}

\bibitem[{{Cantalupo} {et~al.}(2014){Cantalupo}, {Arrigoni-Battaia}, {Prochaska}, {Hennawi}, \& {Madau}}]{cantalupo14}
{Cantalupo}, S., {Arrigoni-Battaia}, F., {Prochaska}, J.~X., {Hennawi}, J.~F., \& {Madau}, P. 2014, \nat, 506, 63, \dodoi{10.1038/nature12898}

\bibitem[{{Capak} {et~al.}(2015){Capak}, {Carilli}, {Jones}, {Casey}, {Riechers}, {Sheth}, {Carollo}, {Ilbert}, {Karim}, {Lefevre}, {Lilly}, {Scoville}, {Smolcic}, \& {Yan}}]{capak15}
{Capak}, P.~L., {Carilli}, C., {Jones}, G., {et~al.} 2015, \nat, 522, 455, \dodoi{10.1038/nature14500}

\bibitem[{{Carilli} \& {Walter}(2013)}]{carilli13}
{Carilli}, C.~L., \& {Walter}, F. 2013, \araa, 51, 105, \dodoi{10.1146/annurev-astro-082812-140953}

\bibitem[{{Carniani} {et~al.}(2018){Carniani}, {Maiolino}, {Smit}, \& {Amor{\'\i}n}}]{carniani18}
{Carniani}, S., {Maiolino}, R., {Smit}, R., \& {Amor{\'\i}n}, R. 2018, \apjl, 854, L7, \dodoi{10.3847/2041-8213/aaab45}

\bibitem[{{Carniani} {et~al.}(2020){Carniani}, {Ferrara}, {Maiolino}, {Castellano}, {Gallerani}, {Fontana}, {Kohandel}, {Lupi}, {Pallottini}, {Pentericci}, {Vallini}, \& {Vanzella}}]{carniani20}
{Carniani}, S., {Ferrara}, A., {Maiolino}, R., {et~al.} 2020, \mnras, 499, 5136, \dodoi{10.1093/mnras/staa3178}

\bibitem[{{CASA Team} {et~al.}(2022){CASA Team}, {Bean}, {Bhatnagar}, {Castro}, {Donovan Meyer}, {Emonts}, {Garcia}, {Garwood}, {Golap}, {Gonzalez Villalba}, {Harris}, {Hayashi}, {Hoskins}, {Hsieh}, {Jagannathan}, {Kawasaki}, {Keimpema}, {Kettenis}, {Lopez}, {Marvil}, {Masters}, {McNichols}, {Mehringer}, {Miel}, {Moellenbrock}, {Montesino}, {Nakazato}, {Ott}, {Petry}, {Pokorny}, {Raba}, {Rau}, {Schiebel}, {Schweighart}, {Sekhar}, {Shimada}, {Small}, {Steeb}, {Sugimoto}, {Suoranta}, {Tsutsumi}, {van Bemmel}, {Verkouter}, {Wells}, {Xiong}, {Szomoru}, {Griffith}, {Glendenning}, \& {Kern}}]{casa22}
{CASA Team}, {Bean}, B., {Bhatnagar}, S., {et~al.} 2022, \pasp, 134, 114501, \dodoi{10.1088/1538-3873/ac9642}

\bibitem[{{Casasola} {et~al.}(2020){Casasola}, {Bianchi}, {De Vis}, {Magrini}, {Corbelli}, {Clark}, {Fritz}, {Nersesian}, {Viaene}, {Baes}, {Cassar{\`a}}, {Davies}, {De Looze}, {Dobbels}, {Galametz}, {Galliano}, {Jones}, {Madden}, {Mosenkov}, {Tr{\v{c}}ka}, \& {Xilouris}}]{casasola20}
{Casasola}, V., {Bianchi}, S., {De Vis}, P., {et~al.} 2020, \aap, 633, A100, \dodoi{10.1051/0004-6361/201936665}

\bibitem[{{Casey} {et~al.}(2014){Casey}, {Narayanan}, \& {Cooray}}]{casey14}
{Casey}, C.~M., {Narayanan}, D., \& {Cooray}, A. 2014, \physrep, 541, 45, \dodoi{10.1016/j.physrep.2014.02.009}

\bibitem[{{Casey} {et~al.}(2023){Casey}, {Kartaltepe}, {Drakos}, {Franco}, {Harish}, {Paquereau}, {Ilbert}, {Rose}, {Cox}, {Nightingale}, {Robertson}, {Silverman}, {Koekemoer}, {Massey}, {McCracken}, {Rhodes}, {Akins}, {Allen}, {Amvrosiadis}, {Arango-Toro}, {Bagley}, {Bongiorno}, {Capak}, {Champagne}, {Chartab}, {Ch{\'a}vez Ortiz}, {Chworowsky}, {Cooke}, {Cooper}, {Darvish}, {Ding}, {Faisst}, {Finkelstein}, {Fujimoto}, {Gentile}, {Gillman}, {Gould}, {Gozaliasl}, {Hayward}, {He}, {Hemmati}, {Hirschmann}, {Jahnke}, {Jin}, {Khostovan}, {Kokorev}, {Lambrides}, {Laigle}, {Larson}, {Leung}, {Liu}, {Liaudat}, {Long}, {Magdis}, {Mahler}, {Mainieri}, {Manning}, {Maraston}, {Martin}, {McCleary}, {McKinney}, {McPartland}, {Mobasher}, {Pattnaik}, {Renzini}, {Rich}, {Sanders}, {Sattari}, {Scognamiglio}, {Scoville}, {Sheth}, {Shuntov}, {Sparre}, {Suzuki}, {Talia}, {Toft}, {Trakhtenbrot}, {Urry}, {Valentino}, {Vanderhoof}, {Vardoulaki}, {Weaver}, {Whitaker}, {Wilkins}, {Yang}, \& {Zavala}}]{casey23}
{Casey}, C.~M., {Kartaltepe}, J.~S., {Drakos}, N.~E., {et~al.} 2023, \apj, 954, 31, \dodoi{10.3847/1538-4357/acc2bc}

\bibitem[{{Ceverino} {et~al.}(2017){Ceverino}, {Glover}, \& {Klessen}}]{ceverino17}
{Ceverino}, D., {Glover}, S. C.~O., \& {Klessen}, R.~S. 2017, \mnras, 470, 2791, \dodoi{10.1093/mnras/stx1386}

\bibitem[{{Chabrier}(2003)}]{chabrier03}
{Chabrier}, G. 2003, \pasp, 115, 763, \dodoi{10.1086/376392}

\bibitem[{{Chen} {et~al.}(2023){Chen}, {Stark}, {Endsley}, {Topping}, {Whitler}, \& {Charlot}}]{chen23}
{Chen}, Z., {Stark}, D.~P., {Endsley}, R., {et~al.} 2023, \mnras, 518, 5607, \dodoi{10.1093/mnras/stac3476}

\bibitem[{{Claeyssens} {et~al.}(2023){Claeyssens}, {Adamo}, {Richard}, {Mahler}, {Messa}, \& {Dessauges-Zavadsky}}]{claeyssens23}
{Claeyssens}, A., {Adamo}, A., {Richard}, J., {et~al.} 2023, \mnras, 520, 2180, \dodoi{10.1093/mnras/stac3791}

\bibitem[{{Clark} {et~al.}(2016){Clark}, {Schofield}, {Gomez}, \& {Davies}}]{clark16}
{Clark}, C. J.~R., {Schofield}, S.~P., {Gomez}, H.~L., \& {Davies}, J.~I. 2016, \mnras, 459, 1646, \dodoi{10.1093/mnras/stw647}

\bibitem[{{Conroy} \& {Gunn}(2010)}]{conroy10}
{Conroy}, C., \& {Gunn}, J.~E. 2010, \apj, 712, 833, \dodoi{10.1088/0004-637X/712/2/833}

\bibitem[{{Conroy} {et~al.}(2009){Conroy}, {Gunn}, \& {White}}]{conroy09}
{Conroy}, C., {Gunn}, J.~E., \& {White}, M. 2009, \apj, 699, 486, \dodoi{10.1088/0004-637X/699/1/486}

\bibitem[{{Curti} {et~al.}(2024){Curti}, {Maiolino}, {Curtis-Lake}, {Chevallard}, {Carniani}, {D'Eugenio}, {Looser}, {Scholtz}, {Charlot}, {Cameron}, {{\"U}bler}, {Witstok}, {Boyett}, {Laseter}, {Sandles}, {Arribas}, {Bunker}, {Giardino}, {Maseda}, {Rawle}, {Rodr{\'\i}guez Del Pino}, {Smit}, {Willott}, {Eisenstein}, {Hausen}, {Johnson}, {Rieke}, {Robertson}, {Tacchella}, {Williams}, {Willmer}, {Baker}, {Bhatawdekar}, {Egami}, {Helton}, {Ji}, {Kumari}, {Perna}, {Shivaei}, \& {Sun}}]{curti24a}
{Curti}, M., {Maiolino}, R., {Curtis-Lake}, E., {et~al.} 2024, \aap, 684, A75, \dodoi{10.1051/0004-6361/202346698}

\bibitem[{{da Cunha} {et~al.}(2015){da Cunha}, {Walter}, {Smail}, {Swinbank}, {Simpson}, {Decarli}, {Hodge}, {Weiss}, {van der Werf}, {Bertoldi}, {Chapman}, {Cox}, {Danielson}, {Dannerbauer}, {Greve}, {Ivison}, {Karim}, \& {Thomson}}]{dacunha15}
{da Cunha}, E., {Walter}, F., {Smail}, I.~R., {et~al.} 2015, \apj, 806, 110, \dodoi{10.1088/0004-637X/806/1/110}

\bibitem[{{De Vis} {et~al.}(2019){De Vis}, {Jones}, {Viaene}, {Casasola}, {Clark}, {Baes}, {Bianchi}, {Cassara}, {Davies}, {De Looze}, {Galametz}, {Galliano}, {Lianou}, {Madden}, {Manilla-Robles}, {Mosenkov}, {Nersesian}, {Roychowdhury}, {Xilouris}, \& {Ysard}}]{devis19}
{De Vis}, P., {Jones}, A., {Viaene}, S., {et~al.} 2019, \aap, 623, A5, \dodoi{10.1051/0004-6361/201834444}

\bibitem[{{D{\'\i}az-Santos} {et~al.}(2013){D{\'\i}az-Santos}, {Armus}, {Charmandaris}, {Stierwalt}, {Murphy}, {Haan}, {Inami}, {Malhotra}, {Meijerink}, {Stacey}, {Petric}, {Evans}, {Veilleux}, {van der Werf}, {Lord}, {Lu}, {Howell}, {Appleton}, {Mazzarella}, {Surace}, {Xu}, {Schulz}, {Sanders}, {Bridge}, {Chan}, {Frayer}, {Iwasawa}, {Melbourne}, \& {Sturm}}]{diaz-santos13}
{D{\'\i}az-Santos}, T., {Armus}, L., {Charmandaris}, V., {et~al.} 2013, \apj, 774, 68, \dodoi{10.1088/0004-637X/774/1/68}

\bibitem[{{Dijkstra} \& {Loeb}(2009)}]{dijkstra09}
{Dijkstra}, M., \& {Loeb}, A. 2009, \mnras, 400, 1109, \dodoi{10.1111/j.1365-2966.2009.15533.x}

\bibitem[{{Dudzevi{\v{c}}i{\={u}}t{\.{e}}} {et~al.}(2020){Dudzevi{\v{c}}i{\={u}}t{\.{e}}}, {Smail}, {Swinbank}, {Stach}, {Almaini}, {da Cunha}, {An}, {Arumugam}, {Birkin}, {Blain}, {Chapman}, {Chen}, {Conselice}, {Coppin}, {Dunlop}, {Farrah}, {Geach}, {Gullberg}, {Hartley}, {Hodge}, {Ivison}, {Maltby}, {Scott}, {Simpson}, {Simpson}, {Thomson}, {Walter}, {Wardlow}, {Weiss}, \& {van der Werf}}]{dudzevivciute20}
{Dudzevi{\v{c}}i{\={u}}t{\.{e}}}, U., {Smail}, I., {Swinbank}, A.~M., {et~al.} 2020, \mnras, 494, 3828, \dodoi{10.1093/mnras/staa769}

\bibitem[{{Dunlop} {et~al.}(2017){Dunlop}, {McLure}, {Biggs}, {Geach}, {Micha{\l}owski}, {Ivison}, {Rujopakarn}, {van Kampen}, {Kirkpatrick}, {Pope}, {Scott}, {Swinbank}, {Targett}, {Aretxaga}, {Austermann}, {Best}, {Bruce}, {Chapin}, {Charlot}, {Cirasuolo}, {Coppin}, {Ellis}, {Finkelstein}, {Hayward}, {Hughes}, {Ibar}, {Jagannathan}, {Khochfar}, {Koprowski}, {Narayanan}, {Nyland}, {Papovich}, {Peacock}, {Rieke}, {Robertson}, {Vernstrom}, {Werf}, {Wilson}, \& {Yun}}]{dunlop17}
{Dunlop}, J.~S., {McLure}, R.~J., {Biggs}, A.~D., {et~al.} 2017, \mnras, 466, 861, \dodoi{10.1093/mnras/stw3088}

\bibitem[{{Dunne} {et~al.}(2000){Dunne}, {Eales}, {Edmunds}, {Ivison}, {Alexander}, \& {Clements}}]{dunne00}
{Dunne}, L., {Eales}, S., {Edmunds}, M., {et~al.} 2000, \mnras, 315, 115, \dodoi{10.1046/j.1365-8711.2000.03386.x}

\bibitem[{{Dutton} \& {Macci{\`o}}(2014)}]{dutton14}
{Dutton}, A.~A., \& {Macci{\`o}}, A.~V. 2014, \mnras, 441, 3359, \dodoi{10.1093/mnras/stu742}

\bibitem[{{Endsley} {et~al.}(2021){Endsley}, {Stark}, {Charlot}, {Chevallard}, {Robertson}, {Bouwens}, \& {Stefanon}}]{endsley21}
{Endsley}, R., {Stark}, D.~P., {Charlot}, S., {et~al.} 2021, \mnras, 502, 6044, \dodoi{10.1093/mnras/stab432}

\bibitem[{{Finkelstein} {et~al.}(2015){Finkelstein}, {Ryan}, {Papovich}, {Dickinson}, {Song}, {Somerville}, {Ferguson}, {Salmon}, {Giavalisco}, {Koekemoer}, {Ashby}, {Behroozi}, {Castellano}, {Dunlop}, {Faber}, {Fazio}, {Fontana}, {Grogin}, {Hathi}, {Jaacks}, {Kocevski}, {Livermore}, {McLure}, {Merlin}, {Mobasher}, {Newman}, {Rafelski}, {Tilvi}, \& {Willner}}]{finkelstein15}
{Finkelstein}, S.~L., {Ryan}, Jr., R.~E., {Papovich}, C., {et~al.} 2015, \apj, 810, 71, \dodoi{10.1088/0004-637X/810/1/71}

\bibitem[{{Foreman-Mackey} {et~al.}(2013){Foreman-Mackey}, {Hogg}, {Lang}, \& {Goodman}}]{Foreman-Mackey13}
{Foreman-Mackey}, D., {Hogg}, D.~W., {Lang}, D., \& {Goodman}, J. 2013, \pasp, 125, 306, \dodoi{10.1086/670067}

\bibitem[{{Fujimoto} {et~al.}(2024{\natexlab{a}}){Fujimoto}, {Ouchi}, {Nakajima}, {Harikane}, {Isobe}, {Brammer}, {Oguri}, {Gim{\'e}nez-Arteaga}, {Heintz}, {Kokorev}, {Bauer}, {Ferrara}, {Kojima}, {Lagos}, {Laura}, {Schaerer}, {Shimasaku}, {Hatsukade}, {Kohno}, {Sun}, {Valentino}, {Watson}, {Fudamoto}, {Inoue}, {Gonz{\'a}lez-L{\'o}pez}, {Koekemoer}, {Knudsen}, {Lee}, {Magdis}, {Richard}, {Strait}, {Sugahara}, {Tamura}, {Toft}, {Umehata}, \& {Walth}}]{fujimoto24}
{Fujimoto}, S., {Ouchi}, M., {Nakajima}, K., {et~al.} 2024{\natexlab{a}}, \apj, 964, 146, \dodoi{10.3847/1538-4357/ad235c}

\bibitem[{{Fujimoto} {et~al.}(2024{\natexlab{b}}){Fujimoto}, {Ouchi}, {Kohno}, {Valentino}, {Gim\textbackslash'enez-Arteaga}, {Brammer}, {Furtak}, {Kohandel}, {Oguri}, {Pallottini}, {Richard}, {Zitrin}, {Bauer}, {Boylan-Kolchin}, {Dessauges-Zavadsky}, {Egami}, {Finkelstein}, {Ma}, {Smail}, {Watson}, {Hutchison}, {Rigby}, {Welch}, {Ao}, {Bradley}, {Caminha}, {Caputi}, {Espada}, {Endsley}, {Fudamoto}, {Gonz\textbackslash'alez-L\textbackslash'opez}, {Hatsukade}, {Koekemoer}, {Kokorev}, {Laporte}, {Lee}, {Magdis}, {Ono}, {Rizzo}, {Shibuya}, {Shimasaku}, {Sun}, {Toft}, {Umehata}, {Wang}, \& {Yajima}}]{fujimoto24_cosmicgrape}
{Fujimoto}, S., {Ouchi}, M., {Kohno}, K., {et~al.} 2024{\natexlab{b}}, arXiv e-prints, arXiv:2402.18543, \dodoi{10.48550/arXiv.2402.18543}

\bibitem[{{Gardner} {et~al.}(2023){Gardner}, {Mather}, {Abbott}, {Abell}, {Abernathy}, {Abney}, {Abraham}, {Abraham}, {Abul-Huda}, {Acton}, {Adams}, {Adams}, {Adler}, {Adriaensen}, {Aguilar}, {Ahmed}, {Ahmed}, {Ahmed}, {Albat}, {Albert}, {Alberts}, {Aldridge}, {Allen}, {Allen}, {Altenburg}, {Altunc}, {Alvarez}, {{\'A}lvarez-M{\'a}rquez}, {Alves de Oliveira}, {Ambrose}, {Anandakrishnan}, {Andersen}, {Anderson}, {Anderson}, {Anderson}, {Anderson}, {Aprea}, {Archer}, {Arenberg}, {Argyriou}, {Arribas}, {Artigau}, {Arvai}, {Atcheson}, {Atkinson}, {Averbukh}, {Aymergen}, {Bacinski}, {Baggett}, {Bagnasco}, {Baker}, {Balzano}, {Banks}, {Baran}, {Barker}, {Barrett}, {Barringer}, {Barto}, {Bast}, {Baudoz}, {Baum}, {Beatty}, {Beaulieu}, {Bechtold}, {Beck}, {Beddard}, {Beichman}, {Bellagama}, {Bely}, {Berger}, {Bergeron}, {Bernier}, {Bertch}, {Beskow}, {Betz}, {Biagetti}, {Birkmann}, {Bjorklund}, {Blackwood}, {Blazek}, {Blossfeld}, {Bluth}, {Boccaletti}, {Boegner}, {Bohlin}, {Boia}, {B{\"o}ker}, {Bonaventura}, {Bond},
  {Bosley}, {Boucarut}, {Bouchet}, {Bouwman}, {Bower}, {Bowers}, {Bowers}, {Boyce}, {Boyer}, {Boyer}, {Boyer}, {Boyer}, {Bradley}, {Brady}, {Brandl}, {Brannen}, {Breda}, {Bremmer}, {Brennan}, {Bresnahan}, {Bright}, {Broiles}, {Bromenschenkel}, {Brooks}, {Brooks}, {Brown}, {Brown}, {Brown}, {Bruce}, {Bryson}, {Bujanda}, {Bullock}, {Bunker}, {Bureo}, {Burt}, {Bush}, {Bushouse}, {Bussman}, {Cabaud}, {Cale}, {Calhoon}, {Calvani}, {Canipe}, {Caputo}, {Cara}, {Carey}, {Case}, {Cesari}, {Cetorelli}, {Chance}, {Chandler}, {Chaney}, {Chapman}, {Charlot}, {Chayer}, {Cheezum}, {Chen}, {Chen}, {Cherinka}, {Chichester}, {Chilton}, {Chittiraibalan}, {Clampin}, {Clark}, {Clark}, {Clark}, {Claybrooks}, {Cleveland}, {Cohen}, {Cohen}, {Col{\'o}n}, {Coleman}, {Colina}, {Comber}, {Comeau}, {Comer}, {Conde Reis}, {Connolly}, {Conroy}, {Contos}, {Contreras}, {Cook}, {Cooper}, {Cooper}, {Correia}, {Correnti}, {Cossou}, {Costanza}, {Coulais}, {Cox}, {Coyle}, {Cracraft}, {Crew}, {Curtis}, {Cusveller}, {Da Costa Maciel}, {Dailey},
  {Daugeron}, {Davidson}, {Davies}, {Davis}, {Davis}, {Day}, {de Chambure}, {de Jong}, {De Marchi}, {Dean}, {Decker}, {Delisa}, {Dell}, {Dellagatta}, {Dembinska}, {Demosthenes}, {Dencheva}, {Deneu}, {DePriest}, {Deschenes}, {Dethienne}, {Detre}, {Diaz}, {Dicken}, {DiFelice}, {Dillman}, {Disharoon}, {Dixon}, {Doggett}, {Dominguez}, {Donaldson}, {Doria-Warner}, {Santos}, {Doty}, {Douglas}, {Doyon}, {Dressler}, {Driggers}, {Driggers}, {Dunn}, {DuPrie}, {Dupuis}, {Durning}, {Dutta}, {Earl}, {Eccleston}, {Ecobichon}, {Egami}, {Ehrenwinkler}, {Eisenhamer}, {Eisenhower}, {Eisenstein}, {El Hamel}, {Elie}, {Elliott}, {Elliott}, {Engesser}, {Espinoza}, {Etienne}, {Etxaluze}, {Evans}, {Fabreguettes}, {Falcolini}, {Falini}, {Fatig}, {Feeney}, {Feinberg}, {Fels}, {Ferdous}, {Ferguson}, {Ferrarese}, {Ferreira}, {Ferruit}, {Ferry}, {Filippazzo}, {Firre}, {Fix}, {Flagey}, {Flanagan}, {Fleming}, {Florian}, {Flynn}, {Foiadelli}, {Fontaine}, {Fontanella}, {Forshay}, {Fortner}, {Fox}, {Framarini}, {Francisco}, {Franck}, {Franx},
  {Franz}, {Friedman}, {Friend}, {Frost}, {Fu}, {Fullerton}, {Gaillard}, {Galkin}, {Gallagher}, {Galyer}, {Garc{\'\i}a Mar{\'\i}n}, {Gardner}, {Garland}, {Garrett}, {Gasman}, {G{\'a}sp{\'a}r}, {Gastaud}, {Gaudreau}, {Gauthier}, {Geers}, {Geithner}, {Gennaro}, {Gerber}, {Gereau}, {Giampaoli}, {Giardino}, {Gibbons}, {Gilbert}, {Gilman}, {Girard}, {Giuliano}, {Gkountis}, {Glasse}, {Glassmire}, {Glauser}, {Glazer}, {Goldberg}, {Golimowski}, {Gonzaga}, {Gordon}, {Gordon}, {Goudfrooij}, {Gough}, {Graham}, {Grau}, {Green}, {Greene}, {Greene}, {Greenfield}, {Greenhouse}, {Greve}, {Greville}, {Grimaldi}, {Groe}, {Groebner}, {Grumm}, {Grundy}, {G{\"u}del}, {Guillard}, {Guldalian}, {Gunn}, {Gurule}, {Gutman}, {Guy}, {Guyot}, {Hack}, {Haderlein}, {Hagan}, {Hagedorn}, {Hainline}, {Haley}, {Hami}, {Hamilton}, {Hammann}, {Hammel}, {Hanley}, {Hansen}, {Hardy}, {Harnisch}, {Harr}, {Harris}, {Hart}, {Hartig}, {Hasan}, {Hashim}, {Hashimoto}, {Haskins}, {Hawkins}, {Hayden}, {Hayden}, {Healy}, {Hecht}, {Heeg}, {Hejal}, {Helm},
  {Hengemihle}, {Henning}, {Henry}, {Henry}, {Henshaw}, {Hernandez}, {Herrington}, {Heske}, {Hesman}, {Hickey}, {Hilbert}, {Hines}, {Hinz}, {Hirsch}, {Hitcho}, {Hodapp}, {Hodge}, {Hoffman}, {Holfeltz}, {Holler}, {Hoppa}, {Horner}, {Howard}, {Howard}, {Huber}, {Hunkeler}, {Hunter}, {Hunter}, {Hurd}, {Hurst}, {Hutchings}, {Hylan}, {Ignat}, {Illingworth}, {Irish}, {Isaacs}, {Jackson}, {Jaffe}, {Jahic}, {Jahromi}, {Jakobsen}, {James}, {James}, {James}, {Jamieson}, {Jandra}, {Jayawardhana}, {Jedrzejewski}, {Jeffers}, {Jensen}, {Joanne}, {Johns}, {Johnson}, {Johnson}, {Johnson}, {Johnson}, {Johnson}, {Johnson}, {Johnstone}, {Jollet}, {Jones}, {Jones}, {Jones}, {Jones}, {Jones}, {Jordan}, {Jordan}, {Jue}, {Jurkowski}, {Justis}, {Justtanont}, {Kaleida}, {Kalirai}, {Kalmanson}, {Kaltenegger}, {Kammerer}, {Kan}, {Kanarek}, {Kao}, {Karakla}, {Karl}, {Kassin}, {Kauffman}, {Kavanagh}, {Kelley}, {Kelly}, {Kendrew}, {Kennedy}, {Kenny}, {Keski-Kuha}, {Keyes}, {Khan}, {Kidwell}, {Kimble}, {King}, {King}, {Kinzel}, {Kirk},
  {Kirkpatrick}, {Klaassen}, {Klingemann}, {Klintworth}, {Knapp}, {Knight}, {Knollenberg}, {Knutsen}, {Koehler}, {Koekemoer}, {Kofler}, {Kontson}, {Kovacs}, {Kozhurina-Platais}, {Krause}, {Kriss}, {Krist}, {Kristoffersen}, {Krogel}, {Krueger}, {Kulp}, {Kumari}, {Kwan}, {Kyprianou}, {Labador}, {Labiano}, {Lafreni{\`e}re}, {Lagage}, {Laidler}, {Laine}, {Laird}, {Lajoie}, {Lallo}, {Lam}, {LaMassa}, {Lambros}, {Lampenfield}, {Lander}, {Langston}, {Larson}, {Larson}, {LaVerghetta}, {Law}, {Lawrence}, {Lee}, {Lee}, {Lee}, {Leisenring}, {Leveille}, {Levenson}, {Levi}, {Levine}, {Lewis}, {Lewis}, {Lewis}, {Libralato}, {Lidon}, {Liebrecht}, {Lightsey}, {Lilly}, {Lim}, {Lim}, {Ling}, {Link}, {Link}, {Lipinski}, {Liu}, {Lo}, {Lobmeyer}, {Logue}, {Long}, {Long}, {Long}, {Long}, {L{\'o}pez-Caniego}, {Lotz}, {Love-Pruitt}, {Lubskiy}, {Luers}, {Luetgens}, {Luevano}, {Lui}, {Lund}, {Lundquist}, {Lunine}, {L{\"u}tzgendorf}, {Lynch}, {MacDonald}, {MacDonald}, {Macias}, {Macklis}, {Maghami}, {Maharaja}, {Maiolino},
  {Makrygiannis}, {Malla}, {Malumuth}, {Manjavacas}, {Marini}, {Marrione}, {Marston}, {Martel}, {Martin}, {Martin}, {Martinez}, {Maschmann}, {Masci}, {Masetti}, {Maszkiewicz}, {Matthews}, {Matuskey}, {McBrayer}, {McCarthy}, {McCaughrean}, {McClare}, {McClare}, {McCloskey}, {McClurg}, {McCoy}, {McElwain}, {McGregor}, {McGuffey}, {McKay}, {McKenzie}, {McLean}, {McMaster}, {McNeil}, {De Meester}, {Mehalick}, {Meixner}, {Mel{\'e}ndez}, {Menzel}, {Menzel}, {Merz}, {Mesterharm}, {Meyer}, {Meyett}, {Meza}, {Midwinter}, {Milam}, {Miller}, {Miller}, {Miskey}, {Misselt}, {Mitchell}, {Mohan}, {Montoya}, {Moran}, {Morishita}, {Moro-Mart{\'\i}n}, {Morrison}, {Morrison}, {Morse}, {Moschos}, {Moseley}, {Mosier}, {Mosner}, {Mountain}, {Muckenthaler}, {Mueller}, {Mueller}, {Muhiem}, {M{\"u}hlmann}, {Mullally}, {Mullen}, {Munger}, {Murphy}, {Murray}, {Muzerolle}, {Mycroft}, {Myers}, {Myers}, {Myers}, {Myers}, {Myrick}, {Nagle}, {Nayak}, {Naylor}, {Neff}, {Nelan}, {Nella}, {Nguyen}, {Nguyen}, {Nickson}, {Nidhiry}, {Niedner},
  {Nieto-Santisteban}, {Nikolov}, {Nishisaka}, {Noriega-Crespo}, {Nota}, {O'Mara}, {Oboryshko}, {O'Brien}, {Ochs}, {Offenberg}, {Ogle}, {Ohl}, {Olmsted}, {Osborne}, {O'Shaughnessy}, {{\"O}stlin}, {O'Sullivan}, {Otor}, {Ottens}, {Ouellette}, {Outlaw}, {Owens}, {Pacifici}, {Page}, {Paranilam}, {Park}, {Parrish}, {Paschal}, {Patapis}, {Patel}, {Patrick}, {Pattishall}, {Paul}, {Paul}, {Pauly}, {Pavlovsky}, {Pe{\~n}a-Guerrero}, {Pedder}, {Peek}, {Pelham}, {Penanen}, {Perriello}, {Perrin}, {Perrine}, {Perrygo}, {Peslier}, {Petach}, {Peterson}, {Pfarr}, {Pierson}, {Pietraszkiewicz}, {Pilchen}, {Pipher}, {Pirzkal}, {Pitman}, {Player}, {Plesha}, {Plitzke}, {Pohner}, {Poletis}, {Pollizzi}, {Polster}, {Pontius}, {Pontoppidan}, {Porges}, {Potter}, {Prescott}, {Proffitt}, {Pueyo}, {Quispe Neira}, {Radich}, {Rager}, {Rameau}, {Ramey}, {Ramos Alarcon}, {Rampini}, {Rapp}, {Rashford}, {Rauscher}, {Ravindranath}, {Rawle}, {Rawlings}, {Ray}, {Regan}, {Rehm}, {Rehm}, {Reid}, {Reis}, {Renk}, {Reoch}, {Ressler}, {Rest},
  {Reynolds}, {Richon}, {Richon}, {Ridgaway}, {Riedel}, {Rieke}, {Rieke}, {Rifelli}, {Rigby}, {Riggs}, {Ringel}, {Ritchie}, {Rix}, {Robberto}, {Robinson}, {Robinson}, {Robinson}, {Rock}, {Rodriguez}, {Rodr{\'\i}guez del Pino}, {Roellig}, {Rohrbach}, {Roman}, {Romelfanger}, {Romo}, {Rosales}, {Rose}, {Roteliuk}, {Roth}, {Rothwell}, {Rouzaud}, {Rowe}, {Rowlands}, {Roy}, {Royer}, {Rui}, {Rumler}, {Rumpl}, {Russ}, {Ryan}, {Ryan}, {Saad}, {Sabata}, {Sabatino}, {Sabbi}, {Sabelhaus}, {Sabia}, {Sahu}, {Saif}, {Salvignol}, {Samara-Ratna}, {Samuelson}, {Sanders}, {Sappington}, {Sargent}, {Sauer}, {Savadkin}, {Sawicki}, {Schappell}, {Scheffer}, {Scheithauer}, {Scherer}, {Schiff}, {Schlawin}, {Schmeitzky}, {Schmitz}, {Schmude}, {Schneider}, {Schreiber}, {Schroeven-Deceuninck}, {Schultz}, {Schwab}, {Schwartz}, {Scoccimarro}, {Scott}, {Scott}, {Seaton}, {Seely}, {Seery}, {Seidleck}, {Sembach}, {Shanahan}, {Shaughnessy}, {Shaw}, {Shay}, {Sheehan}, {Sheth}, {Shih}, {Shivaei}, {Siegel}, {Sienkiewicz}, {Simmons}, {Simon},
  {Sirianni}, {Sivaramakrishnan}, {Slade}, {Sloan}, {Slocum}, {Slowinski}, {Smith}, {Smith}, {Smith}, {Smith}, {Smith}, {Smith}, {Smolik}, {Soderblom}, {Sohn}, {Sokol}, {Sonneborn}, {Sontag}, {Sooy}, {Soummer}, {Southwood}, {Spain}, {Sparmo}, {Speer}, {Spencer}, {Sprofera}, {Stallcup}, {Stanley}, {Stansberry}, {Stark}, {Starr}, {Stassi}, {Steck}, {Steeley}, {Stephens}, {Stephenson}, {Stewart}, {Stiavelli}, {}, {Strada}, {Straughn}, {Streetman}, {Strickland}, {Strobele}, {Stuhlinger}, {Stys}, {Such}, {Sukhatme}, {Sullivan}, {Sullivan}, {Sumner}, {Sun}, {Sunnquist}, {Swade}, {Swam}, {Swenton}, {Swoish}, {Tam Litten}, {Tamas}, {Tao}, {Taylor}, {Taylor}, {te Plate}, {Van Tea}, {Teague}, {Telfer}, {Temim}, {Texter}, {Thatte}, {Thompson}, {Thompson}, {Thomson}, {Thronson}, {Tierney}, {Tikkanen}, {Tinnin}, {Tippet}, {Todd}, {Tran}, {Trauger}, {Trejo}, {Vinh Truong}, {Tsukamoto}, {Tufail}, {Tumlinson}, {Tustain}, {Tyra}, {Ubeda}, {Underwood}, {Uzzo}, {Vaclavik}, {Valenduc}, {Valenti}, {Van Campen}, {van de Wetering},
  {Van Der Marel}, {van Haarlem}, {Vandenbussche}, {van Dishoeck}, {Vanterpool}, {Vernoy}, {Vila Costas}, {Volk}, {Voorzaat}, {Voyton}, {Vydra}, {Waddy}, {Waelkens}, {Wahlgren}, {Walker}, {Wander}, {Warfield}, {Warner}, {Wasiak}, {Wasiak}, {Wehner}, {Weiler}, {Weilert}, {Weiss}, {Wells}, {Welty}, {Wheate}, {Wheeler}, {White}, {Whitehouse}, {Whiteleather}, {Whitman}, {Williams}, {Willmer}, {Willott}, {Willoughby}, {Wilson}, {Wilson}, {Wilson}, {Windhorst}, {Wislowski}, {Wolfe}, {Wolfe}, {Wolff}, {Wondel}, {Woo}, {Woods}, {Worden}, {Workman}, {Wright}, {Wu}, {Wu}, {Wun}, {Wymer}, {Yadetie}, {Yan}, {Yang}, {Yates}, {Yeager}, {Yerger}, {Young}, {Young}, {Yu}, {Yu}, {Zak}, {Zeidler}, {Zepp}, {Zhou}, {Zincke}, {Zonak}, \& {Zondag}}]{gardner23}
{Gardner}, J.~P., {Mather}, J.~C., {Abbott}, R., {et~al.} 2023, \pasp, 135, 068001, \dodoi{10.1088/1538-3873/acd1b5}

\bibitem[{{Garnett}(1992)}]{garnett92}
{Garnett}, D.~R. 1992, \aj, 103, 1330, \dodoi{10.1086/116146}

\bibitem[{{Gim{\'e}nez-Arteaga} {et~al.}(2023){Gim{\'e}nez-Arteaga}, {Oesch}, {Brammer}, {Valentino}, {Mason}, {Weibel}, {Barrufet}, {Fujimoto}, {Heintz}, {Nelson}, {Strait}, {Suess}, \& {Gibson}}]{gimenez'-arteaga23}
{Gim{\'e}nez-Arteaga}, C., {Oesch}, P.~A., {Brammer}, G.~B., {et~al.} 2023, \apj, 948, 126, \dodoi{10.3847/1538-4357/acc5ea}

\bibitem[{{Greene} \& {Ho}(2005)}]{green05}
{Greene}, J.~E., \& {Ho}, L.~C. 2005, \apj, 630, 122, \dodoi{10.1086/431897}

\bibitem[{{Greene} {et~al.}(2024){Greene}, {Labbe}, {Goulding}, {Furtak}, {Chemerynska}, {Kokorev}, {Dayal}, {Volonteri}, {Williams}, {Wang}, {Setton}, {Burgasser}, {Bezanson}, {Atek}, {Brammer}, {Cutler}, {Feldmann}, {Fujimoto}, {Glazebrook}, {de Graaff}, {Khullar}, {Leja}, {Marchesini}, {Maseda}, {Matthee}, {Miller}, {Naidu}, {Nanayakkara}, {Oesch}, {Pan}, {Papovich}, {Price}, {van Dokkum}, {Weaver}, {Whitaker}, \& {Zitrin}}]{greene24}
{Greene}, J.~E., {Labbe}, I., {Goulding}, A.~D., {et~al.} 2024, \apj, 964, 39, \dodoi{10.3847/1538-4357/ad1e5f}

\bibitem[{{Gullberg} {et~al.}(2015){Gullberg}, {De Breuck}, {Vieira}, {Wei{\ss}}, {Aguirre}, {Aravena}, {B{\'e}thermin}, {Bradford}, {Bothwell}, {Carlstrom}, {Chapman}, {Fassnacht}, {Gonzalez}, {Greve}, {Hezaveh}, {Holzapfel}, {Husband}, {Ma}, {Malkan}, {Marrone}, {Menten}, {Murphy}, {Reichardt}, {Spilker}, {Stark}, {Strandet}, \& {Welikala}}]{gullberg15}
{Gullberg}, B., {De Breuck}, C., {Vieira}, J.~D., {et~al.} 2015, \mnras, 449, 2883, \dodoi{10.1093/mnras/stv372}

\bibitem[{{Harikane} {et~al.}(2019){Harikane}, {Ouchi}, {Ono}, {Fujimoto}, {Donevski}, {Shibuya}, {Faisst}, {Goto}, {Hatsukade}, {Kashikawa}, {Kohno}, {Hashimoto}, {Higuchi}, {Inoue}, {Lin}, {Martin}, {Overzier}, {Smail}, {Toshikawa}, {Umehata}, {Ao}, {Chapman}, {Clements}, {Im}, {Jing}, {Kawaguchi}, {Lee}, {Lee}, {Lin}, {Matsuoka}, {Marinello}, {Nagao}, {Onodera}, {Toft}, \& {Wang}}]{harikane19a}
{Harikane}, Y., {Ouchi}, M., {Ono}, Y., {et~al.} 2019, \apj, 883, 142, \dodoi{10.3847/1538-4357/ab2cd5}

\bibitem[{{Harikane} {et~al.}(2023){Harikane}, {Zhang}, {Nakajima}, {Ouchi}, {Isobe}, {Ono}, {Hatano}, {Xu}, \& {Umeda}}]{harikane23b}
{Harikane}, Y., {Zhang}, Y., {Nakajima}, K., {et~al.} 2023, \apj, 959, 39, \dodoi{10.3847/1538-4357/ad029e}

\bibitem[{{Harikane} {et~al.}(2025){Harikane}, {Inoue}, {Ellis}, {Ouchi}, {Nakazato}, {Yoshida}, {Ono}, {Sun}, {Sato}, {Ferrami}, {Fujimoto}, {Kashikawa}, {McLeod}, {P{\'e}rez-Gonz{\'a}lez}, {Sawicki}, {Sugahara}, {Xu}, {Yamanaka}, {Carnall}, {Cullen}, {Dunlop}, {Egami}, {Grogin}, {Isobe}, {Koekemoer}, {Laporte}, {Lee}, {Magee}, {Matsuo}, {Matsuoka}, {Mawatari}, {Nakajima}, {Nakane}, {Tamura}, {Umeda}, \& {Yanagisawa}}]{harikane25}
{Harikane}, Y., {Inoue}, A.~K., {Ellis}, R.~S., {et~al.} 2025, \apj, 980, 138, \dodoi{10.3847/1538-4357/ad9b2c}

\bibitem[{Harris {et~al.}(2020)Harris, Millman, van~der Walt, Gommers, Virtanen, Cournapeau, Wieser, Taylor, Berg, Smith, Kern, Picus, Hoyer, van Kerkwijk, Brett, Haldane, del R{'{\i}}o, Wiebe, Peterson, G{'{e}}rard-Marchant, Sheppard, Reddy, Weckesser, Abbasi, Gohlke, \& Oliphant}]{harris20}
Harris, C.~R., Millman, K.~J., van~der Walt, S.~J., {et~al.} 2020, Nature, 585, 357, \dodoi{10.1038/s41586-020-2649-2}

\bibitem[{{Hashimoto} {et~al.}(2019){Hashimoto}, {Inoue}, {Mawatari}, {Tamura}, {Matsuo}, {Furusawa}, {Harikane}, {Shibuya}, {Knudsen}, {Kohno}, {Ono}, {Zackrisson}, {Okamoto}, {Kashikawa}, {Oesch}, {Ouchi}, {Ota}, {Shimizu}, {Taniguchi}, {Umehata}, \& {Watson}}]{hashimoto19}
{Hashimoto}, T., {Inoue}, A.~K., {Mawatari}, K., {et~al.} 2019, \pasj, 71, 71, \dodoi{10.1093/pasj/psz049}

\bibitem[{{Hayes} {et~al.}(2011){Hayes}, {Scarlata}, \& {Siana}}]{hayes11}
{Hayes}, M., {Scarlata}, C., \& {Siana}, B. 2011, \nat, 476, 304, \dodoi{10.1038/nature10320}

\bibitem[{{Heintz} {et~al.}(2023){Heintz}, {Brammer}, {Gim{\'e}nez-Arteaga}, {Strait}, {del P. Lagos}, {Vijayan}, {Matthee}, {Watson}, {Mason}, {Hutter}, {Toft}, {Fynbo}, \& {Oesch}}]{heintz23}
{Heintz}, K.~E., {Brammer}, G.~B., {Gim{\'e}nez-Arteaga}, C., {et~al.} 2023, Nature Astronomy, 7, 1517, \dodoi{10.1038/s41550-023-02078-7}

\bibitem[{{Herenz} {et~al.}(2020){Herenz}, {Hayes}, \& {Scarlata}}]{herenz20}
{Herenz}, E.~C., {Hayes}, M., \& {Scarlata}, C. 2020, \aap, 642, A55, \dodoi{10.1051/0004-6361/202037464}

\bibitem[{{Hirashita} {et~al.}(2014){Hirashita}, {Ferrara}, {Dayal}, \& {Ouchi}}]{hirashita14}
{Hirashita}, H., {Ferrara}, A., {Dayal}, P., \& {Ouchi}, M. 2014, \mnras, 443, 1704, \dodoi{10.1093/mnras/stu1290}

\bibitem[{{Hirschmann} {et~al.}(2023){Hirschmann}, {Charlot}, \& {Somerville}}]{hirschmann23}
{Hirschmann}, M., {Charlot}, S., \& {Somerville}, R.~S. 2023, \mnras, 526, 3504, \dodoi{10.1093/mnras/stad2745}

\bibitem[{{Holland} {et~al.}(1999){Holland}, {Robson}, {Gear}, {Cunningham}, {Lightfoot}, {Jenness}, {Ivison}, {Stevens}, {Ade}, {Griffin}, {Duncan}, {Murphy}, \& {Naylor}}]{holland99}
{Holland}, W.~S., {Robson}, E.~I., {Gear}, W.~K., {et~al.} 1999, \mnras, 303, 659, \dodoi{10.1046/j.1365-8711.1999.02111.x}

\bibitem[{{Hollenbach} \& {Tielens}(1999)}]{hollenbach99}
{Hollenbach}, D.~J., \& {Tielens}, A.~G.~G.~M. 1999, Reviews of Modern Physics, 71, 173, \dodoi{10.1103/RevModPhys.71.173}

\bibitem[{{Hopkins} {et~al.}(2008){Hopkins}, {Hernquist}, {Cox}, \& {Kere{\v{s}}}}]{hopkins08}
{Hopkins}, P.~F., {Hernquist}, L., {Cox}, T.~J., \& {Kere{\v{s}}}, D. 2008, \apjs, 175, 356, \dodoi{10.1086/524362}

\bibitem[{Hunter(2007)}]{hunter07}
Hunter, J.~D. 2007, Computing in Science \& Engineering, 9, 90, \dodoi{10.1109/MCSE.2007.55}

\bibitem[{{Inami} {et~al.}(2022){Inami}, {Algera}, {Schouws}, {Sommovigo}, {Bouwens}, {Smit}, {Stefanon}, {Bowler}, {Endsley}, {Ferrara}, {Oesch}, {Stark}, {Aravena}, {Barrufet}, {da Cunha}, {Dayal}, {De Looze}, {Fudamoto}, {Gonzalez}, {Graziani}, {Hodge}, {Hygate}, {Nanayakkara}, {Pallottini}, {Riechers}, {Schneider}, {Topping}, \& {van der Werf}}]{inami22}
{Inami}, H., {Algera}, H. S.~B., {Schouws}, S., {et~al.} 2022, \mnras, 515, 3126, \dodoi{10.1093/mnras/stac1779}

\bibitem[{{Ishikawa} {et~al.}(2024){Ishikawa}, {Zakamska}, {Shen}, {Liu}, {Chen}, {Hwang}, {Vayner}, {Veilleux}, {Rupke}, {Wylezalek}, {Gross}, {Sankar}, \& {Diachenko}}]{ishikawa24}
{Ishikawa}, Y., {Zakamska}, N.~L., {Shen}, Y., {et~al.} 2024, arXiv e-prints, arXiv:2403.08098, \dodoi{10.48550/arXiv.2403.08098}

\bibitem[{{Izotov} {et~al.}(2006){Izotov}, {Stasi{\'n}ska}, {Meynet}, {Guseva}, \& {Thuan}}]{izotov06}
{Izotov}, Y.~I., {Stasi{\'n}ska}, G., {Meynet}, G., {Guseva}, N.~G., \& {Thuan}, T.~X. 2006, \aap, 448, 955, \dodoi{10.1051/0004-6361:20053763}

\bibitem[{{Izumi} {et~al.}(2024){Izumi}, {Matsuoka}, {Onoue}, {Strauss}, {Umehata}, {Silverman}, {Nagao}, {Imanishi}, {Kohno}, {Toba}, {Iwasawa}, {Nakanishi}, {Sawamura}, {Fujimoto}, {Kikuta}, {Kawaguchi}, {Aoki}, \& {Goto}}]{izumi24}
{Izumi}, T., {Matsuoka}, Y., {Onoue}, M., {et~al.} 2024, \apj, 972, 116, \dodoi{10.3847/1538-4357/ad57c6}

\bibitem[{{Jaff{\'e}} {et~al.}(2015){Jaff{\'e}}, {Smith}, {Candlish}, {Poggianti}, {Sheen}, \& {Verheijen}}]{jaffe15}
{Jaff{\'e}}, Y.~L., {Smith}, R., {Candlish}, G.~N., {et~al.} 2015, \mnras, 448, 1715, \dodoi{10.1093/mnras/stv100}

\bibitem[{{Jakobsen} {et~al.}(2022){Jakobsen}, {Ferruit}, {Alves de Oliveira}, {Arribas}, {Bagnasco}, {Barho}, {Beck}, {Birkmann}, {B{\"o}ker}, {Bunker}, {Charlot}, {de Jong}, {de Marchi}, {Ehrenwinkler}, {Falcolini}, {Fels}, {Franx}, {Franz}, {Funke}, {Giardino}, {Gnata}, {Holota}, {Honnen}, {Jensen}, {Jentsch}, {Johnson}, {Jollet}, {Karl}, {Kling}, {K{\"o}hler}, {Kolm}, {Kumari}, {Lander}, {Lemke}, {L{\'o}pez-Caniego}, {L{\"u}tzgendorf}, {Maiolino}, {Manjavacas}, {Marston}, {Maschmann}, {Maurer}, {Messerschmidt}, {Moseley}, {Mosner}, {Mott}, {Muzerolle}, {Pirzkal}, {Pittet}, {Plitzke}, {Posselt}, {Rapp}, {Rauscher}, {Rawle}, {Rix}, {R{\"o}del}, {Rumler}, {Sabbi}, {Salvignol}, {Schmid}, {Sirianni}, {Smith}, {Strada}, {te Plate}, {Valenti}, {Wettemann}, {Wiehe}, {Wiesmayer}, {Willott}, {Wright}, {Zeidler}, \& {Zincke}}]{jakobsen22}
{Jakobsen}, P., {Ferruit}, P., {Alves de Oliveira}, C., {et~al.} 2022, \aap, 661, A80, \dodoi{10.1051/0004-6361/202142663}

\bibitem[{{Johnson} {et~al.}(2021){Johnson}, {Leja}, {Conroy}, \& {Speagle}}]{johnson21}
{Johnson}, B.~D., {Leja}, J., {Conroy}, C., \& {Speagle}, J.~S. 2021, \apjs, 254, 22, \dodoi{10.3847/1538-4365/abef67}

\bibitem[{{Jones} {et~al.}(2020){Jones}, {B{\'e}thermin}, {Fudamoto}, {Ginolfi}, {Capak}, {Cassata}, {Faisst}, {Le F{\`e}vre}, {Schaerer}, {Silverman}, {Yan}, {Bardelli}, {Boquien}, {Cimatti}, {Dessauges-Zavadsky}, {Giavalisco}, {Gruppioni}, {Ibar}, {Khusanova}, {Koekemoer}, {Lemaux}, {Loiacono}, {Maiolino}, {Oesch}, {Pozzi}, {Riechers}, {Rodighiero}, {Talia}, {Vallini}, {Vergani}, {Zamorani}, \& {Zucca}}]{jones20}
{Jones}, G.~C., {B{\'e}thermin}, M., {Fudamoto}, Y., {et~al.} 2020, \mnras, 491, L18, \dodoi{10.1093/mnrasl/slz154}

\bibitem[{{Jones} {et~al.}(2024{\natexlab{a}}){Jones}, {{\"U}bler}, {Perna}, {Arribas}, {Bunker}, {Carniani}, {Charlot}, {Maiolino}, {Del Pino}, {Willott}, {Bowler}, {B{\"o}ker}, {Cameron}, {Chevallard}, {Cresci}, {Curti}, {D'Eugenio}, {Kumari}, {Saxena}, {Scholtz}, {Venturi}, \& {Witstok}}]{jones24a}
{Jones}, G.~C., {{\"U}bler}, H., {Perna}, M., {et~al.} 2024{\natexlab{a}}, \aap, 682, A122, \dodoi{10.1051/0004-6361/202347838}

\bibitem[{{Jones} {et~al.}(2024{\natexlab{b}}){Jones}, {Bowler}, {Bunker}, {Arribas}, {Carniani}, {Charlot}, {Perna}, {Rodr{\'\i}guez Del Pino}, {{\"U}bler}, {Willott}, {Chevallard}, {Cresci}, {Parlanti}, {Scholtz}, \& {Venturi}}]{jones24b}
{Jones}, G.~C., {Bowler}, R., {Bunker}, A.~J., {et~al.} 2024{\natexlab{b}}, arXiv e-prints, arXiv:2412.15027, \dodoi{10.48550/arXiv.2412.15027}

\bibitem[{{Jones} {et~al.}(2024{\natexlab{c}}){Jones}, {Bunker}, {Telikova}, {Arribas}, {Carniani}, {Charlot}, {D'Eugenio}, {Maiolino}, {Perna}, {Rodriguez Del Pino}, {Ubler}, {Willott}, {Aravena}, {Boker}, {Cresci}, {Curti}, {Herrera-Camus}, {Lamperti}, {Parlanti}, {Perez-Gonzalez}, \& {Villanueva}}]{jones24c}
{Jones}, G.~C., {Bunker}, A.~J., {Telikova}, K., {et~al.} 2024{\natexlab{c}}, arXiv e-prints, arXiv:2405.12955, \dodoi{10.48550/arXiv.2405.12955}

\bibitem[{{Juod{\v{z}}balis} {et~al.}(2025){Juod{\v{z}}balis}, {Maiolino}, {Baker}, {Lake}, {Scholtz}, {D'Eugenio}, {Trefoloni}, {Isobe}, {Tacchella}, {Bunker}, {Carniani}, {Charlot}, {Jones}, {Parlanti}, {Perna}, {Rinaldi}, {Robertson}, {{\"U}bler}, {Venturi}, \& {Willott}}]{juodvzbalis25}
{Juod{\v{z}}balis}, I., {Maiolino}, R., {Baker}, W.~M., {et~al.} 2025, arXiv e-prints, arXiv:2504.03551, \dodoi{10.48550/arXiv.2504.03551}

\bibitem[{{Kauffmann} {et~al.}(2003){Kauffmann}, {Heckman}, {Tremonti}, {Brinchmann}, {Charlot}, {White}, {Ridgway}, {Brinkmann}, {Fukugita}, {Hall}, {Ivezi{\'c}}, {Richards}, \& {Schneider}}]{kauffmann03}
{Kauffmann}, G., {Heckman}, T.~M., {Tremonti}, C., {et~al.} 2003, \mnras, 346, 1055, \dodoi{10.1111/j.1365-2966.2003.07154.x}

\bibitem[{{Kewley} {et~al.}(2001){Kewley}, {Dopita}, {Sutherland}, {Heisler}, \& {Trevena}}]{kewley01}
{Kewley}, L.~J., {Dopita}, M.~A., {Sutherland}, R.~S., {Heisler}, C.~A., \& {Trevena}, J. 2001, \apj, 556, 121, \dodoi{10.1086/321545}

\bibitem[{{Kikuta} {et~al.}(2023){Kikuta}, {Ouchi}, {Shibuya}, {Liang}, {Umeda}, {Matsumoto}, {Shimasaku}, {Harikane}, {Ono}, {Inoue}, {Yamanaka}, {Kusakabe}, {Momose}, {Kashikawa}, {Matsuda}, \& {Lee}}]{kikuta23}
{Kikuta}, S., {Ouchi}, M., {Shibuya}, T., {et~al.} 2023, \apjs, 268, 24, \dodoi{10.3847/1538-4365/ace4cb}

\bibitem[{{Kimock} {et~al.}(2021){Kimock}, {Narayanan}, {Smith}, {Ma}, {Feldmann}, {Angl{\'e}s-Alc{\'a}zar}, {Bromm}, {Dav{\'e}}, {Geach}, {Hopkins}, \& {Keres̆}}]{kimock21}
{Kimock}, B., {Narayanan}, D., {Smith}, A., {et~al.} 2021, \apj, 909, 119, \dodoi{10.3847/1538-4357/abbe89}

\bibitem[{{Knudsen} {et~al.}(2017){Knudsen}, {Watson}, {Frayer}, {Christensen}, {Gallazzi}, {Micha{\l}owski}, {Richard}, \& {Zavala}}]{knudsen17}
{Knudsen}, K.~K., {Watson}, D., {Frayer}, D., {et~al.} 2017, \mnras, 466, 138, \dodoi{10.1093/mnras/stw3066}

\bibitem[{{Kocevski} {et~al.}(2023){Kocevski}, {Onoue}, {Inayoshi}, {Trump}, {Arrabal Haro}, {Grazian}, {Dickinson}, {Finkelstein}, {Kartaltepe}, {Hirschmann}, {Aird}, {Holwerda}, {Fujimoto}, {Juneau}, {Amor{\'\i}n}, {Backhaus}, {Bagley}, {Barro}, {Bell}, {Bisigello}, {Calabr{\`o}}, {Cleri}, {Cooper}, {Ding}, {Grogin}, {Ho}, {Hutchison}, {Inoue}, {Jiang}, {Jones}, {Koekemoer}, {Li}, {Li}, {McGrath}, {Molina}, {Papovich}, {P{\'e}rez-Gonz{\'a}lez}, {Pirzkal}, {Wilkins}, {Yang}, \& {Yung}}]{kocevski23}
{Kocevski}, D.~D., {Onoue}, M., {Inayoshi}, K., {et~al.} 2023, \apjl, 954, L4, \dodoi{10.3847/2041-8213/ace5a0}

\bibitem[{{Kokubo} \& {Harikane}(2024)}]{kokubo24}
{Kokubo}, M., \& {Harikane}, Y. 2024, arXiv e-prints, arXiv:2407.04777, \dodoi{10.48550/arXiv.2407.04777}

\bibitem[{{Kusakabe} {et~al.}(2019){Kusakabe}, {Shimasaku}, {Momose}, {Ouchi}, {Nakajima}, {Hashimoto}, {Harikane}, {Silverman}, \& {Capak}}]{kusakabe19}
{Kusakabe}, H., {Shimasaku}, K., {Momose}, R., {et~al.} 2019, \pasj, 71, 55, \dodoi{10.1093/pasj/psz029}

\bibitem[{{Lake} {et~al.}(2015){Lake}, {Zheng}, {Cen}, {Sadoun}, {Momose}, \& {Ouchi}}]{lake15}
{Lake}, E., {Zheng}, Z., {Cen}, R., {et~al.} 2015, \apj, 806, 46, \dodoi{10.1088/0004-637X/806/1/46}

\bibitem[{{Lamperti} {et~al.}(2024){Lamperti}, {Arribas}, {Perna}, {Rodr{\'\i}guez Del Pino}, {Circosta}, {P{\'e}rez-Gonz{\'a}lez}, {Bunker}, {Carniani}, {Charlot}, {D'Eugenio}, {Maiolino}, {{\"U}bler}, {Willott}, {Bertola}, {B{\"o}ker}, {Cresci}, {Curti}, {Jones}, {Kumari}, {Parlanti}, {Scholtz}, \& {Venturi}}]{lamperti24}
{Lamperti}, I., {Arribas}, S., {Perna}, M., {et~al.} 2024, \aap, 691, A153, \dodoi{10.1051/0004-6361/202451021}

\bibitem[{{Laporte} {et~al.}(2017){Laporte}, {Nakajima}, {Ellis}, {Zitrin}, {Stark}, {Mainali}, \& {Roberts-Borsani}}]{laporte17}
{Laporte}, N., {Nakajima}, K., {Ellis}, R.~S., {et~al.} 2017, \apj, 851, 40, \dodoi{10.3847/1538-4357/aa96a8}

\bibitem[{{Leclercq} {et~al.}(2017){Leclercq}, {Bacon}, {Wisotzki}, {Mitchell}, {Garel}, {Verhamme}, {Blaizot}, {Hashimoto}, {Herenz}, {Conseil}, {Cantalupo}, {Inami}, {Contini}, {Richard}, {Maseda}, {Schaye}, {Marino}, {Akhlaghi}, {Brinchmann}, \& {Carollo}}]{leclercq17}
{Leclercq}, F., {Bacon}, R., {Wisotzki}, L., {et~al.} 2017, \aap, 608, A8, \dodoi{10.1051/0004-6361/201731480}

\bibitem[{{Leja} {et~al.}(2017){Leja}, {Johnson}, {Conroy}, {van Dokkum}, \& {Byler}}]{leja17}
{Leja}, J., {Johnson}, B.~D., {Conroy}, C., {van Dokkum}, P.~G., \& {Byler}, N. 2017, \apj, 837, 170, \dodoi{10.3847/1538-4357/aa5ffe}

\bibitem[{{Li} {et~al.}(2020){Li}, {Gu}, {Yajima}, {Zhu}, \& {Maji}}]{li20a}
{Li}, Y., {Gu}, M.~F., {Yajima}, H., {Zhu}, Q., \& {Maji}, M. 2020, \mnras, 494, 1919, \dodoi{10.1093/mnras/staa733}

\bibitem[{{Li} {et~al.}(2008){Li}, {Hopkins}, {Hernquist}, {Finkbeiner}, {Cox}, {Springel}, {Jiang}, {Fan}, \& {Yoshida}}]{li08}
{Li}, Y., {Hopkins}, P.~F., {Hernquist}, L., {et~al.} 2008, \apj, 678, 41, \dodoi{10.1086/529364}

\bibitem[{{Liu} {et~al.}(2019){Liu}, {Liu}, {Dong}, {Zhou}, {Wang}, {Lu}, \& {Yuan}}]{liu19}
{Liu}, H.-Y., {Liu}, W.-J., {Dong}, X.-B., {et~al.} 2019, \apjs, 243, 21, \dodoi{10.3847/1538-4365/ab298b}

\bibitem[{{Luridiana} {et~al.}(2015){Luridiana}, {Morisset}, \& {Shaw}}]{luridiana15}
{Luridiana}, V., {Morisset}, C., \& {Shaw}, R.~A. 2015, \aap, 573, A42, \dodoi{10.1051/0004-6361/201323152}

\bibitem[{{Madau}(1995)}]{madau95}
{Madau}, P. 1995, \apj, 441, 18, \dodoi{10.1086/175332}

\bibitem[{{Maiolino} {et~al.}(2023){Maiolino}, {Scholtz}, {Curtis-Lake}, {Carniani}, {Baker}, {de Graaff}, {Tacchella}, {{\"U}bler}, {D'Eugenio}, {Witstok}, {Curti}, {Arribas}, {Bunker}, {Charlot}, {Chevallard}, {Eisenstein}, {Egami}, {Ji}, {Jones}, {Lyu}, {Rawle}, {Robertson}, {Rujopakarn}, {Perna}, {Sun}, {Venturi}, {Williams}, \& {Willott}}]{maiolino23}
{Maiolino}, R., {Scholtz}, J., {Curtis-Lake}, E., {et~al.} 2023, arXiv e-prints, arXiv:2308.01230, \dodoi{10.48550/arXiv.2308.01230}

\bibitem[{{Marconcini} {et~al.}(2024{\natexlab{a}}){Marconcini}, {D'Eugenio}, {Maiolino}, {Arribas}, {Bunker}, {Carniani}, {Charlot}, {Perna}, {Rodr{\'\i}guez Del Pino}, {{\"U}bler}, {P{\'e}rez-Gonz{\'a}lez}, {Willott}, {B{\"o}ker}, {Cresci}, {Curti}, {Lamperti}, {Scholtz}, {Parlanti}, \& {Venturi}}]{marconcini24b}
{Marconcini}, C., {D'Eugenio}, F., {Maiolino}, R., {et~al.} 2024{\natexlab{a}}, arXiv e-prints, arXiv:2411.08627, \dodoi{10.48550/arXiv.2411.08627}

\bibitem[{{Marconcini} {et~al.}(2024{\natexlab{b}}){Marconcini}, {D'Eugenio}, {Maiolino}, {Arribas}, {Bunker}, {Carniani}, {Charlot}, {Perna}, {Rodr{\'\i}guez Del Pino}, {{\"U}bler}, {Willott}, {B{\"o}ker}, {Cresci}, {Curti}, {Jones}, {Lamperti}, {Parlanti}, \& {Venturi}}]{marconcini24a}
---. 2024{\natexlab{b}}, \mnras, 533, 2488, \dodoi{10.1093/mnras/stae1971}

\bibitem[{{Marrone} {et~al.}(2018){Marrone}, {Spilker}, {Hayward}, {Vieira}, {Aravena}, {Ashby}, {Bayliss}, {B{\'e}thermin}, {Brodwin}, {Bothwell}, {Carlstrom}, {Chapman}, {Chen}, {Crawford}, {Cunningham}, {De Breuck}, {Fassnacht}, {Gonzalez}, {Greve}, {Hezaveh}, {Lacaille}, {Litke}, {Lower}, {Ma}, {Malkan}, {Miller}, {Morningstar}, {Murphy}, {Narayanan}, {Phadke}, {Rotermund}, {Sreevani}, {Stalder}, {Stark}, {Strandet}, {Tang}, \& {Wei{\ss}}}]{marrone18}
{Marrone}, D.~P., {Spilker}, J.~S., {Hayward}, C.~C., {et~al.} 2018, \nat, 553, 51, \dodoi{10.1038/nature24629}

\bibitem[{{Marshall} {et~al.}(2024){Marshall}, {Yue}, {Eilers}, {Scholtz}, {Perna}, {Willott}, {Maiolino}, {{\"U}bler}, {Arribas}, {Bunker}, {Charlot}, {Rodr{\'\i}guez Del Pino}, {B{\"o}ker}, {Carniani}, {Cresci}, {D'Eugenio}, {Jones}, {Venturi}, {Bordoloi}, {Kashino}, {Mackenzie}, {Matthee}, {Naidu}, \& {Simcoe}}]{marshall24}
{Marshall}, M.~A., {Yue}, M., {Eilers}, A.-C., {et~al.} 2024, arXiv e-prints, arXiv:2410.11035, \dodoi{10.48550/arXiv.2410.11035}

\bibitem[{{Matsuda} {et~al.}(2004){Matsuda}, {Yamada}, {Hayashino}, {Tamura}, {Yamauchi}, {Ajiki}, {Fujita}, {Murayama}, {Nagao}, {Ohta}, {Okamura}, {Ouchi}, {Shimasaku}, {Shioya}, \& {Taniguchi}}]{matsuda04}
{Matsuda}, Y., {Yamada}, T., {Hayashino}, T., {et~al.} 2004, \aj, 128, 569, \dodoi{10.1086/422020}

\bibitem[{{Matthee} {et~al.}(2015){Matthee}, {Sobral}, {Santos}, {R{\"o}ttgering}, {Darvish}, \& {Mobasher}}]{matthee15}
{Matthee}, J., {Sobral}, D., {Santos}, S., {et~al.} 2015, \mnras, 451, 400, \dodoi{10.1093/mnras/stv947}

\bibitem[{{Matthee} {et~al.}(2017){Matthee}, {Sobral}, {Boone}, {R{\"o}ttgering}, {Schaerer}, {Girard}, {Pallottini}, {Vallini}, {Ferrara}, {Darvish}, \& {Mobasher}}]{matthee17}
{Matthee}, J., {Sobral}, D., {Boone}, F., {et~al.} 2017, \apj, 851, 145, \dodoi{10.3847/1538-4357/aa9931}

\bibitem[{{Matthee} {et~al.}(2020){Matthee}, {Pezzulli}, {Mackenzie}, {Cantalupo}, {Kusakabe}, {Leclercq}, {Sobral}, {Richard}, {Wisotzki}, {Lilly}, {Boogaard}, {Marino}, {Maseda}, \& {Nanayakkara}}]{matthee20}
{Matthee}, J., {Pezzulli}, G., {Mackenzie}, R., {et~al.} 2020, \mnras, 498, 3043, \dodoi{10.1093/mnras/staa2550}

\bibitem[{{Matthee} {et~al.}(2024){Matthee}, {Naidu}, {Brammer}, {Chisholm}, {Eilers}, {Goulding}, {Greene}, {Kashino}, {Labbe}, {Lilly}, {Mackenzie}, {Oesch}, {Weibel}, {Wuyts}, {Xiao}, {Bordoloi}, {Bouwens}, {van Dokkum}, {Illingworth}, {Kramarenko}, {Maseda}, {Mason}, {Meyer}, {Nelson}, {Reddy}, {Shivaei}, {Simcoe}, \& {Yue}}]{matthee24}
{Matthee}, J., {Naidu}, R.~P., {Brammer}, G., {et~al.} 2024, \apj, 963, 129, \dodoi{10.3847/1538-4357/ad2345}

\bibitem[{{Mazzolari} {et~al.}(2024){Mazzolari}, {{\"U}bler}, {Maiolino}, {Ji}, {Nakajima}, {Feltre}, {Scholtz}, {D'Eugenio}, {Curti}, {Mignoli}, \& {Marconi}}]{mazzolari24}
{Mazzolari}, G., {{\"U}bler}, H., {Maiolino}, R., {et~al.} 2024, \aap, 691, A345, \dodoi{10.1051/0004-6361/202450407}

\bibitem[{McKinney {et~al.}(2010)}]{mckinney2010data}
McKinney, W., {et~al.} 2010, in Proceedings of the 9th Python in Science Conference, Vol. 445, Austin, TX, 51--56

\bibitem[{{McLure} \& {Dunlop}(2004)}]{mclure04}
{McLure}, R.~J., \& {Dunlop}, J.~S. 2004, \mnras, 352, 1390, \dodoi{10.1111/j.1365-2966.2004.08034.x}

\bibitem[{{Mo} \& {White}(2002)}]{mo02}
{Mo}, H.~J., \& {White}, S.~D.~M. 2002, \mnras, 336, 112, \dodoi{10.1046/j.1365-8711.2002.05723.x}

\bibitem[{{Momose} {et~al.}(2014){Momose}, {Ouchi}, {Nakajima}, {Ono}, {Shibuya}, {Shimasaku}, {Yuma}, {Mori}, \& {Umemura}}]{momose14}
{Momose}, R., {Ouchi}, M., {Nakajima}, K., {et~al.} 2014, \mnras, 442, 110, \dodoi{10.1093/mnras/stu825}

\bibitem[{{Momose} {et~al.}(2016){Momose}, {Ouchi}, {Nakajima}, {Ono}, {Shibuya}, {Shimasaku}, {Yuma}, {Mori}, \& {Umemura}}]{momose16}
---. 2016, \mnras, 457, 2318, \dodoi{10.1093/mnras/stw021}

\bibitem[{{Morishita} {et~al.}(2025){Morishita}, {Stiavelli}, {Vanzella}, {Bergamini}, {Boyett}, {Chiaberge}, {Grillo}, {Leethochawalit}, {Messa}, {Roberts-Borsani}, {Rosati}, \& {Shajib}}]{morishita25}
{Morishita}, T., {Stiavelli}, M., {Vanzella}, E., {et~al.} 2025, \apj, 985, 83, \dodoi{10.3847/1538-4357/adc4c3}

\bibitem[{{Moster} {et~al.}(2013){Moster}, {Naab}, \& {White}}]{moster13}
{Moster}, B.~P., {Naab}, T., \& {White}, S. D.~M. 2013, \mnras, 428, 3121, \dodoi{10.1093/mnras/sts261}

\bibitem[{{Moster} {et~al.}(2018){Moster}, {Naab}, \& {White}}]{moster18}
---. 2018, \mnras, 477, 1822, \dodoi{10.1093/mnras/sty655}

\bibitem[{{Mowla} {et~al.}(2024){Mowla}, {Iyer}, {Asada}, {Desprez}, {Tan}, {Martis}, {Sarrouh}, {Strait}, {Abraham}, {Brada{\v{c}}}, {Brammer}, {Muzzin}, {Pacifici}, {Ravindranath}, {Sawicki}, {Willott}, {Estrada-Carpenter}, {Jahan}, {Noirot}, {Matharu}, {Rihtar{\v{s}}i{\v{c}}}, \& {Zabl}}]{mowla24}
{Mowla}, L., {Iyer}, K., {Asada}, Y., {et~al.} 2024, \nat, 636, 332, \dodoi{10.1038/s41586-024-08293-0}

\bibitem[{{Nakajima} {et~al.}(2016){Nakajima}, {Ellis}, {Iwata}, {Inoue}, {Kusakabe}, {Ouchi}, \& {Robertson}}]{nakajima16}
{Nakajima}, K., {Ellis}, R.~S., {Iwata}, I., {et~al.} 2016, \apjl, 831, L9, \dodoi{10.3847/2041-8205/831/1/L9}

\bibitem[{{Nakajima} \& {Maiolino}(2022)}]{nakajima22}
{Nakajima}, K., \& {Maiolino}, R. 2022, \mnras, 513, 5134, \dodoi{10.1093/mnras/stac1242}

\bibitem[{{Nakajima} {et~al.}(2023){Nakajima}, {Ouchi}, {Isobe}, {Harikane}, {Zhang}, {Ono}, {Umeda}, \& {Oguri}}]{nakajima23}
{Nakajima}, K., {Ouchi}, M., {Isobe}, Y., {et~al.} 2023, \apjs, 269, 33, \dodoi{10.3847/1538-4365/acd556}

\bibitem[{{Nakazato} {et~al.}(2024){Nakazato}, {Ceverino}, \& {Yoshida}}]{nakazato24}
{Nakazato}, Y., {Ceverino}, D., \& {Yoshida}, N. 2024, \apj, 975, 238, \dodoi{10.3847/1538-4357/ad7d0b}

\bibitem[{{Navarro} {et~al.}(1996){Navarro}, {Frenk}, \& {White}}]{navarro96}
{Navarro}, J.~F., {Frenk}, C.~S., \& {White}, S. D.~M. 1996, \apj, 462, 563, \dodoi{10.1086/177173}

\bibitem[{{Netzer}(2009)}]{netzer09}
{Netzer}, H. 2009, \mnras, 399, 1907, \dodoi{10.1111/j.1365-2966.2009.15434.x}

\bibitem[{{Oke} \& {Gunn}(1983)}]{oke83}
{Oke}, J.~B., \& {Gunn}, J.~E. 1983, \apj, 266, 713, \dodoi{10.1086/160817}

\bibitem[{{Onken} {et~al.}(2004){Onken}, {Ferrarese}, {Merritt}, {Peterson}, {Pogge}, {Vestergaard}, \& {Wandel}}]{onken04}
{Onken}, C.~A., {Ferrarese}, L., {Merritt}, D., {et~al.} 2004, \apj, 615, 645, \dodoi{10.1086/424655}

\bibitem[{{Osterbrock} \& {Ferland}(2006)}]{osterbrock06}
{Osterbrock}, D.~E., \& {Ferland}, G.~J. 2006, {Astrophysics of gaseous nebulae and active galactic nuclei} (University Science Books)

\bibitem[{{Ota} {et~al.}(2014){Ota}, {Walter}, {Ohta}, {Hatsukade}, {Carilli}, {da Cunha}, {Gonz{\'a}lez-L{\'o}pez}, {Decarli}, {Hodge}, {Nagai}, {Egami}, {Jiang}, {Iye}, {Kashikawa}, {Riechers}, {Bertoldi}, {Cox}, {Neri}, \& {Weiss}}]{ota14}
{Ota}, K., {Walter}, F., {Ohta}, K., {et~al.} 2014, \apj, 792, 34, \dodoi{10.1088/0004-637X/792/1/34}

\bibitem[{{Ouchi} {et~al.}(2009){Ouchi}, {Ono}, {Egami}, {Saito}, {Oguri}, {McCarthy}, {Farrah}, {Kashikawa}, {Momcheva}, {Shimasaku}, {Nakanishi}, {Furusawa}, {Akiyama}, {Dunlop}, {Mortier}, {Okamura}, {Hayashi}, {Cirasuolo}, {Dressler}, {Iye}, {Jarvis}, {Kodama}, {Martin}, {McLure}, {Ohta}, {Yamada}, \& {Yoshida}}]{ouchi09}
{Ouchi}, M., {Ono}, Y., {Egami}, E., {et~al.} 2009, \apj, 696, 1164, \dodoi{10.1088/0004-637X/696/2/1164}

\bibitem[{{Ouchi} {et~al.}(2010){Ouchi}, {Shimasaku}, {Furusawa}, {Saito}, {Yoshida}, {Akiyama}, {Ono}, {Yamada}, {Ota}, {Kashikawa}, {Iye}, {Kodama}, {Okamura}, {Simpson}, \& {Yoshida}}]{ouchi10}
{Ouchi}, M., {Shimasaku}, K., {Furusawa}, H., {et~al.} 2010, \apj, 723, 869, \dodoi{10.1088/0004-637X/723/1/869}

\bibitem[{{Ouchi} {et~al.}(2013){Ouchi}, {Ellis}, {Ono}, {Nakanishi}, {Kohno}, {Momose}, {Kurono}, {Ashby}, {Shimasaku}, {Willner}, {Fazio}, {Tamura}, \& {Iono}}]{ouchi13}
{Ouchi}, M., {Ellis}, R., {Ono}, Y., {et~al.} 2013, \apj, 778, 102, \dodoi{10.1088/0004-637X/778/2/102}

\bibitem[{{Ouchi} {et~al.}(2018){Ouchi}, {Harikane}, {Shibuya}, {Shimasaku}, {Taniguchi}, {Konno}, {Kobayashi}, {Kajisawa}, {Nagao}, {Ono}, {Inoue}, {Umemura}, {Mori}, {Hasegawa}, {Higuchi}, {Komiyama}, {Matsuda}, {Nakajima}, {Saito}, \& {Wang}}]{ouchi18}
{Ouchi}, M., {Harikane}, Y., {Shibuya}, T., {et~al.} 2018, \pasj, 70, S13, \dodoi{10.1093/pasj/psx074}

\bibitem[{{Pacucci} {et~al.}(2023){Pacucci}, {Nguyen}, {Carniani}, {Maiolino}, \& {Fan}}]{pacucci23}
{Pacucci}, F., {Nguyen}, B., {Carniani}, S., {Maiolino}, R., \& {Fan}, X. 2023, \apjl, 957, L3, \dodoi{10.3847/2041-8213/ad0158}

\bibitem[{{Palla} {et~al.}(2024){Palla}, {De Looze}, {Rela{\~n}o}, {van der Giessen}, {Dayal}, {Ferrara}, {Schneider}, {Graziani}, {Algera}, {Aravena}, {Bowler}, {Hygate}, {Inami}, {van Leeuwen}, {Bouwens}, {Hodge}, {Smit}, {Stefanon}, \& {van der Werf}}]{palla24}
{Palla}, M., {De Looze}, I., {Rela{\~n}o}, M., {et~al.} 2024, \mnras, 528, 2407, \dodoi{10.1093/mnras/stae160}

\bibitem[{{Parlanti} {et~al.}(2025){Parlanti}, {Carniani}, {Venturi}, {Herrera-Camus}, {Arribas}, {Bunker}, {Charlot}, {D'Eugenio}, {Maiolino}, {Perna}, {{\"U}bler}, {B{\"o}ker}, {Cresci}, {Curti}, {Jones}, {Lamperti}, {P{\'e}rez-Gonz{\'a}lez}, {Del Pino}, \& {Zamora}}]{parlanti25}
{Parlanti}, E., {Carniani}, S., {Venturi}, G., {et~al.} 2025, \aap, 695, A6, \dodoi{10.1051/0004-6361/202451692}

\bibitem[{{Perna}(2023)}]{perna23IAU}
{Perna}, M. 2023, in IAU Symposium, Vol. 373, Resolving the Rise and Fall of Star Formation in Galaxies, ed. T.~{Wong} \& W.-T. {Kim}, 60--62, \dodoi{10.1017/S174392132200374X}

\bibitem[{{Perna} {et~al.}(2023){Perna}, {Arribas}, {Marshall}, {D'Eugenio}, {{\"U}bler}, {Bunker}, {Charlot}, {Carniani}, {Jakobsen}, {Maiolino}, {Rodr{\'\i}guez Del Pino}, {Willott}, {B{\"o}ker}, {Circosta}, {Cresci}, {Curti}, {Husemann}, {Kumari}, {Lamperti}, {P{\'e}rez-Gonz{\'a}lez}, \& {Scholtz}}]{perna23}
{Perna}, M., {Arribas}, S., {Marshall}, M., {et~al.} 2023, \aap, 679, A89, \dodoi{10.1051/0004-6361/202346649}

\bibitem[{{Perna} {et~al.}(2025){Perna}, {Arribas}, {Lamperti}, {Circosta}, {Bertola}, {P{\'e}rez-Gonz{\'a}lez}, {D'Eugenio}, {{\"U}bler}, {Cresci}, {Volonteri}, {Mannucci}, {Maiolino}, {Rodr{\'\i}guez Del Pino}, {B{\"o}ker}, {Bunker}, {Charlot}, {Willott}, {Carniani}, {Curti}, {Jones}, {Kumari}, {Marshall}, {Venturi}, {Saxena}, {Scholtz}, \& {Witstok}}]{perna25}
{Perna}, M., {Arribas}, S., {Lamperti}, I., {et~al.} 2025, \aap, 696, A59, \dodoi{10.1051/0004-6361/202453430}

\bibitem[{{P{\'e}roux} \& {Howk}(2020)}]{peroux20}
{P{\'e}roux}, C., \& {Howk}, J.~C. 2020, \araa, 58, 363, \dodoi{10.1146/annurev-astro-021820-120014}

\bibitem[{{Planck Collaboration} {et~al.}(2020){Planck Collaboration}, {Aghanim}, {Akrami}, {Ashdown}, {Aumont}, {Baccigalupi}, {Ballardini}, {Banday}, {Barreiro}, {Bartolo}, {Basak}, {Battye}, {Benabed}, {Bernard}, {Bersanelli}, {Bielewicz}, {Bock}, {Bond}, {Borrill}, {Bouchet}, {Boulanger}, {Bucher}, {Burigana}, {Butler}, {Calabrese}, {Cardoso}, {Carron}, {Challinor}, {Chiang}, {Chluba}, {Colombo}, {Combet}, {Contreras}, {Crill}, {Cuttaia}, {de Bernardis}, {de Zotti}, {Delabrouille}, {Delouis}, {Di Valentino}, {Diego}, {Dor{\'e}}, {Douspis}, {Ducout}, {Dupac}, {Dusini}, {Efstathiou}, {Elsner}, {En{\ss}lin}, {Eriksen}, {Fantaye}, {Farhang}, {Fergusson}, {Fernandez-Cobos}, {Finelli}, {Forastieri}, {Frailis}, {Fraisse}, {Franceschi}, {Frolov}, {Galeotta}, {Galli}, {Ganga}, {G{\'e}nova-Santos}, {Gerbino}, {Ghosh}, {Gonz{\'a}lez-Nuevo}, {G{\'o}rski}, {Gratton}, {Gruppuso}, {Gudmundsson}, {Hamann}, {Handley}, {Hansen}, {Herranz}, {Hildebrandt}, {Hivon}, {Huang}, {Jaffe}, {Jones}, {Karakci}, {Keih{\"a}nen},
  {Keskitalo}, {Kiiveri}, {Kim}, {Kisner}, {Knox}, {Krachmalnicoff}, {Kunz}, {Kurki-Suonio}, {Lagache}, {Lamarre}, {Lasenby}, {Lattanzi}, {Lawrence}, {Le Jeune}, {Lemos}, {Lesgourgues}, {Levrier}, {Lewis}, {Liguori}, {Lilje}, {Lilley}, {Lindholm}, {L{\'o}pez-Caniego}, {Lubin}, {Ma}, {Mac{\'\i}as-P{\'e}rez}, {Maggio}, {Maino}, {Mandolesi}, {Mangilli}, {Marcos-Caballero}, {Maris}, {Martin}, {Martinelli}, {Mart{\'\i}nez-Gonz{\'a}lez}, {Matarrese}, {Mauri}, {McEwen}, {Meinhold}, {Melchiorri}, {Mennella}, {Migliaccio}, {Millea}, {Mitra}, {Miville-Desch{\^e}nes}, {Molinari}, {Montier}, {Morgante}, {Moss}, {Natoli}, {N{\o}rgaard-Nielsen}, {Pagano}, {Paoletti}, {Partridge}, {Patanchon}, {Peiris}, {Perrotta}, {Pettorino}, {Piacentini}, {Polastri}, {Polenta}, {Puget}, {Rachen}, {Reinecke}, {Remazeilles}, {Renzi}, {Rocha}, {Rosset}, {Roudier}, {Rubi{\~n}o-Mart{\'\i}n}, {Ruiz-Granados}, {Salvati}, {Sandri}, {Savelainen}, {Scott}, {Shellard}, {Sirignano}, {Sirri}, {Spencer}, {Sunyaev}, {Suur-Uski}, {Tauber}, {Tavagnacco},
  {Tenti}, {Toffolatti}, {Tomasi}, {Trombetti}, {Valenziano}, {Valiviita}, {Van Tent}, {Vibert}, {Vielva}, {Villa}, {Vittorio}, {Wandelt}, {Wehus}, {White}, {White}, {Zacchei}, \& {Zonca}}]{planck20}
{Planck Collaboration}, {Aghanim}, N., {Akrami}, Y., {et~al.} 2020, \aap, 641, A6, \dodoi{10.1051/0004-6361/201833910}

\bibitem[{{Prescott} {et~al.}(2012){Prescott}, {Dey}, {Brodwin}, {Chaffee}, {Desai}, {Eisenhardt}, {Le Floc'h}, {Jannuzi}, {Kashikawa}, {Matsuda}, \& {Soifer}}]{prescott12}
{Prescott}, M. K.~M., {Dey}, A., {Brodwin}, M., {et~al.} 2012, \apj, 752, 86, \dodoi{10.1088/0004-637X/752/2/86}

\bibitem[{{Rauscher}(2024)}]{rauscher24}
{Rauscher}, B.~J. 2024, \pasp, 136, 015001, \dodoi{10.1088/1538-3873/ad1b36}

\bibitem[{{Reines} {et~al.}(2013){Reines}, {Greene}, \& {Geha}}]{reines13}
{Reines}, A.~E., {Greene}, J.~E., \& {Geha}, M. 2013, \apj, 775, 116, \dodoi{10.1088/0004-637X/775/2/116}

\bibitem[{{Reines} \& {Volonteri}(2015)}]{reines15}
{Reines}, A.~E., \& {Volonteri}, M. 2015, \apj, 813, 82, \dodoi{10.1088/0004-637X/813/2/82}

\bibitem[{{Rhee} {et~al.}(2017){Rhee}, {Smith}, {Choi}, {Yi}, {Jaff{\'e}}, {Candlish}, \& {S{\'a}nchez-J{\'a}nssen}}]{rhee17}
{Rhee}, J., {Smith}, R., {Choi}, H., {et~al.} 2017, \apj, 843, 128, \dodoi{10.3847/1538-4357/aa6d6c}

\bibitem[{{Richards} {et~al.}(2006){Richards}, {Lacy}, {Storrie-Lombardi}, {Hall}, {Gallagher}, {Hines}, {Fan}, {Papovich}, {Vanden Berk}, {Trammell}, {Schneider}, {Vestergaard}, {York}, {Jester}, {Anderson}, {Budav{\'a}ri}, \& {Szalay}}]{richards06}
{Richards}, G.~T., {Lacy}, M., {Storrie-Lombardi}, L.~J., {et~al.} 2006, \apjs, 166, 470, \dodoi{10.1086/506525}

\bibitem[{{Riechers} {et~al.}(2013){Riechers}, {Bradford}, {Clements}, {Dowell}, {P{\'e}rez-Fournon}, {Ivison}, {Bridge}, {Conley}, {Fu}, {Vieira}, {Wardlow}, {Calanog}, {Cooray}, {Hurley}, {Neri}, {Kamenetzky}, {Aguirre}, {Altieri}, {Arumugam}, {Benford}, {B{\'e}thermin}, {Bock}, {Burgarella}, {Cabrera-Lavers}, {Chapman}, {Cox}, {Dunlop}, {Earle}, {Farrah}, {Ferrero}, {Franceschini}, {Gavazzi}, {Glenn}, {Solares}, {Gurwell}, {Halpern}, {Hatziminaoglou}, {Hyde}, {Ibar}, {Kov{\'a}cs}, {Krips}, {Lupu}, {Maloney}, {Martinez-Navajas}, {Matsuhara}, {Murphy}, {Naylor}, {Nguyen}, {Oliver}, {Omont}, {Page}, {Petitpas}, {Rangwala}, {Roseboom}, {Scott}, {Smith}, {Staguhn}, {Streblyanska}, {Thomson}, {Valtchanov}, {Viero}, {Wang}, {Zemcov}, \& {Zmuidzinas}}]{riechers13}
{Riechers}, D.~A., {Bradford}, C.~M., {Clements}, D.~L., {et~al.} 2013, \nat, 496, 329, \dodoi{10.1038/nature12050}

\bibitem[{{Rieke} {et~al.}(2023){Rieke}, {Kelly}, {Misselt}, {Stansberry}, {Boyer}, {Beatty}, {Egami}, {Florian}, {Greene}, {Hainline}, {Leisenring}, {Roellig}, {Schlawin}, {Sun}, {Tinnin}, {Williams}, {Willmer}, {Wilson}, {Clark}, {Rohrbach}, {Brooks}, {Canipe}, {Correnti}, {DiFelice}, {Gennaro}, {Girard}, {Hartig}, {Hilbert}, {Koekemoer}, {Nikolov}, {Pirzkal}, {Rest}, {Robberto}, {Sunnquist}, {Telfer}, {Wu}, {Ferry}, {Lewis}, {Baum}, {Beichman}, {Doyon}, {Dressler}, {Eisenstein}, {Ferrarese}, {Hodapp}, {Horner}, {Jaffe}, {Johnstone}, {Krist}, {Martin}, {McCarthy}, {Meyer}, {Rieke}, {Trauger}, \& {Young}}]{rieke23}
{Rieke}, M.~J., {Kelly}, D.~M., {Misselt}, K., {et~al.} 2023, \pasp, 135, 028001, \dodoi{10.1088/1538-3873/acac53}

\bibitem[{Rigby {et~al.}(2024)Rigby, Hutchison, Rivera-Thorsen, phadkekd, Spilker, \& Welch}]{jane_rigby_2024_10933642}
Rigby, J., Hutchison, T.~A., Rivera-Thorsen, T.~E., {et~al.} 2024, {JWST-Templates/Notebooks: minor update, added DOI tracker}, v1.0.2,  Zenodo, \dodoi{10.5281/zenodo.10933642}

\bibitem[{{Rigby} {et~al.}(2023){Rigby}, {Vieira}, {Phadke}, {Hutchison}, {Welch}, {Cathey}, {Spilker}, {Gonzalez}, {Adhikari}, {Aravena}, {Bayliss}, {Birkin}, {Bursk}, {Chapman}, {Dahle}, {Elicker}, {Fischer}, {Florian}, {Gladders}, {Hayward}, {Hewald}, {Kettler}, {Khullar}, {Kim}, {Law}, {Mahler}, {Malhotra}, {Murphy}, {Narayanan}, {Olivier}, {Rhoads}, {Sharon}, {Solimano}, {Thiruvengadam}, {Vizgan}, \& {Younker}}]{rigby23}
{Rigby}, J.~R., {Vieira}, J.~D., {Phadke}, K.~A., {et~al.} 2023, arXiv e-prints, arXiv:2312.10465, \dodoi{10.48550/arXiv.2312.10465}

\bibitem[{Robitaille {et~al.}(2024)Robitaille, Ginsburg, Mumford, Sipőcz, Kooten, Deil, Lim, Biscani, Stansby, Craig, jimboH, Williams, Barentsen, Singer, AlistairSymonds, Robert, Tollerud, Jankowski, Davies, Streicher, \& Maret}]{robitaille_2024_10931886}
Robitaille, T., Ginsburg, A., Mumford, S., {et~al.} 2024, astropy/reproject: v0.13.1, v0.13.1,  Zenodo, \dodoi{10.5281/zenodo.10931886}

\bibitem[{{Santini} {et~al.}(2017){Santini}, {Fontana}, {Castellano}, {Di Criscienzo}, {Merlin}, {Amorin}, {Cullen}, {Daddi}, {Dickinson}, {Dunlop}, {Grazian}, {Lamastra}, {McLure}, {Micha{\l}owski}, {Pentericci}, \& {Shu}}]{santini17}
{Santini}, P., {Fontana}, A., {Castellano}, M., {et~al.} 2017, \apj, 847, 76, \dodoi{10.3847/1538-4357/aa8874}

\bibitem[{{Scholtz} {et~al.}(2024){Scholtz}, {Curti}, {D'Eugenio}, {{\"U}bler}, {Maiolino}, {Marconcini}, {Smit}, {Perna}, {Witstok}, {Arribas}, {B{\"o}ker}, {Bunker}, {Carniani}, {Charlot}, {Cresci}, {P{\'e}rez-Gonz{\'a}lez}, {Lamperti}, {Rodr{\'\i}guez Del Pino}, {Parlanti}, \& {Venturi}}]{scholtz24}
{Scholtz}, J., {Curti}, M., {D'Eugenio}, F., {et~al.} 2024, arXiv e-prints, arXiv:2411.07695, \dodoi{10.48550/arXiv.2411.07695}

\bibitem[{{Shapley} {et~al.}(2020){Shapley}, {Cullen}, {Dunlop}, {McLure}, {Kriek}, {Reddy}, \& {Sanders}}]{shapley20}
{Shapley}, A.~E., {Cullen}, F., {Dunlop}, J.~S., {et~al.} 2020, \apjl, 903, L16, \dodoi{10.3847/2041-8213/abc006}

\bibitem[{{Shen} {et~al.}(2011){Shen}, {Richards}, {Strauss}, {Hall}, {Schneider}, {Snedden}, {Bizyaev}, {Brewington}, {Malanushenko}, {Malanushenko}, {Oravetz}, {Pan}, \& {Simmons}}]{shen11}
{Shen}, Y., {Richards}, G.~T., {Strauss}, M.~A., {et~al.} 2011, \apjs, 194, 45, \dodoi{10.1088/0067-0049/194/2/45}

\bibitem[{{Shen} {et~al.}(2019){Shen}, {Wu}, {Jiang}, {Ba{\~n}ados}, {Fan}, {Ho}, {Riechers}, {Strauss}, {Venemans}, {Vestergaard}, {Walter}, {Wang}, {Willott}, {Wu}, \& {Yang}}]{shen19}
{Shen}, Y., {Wu}, J., {Jiang}, L., {et~al.} 2019, \apj, 873, 35, \dodoi{10.3847/1538-4357/ab03d9}

\bibitem[{{Shibuya} {et~al.}(2018){Shibuya}, {Ouchi}, {Konno}, {Higuchi}, {Harikane}, {Ono}, {Shimasaku}, {Taniguchi}, {Kobayashi}, {Kajisawa}, {Nagao}, {Furusawa}, {Goto}, {Kashikawa}, {Komiyama}, {Kusakabe}, {Lee}, {Momose}, {Nakajima}, {Tanaka}, {Wang}, \& {Yuma}}]{shibuya18}
{Shibuya}, T., {Ouchi}, M., {Konno}, A., {et~al.} 2018, \pasj, 70, S14, \dodoi{10.1093/pasj/psx122}

\bibitem[{{Shuntov} {et~al.}(2025){Shuntov}, {Akins}, {Paquereau}, {Casey}, {Ilbert}, {Arango-Toro}, {McCracken}, {Franco}, {Harish}, {Kartaltepe}, {Koekemoer}, {Yang}, {Huertas-Company}, {Berman}, {McCleary}, {Toft}, {Gavazzi}, {Achenbach}, {Bertin}, {Brinch}, {Champagne}, {Chartab}, {Drakos}, {Egami}, {Endsley}, {Faisst}, {Fan}, {Flayhart}, {Hartley}, {Hatamnia}, {Gozaliasl}, {Gentile}, {Jermann}, {Jin}, {Kakiichi}, {Khostovan}, {K{\"u}mmel}, {Laigle}, {Laishram}, {Lambrides}, {Liu}, {Lyu}, {Magdis}, {Mobasher}, {Moutard}, {Renzini}, {Robertson}, {Schefer}, {Scognamiglio}, {Scoville}, {Sattari}, {Sanders}, {Taamoli}, {Trakhtenbrot}, {Valentino}, {Wang}, {Weaver}, \& {Yang}}]{shuntov25}
{Shuntov}, M., {Akins}, H.~B., {Paquereau}, L., {et~al.} 2025, arXiv e-prints, arXiv:2506.03243, \dodoi{10.48550/arXiv.2506.03243}

\bibitem[{{Smit} {et~al.}(2018){Smit}, {Bouwens}, {Carniani}, {Oesch}, {Labb{\'e}}, {Illingworth}, {van der Werf}, {Bradley}, {Gonzalez}, {Hodge}, {Holwerda}, {Maiolino}, \& {Zheng}}]{smit18}
{Smit}, R., {Bouwens}, R.~J., {Carniani}, S., {et~al.} 2018, \nat, 553, 178, \dodoi{10.1038/nature24631}

\bibitem[{{Sobral} {et~al.}(2015){Sobral}, {Matthee}, {Darvish}, {Schaerer}, {Mobasher}, {R{\"o}ttgering}, {Santos}, \& {Hemmati}}]{sobral15}
{Sobral}, D., {Matthee}, J., {Darvish}, B., {et~al.} 2015, \apj, 808, 139, \dodoi{10.1088/0004-637X/808/2/139}

\bibitem[{{Sobral} {et~al.}(2018){Sobral}, {Matthee}, {Darvish}, {Smail}, {Best}, {Alegre}, {R{\"o}ttgering}, {Mobasher}, {Paulino-Afonso}, {Stroe}, \& {Oteo}}]{sobral18b}
---. 2018, \mnras, 477, 2817, \dodoi{10.1093/mnras/sty782}

\bibitem[{{Sobral} {et~al.}(2019){Sobral}, {Matthee}, {Brammer}, {Ferrara}, {Alegre}, {R{\"o}ttgering}, {Schaerer}, {Mobasher}, \& {Darvish}}]{sobral19}
{Sobral}, D., {Matthee}, J., {Brammer}, G., {et~al.} 2019, \mnras, 482, 2422, \dodoi{10.1093/mnras/sty2779}

\bibitem[{{Solomon} \& {Vanden Bout}(2005)}]{solomon05}
{Solomon}, P.~M., \& {Vanden Bout}, P.~A. 2005, \araa, 43, 677, \dodoi{10.1146/annurev.astro.43.051804.102221}

\bibitem[{{Sorba} \& {Sawicki}(2015)}]{sorba15}
{Sorba}, R., \& {Sawicki}, M. 2015, \mnras, 452, 235, \dodoi{10.1093/mnras/stv1235}

\bibitem[{{Sorba} \& {Sawicki}(2018)}]{sorba18}
---. 2018, \mnras, 476, 1532, \dodoi{10.1093/mnras/sty186}

\bibitem[{{Steidel} {et~al.}(2000){Steidel}, {Adelberger}, {Shapley}, {Pettini}, {Dickinson}, \& {Giavalisco}}]{steidel00}
{Steidel}, C.~C., {Adelberger}, K.~L., {Shapley}, A.~E., {et~al.} 2000, \apj, 532, 170, \dodoi{10.1086/308568}

\bibitem[{{Storey} \& {Zeippen}(2000)}]{storey00}
{Storey}, P.~J., \& {Zeippen}, C.~J. 2000, \mnras, 312, 813, \dodoi{10.1046/j.1365-8711.2000.03184.x}

\bibitem[{{Sugahara} {et~al.}(2025){Sugahara}, {{\'A}lvarez-M{\'a}rquez}, {Hashimoto}, {Colina}, {Inoue}, {Costantin}, {Fudamoto}, {Mawatari}, {Ren}, {Arribas}, {Bakx}, {Blanco-Prieto}, {Ceverino}, {Crespo G{\'o}mez}, {Hagimoto}, {Hashigaya}, {Marques-Chaves}, {Matsuo}, {Nakazato}, {Pereira-Santaella}, {Tamura}, {Usui}, \& {Yoshida}}]{sugahara25}
{Sugahara}, Y., {{\'A}lvarez-M{\'a}rquez}, J., {Hashimoto}, T., {et~al.} 2025, \apj, 981, 135, \dodoi{10.3847/1538-4357/adb02a}

\bibitem[{{Tacconi} {et~al.}(2008){Tacconi}, {Genzel}, {Smail}, {Neri}, {Chapman}, {Ivison}, {Blain}, {Cox}, {Omont}, {Bertoldi}, {Greve}, {F{\"o}rster Schreiber}, {Genel}, {Lutz}, {Swinbank}, {Shapley}, {Erb}, {Cimatti}, {Daddi}, \& {Baker}}]{tacconi08}
{Tacconi}, L.~J., {Genzel}, R., {Smail}, I., {et~al.} 2008, \apj, 680, 246, \dodoi{10.1086/587168}

\bibitem[{{Tadaki} {et~al.}(2018){Tadaki}, {Iono}, {Yun}, {Aretxaga}, {Hatsukade}, {Hughes}, {Ikarashi}, {Izumi}, {Kawabe}, {Kohno}, {Lee}, {Matsuda}, {Nakanishi}, {Saito}, {Tamura}, {Ueda}, {Umehata}, {Wilson}, {Michiyama}, {Ando}, \& {Kamieneski}}]{tadaki18}
{Tadaki}, K., {Iono}, D., {Yun}, M.~S., {et~al.} 2018, \nat, 560, 613, \dodoi{10.1038/s41586-018-0443-1}

\bibitem[{{Toft} {et~al.}(2014){Toft}, {Smol{\v{c}}i{\'c}}, {Magnelli}, {Karim}, {Zirm}, {Michalowski}, {Capak}, {Sheth}, {Schawinski}, {Krogager}, {Wuyts}, {Sanders}, {Man}, {Lutz}, {Staguhn}, {Berta}, {Mccracken}, {Krpan}, \& {Riechers}}]{toft14}
{Toft}, S., {Smol{\v{c}}i{\'c}}, V., {Magnelli}, B., {et~al.} 2014, \apj, 782, 68, \dodoi{10.1088/0004-637X/782/2/68}

\bibitem[{{Toshikawa} {et~al.}(2018){Toshikawa}, {Uchiyama}, {Kashikawa}, {Ouchi}, {Overzier}, {Ono}, {Harikane}, {Ishikawa}, {Kodama}, {Matsuda}, {Lin}, {Onoue}, {Tanaka}, {Nagao}, {Akiyama}, {Komiyama}, {Goto}, \& {Lee}}]{toshikawa18}
{Toshikawa}, J., {Uchiyama}, H., {Kashikawa}, N., {et~al.} 2018, \pasj, 70, S12, \dodoi{10.1093/pasj/psx102}

\bibitem[{{Trakhtenbrot} {et~al.}(2011){Trakhtenbrot}, {Netzer}, {Lira}, \& {Shemmer}}]{trakhtenbrot11}
{Trakhtenbrot}, B., {Netzer}, H., {Lira}, P., \& {Shemmer}, O. 2011, \apj, 730, 7, \dodoi{10.1088/0004-637X/730/1/7}

\bibitem[{{{\"U}bler} {et~al.}(2024){{\"U}bler}, {Maiolino}, {P{\'e}rez-Gonz{\'a}lez}, {D'Eugenio}, {Perna}, {Curti}, {Arribas}, {Bunker}, {Carniani}, {Charlot}, {Rodr{\'\i}guez Del Pino}, {Baker}, {B{\"o}ker}, {Cresci}, {Dunlop}, {Grogin}, {Jones}, {Kumari}, {Lamperti}, {Laporte}, {Marshall}, {Mazzolari}, {Parlanti}, {Rawle}, {Scholtz}, {Venturi}, \& {Witstok}}]{ubler24}
{{\"U}bler}, H., {Maiolino}, R., {P{\'e}rez-Gonz{\'a}lez}, P.~G., {et~al.} 2024, \mnras, 531, 355, \dodoi{10.1093/mnras/stae943}

\bibitem[{{Umehata} {et~al.}(2021){Umehata}, {Smail}, {Steidel}, {Hayes}, {Scott}, {Swinbank}, {Ivison}, {Nagao}, {Kubo}, {Nakanishi}, {Matsuda}, {Ikarashi}, {Tamura}, \& {Geach}}]{umehata21}
{Umehata}, H., {Smail}, I., {Steidel}, C.~C., {et~al.} 2021, \apj, 918, 69, \dodoi{10.3847/1538-4357/ac1106}

\bibitem[{{Valentino} {et~al.}(2023){Valentino}, {Brammer}, {Gould}, {Kokorev}, {Fujimoto}, {Jespersen}, {Vijayan}, {Weaver}, {Ito}, {Tanaka}, {Ilbert}, {Magdis}, {Whitaker}, {Faisst}, {Gallazzi}, {Gillman}, {Gim{\'e}nez-Arteaga}, {G{\'o}mez-Guijarro}, {Kubo}, {Heintz}, {Hirschmann}, {Oesch}, {Onodera}, {Rizzo}, {Lee}, {Strait}, \& {Toft}}]{valentino23}
{Valentino}, F., {Brammer}, G., {Gould}, K. M.~L., {et~al.} 2023, \apj, 947, 20, \dodoi{10.3847/1538-4357/acbefa}

\bibitem[{{Villanueva} {et~al.}(2024){Villanueva}, {Herrera-Camus}, {Gonz{\'a}lez-L{\'o}pez}, {Aravena}, {Assef}, {Baeza-Garay}, {Barcos-Mu{\~n}oz}, {Bovino}, {Bowler}, {da Cunha}, {De Looze}, {Diaz-Santos}, {Ferrara}, {F{\"o}rster Schreiber}, {Algera}, {Ikeda}, {Killi}, {Mitsuhashi}, {Naab}, {Relano}, {Spilker}, {Solimano}, {Palla}, {Price}, {Posses}, {Tadaki}, {Telikova}, \& {{\"U}bler}}]{villanueva24}
{Villanueva}, V., {Herrera-Camus}, R., {Gonz{\'a}lez-L{\'o}pez}, J., {et~al.} 2024, \aap, 691, A133, \dodoi{10.1051/0004-6361/202451490}

\bibitem[{Virtanen {et~al.}(2020)Virtanen, Gommers, Oliphant, Haberland, Reddy, Cournapeau, Burovski, Peterson, Weckesser, Bright, {van der Walt}, Brett, Wilson, Millman, Mayorov, Nelson, Jones, Kern, Larson, Carey, Polat, Feng, Moore, {VanderPlas}, Laxalde, Perktold, Cimrman, Henriksen, Quintero, Harris, Archibald, Ribeiro, Pedregosa, {van Mulbregt}, \& {SciPy 1.0 Contributors}}]{virtanen20}
Virtanen, P., Gommers, R., Oliphant, T.~E., {et~al.} 2020, Nature Methods, 17, 261, \dodoi{10.1038/s41592-019-0686-2}

\bibitem[{{Walter} {et~al.}(2008){Walter}, {Brinks}, {de Blok}, {Bigiel}, {Kennicutt}, {Thornley}, \& {Leroy}}]{walter08}
{Walter}, F., {Brinks}, E., {de Blok}, W.~J.~G., {et~al.} 2008, \aj, 136, 2563, \dodoi{10.1088/0004-6256/136/6/2563}

\bibitem[{{Walter} {et~al.}(2012){Walter}, {Decarli}, {Carilli}, {Riechers}, {Bertoldi}, {Wei{\ss}}, {Cox}, {Neri}, {Maiolino}, {Ouchi}, {Egami}, \& {Nakanishi}}]{walter12}
{Walter}, F., {Decarli}, R., {Carilli}, C., {et~al.} 2012, \apj, 752, 93, \dodoi{10.1088/0004-637X/752/2/93}

\bibitem[{{Watson} {et~al.}(2015){Watson}, {Christensen}, {Knudsen}, {Richard}, {Gallazzi}, \& {Micha{\l}owski}}]{watson15}
{Watson}, D., {Christensen}, L., {Knudsen}, K.~K., {et~al.} 2015, \nat, 519, 327, \dodoi{10.1038/nature14164}

\bibitem[{{Witstok} {et~al.}(2022){Witstok}, {Smit}, {Maiolino}, {Kumari}, {Aravena}, {Boogaard}, {Bouwens}, {Carniani}, {Hodge}, {Jones}, {Stefanon}, {van der Werf}, \& {Schouws}}]{witstok22}
{Witstok}, J., {Smit}, R., {Maiolino}, R., {et~al.} 2022, \mnras, 515, 1751, \dodoi{10.1093/mnras/stac1905}

\bibitem[{{Witten} {et~al.}(2024){Witten}, {Laporte}, {Martin-Alvarez}, {Sijacki}, {Yuan}, {Haehnelt}, {Baker}, {Dunlop}, {Ellis}, {Grogin}, {Illingworth}, {Katz}, {Koekemoer}, {Magee}, {Maiolino}, {McClymont}, {P{\'e}rez-Gonz{\'a}lez}, {Pusk{\'a}s}, {Roberts-Borsani}, {Santini}, \& {Simmonds}}]{witten24}
{Witten}, C., {Laporte}, N., {Martin-Alvarez}, S., {et~al.} 2024, Nature Astronomy, 8, 384, \dodoi{10.1038/s41550-023-02179-3}

\bibitem[{{Wong} {et~al.}(2022){Wong}, {Wang}, {Hashimoto}, {Takagi}, {Goto}, {Kim}, {Wu}, {On}, {Santos}, {Lu}, {Kilerci-Eser}, {Ho}, \& {Hsiao}}]{wong22}
{Wong}, Y. H.~V., {Wang}, P., {Hashimoto}, T., {et~al.} 2022, \apj, 929, 161, \dodoi{10.3847/1538-4357/ac5cc7}

\bibitem[{{Yajima} {et~al.}(2023){Yajima}, {Abe}, {Fukushima}, {Ono}, {Harikane}, {Ouchi}, {Hashimoto}, \& {Khochfar}}]{yajima23}
{Yajima}, H., {Abe}, M., {Fukushima}, H., {et~al.} 2023, \mnras, 525, 4832, \dodoi{10.1093/mnras/stad2497}

\bibitem[{{Yajima} {et~al.}(2013){Yajima}, {Li}, \& {Zhu}}]{yajima13}
{Yajima}, H., {Li}, Y., \& {Zhu}, Q. 2013, \apj, 773, 151, \dodoi{10.1088/0004-637X/773/2/151}

\bibitem[{{Yajima} {et~al.}(2012){Yajima}, {Umemura}, \& {Mori}}]{yajima12c}
{Yajima}, H., {Umemura}, M., \& {Mori}, M. 2012, \mnras, 420, 3381, \dodoi{10.1111/j.1365-2966.2011.20261.x}

\bibitem[{{Yajima} {et~al.}(2022){Yajima}, {Abe}, {Khochfar}, {Nagamine}, {Inoue}, {Kodama}, {Arata}, {Dalla Vecchia}, {Fukushima}, {Hashimoto}, {Kashikawa}, {Kubo}, {Li}, {Matsuda}, {Mawatari}, {Ouchi}, \& {Umehata}}]{yajima22}
{Yajima}, H., {Abe}, M., {Khochfar}, S., {et~al.} 2022, \mnras, 509, 4037, \dodoi{10.1093/mnras/stab3092}

\bibitem[{{Yoon} {et~al.}(2017){Yoon}, {Chung}, {Smith}, \& {Jaff{\'e}}}]{yoon17}
{Yoon}, H., {Chung}, A., {Smith}, R., \& {Jaff{\'e}}, Y.~L. 2017, \apj, 838, 81, \dodoi{10.3847/1538-4357/aa6579}

\bibitem[{{Zabl} {et~al.}(2015){Zabl}, {N{\o}rgaard-Nielsen}, {Fynbo}, {Laursen}, {Ouchi}, \& {Kj{\ae}rgaard}}]{zabl15}
{Zabl}, J., {N{\o}rgaard-Nielsen}, H.~U., {Fynbo}, J.~P.~U., {et~al.} 2015, \mnras, 451, 2050, \dodoi{10.1093/mnras/stv1019}

\bibitem[{{Zamora} {et~al.}(2024){Zamora}, {Venturi}, {Carniani}, {Bertola}, {Parlanti}, {Perna}, {Arribas}, {B{\"o}ker}, {Bunker}, {Charlot}, {D'Eugenio}, {Maiolino}, {Rodr{\'\i}guez Del Pino}, {{\"U}bler}, {Cresci}, {Jones}, \& {Lamperti}}]{zamora24}
{Zamora}, S., {Venturi}, G., {Carniani}, S., {et~al.} 2024, arXiv e-prints, arXiv:2412.02751, \dodoi{10.48550/arXiv.2412.02751}

\bibitem[{{Zana} {et~al.}(2022){Zana}, {Gallerani}, {Carniani}, {Vito}, {Ferrara}, {Lupi}, {Di Mascia}, \& {Barai}}]{zana22}
{Zana}, T., {Gallerani}, S., {Carniani}, S., {et~al.} 2022, \mnras, 513, 2118, \dodoi{10.1093/mnras/stac978}

\bibitem[{{Zanella} {et~al.}(2018){Zanella}, {Daddi}, {Magdis}, {Diaz Santos}, {Cormier}, {Liu}, {Cibinel}, {Gobat}, {Dickinson}, {Sargent}, {Popping}, {Madden}, {Bethermin}, {Hughes}, {Valentino}, {Rujopakarn}, {Pannella}, {Bournaud}, {Walter}, {Wang}, {Elbaz}, \& {Coogan}}]{zanella18}
{Zanella}, A., {Daddi}, E., {Magdis}, G., {et~al.} 2018, \mnras, 481, 1976, \dodoi{10.1093/mnras/sty2394}

\end{thebibliography}
\bibliographystyle{aasjournal}

\end{document}